\documentclass[a4paper,12pt,times,numbered,oneside,print,index,custommargin]{PhDThesisPSnPDF}

\usepackage{comment}
\usepackage[utf8]{inputenc}
\usepackage{float}
\usepackage{alphabeta}
\usepackage{indentfirst}
\usepackage{mathtools}
\usepackage{amsmath}
\usepackage{amssymb}
\usepackage{amsthm}

\usepackage{nicematrix,tikz}
\usepackage{array}
\usepackage{cleveref}
\usepackage{physics}
\usepackage{thmtools}
\newtheorem{theorem}{Theorem}[section]
\newtheorem{prop}[theorem]{Proposition}
\newtheorem{lem}[theorem]{Lemma}
\newtheorem{corollary}[theorem]{Corollary}
\newtheorem{defn}[theorem]{Definition}
\newtheorem{remark}[theorem]{Remark}
\newtheorem{example}[theorem]{Example}

\newtheorem{property}[theorem]{Property}
\usepackage{bm}
\usepackage{dsfont}

\input{Preamble/preamble}

\ifdefineAbstract
\fi
\newcommand{\ri}{{\rm i}}
\newcommand{\re}{{e}}

\def\bs#1{\boldsymbol{#1}}
\newcommand{\no}{\nonumber}
\newcommand{\n}{\nonumber\\}
\newcommand{\bc}{\begin{center}}
\newcommand{\ec}{\end{center}}
\def\ba#1{\begin{array}{#1}\displaystyle}
\newcommand{\ea}{\end{array}}

\newcommand{\beq}{\begin{equation}}
\newcommand{\eeq}{\end{equation}}
\newcommand{\beqa}{\begin{eqnarray}}
\newcommand{\eeqa}{\end{eqnarray}}
\newcommand{\bi}{\begin{itemize}}
\newcommand{\ei}{\end{itemize}}

\def\ba#1{\begin{array}{#1}\displaystyle}

\def\lt#1{\left#1}
\def\rt#1{\right#1}
\def\t#1{\tilde{#1}}

\def\b#1{\bar{#1}}
\def\frc#1#2{\frac{#1}{#2}}

\newcommand{\p}{\partial}

\newcommand{\varep}{\varepsilon}

\def\bs#1{\boldsymbol{#1}}

\title{Application of the theory of C* algebras  to the emergence of hydrodynamics in quantum many-body systems }
\subtitle{}
\author{Dimitrios Ampelogiannis}
\university{King's College London}
\dept{Department of Mathematics}
\crest{\includegraphics[width=0.4\textwidth]{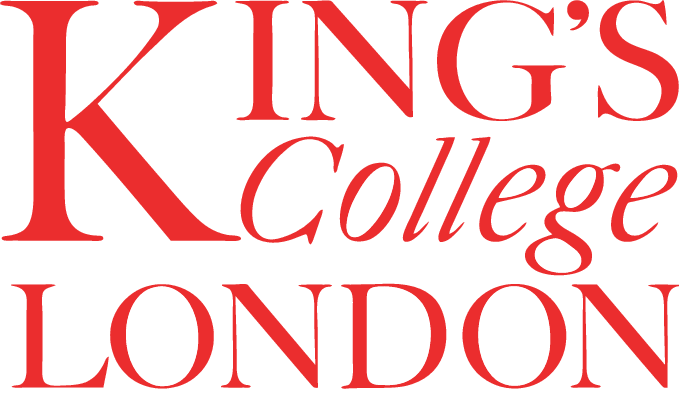}}
\supervisor{Benjamin Doyon}
\degreedate{April 2025}
\degreetitle{Doctor of Philosophy}

\begin{document}

\mainmatter

 \maketitle

% ******************************* Thesis Dedidcation ********************************

\begin{dedication} 

Dedicated to the memory of Petros Meramveliotakis

\end{dedication}

% ******************************* Thesis Declaration ***************************

\begin{declaration}

I hereby declare that except where specific reference is made to the work of 
others, the contents of this dissertation are original and have not been 
submitted in whole or in part for consideration for any other degree or 
qualification in this, or any other university. This dissertation is my own 
work and contains nothing which is the outcome of work done in collaboration 
with others, except as specified in the text and Acknowledgements. This 
dissertation contains fewer than 100,000 words including appendices, 
bibliography, footnotes, tables and equations and has fewer than 150 figures.

% Author and date will be inserted automatically from thesis.tex \author \degreedate

\end{declaration}

% ************************** Thesis Acknowledgements **************************

\begin{acknowledgements}

This thesis wouldn't have been possible without my dear friends Petros, Thodoris and Kostakis, who all helped shape me during my formative years as a physicist  in Athens. They are responsible for making me realise the beauty of mathematics.
\\

I am proud to have done my Ph.D. next to Friedrich, who was always there with the right answers, and the right questions. 
\\

On a more personal level, I would  like to thank my family, Angie, Mom and Dad, for always being there for me. Melina, for understanding me and making me a better person. 
\\

Last but not least, I would like to thank Benjamin, for not just being a supervisor, but also a friend, and guiding my mind to the right places, while at the same time giving me the freedom to explore my own ideas.
%And I would like to acknowledge ... Benj, Thodoris, Petros, Kostakis, Friedrich

\end{acknowledgements}

% ************************** Thesis Abstract *****************************
% Use `abstract' as an option in the document class to print only the titlepage and the abstract.
\begin{abstract}

%%%%%%%%%%%%
The dynamics of many-body, extended, interacting quantum systems has been the focus of much recent theoretical and experimental research, with particular interest in their universal, large-scale behaviours away from equilibrium. Given the panoply of phenomena observed and investigated, it is crucial to rigorously establish the fundamental physical principles that guide them.

This Ph.D. thesis reports on progress in the problem of rigorously establishing hydrodynamic principles from the microscopic Hamiltonian dynamics of quantum many-body systems, in a general, not model-specific, manner. We use the algebraic (C* algebra) framework of statistical mechanics, to treat systems directly in the large volume limit, and mostly study quantum lattice models where one can take advantage of powerful tools, such as Lieb-Robinson bounds, to obtain rigorous statements. Hence, we obtain a proof-of-principle that large-scale behaviours can indeed be seen as emerging from microscopic dynamics, with mathematical proof. We first report on ergodicity results  in short range models, with exponentially decaying or finite range interactions, whereby we show that time-averaged observables converge to their ensemble averages and decorrelate from all other observables, almost everywhere within the light-cone defined by the Lieb-Robinson bounds. These are developed in \cite{ampelogiannis_2023_almost}. This is a relaxation property that indicates the loss of information at large scales and from it one is able to prove a Boltzmann-Gibbs principle: observables project onto hydrodynamic modes (the extensive conserved quantities of the system) in correlation functions, at the Euler scaling limit of large time and distances \cite{ampelogiannis_long-time_2023}. These results hold independently from the details of the microscopic dynamics, capturing the physical idea that the microscopic details are lost at large space-time scales. 

Going to finer scales of hydrodynamics, we discuss rigorous lower bounds on the strength of diffusion, developed in \cite{ampelogiannis_2024_diffusion}. We show a general result on clustering (correlation decay) of n-th order connected correlations, in the context of C* dynamical systems, proved in \cite{ampelogiannis_2024_clustering}. We then use these results to obtain a strictly positive lower bound on the Diffusion constant of chaotic open quantum spin chains with nearest neighbor interactions. 

This thesis underlines the large universality of hydrodynamic principles, provides a framework for establishing them rigorously and sets the stage for many open problems towards the final goal of proving the hydrodynamic equations.
\end{abstract}

% *********************** Adding TOC and List of Figures ***********************

\tableofcontents

%\listoffigures
%\listoftables
% \printnomenclature[space] space can be set as 2em between symbol and description
%\printnomenclature[3em]

\printnomenclature

% ******************************** Main Matter *********************************

\part{Set up}
\chapter{Introduction} 

\section{Motivation}

At the Paris conference of the International Congress of Mathematicians in 1900, Hilbert presented a list of problems, all unsolved at the time, as the most important mathematical challenges of the new century. Hilbert's sixth problem, to which he devoted much of his research \cite{corry_1997_hilbert_axiomatisation}, concerns the axiomatic derivation of the laws of physics, in Hilbert's description \cite{hilbert_mathematical_1902}:
%\vspace*{\fill} 
\begin{quote} 
\centering 
``The investigations on the foundations of geometry suggest the problem: To treat in the
same manner, by means of axioms, those physical sciences in which already today
mathematics plays an important part; in the first rank are the theory of probabilities
and mechanics.''
\end{quote}
%\vspace*{\fill}
Hilbert further explained the problem:
%\vspace*{\fill} 
\begin{quote} 
\centering 
``As to the axioms of the theory of probabilities, it seems to me desirable that their logical investigation should be accompanied by a rigorous and satisfactory development of the method of mean values in mathematical physics, and in particular in the kinetic theory of gases. ... Boltzmann's
work on the principles of mechanics suggests the problem
of developing mathematically the limiting processes, there
merely indicated, which lead from the atomistic view to the
laws of motion of continua. ''
\end{quote}
%\vspace*{\fill}

The first part was resolved in the 1930s by the work of  Andrey Kolmogorov, who axiomatised probability theory by using measure theory \cite{kolmogorov_foundations}. The second part remains unsolved until today, and encompasses  most of modern mathematical physics. Indeed, understanding how macroscopic physical laws—such as those governing thermodynamics, fluid dynamics, or other large-scale phenomena—emerge from the underlying microscopic dynamics is one of the most profound and enduring challenges in both mathematics and physics. Bridging this gap not only deepens our conceptual understanding of nature’s hierarchy but also forms the cornerstone of statistical mechanics and the theory of emergent phenomena. Consider the flow of a river, as we know described by the partial differential equations called Navier–Stokes equations. Each Liter of water in the river consists of $10^{25}$ molecules; if we zoom in, then, at the most fundamental level these are described by quantum mechanics. Of course, we do not need $10^{25}$ degrees of freedom to describe the flow of water, at large-scales only a small set of collective variables—such as density, flow velocity, and temperature—are needed to capture the essential dynamics of the system. This striking reduction in complexity prompts several fundamental questions: How can the collective, large-scale behavior of such an enormous assembly of particles be systematically derived from the underlying microscopic theory? Is it possible to rigorously justify the emergence of effective macroscopic equations, like the Navier–Stokes equations, from first principles? More specifically, can one identify which degrees of freedom are relevant at macroscopic scales, and precisely describe their dynamics within a mathematically rigorous framework? Addressing these deep and challenging questions concerning the emergence and justification of macroscopic physical laws from microscopic dynamics forms the central motivation for this Ph.D. project.

The effective theory that describes the large-scale, long-time behavrior of inhomogeneous many-body systems, at the presence of nonzero currents -of particles, charge, spin- between various parts of the system, in terms of a reduced set of emergent quantities, is called hydrodynamics \cite{kadanoff_1963_hydrodynamics,pines_1967_theory,spohn_large_1991}. Unlike what the word ``hydrodynamics'' suggests (``hydro'' from the greek ``ύδωρ'' meaning water), the theory isn't just confined to fluids, but instead is a universal description of emergent  large-scale properties. For example, it can be used to described the experimentally observed \cite{moll_2016_graphene} electron transport in graphene \cite{muller_2009_graphene,narozhny_2015_hydro_graphene,lucas_2018_electrons_graphene,torre_2015_viscosity_graphene}, or in the context of high-energy physics \cite{jeon_1996_qft_hydro,pavel_2003_holo_hydro}. The Navier-Stokes equation is a specific type of hydrodynamic equation, that describes the motion of viscous fluids.  Given the panoply of phenomena
observed, it is crucial to rigorously establish the physical principles that guide them, in a universal manner.

The goal of this Ph.D. project was to contribute to the development of a universal method for rigorously proving the emergence of hydrodynamics, strating from the microscopic Hamiltonian description of the dynamics and guided by the principles of statistical mechanics. This is a monumental task, and we are aware of only a few rigorous derivations of the hydrodynamic equations, such as the hard rod gas \cite{boldrighini_1997_hard_rod}, the Rule 54 cellular automaton \cite{Klobas:2018ldp}, for disordered harmonic chains \cite{bernardin_2019_hydro_chain,hannani_2021_hydro_harmonic}, for  systems with conservative noise \cite{ollahydro,even_2014_hydrolimit}, and with significant progress in the domain of stochastic systems \cite{demasi_mathematical_2006,kipnis_1988_scaling}. More recently, in \cite{doyon_hydrodynamic_2022} the Euler hydrodynamic equations (corresponding to the Navier-Stokes with zero viscosity) and  ``... a precise and universal theory for the emergence of ballistic waves
at the Euler scale and how they propagate within homogeneous, stationary states.'' were rigorously obtained in the context of quantum spin chains.

\section{Hydrodynamics} \label{section:intro_hydro}
\begin{figure}
    \centering
    \includegraphics[width=0.7\linewidth]{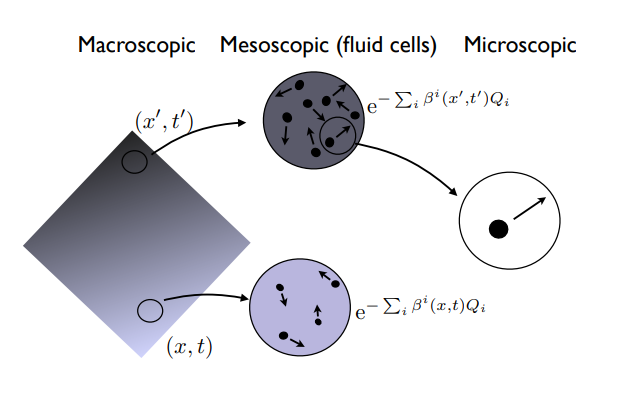}
    \caption{Fluid cell approximation \cite{doyon_lecture_2020} }
    \label{fig:fluid}
\end{figure}

Let us give a heuristic review of the basic principles of hydrodynamics, which this thesis aims to put on a rigorous basis. These principles are generally independent of the microscopic details of the system under consideration; it can be quantum or classical, a spin lattice or a gas of particles, integrable or chaotic. The key idea is that an inhomogeneous  system approaches a state that is locally in equilibrium, after an appropriate relaxation time. That means, locally the system is at a state which has maximised entropy with respect to the conservation laws. This is called the fluid cell approximation \cite{landau_1987_fluid}; imagine at (macroscopic)  point $x$ a ``bubble'' which is infinitesimally small with respect to the full system, but infinitely large with respect to the microscopic scale (characterised by the dynamics) so that it can be described by statistical mechanics, see \Cref{fig:fluid}. 

Suppose $q_i(x,t)$ are conserved densities and $j_i(x,t)$ their respective currents, with $i\in I$ some index set, satisfying the conservation laws:
\begin{equation} \label{eq:intro_conservation}
    \partial_t q_i(x, t)+\partial_x j_i(x, t)=0
\end{equation}
with associated extensive conserved quantities 
\begin{eqnarray}
    Q_i=\int \mathrm{d} x \mathfrak{q}_i(x, t)
\end{eqnarray}
Associated with these, is the maximal entropy state written formally in the Gibbs form as
\begin{equation} \label{eq:intro_max_entr}
    \langle \ldots\rangle = \frac{\Tr( \ldots e^{-\sum_i \beta_i Q_i})}{Z}
\end{equation}
where for each conserved density $q_i$ there's a corresponding Lagrange parameter $β_i$ (generalised temperature). For example, in the case of a system with  energy and particle number conservation, we have the inverse  temperature $β=1/k_BT$ and the chemical potential $μ$. Let us denote the mean values of the densities and currents as:
\begin{eqnarray}
    \overline q_i(x,t) = \langle q_i(x,t)\rangle , \quad     \overline j_i(x,t) = \langle j_i(x,t)\rangle
\end{eqnarray}

In fact, a maximal entropy state is fully characterised by either the set $\{β_i \}_{i \in I}$ or equivalently by the the averages of the conserved densities $\{  \overline q_i \}_{i \in I}$. Consequently, the average of any observable $A$ in a maximal entropy state can be seen as a function of the averages $\langle q_i\rangle$. This means that for each observable $A$ there's a corresponding functional, denoted $A^{\rm av}$, which maps the set of conserved densities $\overline q_i=\langle q_i\rangle$ to the average of $A$ in the maximal entropy state that these characterise:
\begin{eqnarray}
    \{ \overline q_i \}_{i \in I} \mapsto A^{\rm av}(  \{ \overline q_i \}_{i \in I}) = \langle A \rangle
\end{eqnarray}
For simplicity we write $A^{\rm av}( \overline q_i ) = \langle A \rangle$. 
In particular, the dependence of the average currents $ \overline j_i= \langle j_i\rangle$ on the  $ \overline q_i=\langle q_i\rangle$ defines the equation of state:
\begin{eqnarray}
    \overline j_i = j_i^{\rm av}(   \overline q_i)
\end{eqnarray}
and the flux Jacobian as:
\begin{equation} \label{eq:intro_flux}
    \mathrm{A}_i^j=\frac{\partial \overline j_i}{\partial \overline q_i}=\frac{\partial j_i^{\rm av}}{\partial \overline q_i}
\end{equation}

The discussion between \Cref{eq:intro_conservation} and \Cref{eq:intro_flux} describes what happens inside the fluid cell, \Cref{fig:fluid}. The assumption of local equilibrium posits the existence of slowly varying fields $β^i(x,t)$, these describe the equilibrium (maximal entropy) state of each fluid cell. It is with respect to the variation lengths of the $β^i(x,t)$ that the fluid cell is considered small enough.  The expectation value, in the global state $\langle \ldots\rangle_{\rm global}$ of an observable $A$ at point $(x,t)$ is then
\begin{equation} 
    \langle A(x,t) \rangle_{\rm global} \approx \langle A(0,0)\rangle_{β_i(x,t)}
\end{equation}
where $\langle A(0,0)\rangle_{β_i(x,t)}$ is the maximal entropy state with with $x,t$-dependent Lagrange parameters. 

The average currents $\overline j_i(x,t)$ will depend on all the densities $q_k(y,t)$, $k \in I$, at all points $y$, for any given time slice $t$. Since local equilibration is assumed to occur within fluid cells, with large variation length, it is natural to assume that $\overline j_i(x,t)$ only depends on the values of $\overline q_i(y,t)$ for $y$ near $x$.
One then expresses $\overline j_i(x,t)= \langle j_i(x,t) \rangle_{\rm global} $ by a derivative expansion
\begin{equation} \label{eq:intro_current_expansion}
    \overline j_i(x,t) =  j_i^{\rm av}(   \overline q(x,t)) - \frac12 \sum_k \mathcal{D}_i^k( \overline q(x,t)) \partial_x \overline q_k(x,t) + \mathcal{O} ( \partial^2_x)
\end{equation}
where the factor $\frac12$ is by convention, and $\mathcal{D}_i^k$ is called the diffusion matrix. Combining the current expansion \Cref{eq:intro_current_expansion} with the conservation law for the averages:
\begin{equation} \label{eq:intro_conservation_av}
    \partial_t \overline q_i(x, t)+\partial_x \overline j_i(x, t)=0
\end{equation}
and ignoring higher orders, yields the hydrodynamic equation:
\begin{equation} \label{eq:hydrodynamic_eq}
    \partial_t \overline q_i + \sum_k A_i^k \partial_x \overline q_k - \frac12 \sum_k \mathcal{D}_i^k \partial_x \overline q_k = 0
\end{equation}
The first two terms correspond to the Euler equation, and $A_i^k$ is the flux Jacobian, \Cref{eq:intro_flux}, describing ballistic transport. The third term is of Navier-Stokes type, with the diffusion matrix $\mathcal{D}_i^k$ depending on the microscopic details of the system. See also the discussions in \cite{doyon_lecture_2020, nardis_2019_diffusion_ghd}.

The above discussion is formal and making it mathematically precise  is clearly a difficult task. Firstly,  the maximal entropy states as expressed in \Cref{eq:intro_max_entr} are not well-defined. What types of charges need to be considered? How are the charges $Q_i$ defined and  what is the meaning of the sum $\sum β_i Q_i$ in the infinite volume limit? Further, the validity of the fluid cell approximation is not generally established, with an exception being the hard rod gas \cite{Boldrighini_HardRodGas}.  Rigorously constructing hydrodynamics, from the microscopic description of the system, is the goal of the present thesis.

\section{Organisation of the Thesis}
This thesis reports on recent advances \cite{ampelogiannis_2023_almost,ampelogiannis_long-time_2023,ampelogiannis_2024_clustering,ampelogiannis_2024_diffusion} in rigorously proving the emergence of hydrodynamics in quantum lattice models, which have their starting point in \cite{doyon_hydrodynamic_2022}, and is organised as follows. In \Cref{chapter:math} we provide the precise mathematical set up we use in our work: the algebraic ($C^*$ algebra) approach to statistical mechanics, and review some of the key results in current literature. In \Cref{chapter:ergo} we discuss ergodicity  results obtained in \cite{ampelogiannis_2023_almost}, which provide the necessary mechanism for the loss of information  and subsequent reduction of the number of degrees of freedom at large scales; this is captured by  a ``hydrodynamic projection'' principle, shown in \cite{ampelogiannis_long-time_2023} and discussed here in \Cref{chapter:hydroproj}. Finally, in \Cref{chapter:higher_clustering} we present the results of \cite{ampelogiannis_2024_clustering} on the decay of higher order correlation functions, which are used in \cite{ampelogiannis_2024_diffusion} to obtain a lower bound on the strength of diffusion in open quantum spin chains, presented here in \Cref{chapter:diffusion}.

\chapter{Mathematical Set-up: the algebraic approach} \label{chapter:math}

Statistical mechanics seeks to describe the macroscopic physical properties of matter—such as pressure, temperature, and magnetization—in terms of the collective behavior of a vast number of microscopic components, like atoms in a gas or spins in a magnet. When the number of constituents $N$, is extremely large, contributions from terms of order $N^{-1}$ or smaller become negligible, allowing one to approximate the finite system by an idealised infinite system. Consider the example of a gas confined within a finite box. The physical expectation is that its thermodynamic properties should be independent of the properties of the walls, and the same should be true for hydrodynamics; we are interested in the bulk properties. We can then either follow the usual description of the system by  ordinary quantum (or classical) mechanics, and ignore terms that are inverse powers of the system size, or we could instead treat an infinitely extended system. The latter is called the ``algebraic approach'', as the algebra of observables is the central object of study and not the Hilbert space of wavefunctions. The algebraic approach provides a powerful and flexible language for treating infinite systems, offering a natural generalisation of ordinary quantum mechanics. Importantly, for systems with a finite number of degrees of freedom, the algebraic and conventional approaches are mathematically equivalent, but only the algebraic formalism can rigorously accommodate the complexities arising in infinitely large systems—such as the definition of Gibbs (equilibrium) states, the emergence of distinct thermodynamic phases and long-range correlations \cite{hugenholtz_1972_why_algebras, bratteli_operator_1987,bratteli_operator_1997}.

The main obstacle stems from the fact that taking the thermodynamic limit ($N\to\infty$) of a sequence of finite quantum systems can lead to profound changes in the mathematical structure of the theory. For finite systems, the algebra of observables is always of type I, meaning that  the Gelfand–Naimark–Segal construction (GNS) representation associated with any state is unitarily equivalent to the familiar Schrödinger representation \cite[Section 2.7]{bratteli_operator_1987}. However, in the infinite limit, new features emerge: the algebra of observables becomes a genuine (often simple) $C^*$-algebra, and the space of physically relevant states may comprise mutually inequivalent (i.e., non-unitarily equivalent) Hilbert space representations. This is a crucial distinction, as different thermodynamic phases (such as those with spontaneous symmetry breaking) are associated with distinct representations, which cannot be accommodated within a single Hilbert space in the Schrödinger picture. From a more physical point of view, consider a liquid at fixed pressure with its temperature being raised. At a critical value of the temperature $T_c$ a phase transition occurs, whereby small variations of temperature around $T_c$ lead to large changes in  quantities like the density. At $T_c$ multiple phases, mixtures of liquid and gas, can coexist; the equilibrium state is not unique, hence the specific heat per gram $^\circ K$, or the density per unit volume, are not well-defined at these critical values. Thus, one anticipates that the theoretical counterparts of these physical quantities will exhibit pronounced changes with respect to temperature or other control parameters at the critical points. The thermodynamic limit once again proves advantageous here: in finite systems, such rapid variations manifest over a narrow range but, upon taking the infinite-volume limit, these become true discontinuities. Indeed, the emergence of these discontinuities is often regarded as a key motivation for considering the thermodynamic limit. 

The present thesis deals with the problem of emergence of hydrodynamics in quantum many-body systems, which refers to the derivation of macroscopic, effective equations—such as the continuity and Navier-Stokes equations—for conserved quantities (e.g., energy, momentum, particle number) from the underlying quantum-mechanical description, particularly in the thermodynamic and long-time limits. This task poses several major conceptual and technical challenges that are most naturally addressed within the algebraic  framework. In fact, we believe our present work further solidifies the need for the algebraic approach to statistical mechanics. Hydrodynamics is inherently a theory of the bulk: it concerns local densities and currents in the limit of infinite system size. The algebraic approach allows one to define quasi-local algebras of observables for infinite quantum systems, rigorously capturing the notion of spatial locality.  In contrast, the conventional Hilbert space formalism, which ties all observables to a single global space, lacks a natural way of encoding spatially local degrees of freedom in the thermodynamic limit. The hydrodynamic description emerges from the nonequilibrium dynamics of conserved quantities. The algebraic approach supplies the tools for constructing and analyzing nonequilibrium steady states, entropy production, and linear response, via, for example, the use of KMS states and the theory of modular automorphisms. These structures are vital for making sense of macroscopic transport coefficients and response functions in a rigorous way. 
 Finally, derivation of hydrodynamics requires careful control over various limits: large system size, long time, and long wavelength \cite{spohn_2014_nonlinear}. The C$^*$-algebra approach provides a solid backround on which to build these notions, where the thermodynamic limit has already been taken and various properties of the state can be used to show existence of the long time and/or wavelength limits, while local observables can be bundled up in appropriate quotient spaces of the C$^*$-algebra to construct Hilbert spaces which only carry the relevant information necessary to describe hydrodynamics \cite{doyon_hydrodynamic_2022,ampelogiannis_long-time_2023}.

In this chapter we provide a short introduction to the algebraic (C$^*$-algebra) description of quantum statistical mechanics, organised as follows. \Cref{section:c*} introduces C$^*$-algebras as the main mathematical structure that describes our physical observables, states as linear functionals over this algebra and the important Gelfand–Naimark–Segal  (GNS) representation, which will be used throughout our work. In \Cref{section:quantum_spin_lattice} we specify the description of quantum spin lattice models, described by quasi-local algebras, and their dynamics. One of the most pivotal results in quantum spin lattices, the Lieb-Robinson bound, is discussed in \Cref{section:LR}, along with a brief historical overview. The Kubo-Martin-Schwinger (KMS) states are defined in \Cref{section:KMS}. Finally, in \Cref{section:clustering} we discuss clustering of correlations, with an overview of some key results.

%gas in a box, fock space inequivalent rep, bratteli

%\section{Introduction Title with math \texorpdfstring{$\sigma$}{[sigma]}}x

\section{Introduction to the C*-algebra framework} \label{section:c*}
Throughout the present thesis we consider quantum many-body systems described by C$^*$ dynamical systems. We refer to the textbooks of Ola Bratteli and  Derek W. Robinson \cite{bratteli_operator_1987,bratteli_operator_1997} for the detailed mathematical framework, on which we base a lot of the work of this Ph.D. thesis. We assume that the reader is familiar with basic notions of functional analysis, operator theory, and quantum statistical mechanics.

To start, the observables of our system will be elements of a C$^*$-algebra, which is a Banach space that has an algebra structure and an involution which   satisfies the properties of the adjoint. Any non-commutative C$^*$-algebra is isomoprhic to a space of bounded operators acting on a Hilbert space.

\begin{defn}[C$^*$-algebra]
    A Banach algebra is a pair $(\mathfrak{A},\norm{\cdot})$ where $\mathfrak{A}$  an algebra  (over $\C$) equipped with a norm $\norm{\cdot}$ such that $(\mathfrak{A},\norm{\cdot})$ is a Banach space (complete) and for all $A,B\in \mathfrak{A}$
    \begin{equation}
        \norm{AB} \leq \norm{A} \norm{B}
    \end{equation}
 An involution, or adjoint operation, on an algebra $\mathfrak{U}$ is a map $*: \mathfrak{U} \to \mathfrak{U}$, denoted $A \mapsto A^*$, such that $A^{**}=A$, $(AB)^*=B^*A^*$ and $(aA+bB)^*= \overline{a}A^*+\overline{b}B^*$ for all $A,B \in \mathfrak{U}$, $a,b\in \C$.

If  $\mathfrak{U}$  is a Banach algebra and $^*$ an involution on it, such that 
\begin{equation}
    \norm{A^*}= \norm{A} , \quad \forall A\in \mathfrak{U}
\end{equation}
then $(\mathfrak{U},^*)$ is a Banach $^*$-algebra. 

A C$^*$-algebra is a Banach $^*$-algebra such that 
\begin{equation}
    \norm{AA^*} =\norm{A}^2 , \quad \forall A \in \mathfrak{U}
\end{equation}
We will only consider $C^*$ algebras that possess a unit element $\mathds{1}$.
\end{defn}

In quantum mechanics, the quantities that one is interested in, and can measure experimentally, are the expectation values of observables in some quantum state $Ψ$
\begin{equation}
    \langle A \rangle = \bra{Ψ}A\ket{Ψ}
\end{equation}
where $Ψ$ is a pure state: an element of the underlying Hilbert space. In statistical mechanics we consider an ensemble of possible quantum states $Ψ_i$, each occuring with classical probability $ρ_i$, therefore the expectation values of observables are
\begin{equation}
   \langle A\rangle=\sum_i \rho_i\left\langle\Psi_i\right| A\left|\Psi_i\right\rangle
\end{equation}
This ensemble of pure states is called a mixed state and is described by a density matrix, see also the textbook \cite{greiner2012thermodynamics}. In our case, the density matrix is $\rho=\sum_i \rho_i\left|\Psi_i\right\rangle\left\langle\Psi_i\right|$, and the expected value is then written as
\begin{equation}
    \langle A\rangle=\operatorname{Tr}(\rho A)
\end{equation}

In both cases, $\langle A \rangle$ is a linear functional on the set of observables. It is worth pointing out that density matrices can be problematic in the thermodynamic limit, where one takes the system size to be infnite. Indeed, the trace of operators in infinite dimensional Hilbert spaces is not always defined. The C$^*$-algebra framework deals with this issue by defining states directly as positive linear functionals of the C$^*$-algebra.

\begin{defn}[State] \label{def:state}
    A state of a C$^*$-algebra $\mathfrak{U}$ is a positive linear functional $ω: \mathfrak{U} \to \C$; $ω(AA^*) \geq 0$, $\forall A\in \mathfrak{U}$, with $\norm{ω}=1$. The set of states of $\mathfrak{U}$ is denoted $E_{\mathfrak{U}}$.
\end{defn}
\nomenclature[ω]{$ω( \cdot)$}{State of C$^*$-algebra}     
\nomenclature[EU]{$E_{\mathfrak{U}}$}{Set of states}     
In quantum mechanics time evolution is implemented either in the Shrodinger picture, by evolving the state which is an element of the Hilbert space, or in the Heisenberg picture where one evolves the operators. In the C$^*$-algebra framework, we have ``forgotten'' about the underlying Hilbert space, so we define define time evolution directly on the observables, by automorphisms of the C$^*$-algebra. A C$^*$-algebra together with such an automorphism is called a C$^*$ dynamical system. We first give a general definition, which will be used in \Cref{chapter:higher_clustering}; the automorphism needs not be time evolution, but can instead describe symmetries of the system, or, for example, space translations. In the next section, we will give a specific definition of the dynamical systems that we will use for quantum lattice models.

\begin{defn}[C$^*$ dynamical system] \label{def:dynamical_system}
    A C$^*$ dynamical system is a triple $(\mathfrak{U}, G, a)$ consisting of a C$^*$-algebra $\mathfrak{U}$, a locally compact group $G$ and a strongly continuous representation of $G$ by $^*$-automorphisms  $\{ a_g: \mathfrak{U} \xrightarrow{\sim} \mathfrak{U} \}_{g \in G}$.
\end{defn}

Given a state $ω$ we can construct a Hilbert space $H_ω$ and a representation of $\mathfrak{U}$ into the bounded operators $B(H_ω)$ acting on $H_ω$. This is done by the Gelfand–Naimark–Segal (GNS) construction \cite[Theorem 2.3.16]{bratteli_operator_1987}. Further, if we have a group of automorphisms of $\mathfrak{U}$, then these can be represented by a unitary group of operators acting on $H_ω$ \cite[Corollary 2.3.17]{bratteli_operator_1987}. We summarise these two central results of the theory in the following Proposition, which will be used extensively in our proofs.
\begin{prop}[GNS representation] \label{prop:gns}
Given a state $ω \in E_{\mathfrak{U}}$ of a unital C$^*$ algebra $\mathfrak{U}$ there exists  a (unique, up to unitary equivalence)  triple $(H_ω,π_ω,Ω_ω)$ where $H_ω$ is a Hilbert space with inner product $\langle \cdot, \cdot \rangle$, $π_ω$ is a representation of the C$^*$ algebra by bounded operators acting on $H_ω$ and $Ω_ω$ is a cyclic vector for $π_ω$, i.e.\ the $\operatorname{span} \{ π_ω(A)Ω_ω : A \in \mathfrak{U}\}$ is dense in $H_ω$, such that
\begin{equation}
    ω(A) = \langle Ω_ω , π_ω(A) Ω_ω \rangle ,  \ \ A \in \mathfrak{U}
\end{equation}
If additionally we have a group $G$ of automorphisms $\{ τ_g \}_{g \in G}$ of $\mathfrak{U}$ and $ω$ is $τ$-invariant, then there exists a representation of $G$ by unitary operators $U_ω(g)$ acting on $H_ω$. This representation is uniquely determined by 
\begin{equation}
    U_ω(g) π_ω(A) U_ω(g)^* = π_ω (τ_g (A)) \ , \ \ A \in \mathfrak{U}, g \in G
\end{equation}
and invariance of the cyclic vector
\begin{equation}
    U_ω(g)Ω_ω = Ω_ω \ , \forall g \in G
\end{equation}
\end{prop}
\nomenclature[BH]{$B(H)$}{Set of bounded operators over a hilbert space $H$}     
\section{Quantum spin lattice description} \label{section:quantum_spin_lattice}

The results presented in this Ph.D. thesis mainly focus on quantum spin models on a square lattice. The basic theory of quantum spin models, in the C$^*$-algebra framework, is treated in the second book of Bratteli and Robinson \cite{bratteli_operator_1987}, but we also refer to \cite{naaijkens_quantum_2017} for a more concise and modern introduction. The kinematics are described by a quasi-local uniformly hyperfinite (UHF) C$^*$-algebra \cite{glimm_certain_1960}, which is built up as a union of finite-dimensional  matrix algebras, with an added locality structure. The dynamics are described by a Hamiltonian, which defines the $^*$-automoprhism of time evolution via the Heisenberg picture. 

\subsection{Quasi-local C*-algebras} \label{section:quasi}

We consider a lattice $\Z^D$, as a $\Z-$module, equipped with the $l_1$ norm denoted . We denote the distance between two lattice points $x,y\in \Z^D$ by $d(x,y)$. On each site $x \in \Z^D$ we associate a quantum spin described by the matrix algebra of observables $\mathfrak{U}_{x} \coloneqq M_{N_{x}} ( \C)$,  $N_{x} \in \N$, of $N_x \times N_x$ matrices, equipped with the operator norm $||\cdot ||$.  We assume the dimension of the spin matrix algebra to be uniformly bounded, $N_{x} \leq N$ for some $N \in \N$. To each finite $Λ \subset \Z^D$ we associate the algebra 
\begin{equation}
    \mathfrak{U}_Λ\coloneqq \bigotimes_{x \in Λ} \mathfrak{U}_{x}
\end{equation}
If $Λ_1 \cap Λ_2= \emptyset$, then $\mathfrak{U}_{Λ_1 \cap Λ_2} =\mathfrak{U}_{Λ_1} \otimes \mathfrak{U}_{Λ_2}$ and $\mathfrak{U}_{Λ_1}$ is isomorphically embedded in $\mathfrak{U}_{Λ_1} \otimes \mathds{1}_{Λ_2}$, where $\mathds{1}_{Λ_2}$ is the unit of $\mathfrak{U}_{Λ_2}$. The algebra of local observables is then defined  as the direct limit of the increasing net of algebras $\{ \mathfrak{U}_Λ \}_{Λ \in \mathbb{P}_f(\Z^D)}$, where $\mathbb{P}_f(\Z^D)$ denotes the set of finite subsets of $\Z^D$,
\begin{equation}
    \mathfrak{U}_{\rm loc} \coloneqq \lim_{\longrightarrow} \mathfrak{U}_Λ
\end{equation}
\nomenclature[PF]{$\mathbb{P}_F(A)$}{Set of finite subsets of A}        
The norm completion of $\mathfrak{U}_{\rm loc}$ defines the quasi-local C$^*$-algebra of the quantum spin lattice:
\begin{equation}
    \mathfrak{U} \coloneqq \overline{\mathfrak{U}_{\rm loc}}
\end{equation}
with the following structure:
\begin{itemize}
    \item $\mathfrak{U}_{Λ_2} \subset \mathfrak{U}_{Λ_1}$, if $Λ_2 \subset Λ_1$,
    \item $\mathfrak{U} = \overline{\bigcup_Λ \mathfrak{U}_Λ }$,
    \item $\mathfrak{U}$ and $\mathfrak{U}_Λ$ have common unit $\mathds{1}$, for any $Λ$,
    \item $[\mathfrak{U}_{Λ_1}, \mathfrak{U}_{Λ_2}] =0$ whenever  $Λ_2 \cap Λ_1= \emptyset$,
\end{itemize}
\nomenclature[ι]{$ι_x$}{Space translations acting as group of automorphisms on C$^*$-algebra}   
\nomenclature[τ]{$τ_t$}{Time evolution acting as group of automorphisms on C$^*$-algebra}   
\nomenclature[A]{$A(x,t)= ι_x τ_t(A)$}{Space-time translation of observable $A$}   
Space translations are naturally defined on $\mathfrak{U}$, as an action  of the group $\Z^D$ by $^*$-automorphisms $\{ ι_x :\mathfrak{U} \xrightarrow{\sim} \mathfrak{U} \}_{x \in \Z^D}$, see \cite{bratteli_operator_1987}, with the following properties
\begin{itemize}
    \item $A \in \mathfrak{U}_Λ$, then $ι_x(A) \in \mathfrak{U}_{Λ+x}$,
    \item $\lim_{x \to \infty} \norm{ [ι_x (A), B ]} =0$
\end{itemize}
The latter property is called asymptotic abeliannes of the algebra with respect to the automorhpism group of space translations, and plays an important role in the development of the general theory, as well as our present results. It conveys the intuitively anticipated principle that distant observations remain unaffected by each other, which stems from the quasi-local structure.

Throughout this thesis we use the following quantities. The support of an observable $A \in \mathfrak{U}_{\rm loc}$, and the distance of the supports of $A,B \in \mathfrak{U}_{\rm loc}$ are defined as
\begin{equation}
    \supp(A) = \bigcap \{ X \subset \mathbb{Z}^D : A \in \mathfrak{U}_X \} 
\end{equation}
and
\begin{equation}
    \dist(A,B) \coloneqq \dist( \supp(A), \supp(B))
\end{equation}
respectively, where the distance between two subsets $X,Y$ of $\mathbb{Z}^D$ is 
\begin{equation}
    \dist(X,Y) = \min \{ |x-y|_1 : x \in X , y \in Y \}.
\end{equation}
The diameter of subsets $X$ of $\mathbb{Z}^D$ is
\begin{equation}
    \diam(X) = \max \{ |x-y|_1: x,y \in X \}
\end{equation}
%Finally, the ball around $x_0$ of radius $d$ is \begin{equation}
%   B_{x_0}(d) = \{ x \in \mathbb{Z}^D : |x-x_0| \leq d \}
%\end{equation}
and, finally,  $|X|$ denotes the number of elements in $X$.
\subsection{Interactions and Dynamics} \label{section:dynamics}
The dynamics are defined from the Hamiltonian using the Heisenberg picture. To define the Hamiltonian, we introduce a notion of interactions for finite subsets of the lattice, by associated with each such subset a self-adjoint element of the C$^*$-algebra, interpreted as the total interaction energy.

\begin{defn}[Interaction] \label{def:interaction}
    An interaction is a map $Φ: \mathbb{P}_f(\Z^D) \to \mathfrak{U}_{loc}$ such that $Φ(Λ) \in \mathfrak{U}_Λ$ is hermitian $Φ(Λ)= Φ^*(Λ)$, $\forall Λ \in \mathbb{P}_f(\Z^D)$. 
\end{defn}
Given an interaction, the Hamiltonian associated with each finite subset $Λ$ is defined as the total interaction energy: 
\begin{equation}
        H_Λ = \sum_{X \subset Λ} Φ(X)
    \end{equation}
which defines the local time evolution 
\begin{equation}
    τ_t^Λ(A) = e^{itH_Λ} Ae^{-itH_Λ} , \, t\in \R, \, A\in \mathfrak{U}_Λ
\end{equation}
Time evolution of the infinite system is defined when the limit $\lim_{Λ\to \Z^D} τ_t(A) $
exists in the norm for all $A \in \mathfrak{U}_{\rm loc}$ and can be uniquely extended to a strongly continuous $^*$-automorphism $τ_t$ of $\mathfrak{U}$. This can be proved for a large class of interactions, including finite range and exponentially decaying ones \cite[Chapter 6]{bratteli_operator_1997}, as discussed below.

We summarise the mathematical framework used to decribe quantum spin lattice systems in the following definition, where we also assume time evolution to be homogeneous, that is we impose space translations to commute with time evolution.

\begin{defn}[Quantum lattice C$^*$-dynamical system] \label{def:qsl_dynamical_system}
A C$^*$-dynamical system of a  quantum spin lattice is a triple $(\mathfrak{U},ι,τ)$ where $\mathfrak{U}$ is a quasi-local UHF $C^*$-algebra, $τ$ is a strongly continuous representation of the group $\R$ by $^*$-automorphisms $\{ τ_t : \mathfrak{U} \xrightarrow{\sim} \mathfrak{U} \}_{t \in \R}$, and $ι$ is a representation of the translation group $\Z^D$ by $^*$-automorphisms $\{ ι_{x}: \mathfrak{U} \xrightarrow{\sim} \mathfrak{U} \}_{x \in \Z^D}$, such that for any $Λ \in \mathbb{P}_f(\Z^D)$: $A \in \mathfrak{U}_Λ \implies ι_{x }(A) \in \mathfrak{U}_{Λ+x }$ for all $x \in \Z^D$. We further assume that $τ$ is such that $τ_t ι_{x} = ι_{x} τ_t  $, $\forall t \in \R$, $x \in \Z^D$, i.e.\ time evolution is homogeneous. 
\end{defn}

The results presented in this  thesis  will be for quantum lattice systems which we call ``short-range''; they have interactions with finite, or exponentially decaying, range. To describe such interactions, consider a function $F:[0,\infty) \to (0,\infty)$ such that:
\begin{equation}
    \norm{F} \coloneq \sup_{x \in \Z^D} \sum_{y \in \Z^D} F( d(x,y)) < \infty
\end{equation}
and 
\begin{equation}
  C \coloneqq  \sup_{x,y \in \Z^D}\sum_{z \in \Z^D}  \frac{F(d(x, z)) F(d(z, y))}{F(d(x, y))}<\infty 
\end{equation}
Given any $F$ that satisfies these two conditions, we define for $λ>0$
\begin{equation}
    F_λ(x) \coloneqq e^{-λx} F(x)
\end{equation}
which also satisfies the conditions. See also  the discussion in \cite{nachtergaele_2006_propagation}.

Using the functions $F_a$ we characterise the interactions that we consider. We summarise the full description of the systems under consideration in the following definition.

\begin{defn}[Short-range quantum spin lattice] \label{def:short_range_qsl}
A short-range quantum spin lattice is a C$^*$ dynamical system, \Cref{def:qsl_dynamical_system}, with $τ_t$ defined from an interaction $Φ$ such that there exists a $λ>0$ and function $F_λ$, with
\begin{equation} \label{eq:interaction}
   \|\Phi\|_λ:=\sup _{x, y \in \Lambda} \sum_{X \ni x, y} \frac{\|\Phi(X)\|}{F_λ(d(x, y))}<\infty .
\end{equation}
\end{defn}

We call any  interaction with strength that decays exponentially with distance ``short-range''. Note that an interaction is called finite range when $Φ(X)=0$ whenever $\diam(X)> r_0$, for some finite $r_0>0$.
\begin{example}
    A simple example of the above construction is the Heisenberg XYZ model. Consider a chain $\Z$, with spin $1/2$ at each site $x\in \Z$. The infinite chain is described by the algebra
    \begin{equation}
     \mathfrak{U}= \overline{\mathfrak{U}_{\rm loc}} , \quad \mathfrak{U}_{\rm loc}=  \lim_{\longrightarrow} \bigotimes_{x=-N}^N \C^2
    \end{equation}
 and the interaction contains one and two-body  translationally invariant    terms:
 \begin{equation}
     Φ( \{0 \} )= h σ^z_0 , \quad Φ( \{0,1 \})= J_x σ_0^x σ_1^x+J_yσ_0^y σ_1^y+J_z σ_0^z σ_1^z
 \end{equation}
with $Φ(\{ i\})= ι_x Φ( \{0\})$, $Φ( \{i,i+1\} ) = ι_x Φ( \{0,1 \}) $ and $Φ(X)=0$ if $|X|>2$ or $\diam(X)>1$. Thus, the local Hamiltonian has the familiar form
\begin{equation}
    H_Λ^{\rm    XYZ}  = \sum_{i \in Λ} ( J_x σ_i^x σ_{i+1}^x+J_yσ_i^y σ_{i+1}^y+J_z σ_i^z σ_{i+1}^z +h σ^z_i )
\end{equation}
and the infinite volume limit of time evolution exists, since $H_Λ^{XYZ}$ has finite range interactions.
\end{example}

\section{Lieb-Robinson bounds and quasi-locality} \label{section:LR}
How fast information propagates in a many-body system is a pivotal question, with far
reaching consequences, such as  on thermalization rates \cite{polkovnikov_2011_nonequilibrium, gogolin_equilibration_2016}, determining the strength and range of correlations in equilibrium \cite{Hastings_2006_spectralgap}, imposing speed limits on quantum-state transfer \cite{Bose_2007_qtransfer}, and of great interest for digital quantum simulation of many-body systems \cite{Tran_2019_qsimulation,Haah_2021_qalgorithm} and quantum information scrambling \cite{Roberts_2016_butterfly, maldacena_bound_2016}.

In relativistic systems, information cannot propagate outside the light cone defined by the speed of light, similarly in non-relativistic  quantum systems  a central result is the Lieb-Robinson bound, which provides an upper bound on the speed of information propagation through the system, up to exponentially small corrections. The Lieb-Robinson bound implies that the time evolution of a local observable, under a local Hamiltonian, has an approximately (up to exponential corrections) finite support that grows linearly in time, with velocity that depends on the interaction and geometry of the underlying space. It was first proven by Lieb and Robinson in 1972 \cite{Lieb:1972wy}, in the context of quantum spin lattice models with short-range interactions, and remained without many applications or improvements until 2004 when Hastings used it to  obtain a generalization  of the Lieb-Schultz-Mattis theorem \cite{Lieb_1961_Lieb-Schultz-Mattis} to higher dimensions \cite{Hastings_2004_lieb-schultz-mattis}. This sparked many new extensions and applications. New Lieb-Robinson bounds and results on exponential clustering appeared in 2006, in the works of Hastings and Koma \cite{Hastings_2006_spectralgap},  Nachtergaele, Ogata and Sims \cite{nachtergaele_2006_propagation}, Nachtergaele and Sims \cite{nachtergaele_2006_LR}. In disordered systems, bounds where obtained by  Burrell and  Osborne in 2007 \cite{Burrell_2007_LR_disorder}. 
Initially Lieb-Robinson bounds where considered for, and applied to, systems with short-range interactions (finite or exponentially decaying), where the effective light-cone is linear. characterised by a finite constant velocity.  The case is more subtle for long-range (power-law decaying) interactions, with the first extension found in \cite{Hastings_2006_spectralgap}, where the upper bound for the group velocity grows exponentially with time leading to a curved light-``cone''. Various improvements have been made to these long-range results, importantly in \cite{foss-feig_2015_nearly_linear} where it is shown that light-cones of power-law $1/r^a$ interacting systems are polynomial for $α > 2D$, and in \cite{Kuwahara_2020_linear_LR} a linear light-cone is obtained by $a>2D+1$, with $D$ the lattice dimension. Lieb-Robinson bounds  were extended in 2010 to the case of dissipative systems, described by Lindbladian dynamics,  see \cite{poulin_2010_LR_markovian,nachtergaele_2011_LR_irreversible}. Finally, it is worth noting that Lieb-Robinson bounds are generally loose, and improvements can be made in the estimates of the velocities, see for example the recent work \cite{Wang_2020_tightening_LR}. We also refer to the recent review \cite{chen2023speed} for a more complete list of results.

The results presented here are in the context of short-range interactions and rely heavily on the Lieb-Robinson bound,   which we state below as proven in \cite{nachtergaele_2006_propagation}. 

\begin{theorem}[Lieb-Robinson bound] \label{th:lieb_robinson}
Consider a quantum lattice dynamical system with interaction $Φ $ satisfying \Cref{eq:interaction}, and local time evolution $τ^Λ_t$ associated with any finite $Λ\subset \Z^D$. There exist  $L,υ_{LR}>0$ such that for any $Λ$ and $A,B \in \mathfrak{U}_{\rm loc}$
   \begin{equation} \label{eq:lieb_robinson}
        \norm{ [τ_t^ΛA,B]} \leq L \norm{A} \norm{B} \min(|A|, |B|) e^{-λ(\dist(A,B)- υ_{LR}t)}
   \end{equation}
for any $t \in \R$. 
\end{theorem}

An important implication of the Lieb-Robinson bound is the quasi-locality of the time evolution of local observables. Lieb-Robinson bounds can be applied  to approximate such timve evolved observables by strictly local ones with an error bound, see for example \cite{nachtergaele_2011_local_approx, nachtergaele_quasi-locality_2019}. Following \cite[Section 4]{nachtergaele_quasi-locality_2019}, we first define a localizing family for the algebra of observables $\mathfrak{U}$. Consider a finite $Λ\subset \Z^D$ and
 a locally normal\footnote{A normal state is one which is defined by a density matrix} product state $\rho$ : i.e. for each site $x \in \Z^D$ we fix a normal state $\rho_x$ on $\mathfrak{U}_x={B}\left(\mathcal{H}_x\right)$, and take the unique state $\rho$ on $\mathfrak{U}$ such that the restriction of $ρ$ to $\mathfrak{U}_Λ$ is $\rho_{\left\lceil\mathfrak{U}_{\Lambda}\right.}=\bigotimes_{x \in \Lambda} \rho_x$ for all finite $\Lambda \subset \Z^D$. Here we use the notation $\rho_{\left\lceil\mathfrak{U}_{\Lambda}\right.}$ for the restriction of $ρ$ to $\mathfrak{U}_Λ$. Then, given $X \subset \Lambda \in \mathcal{P}_F(\Z^D)$, we define the maps $\Pi_\rho^{X, \Lambda}: \mathfrak{U}_{\Lambda} \rightarrow \mathfrak{U}_X$ by
\begin{equation}
\Pi_\rho^{X, \Lambda}(A)=\left(\mathds{1}_X \otimes \rho_{\Lambda \backslash X}\right)(A) , \quad  A \in \mathfrak{U}_{\Lambda}
\end{equation}
where $\mathds{1}_X$ is the  unit of $\mathfrak{U}_X$. Making the dependence on $ρ$ implicit, define $\Pi_X^{\Lambda}: \mathfrak{U}_{\Lambda} \rightarrow \mathfrak{U}_{\Lambda}$ by
\begin{equation}
\Pi_X^{\Lambda}(A)=\Pi_\rho^{X, \Lambda}(A) \otimes \mathds{1}_{\Lambda \backslash X}
\end{equation}
For fixed $X$, these projections are compatible in the sense that if $X \cup \Lambda^{\prime} \subset \Lambda$ and $\Lambda \in \mathcal{P}_F(\Z^D)$, then

\begin{equation} \label{eq:localising_consistency}
\Pi_X^{\Lambda}(A \otimes \mathds{1})=\Pi_{X \cap \Lambda^{\prime}}^{\Lambda^{\prime}}(A) \otimes \mathds{1}, \quad \forall A \in \mathfrak{U}_{\Lambda^{\prime}}
\end{equation}

For any $\Lambda^{\prime} \subseteq \Lambda$, note that we can identify $\mathfrak{U}_{\Lambda^{\prime}}$ as a sub-algebra of $\mathfrak{U}_{\Lambda}$, as discussed in \Cref{section:c*}, hence we  write \Cref{eq:localising_consistency} as $\Pi_X^{\Lambda} \upharpoonright_{\mathfrak{U}_{\Lambda^{\prime}}}=\Pi_{X \cap \Lambda^{\prime}}^{\Lambda^{\prime}}$. In particular, if $X \subseteq \Lambda^{\prime} \subseteq \Lambda$, one has that $\Pi_X^{\Lambda} \upharpoonright_{\mathfrak{U}_{\Lambda^{\prime}}}=\Pi_X^{\Lambda^{\prime}}$, from which we see that the  map $\Pi_X: \mathfrak{U}_{\rm{loc}} \rightarrow \mathfrak{U}_{\rm{loc}}$ is well-defined as:

\begin{equation}
\Pi_X(A):=\Pi_X^{\operatorname{supp}(A) \cup X}(A)
\end{equation}

This map is bounded with norm one, therefore $\Pi_X$ has a unique extension to $\mathfrak{U}$ which we also denote by $\Pi_X$. Note that $\Pi_X(A)=A$ if $A \in \mathfrak{U}_X$. The $\Pi_X, X \in \mathbb{P}_f(\Z^D)$ is called a localizing family, which can be used to define local approximations of time evolved observables, in combination with the following Lemma from \cite{nachtergaele_quasi-locality_2019}:
\begin{lem} \label{lem:localise}
    Consider a quantum lattice dynamical system, as in \Cref{defn:dynamicalsystem} and the localising map $\Pi_X$ defined above. Let $ε>0$ and $A \in \mathfrak{U}$ such that 
    \begin{equation} \label{eq:localise1}
        \norm{ [A,B]} \leq ε \norm{B} , \quad \forall B \in \mathfrak{U}_{ \Z^D \setminus Λ }
    \end{equation}
then we can approximate $A$ by $Π_Λ(A) \in \mathfrak{U}_Λ$:
\begin{equation} \label{eq:localise2}
    \norm{ Π_Λ(A) - A} \leq 2ε
\end{equation}
\end{lem}

Consider now a local observable $A \in \mathfrak{U}_{\rm loc}$  with support $\supp A = Λ$ and a sequence of finite sets $Λ_r = \cup_{x \in Λ}B_{ x}(r)$, $r=1,2,3, \dots$ extending a distance $r$ around $Λ$.  By the Lieb-Robinson bound \Cref{eq:lieb_robinson}, for any $B \in \mathfrak{U}_{Z^D \setminus Λ_r}$ we have 
\begin{equation}
    \norm{ [τ_t(A), B]} \leq L \norm{A} \norm{B} \min(|Λ|, |\supp B|) e^{-λ(r- υ_{LR}t)}
\end{equation}
and applying the above Lemma, we can then approximate $τ_t(A)$ by the sequence $Π_{Λ_r}(τ_tA)$

\begin{theorem}
    Consider a quantum spin lattice dynamical system, as in \Cref{defn:dynamicalsystem}, with finite range or exponentially decaying interactions so that it satisfies the Lieb-Robinson bound \Cref{eq:lieb_robinson}. It follows that time evolution is quasi-local, in the sense that for any $A \in \mathfrak{U}_{\rm loc}$, with finite support $Λ$, its time evolution for any $t\in \R$  can be approximated by a sequence of local observables $Π_{Λ_r}(τ_tA)$:
 \begin{equation}\label{eq:time_evolution_local_approx}
    \norm{Π_{Λ_r}(τ_tA) - τ_tA} \leq  2 L \norm{A} |Λ| e^{-λ(r- υ_{LR}t)} , \quad r \in \N
\end{equation}
\end{theorem}

\section{Equilibrium: KMS states} \label{section:KMS}
Ιn quantum systems with finite dimensional Hilbert space, thermal equilibrium at inverse temperature $β$ is described by  the Gibbs  state:
\begin{equation} 
    ω(A) = \frac{\Tr e^{-βH} A}{\Tr e^{-βH}}
\end{equation}
However, in infinite systems we run into problems, with the most obvious one being that operators might not be trace-class, or the partition function can diverge, making the Gibbs state undefined. The situation can get worse, if we consider the presence of phase transitions, where there exist multiple equilibria.

When the Gibbs state is well-defined, we can easily see that it satisfies the Kubo-Martin-Schwinger (KMS) condition: $\left.\omega\left(A \tau_t(B)\right)\right|_{t=i \beta}=\omega(B A)$.
This condition is described in detail in \cite[Section 5.3.1]{bratteli_operator_1997} and is utilized in the C$^*$-algebra framework as a definition of thermal equilibrium, serving as a generalisation of the Gibbs state. In fact, it is equivalent to the Gibbs condition in finite systems. The KMS condition was introduced by Ryogo Kubo  in 1957 \cite{kubo_1957_statistical} and   was further employed by Paul C. Martin  and Julian Schwinger in 1959 to define thermodynamic Green's functions \cite{martin_1959_theory}. It was then used as a defining property of thermal equilibrium in 1967 by Rudolf Haag, Marinus Winnink and Nico Hugenholtz \cite{haag_1967_equilibrium}.

\begin{defn}[KMS state] \label{defn:KMS}
Consider a dynamical system $(\mathfrak{U},ι,τ)$. A state $ω$ is called a $(τ,β)$-KMS state, at inverse temperature $β$, if 
\begin{equation}
    ω(A τ_{iβ}B) = ω(BA) \label{eq:KMS_state}
\end{equation}
for all $A,B$ in a dense (with respect to the norm topology) $τ$-invariant $^*$-subalgebra of $\mathfrak{U}_τ$, where $\mathfrak{U}_τ$ is the set of entire analytic elements for $τ$.
\end{defn}

One can easily prove, directly from the above definition, that a  $(τ,β)$-KMS state is $τ-$invariant. 

It is established that the set of analytic elements $\mathfrak{U}_τ$ is (norm) dense in $\mathfrak{U}$ for any strongly continuous one-parameter group of automorphisms $τ$, see \cite[Proposition IV.4.6]{simon_statistical_2014}.  Hence, any dense subset of $\mathfrak{U}_τ$ is again dense in $\mathfrak{U}$. 

Two important questions that arise from the above definition have to do with existence and uniqueness of KMS states. Firstly, one sees that the set of inverse temperatures $β$ such that  $(τ,β)$-KMS state exists is a closed subset of $\R \cup \{ \pm \infty \}$, \cite[Proposition 5.3.25]{bratteli_operator_1987}. We denote by $K_β$ the set of $(τ,β)-$KMS states and interpret it as the set of equilibrium states of the system at inverse temperature $β$. Physically, one expects that $K_β$ contains a unique element at higher temperatures, while at lower temperatures it contains multiple elements corresponding to different thermodynamic phases and their mixtures. One can easily see that $K_β$ is convex \cite[Theorem 5.3.30]{bratteli_operator_1997} and its extremal points are interpreted as the pure thermodynamic phases of the system. 

An important property of the extremal KMS states is that they are factorial, as a consequence of \cite[Theorem 5.3.30]{bratteli_operator_1997}. As such, when there is a unique KMS state at a given temperature, this state is factorial. A state $ω$ is called factorial when the bicommutant $π''_ω(\mathfrak{U})$ of the respective GNS representation $π_ω$ has trivial center $π''_ω(\mathfrak{U}) \cap π'_ω(\mathfrak{U})= \mathds{1} \C$. While we do not make use of the defining property of factorial states here, they are important because they have clustering properties \cite[Example 4.3.24]{bratteli_operator_1987}, which are essential for the results obtained throughout  the present Ph.D. project. 

\section{Clustering} \label{section:clustering}
In the context of C$^*$-dynamical systems, acted on by a group such as that of space or time translations, clustering is  the vanishing of the two-point connected correlation function when one observable is translated under the group action. 

The decay of 2-point connected correlations, for space-translations, has been established rigorously in many physical systems. In classical systems see \cite{ruelle_1974_rigorous}; in particular a one-dimensional classical lattice gas satisfies the clustering property, as shown by Ruelle  \cite{Ruelle_1968_classical_gas}. For quantum spin chains (QSC), Araki \cite{Araki:1969bj} showed exponential clustering in Gibbs states, for finite-range interactions. In QSC with short-range interactions the correlations also decay exponentially in Gibbs states \cite{perez_2023_locality}. Clustering for quantum gases was established by Ginibre \cite{Gibibre_1965_quant_gases}. Cluster expansion methods can be used quite generally, to show decay of correlations for both classical and quantum systems \cite{park_1982_clusterexpansion,Ueltschi2003Cluster, park_1995_clustering_gibbs}.

We begin this section with a general definition and basic results, and then specify to the clustering properties with respect to space translations in quantum lattice models. 

\begin{defn}[Clustering] %\label{defn:clustering}
A state $ω$ of a $C^*$ dynamical system  $(\mathfrak{U},G,α)$ is called $G$-clustering if there exists a net $ g_i \in G$ such that 
\begin{equation}\label{clustering}
    \lim_i \big(ω( α_{g^i}(A) B) - ω(α_{g^i}A) ω(B) \big)= 0 , \ \forall A,B \in \mathfrak{U}.
\end{equation} 

\end{defn}

In the non-commutative setup, it can be the case that the algebra is asymptotically abelian under the action of $G$, i.e.\ there exists a net $g_i \in G$:
\begin{equation}
    \lim_i \norm{ [ α_{g^i} A, B] }=0 , \ \forall A,B \in \mathfrak{U}.
\end{equation}
It follows from asymptotic abelinanness that every factor state will satisfy the cluster property \Cref{clustering}, see \cite[Example 4.3.24]{bratteli_operator_1987}.  

\begin{prop} \label{theorem:asymptotic_clustering}
    Consider a $C^*$ dynamical system  $(\mathfrak{U},G,α)$ that is asymptotically abelian. If $ω$ is a factor state, then ω is $G$-clustering. 
\end{prop}

\begin{corollary}
Consider a $C^*$ dynamical system  $(\mathfrak{U},G,α)$ that is asymptotically abelian and $τ_t$ is the strongly continuous group of time evolution. If $ω_β$ is an extremal $(τ,β)$-KMS state, then $ω$ is factor and hence $G$-clustering.
\end{corollary}

Let us go back to quantum spin lattice systems. It is immediate by construction that $\mathfrak{U}$ is asymptotically abelian for space translations $ι$. As such, factor states, and in particular extremal KMS states, i.e.\ pure thermodynamic phases, satisfy the following definition:

\begin{defn}[Clustering in space] \label{def:clustering_space}
A state $ω$ of a dynamical system  $(\mathfrak{U},ι,τ)$  is called clustering with respect to space translations $ι$,  if for any $A,B \in \mathfrak{U}$
    \begin{equation}
        \lim_{ x\to\infty}\big( ω(  ι_xA B)  - ω( ι_xA)ω( B) \big)=0
    \end{equation}
\end{defn}

Of great interest is the rate with which correlations between observables cluster.  Araki showed, in his pivotal 1969 work \cite{Araki:1969bj}, that KMS states of a quantum spin chain are unique for any $β>0$ and exponentially clustering in space.  These results have been improved in the last decade, to include more general interactions and quantum spin systems over general graphs. A very general result, which uses cluster expansions to show 2-point clustering is obtained in \cite{Ueltschi2003Cluster} and subsequently applied in \cite{frohlich_properties_2015} to obtain exponential clustering bounds in high temperature thermal states of quantum spin lattice models. We rely on these results to have the appropriate cluster properties we need in our proofs.

\begin{defn}[Exponential clustering in space] \label{defn:exp_clustering_space}
A state $ω$ of a dynamical system  $(\mathfrak{U},ι,τ)$  is called exponentially clustering, with respect to space translations $ι$,  if there exist constants $E,r,μ>0$ such that for any  $A,B \in \mathfrak{U}_{\rm loc}$ 
\begin{equation}
    |ω(ι_x AB) - ω(A)ω(B) | \leq E_{A,B} e^{-μ | x|} \text{, for all $ x \in \Z^D$.}
\end{equation}
and
\begin{equation}
    E_{A,B}< E \norm{A} \norm{B} |Λ_A|^r |Λ_B|^r
\end{equation}
\end{defn}

To keep our assumptions as minimal as possible, we will also consider states that cluster polynomially, as defined below.

\begin{defn}[Inverse power clustering in space] \label{def:power_clustering_space}
A state $ω$ of a dynamical system  $(\mathfrak{U},ι,τ)$  is called $r$-sizeably $p$-clustering for $p>r>0$, with respect to space translations $ι$, if there exists a constant $u$ such that  for any  $A,B \in \mathfrak{U}_{\rm loc}$ 
\begin{equation}
    |ω(ι_x A B) - ω(A)ω(B) | \leq C_{A,B} (1+ | x|)^{-p} , \text{ for all $ x \in \Z^D$}
\end{equation}
and $C_{A,B}$ is bounded as:
\begin{equation}
   C_{A,B} \leq u \norm{A} \norm{B} |Λ_A|^r |Λ_B|^r.
\end{equation}
%and $C_{A,B} \leq u|Λ_A|^r |Λ_B|^r$, $r>0$,$r,u>0$

A subset $\mathfrak{C} \subset \mathfrak{U}$ is called uniformly $p$-clustering for $p>D$ if there exist a $C>0$ s.t. for all $A,B \in \mathfrak{C}$:
\begin{equation} \label{eq:uniformclustering}
    | ω( ι_xA B )- ω(A) ω(B)| \leq C (1 + | x|)^{-p} , \ \ \forall x \in \Z^D.
\end{equation}
\end{defn}
Of course $C_{A,B}$ will, in general, depend on the pair of observables, i.e.\ on their norm and the size of their support.   This notion originates from \cite[Definition 4.2]{doyon_thermalization_2017}. % In that case, we can easily extend the arguments in \cite[Section 8]{doyon_hydrodynamic_2022} to arbitrary lattice dimension $D$, to show that whenever $p>D(r+1)$, a  $p$-clustering state will be space-like $p_c$-clustering for $p_c>D$, in any system with interaction that satisfies \Cref{eq:interaction}. 

\part{Results}

\chapter{Relaxation properties: Clustering and Ergodicity} \label{chapter:ergo}
Hydrodynamics operates under the assumption that the physical system locally relaxes towards a maximum entropy state, at least at the largest scales of space and time \cite{doyon_2023_mft}. To hope to be able to extract the large scale hydrodynamic behaviours, one first has to examine the relaxation properties of the system, such as clustering and ergodicity. In this chapter we discuss these properties and report on many new results obtained during this Ph.D. project in \cite{ampelogiannis_2023_almost} and \cite{ampelogiannis_2024_clustering}.

Clustering, as discussed in \Cref{section:clustering}, is a property of the state of the system, with respect to the group action of space or time translations, whereby translating an observable infinitely far it decorrelates from all other observables. Thermal states of pure thermodynamic phases, described by extremal KMS states, are generally clustering with respect to space translations. In combination with the Lieb-Robinson bound, one can see that observables will decorrelate when their space-time distance falls outisde the Lieb-Robinson light-cone. This property we call  space-like clustering, and it indicates very strong relaxation outisde the light-cone. However, one then has to ask about what happens inside the Lieb-Robinson light-cone, where very little is known rigorously, and where all interesting physical phenomena, including hydrodynamics, are expected to take place. 

A weaker form of clustering is ergodicity, often considered in the context of C$^*$-algebras as clustering of the mean of covariance \cite[Section 4.3]{bratteli_operator_1987}.
Traditionally, ergodicity is the idea that over long times, the system's state covers uniformly enough the manifold of states, or at least the part of it that is dynamically accessible, such as the energy shell. In the language of Gibbs' statistical mechanics, this translates into the equivalence of time averages with ensemble averages. See for example Khinchin's discussion in \cite[Chapter III]{Khinchin}, for a very instructive introduction to the subject of ergodicity, from a mathematical point of view. A central question is in what situations, and in what specific form, does this happen in extended quantum systems? Can we expect some form of ergodicity inside the Lieb-Robinson light-cone? In particular, a natural idea is that this requires the presence of chaos: classically, the exponential separation, over time, of initially nearby trajectories, or quantum mechanically, certain conditions on the distribution of energies, or on out-of-time-ordered correlators (OTOC) \cite{shenker_black_2014, hosur_chaos_2016, maldacena_bound_2016,hashimoto_out--time-order_2017}. Is there such a link between short-time (chaotic) and long-time (ergodic) behaviours? %
These are some of the questions we aim to answer in \cite{ampelogiannis_2023_almost}, as well as \cite[Section 2]{ampelogiannis_long-time_2023}, and that we address in this chapter.

Von Neumann's work  on the quantum ergodic theorem \cite{neumann_1929_beweis} offers some insight, and we refer to \cite{goldstein_long-time_2010,goldstein_normal_2009} for some very interesting discussions on the matter. In \cite{ampelogiannis_long-time_2023} we also review von Neumann's work in the context of C$^*$ algebras and quantum spin lattices, and highlight some important drawbacks; essentially Von Neumann's quantum ergodic theorem is not formulated directly in the thermodynamic limit and  relies on contrived and hard-to-check conditions. Can something more universal be said?  %In short, von Neumann proved given the time average of an observable $\overline{A(t)}$ and a state $Ψ$ in the microcanincal shell

By taking advantage of the finiteness of local spaces of observables, which yields a countability argument, we will extend space-like clustering inside the Lieb-Robinson light-cone; the drawback is that this will only hold for almost every space-time ray and for the time average of correlations. Strong clustering $ω(ι_xτ_tAB)\to ω(A)ω(B)$ for $x>υ_{LR}t$ becomes mean clustering $\frac{1}{T}\int_0^Tω(ι_{υt}τ_tAB) dt\to ω(A)ω(B)$ for almost every velocity at $υ<υ_{LR}$. This was first considered in spin chains \cite{doyon_hydrodynamic_2022}, while ergodicity results in the C$^*$-algebra framework generally assume asymptotic abelianness \Cref{eq:asympotic_abeliaN} \cite{bratteli_operator_1987,ruelle_1974_rigorous, kastler_invariant_1966,robinson_1967_extremal}. Note that, it is non-trivial to establish whether time evolution is asymptotically abelian, in contrast to space translations.

In fact, the main theorem that we will prove states that the long-time averaging of any observable $A$ converges, in the strong operator topology of the Gelfand-Naimark-Segal (GNS) representation of a state $\omega$, to its expectation in the state $\omega(A)$, times the identity operator. This holds for almost every speed with respect to the Lebesgue measure, and all rational directions on the lattice; we will refer to this as ``almost everywhere" or ``almost every space-time ray". The result holds in every space invariant (that is, invariant under spatial translations) factor state $\omega$ satisfying the Kubo-Martin-Schwinger (KMS) condition, \Cref{defn:KMS}. These include any KMS state of quantum spin lattice models with sufficiently fast-decaying interactions, at high enough temperature.

The result implies that convergence holds in every correlation function. Thus, connected correlation functions (Ursell functions) vanish in the limit, and the $n$-th moment of the ray-average of an observable converges to the $n$-th power of its expectation in the state. We will show, by using the Lieb-Robinson bound, that in fact, the results for two-point correlation functions and for the $n$-th moments hold in every space invariant factor state, not necessarily KMS.

An important new result we obtained in \cite{ampelogiannis_2023_almost} is that the averaging can also be taken with respect to oscillatory factors, in a way that accounts for oscillations, for any wavelength and frequency. We show that ray averaged correlations, weighed by oscillatory factors, vanish in the long time limit for almost every ray, hence precluding persistent oscillatory behaviours.  This oscillatory version of ergodicity will be applied in \Cref{chapter:hydroproj} to obtain an oscillatory hydrodynamic projection theorem. This serves as a demonstration that recent ideas on dynamical symmetries \cite{buca2019nonstationary,buca2020quantum,medenjak2020rigo,buca2021local,gunawardana2021dynamical} can be extended to the hydrodynamic regime of correlation functions.

This Chapter is organised as follows. First, we consider a quantum spin lattice with short-range interactions in a spatially clustering state, and show that the state is also clustering for space-time translation outside the Lieb-Robinson light cone in \Cref{section:space-like-clustering}. In \Cref{section:ergodicity} we state our main theorem of the vanishing of ray-averaged correlations in the long-time limit, for almost every ray. \Cref{section:meansquared} discusses ``mean-square'' ergodicity, which is the vanishing of the variance of ray-averages. Further, if we assume the state to be KMS, we show a strong result of convergence of the ray-average to the ensemble average in \Cref{section:ergodicity_kms}. Finally, in \Cref{section:oscillatory_ergo} we obtain versions of the aforementioned results in which the ray-average is weighed by oscillatory factors. In each Section we include a subsection with the relevant proofs, except for some cases where the proofs are omitted due to similarity with previous ones. All the Theorems presented here, with full proofs, are found in \cite{ampelogiannis_2023_almost}.

\section{Space-like clustering} \label{section:space-like-clustering}
As discussed in \ref{section:KMS}, thermal states -towards which we expect our system to locally relax- exhibit clustering properties with respect to space translations,  reflecting the statistical independence of distant subsystems at equilibrium. However, we should also need relaxation with respect to time evolution in order to be able to extract hydrodynamic behaviours from the microscopic dynamics. The first results that come to mind use Lieb-Robinson bounds, to show exponential clustering, with respect to time evolution, in gapped ground states. These were first shown in  2006 in the works of Hastings and Koma \cite{Hastings_2006_spectralgap}, and Nachtergaele and Sims \cite{nachtergaele_2006_LR}. However, there are two main limitations that prevent direct application of these results to the study of hydrodynamics. First, ground states are not relevant for describing the hydrodynamics of a physical system; we are interested in systems at nonzero temperature, characterized by local equilibrium with slowly varying parameters such as temperature or chemical potential. Second, the existence of a uniform spectral gap—crucial for these clustering theorems—is generally difficult to establish and tends to be a model-specific property. For our purposes, it is essential to develop tools that apply to a broad and physically relevant class of systems, ideally under conditions that are as universal and model-independent as possible.

The Lieb-Robinson bound can also be thought of as asymptotic abelianness  of the algebra of  observables $\mathfrak{U}$ with respect to space time translations $ι_{υt} τ_t$ outside the Lieb-Robinson light-cone $υ>υ_{LR}$ \cite[Theorem 4.2]{ampelogiannis_2023_almost}. More specifically, we have asymptotic abelianness of $\mathfrak{U}$ with respect to the subgroup of space-time translations along a ray $G_{υ,x} \coloneqq  \{ (rx, rυ^{-1}|x|) : r \in \Z \} \subset \Z^D \times \R$, for any $υ\in \R : |υ|>υ_{LR}$ and any $n \in \Z^D$.

\begin{theorem}[Space-like asymptotic abelianness] \label{th:asymptotic_abelianness}
Consider a short-range quantum spin lattice, \Cref{def:short_range_qsl}. It follows that the algebra $\mathfrak{U}$ is $G_{υ,x}$-asymptotically abelian for every $υ \in \R$ s.t.\ $|υ|>υ_{LR}$, and $x \in \Z^D$, i.e.\ 
\begin{equation} \label{eq:asympotic_abeliaN}
   \lim_{r \to \infty} \norm{ [ι_{rx} τ_{rυ^{-1}|{x}|}(A),B]} \rightarrow 0 
\end{equation}
for all $A,B \in \mathfrak{U}$.
\end{theorem}

It is well established that given asymptotic abelianness of a C$^*$-algebra  under a certain group action $τ_g$, it follows that any factor state $ω$ satisfies clustering properties, as discussed in \Cref{section:clustering}.  From these results we  obtain the following in \cite{ampelogiannis_2023_almost}:

\begin{theorem}[Space-like clustering] \label{theor:space-like-clustering}
Consider a short-range quantum spin lattice, \Cref{def:short_range_qsl}, and a factor invariant state $ω$ of $\mathfrak{U}$. It follows that for every $x\in \Z^D$, $|υ|>υ_{LR}$ and all observables $A,B \in \mathfrak{U}$ we have space-like clustering of correlations:
\begin{equation} \label{eq:space-like_clustering}
    \lim_{r \to \infty} | ω\big( ι_{r x} τ_{rυ^{-1}|x|} (A) B \big) - ω(A) ω(B) | = 0 .
\end{equation}
%This immediately implies that $ω$ is also space-like ergodic (\Cref{def:spacelike-ergodic}) and space-like non-oscillating (\Cref{def:phase-spacelike-ergodic}).
\end{theorem}

%In \cite{ampelogiannis_2023_almost} we also discuss a different method of showing space-like clustering
Next, it is natural to ask what the rate of space-like clustering is. Suppose we have a state $ω$ that is exponentially clustering in space, as per \Cref{defn:exp_clustering_space}. Given that the Lieb-Robinson bound is exponential, for short-range interactions, we should be able to show that $ω$ is also exponentially space-like clustering. This was first considered in \cite{doyon_hydrodynamic_2022}, for spin chains $D=1$ with finite range interactions. In \cite{ampelogiannis_long-time_2023} we consider the more general case $D>1$ and short-range interactions and prove that given an exponentially space-clustering state $ω$ and the Lieb-Robinson bound, we have exponential space-like clustering of $ω$. 

The method of proving space-like clustering is straightforward. Using the quasi-locality  of time evolved obvervables $τ_t(A)$ , we approximate $τ_t(A)$ by a sequence of local observables, using \Cref{eq:time_evolution_local_approx}. The connected correlation $ω\big( ι_{r x} τ_{rυ^{-1}|x|} (A) B \big) - ω(A) ω(B)$ is then approximated by a connected correlation between the local approximation of $τ_{rυ^{-1}|x|} (A)$ and $B$, which is bounded by the spatial clustering, plus exponentially small corrections given by the Lieb-Robinson bound.  The following Theorem is proved in \Cref{appC}, see also\cite[Appendix C]{ampelogiannis_long-time_2023}.

\begin{theorem}[Exponential space-like clustering]
   Consider a short-range quantum spin lattice, \Cref{def:short_range_qsl}, in an exponentially clustering state. It follows that the system is space-like exponentially clustering with respect to $ω$ for the Lieb-Robison velocity $υ_{LR}$. For any $A,B\in \mathfrak{U}_{\rm loc}$ and $ t \in \R, x \in \Z^D$ with $|x|> υ_{LR}|t|$:
    \begin{equation}
    |ω(ι_x τ_t A B) - ω(A)ω(B) | \leq k_{A,B} e^{-λ | x|}
\end{equation}
and
\begin{equation}
    k_{A,B}< u \norm{A} \norm{B} |Λ_A|^r |Λ_B|^r
\end{equation}
\end{theorem}

We can also consider a more general case, where the rate of spatial clustering of $ω$ is  polynomial of power $p$. In that case, space-like clustering will be polynomial, but the power will be $p-rD$, where $D$ is the lattice dimension and $r$ is  the rate of polynomial growth of the space-clustering constant with respect to the support sizes.

\begin{theorem}[Power-law space-like clustering]
   Consider a short-range quantum spin lattice, \Cref{def:short_range_qsl}, in an $r-$sizeable $p-$clustering state, \Cref{def:power_clustering_space}. It follows that the system is space-like $p-rD$ clustering with respect to $ω$ for the Lieb-Robison velocity $υ_{LR}$. For any $A,B\in \mathfrak{U}_{\rm loc}$ and $ t \in \R, x \in \Z^D$ with $|x|> υ_{LR}|t|$:
    \begin{equation}
    |ω(ι_x τ_t A B) - ω(A)ω(B) | \leq k_{A,B} \frac{1}{(1+|x|)^{p-rD}}
\end{equation}
and
\begin{equation}
    k_{A,B}< u \norm{A} \norm{B} |Λ_A|^r |Λ_B|^r
\end{equation}
\end{theorem}

The proof of the above theorem is very similar to the exponential case, and is omitted. See also the proofs in \cite[Appendix C]{ampelogiannis_2023_almost}, \cite[Appendix C]{ampelogiannis_long-time_2023}.

\section{Almost everywhere ergodicity: mean-clustering of correlations} \label{section:ergodicity}
While space-like clustering is a strong result that holds in physically relevant states, it still does not reveal anything about what happens inside the Lieb-Robinson light cone. By taking advantage of the finiteness of the local degrees of freedom, which results in seperability of the GNS Hilbert space, we can show that clustering  still holds almost everywhere within the light-cone, but on average. That it, the time average over connected correlations between observables with one translated over a space time ray vanishes in the long-time limit. This was first shown in \cite{doyon_hydrodynamic_2022} in quantum spin chains, $D=1$. Here we present a more general version for $D>1$, with a refined and generalised proof that utilises the GNS representation, shown in \cite{ampelogiannis_2023_almost}.

\begin{theorem}[Almost Everywhere Ergodicity]
Consider a dynamical system $(\mathfrak{U}, ι, τ)$, with interaction that satisfies \Cref{eq:interaction}  and a space-like clustering state $ω\in E_{\mathfrak{U}}$, as defined in \Cref{eq:space-like_clustering}. It follows that for all $A,B \in \mathfrak{U}$, any rational direction $\boldsymbol{q} \in \Srat^{D-1} \coloneqq   \{\boldsymbol x / |\boldsymbol x|:\boldsymbol x\in \mathbb{Z}^D \setminus \{\boldsymbol0\} \}$ and \textbf{almost every speed} $υ \in \mathbb{R}$ (with respect to the Lebesgue measure):
\begin{equation}
    \lim_{T \to \infty} \frac{1}{T} \int_0^T  ω \bigg( ι_{\floor{ \boldsymbol{v}t}}τ_t (A) B \bigg)  \,dt = ω(A) ω(B) \label{eq:maintheorem}
\end{equation}
where $v= υ \boldsymbol{q}$, and for any vector $\boldsymbol{a}=(a_1,a_2,...,a_D)$ we denote $\floor{\boldsymbol{a}} \coloneqq ( \floor{a_1},...,\floor{a_D})$.
\label{th:maintheorem}
\end{theorem}

It is simple to see that \Cref{eq:maintheorem} will hold for any $υ>υ_{LR}$, by the assumption of space-like clustering. However, for $υ<υ_{LR}$ this is non-trivial and we are not aware of any previous results that deal with this situation, besides \cite{doyon_hydrodynamic_2022} in the case of one-dimensional, finite-range interactions.

A few remarks about the physics of the above theorem, and of our other main results below, are in order. The equality of time-averaged quantities with their ensemble averages is a notion of ergodicity. However, conventionally ergodicity does not involve space translation: the ray is at velocity 0. The above theorem, and all the main theorems in this chapter, are notions of ergodicity along almost every ray, but not necessarily at zero velocity. What is the physical meaning or motivation for ergodicity at other velocities?
First, our ergodic theorems can be seen in a somewhat natural fashion as a ``thinning'' of the time evolved observable $τ_t (A)$, as it spatially spreads, over time, within the Lieb-Robinson light-cone. Indeed, looking within the light-cone at different velocities, $ ι_{\floor{ \boldsymbol{υ}t}}τ_t (A)$, we see that the time-average does not correlate with, hence is ``not seen" by, any other observable within the state $ω$, for almost every velocity; it may only be seen on a set of velocities of measure zero. This thinning property is made stronger by the operatorial version of the result , \Cref{th:KMS_average_strong_convergence} in \Cref{section:ergodicity_kms} . This sheds light on the problem of operator spreading discussed in the literature (see e.g.~\cite{nahum_2018_operator,keyserlingk_2018_operator,khemani_2018_operator,rakovszky_2018_diffusive}): we see that under time averaging, much of the structure of operator growth is washed out.

Second, the notion that time averages reproduce ensemble averages is based on the physical idea that a measure of some observable, somewhere in space, in fact will naturally be time averaged, because of the finite measurement time. From this viewpoint, one may argue that time averaging does not have to be in pure-time direction: for instance, a gas may be in a Galilean-boosted thermal state (there may be a nonzero overall velocity), and time averaging is equivalent to ray averaging of the non-boosted gas. Then, the question of ergodicity at almost every velocity is the relevant one. Our results are for lattice gases; it will be interesting to study systems on continuous space, and almost-every-boost ergodicity results.

Third, one of the most important applications of Theorem \ref{th:maintheorem} is to hydrodynamic projections \cite{doyon_hydrodynamic_2022, ampelogiannis_long-time_2023}. Because hydrodynamics only makes predictions for fluid-cell averaged observables, it turns out that it is sufficient to have ergodicity at almost every velocity in order to derive general hydrodynamic projection results. This is made clear in \Cref{chapter:hydroproj}, which is based on our work in \cite{ampelogiannis_long-time_2023}.

Finally, one may ask about the meaning of the measure-zero sets $\mathcal{K}_{\boldsymbol q}$ (one for every direction $\boldsymbol{q}\in\Srat^{D-1})$ of velocities $v$ where \eqref{eq:maintheorem} fails. For generic interacting models, we expect these sets to be empty -- all local observables decorrelate in time, either exponentially or, at the speeds of hydrodynamic modes, algebraically. But examples of models where a $\mathcal{K}_{\boldsymbol q}$ is not empty are the trivial lattice models without inter-site interaction (only site self-energy): one can easily construct such models where some single-site observable $A$ is time invariant. In fact, these are particular examples of {\em constrained models}, and we believe more general, interacting constrained models, actively studied recently (see e.g.~\cite{pancotti_2020_quantum}), may have non-trivial $\mathcal{K}_{\boldsymbol q}$. It would be interesting to investigate this further.

\begin{remark}
It is worth noting that the restriction to the rational directions, which has  measure $0$ in $\mathbb{R}^{D-1}$, is rather natural: as we are on a discrete space, the relevant directions are the ones in $\Srat^{D-1}$, not $\mathbb{S}^{D-1}$ (the unit sphere in $\mathbb{R}^D$). Hence, ``almost everywhere'' indeed means for almost every space-time ray of the lattice: space is discrete. Further, for applications to hydrodynamic projection theorems in \Cref{chapter:hydroproj}, it turns out that considering rational directions is sufficient, as these are dense in $\mathbb S^{D-1}$. One may nevertheless ask what happens along irrational directions, which we leave for future work.
\end{remark}

\subsection{Proof} \label{section:ergodicityproof}
Ergodic theorems usually involve the convergence of series (or integrals) of operators into projections. In our case, we will express  \Cref{eq:maintheorem} in the GNS representation (\Cref{prop:gns}) and use von Neumann's mean ergodic Theorem \ref{th:neumann} ( see also \cite[Theorem II.11]{simon_reed_functional_1}) to get a projection.  This projection will be shown  to have rank one almost everywhere, by use of space-like erogidicity, the group properties of time and space translations and the separability of the GNS Hilbert space of a quantum spin lattice:

\begin{theorem}[Projection]\label{th:rankone}
Consider a quantum spin dynamical system $(\mathfrak{U}, ι, τ)$ with interaction that satisfies \Cref{eq:interaction} and a space-like clustering state $ω\in E_{\mathfrak{U}}$, as defined in \Cref{eq:space-like_clustering}. Let $(H_ω,π_ω,Ω_ω)$ be the GNS representation associated to $ω$ and consider the unitary representation $U_ω(\boldsymbol n,t)$ of the group $G=\mathbb{Z}^D \times \mathbb{R}$  acting on $H_ω$ (\Cref{prop:gns}). It follows for any $\boldsymbol n \in \mathbb{Z}^D$, that for almost all $υ \in \mathbb{R}$ the orthogonal projection $P_{υ,\boldsymbol n}$ on the subspace of  $H_ω$ formed by vectors invariant under $U_ω(\boldsymbol n,υ^{-1}|\boldsymbol n|)$ is the rank one projection on $Ω_ω$.
\end{theorem}

In fact there is an equivalence between Eq. \ref{eq:maintheorem} and $P_{υ,\boldsymbol n}$ being rank one, as will be made clear by the proof, see also \cite[Section 2]{fidaleo2014Nonconventional}. Relevant results include \cite[Theorems 4.3.17, 4.3.20, 4.3.22]{bratteli_operator_1997} and \cite[Proposition 6.3.5]{ruelle_1974_rigorous}, which, however,  assume some form of asymptotic abelianness or clustering for the whole group, while we have it only in a space-like region. 

We begin the proof by showing two Lemmata. 
Consider a space-like clustering $ω \in E_{\mathfrak{U}}$ and its respective GNS representation $(H_ω, π_ω, Ω_ω)$. Since $ω$ is $ι,τ$-invariant there exists a representation of $\mathbb{Z}^D \times \R$ by unitary operators $U(\boldsymbol n,t)$ acting on $H_ω$, see \Cref{prop:gns}. This representation is fully determined by the properties
\begin{equation}
    U_{\omega}(\boldsymbol n,t) \pi_{\omega}(A) U_{\omega}(\boldsymbol n,t)^{*}=\pi_{\omega}\left(ι_{\boldsymbol n} τ_t (A)\right)
\end{equation}
 and invariance of the cyclic vector $Ω_ω$
 \begin{equation}
     U_{\omega}(\boldsymbol n,t) \Omega_{\omega}=\Omega_{\omega}
 \end{equation}
for all $A \in \mathfrak{U}$, $(\boldsymbol n,t) \in \Z^D \times \R$.

Let $P_{υ,\boldsymbol n}$ be the orthogonal projection on the subspace of $H_ω$ that is invariant under $U_\omega(\boldsymbol n, υ^{-1}|\boldsymbol n|)$. This subspace will contain at least $Ω_ω$. 

\begin{lem}
Consider the assumptions of Theorem \ref{th:rankone} and let $P_{υ,\boldsymbol n}^{(r)}$, $r= 1,2,...$, denote the projection onto the subspace of $H_ω$ spanned by vectors invariant under $U_ω^r (\boldsymbol n, υ^{-1}|\boldsymbol n|)$, with $P_{υ,\boldsymbol n}^{(1)}= P_{υ,\boldsymbol n}$. It follows that  $P_{υ,\boldsymbol n}^{(r)}$ is the rank one projection onto $Ω_ω$ for any $υ \in \hat{\mathbb{R}}$ with $|υ|>υ_c$, and all $\boldsymbol n \in \mathbb{Z}^D$, $r=1,2,...$ . 
\label{lem:1}
\end{lem}

\begin{proof}
Space-like clustering, expressed in the GNS representation (\Cref{prop:gns}) implies, for $A,B\in\mathfrak U_{\rm loc}$:
\[\arraycolsep=1.4pt\def\arraystretch{2.5}
\begin{array}{*3{>{\displaystyle}lc}p{5cm}}
\lim_{N \to \infty} \frac{1}{N} \sum_{m=0} ^{N-1} ω\big(  ι_{\boldsymbol n}^{rm} τ_{υ^{-1}|\boldsymbol n|}^{rm} (A) B \big)= ω(A)ω(B) \implies \\ %ω( ι_{x}^{n} τ_{ υ^{-1} |x|}^{n} ab )= ω(A)ω(B) 
\lim_{N \to \infty} \frac{1}{N} \sum_{m=0} ^{N-1} \langle Ω_ω ,  π_ω(ι_{\boldsymbol n}^{rm} τ_{υ^{-1}|\boldsymbol n|}^{rm} (A)B) Ω_ω \rangle = \langle Ω_ω, π_ω(A)Ω_ω \rangle \langle Ω_ω, π_ω(B)Ω_ω \rangle \implies \\
\lim_{N \to \infty} \frac{1}{N} \sum_{m=0} ^{N-1} \langle Ω_ω, U^{rm}_ω(\boldsymbol n, υ^{-1}|\boldsymbol n|)π_ω(A) \big(U^{rm}_ω(\boldsymbol n, υ^{-1}|\boldsymbol n|)\big)^* π_ω(B) Ω_ω \rangle = \\
=\langle Ω_ω, π_ω(A)Ω_ω \rangle \langle Ω_ω, π_ω(B)Ω_ω \rangle. 
\end{array} \]
Taking advantage of the invariance of $Ω_ω$ we get:
\begin{equation}
 \begin{array}{*3{>{\displaystyle}lc}p{5cm}}
\lim_{N \to \infty} \frac{1}{N} \sum_{m=0} ^{N-1} \langle Ω_ω,π_ω(A) \big(U^{rm}_ω(\boldsymbol n, υ^{-1}|\boldsymbol n|)\big)^* π_ω(B) Ω_ω \rangle = \\
=\langle Ω_ω, π_ω(A)Ω_ω \rangle \langle Ω_ω, π_ω(B)Ω_ω \rangle. 
\end{array} 
\end{equation}

Then we move the limit of the sum inside the inner product and the adjoint operation by  continuity and use von Neumann's mean ergodic \Cref{th:neumann} (\cite[Theorem II.11]{simon_reed_functional_1}) which states that this limit will be the projection $P^{(r)}_{υ,\boldsymbol n}$ (in the strong operator topology). Thus, we get
 \begin{equation}
 \langle Ω_ω, π_ω(A) P^{(r)}_{υ, \boldsymbol n} π_ω(B) Ω_ω \rangle = \langle Ω_ω, π_ω(A)Ω_ω \rangle \langle Ω_ω, π_ω(B)Ω_ω \rangle \ , \ \ \forall A,B \in \mathfrak{U}_{\rm loc}.
\end{equation}
 By continuity of the inner product and $π_ω$ this will hold for all $A,B \in \mathfrak{U}$. By cyclicity of $Ω_ω$ the set $\operatorname{span} \{π_ω(B)Ω_ω : B \in \mathfrak{U} \}$ is dense in $H_ω$. Hence, this proves the Lemma.
%maybe explanation in between the lines
\end{proof}

\begin{lem}
Consider the assumptions of Theorem \ref{th:rankone}  and let $υ,w \in \hat{\mathbb{R}}$ with $υ \neq w$ and consider the projections $P_{υ,\boldsymbol n}$, $P_{w,\boldsymbol n}$ for some arbitrary $\boldsymbol n \in \mathbb{Z}^D$. If $Ψ \in P_{υ,\boldsymbol n}H_ω$, $Ψ \bot Ω_ω$ and $Ψ^{\prime} \in P_{w,\boldsymbol n}H_ω$, $Ψ^{\prime} \bot Ω_ω$, then $\langle Ψ,Ψ^{\prime}\rangle =0$. \label{lem:orthogonal}
\end{lem}

\begin{proof}
If $P_{υ,\boldsymbol n}$ (or $P_{w,\boldsymbol n}$) is rank one, then obviously $Ψ$ (resp. $Ψ^{\prime}$) will be $0$ and we are done. Suppose that $Ψ,Ψ^{\prime} \neq 0$ and without loss of generality $υ \neq 0$, $w \neq \infty$. 

\textit{Case 1: $w = 0$}. Then $Ψ^{\prime}$ invariant under time translations, so $U_ω(0,r)Ψ^{\prime}=Ψ^{\prime}$ for any $r \in \mathbb{Z}$. Hence, choosing $r=p|\boldsymbol n| \in \mathbb{Z}$ (note that $|\boldsymbol n|\in \N$ for the $\ell_1$ norm) for any arbitrary $p \in \mathbb{Z}$ we can write
\begin{equation}
\langle Ψ, Ψ^{\prime} \rangle = \langle Ψ, U_ω(0,p|\boldsymbol n|) Ψ^{\prime} \rangle   = \langle U_ω(0,-p|\boldsymbol n|)Ψ, Ψ^{\prime} \rangle .
\end{equation}
 By assumption $U_\omega(\boldsymbol n, υ^{-1} |\boldsymbol n|)Ψ=Ψ$, hence also  $U_\omega^q(\boldsymbol n, υ^{-1} |\boldsymbol n|)Ψ=Ψ$ for any integer $q$. Using this, and the fact that $[ U_ω(\boldsymbol n,t) , U_ω(\boldsymbol n ^{\prime},t^{\prime})]=0 , \forall \boldsymbol n,\boldsymbol n ^{\prime} \in \mathbb{Z}^D$ and $ t,t^{\prime} \in \mathbb{R}$, we have for any $q \in \mathbb{Z}$
\begin{equation}
\langle Ψ, Ψ^{\prime} \rangle =  \langle U_ω^q(\boldsymbol n, υ^{-1}|\boldsymbol n|)U_ω(0,-p|\boldsymbol n|)Ψ, Ψ^{\prime} \rangle = \langle U_ω^q\big(\boldsymbol n, (υ^{-1}- \frac{p}{q})  |\boldsymbol n| \big)Ψ, Ψ^{\prime} \rangle 
\end{equation}
where in the last equality we used the group properties.

For any $υ^{-1} \in \mathbb{R}$, we can find $p \in \mathbb{Z}$ and $q \in \mathbb{Z}_+$ such that \begin{equation}
|qυ^{-1} -p| < qυ_c ^{-1}.
\end{equation}
This is true for any $q$ sufficiently large so that $qυ^{-1}_c >1$, and $p = \floor{qυ^{-1}}$.
Thus, by choosing such $p,q$ we have:
\begin{equation}
    \langle Ψ, Ψ^{\prime} \rangle  = \langle U_ω^q(\boldsymbol n, z^{-1}|\boldsymbol n|)Ψ, Ψ^{\prime} \rangle 
\end{equation}
for $z^{-1} = υ^{-1} - p/q$ with $p,q$ such that $|z| > υ_c$. We can repeat the same process for $\langle U_ω^q(\boldsymbol n, z^{-1}|\boldsymbol n|)Ψ, Ψ^{\prime} \rangle$ to get:
\begin{equation}
 \begin{array}{*3{>{\displaystyle}lc}p{5cm}}
   \langle Ψ, Ψ^{\prime} \rangle&=&   \langle U_ω^q(\boldsymbol n, z^{-1}|\boldsymbol n|)Ψ, Ψ^{\prime} \rangle = \\ 
   &=&  \langle \big(U_ω^q(\boldsymbol n, z^{-1}|\boldsymbol n|) \big)^2Ψ, Ψ^{\prime} \rangle = ...=  \langle \big(U_ω^q(\boldsymbol n, z^{-1}|\boldsymbol n|)\big)^NΨ, Ψ^{\prime} \rangle 
 \end{array}
\end{equation}
for any $ N \in \mathbb{N}$. We then split  $\langle Ψ, Ψ^{\prime} \rangle$ as $\frac{1}{N}$ times a sum of $N$ equal terms:
\begin{equation}
     \langle Ψ, Ψ^{\prime} \rangle = \frac{1}{N} \sum_{m=0}^{N-1}  \langle \big(U_ω^q(\boldsymbol n, z^{-1}|\boldsymbol n|)\big)^mΨ, Ψ^{\prime} \rangle \ , \forall N \in \mathbb{N}.
     \end{equation}
By von Neumann's ergodic theorem the sum converges to the projection $P^{(q)}_{z,\boldsymbol n}$. Since $|z|>υ_c$ Lemma \ref{lem:1} gives that $P^{(q)}_{z,\boldsymbol n}$ is the rank one projection on $Ω_ω$ and by assumption $Ψ \bot Ω_ω$, hence 
\begin{equation}
     \langle Ψ, Ψ^{\prime} \rangle = \langle P_{z,\boldsymbol n}^{(q)}Ψ, Ψ^{\prime} \rangle = 0.
\end{equation}
\textit{Case 2: $w \neq 0$}. For any real $η>0$ we can find $p,q \in \mathbb{Z}$ so that $p/q \in [wυ^{-1}, w(υ^{-1} +η)]$, since the rationals are dense in $\mathbb{R}$. Choose $0<ε < \frac{ |1- wυ^{-1}|}{υ_c + |w|}$ such that \begin{equation}
w(υ^{-1}+ε) = \frac{p}{q} \in \mathbb{Q}. \label{eq:wυ}
\end{equation}
Taking advantage of invariance of $Ψ$ under $U_ω^q(\boldsymbol n,υ^{-1}|\boldsymbol n|)=U_ω(q\boldsymbol n,qυ^{-1}|\boldsymbol n|)$ and $Ψ^{\prime}$ under $U_ω^p(\boldsymbol n,w^{-1}|\boldsymbol n|)$ and the group property, like in the previous case, we can write:
\begin{equation}
\begin{array}{*3{>{\displaystyle}lcl}p{5cm}}
\langle Ψ, Ψ^{\prime} \rangle &=& \langle U_ω^q(\boldsymbol n,υ^{-1}|\boldsymbol n|)Ψ, Ψ^{\prime} \rangle \\
&= &\langle U_ω(0,-qε|\boldsymbol n|)U_ω(q\boldsymbol n,q(υ^{-1}+ε)|\boldsymbol n|)Ψ, Ψ^{\prime} \rangle .
\end{array}\end{equation}
Now, using Eq. \ref{eq:wυ} we have
\begin{equation}
\arraycolsep=1.4pt\def\arraystretch{1.3}
\begin{array}{*3{>{\displaystyle}lc}p{5cm}}
    \langle Ψ, Ψ^{\prime} \rangle&=&  \langle U_ω\big(0,-qε|\boldsymbol n|\big)U_ω\big(q\boldsymbol n,pw^{-1}|\boldsymbol n|\big) Ψ, Ψ^{\prime} \rangle \\
   &=&  \langle U_ω\big((q-p)\boldsymbol n,0\big)U_ω\big(0,-qε|\boldsymbol n|\big)U_ω\big(p\boldsymbol n,pw^{-1}|\boldsymbol n|\big) Ψ, Ψ^{\prime} \rangle \\
  &=&   \langle U_ω\big((q-p)\boldsymbol n,0\big)U_ω\big(0,-qε|\boldsymbol n|\big) Ψ, U_ω^{-p}\big(\boldsymbol n,w^{-1}|\boldsymbol n|\big)Ψ^{\prime} \rangle \\
   &=&  \langle U_ω\big((q-p)\boldsymbol n,0 \big)U_ω\big(0,-qε|\boldsymbol n|\big) Ψ, Ψ^{\prime} \rangle
    \end{array}
\end{equation}

where in the last equality we used invariance of $Ψ^{\prime}$. We can now define a $z \in \mathbb{R}$ by $-qε= z^{-1}(q-p)$, that is
\begin{equation}
    z= - \frac{q-p}{qε} = - \frac{1}{ε} + \frac{p}{q} \frac{1}{ε} =   - \frac{1}{ε}(1- \frac{w}{υ}) + w
\end{equation}
where in the last equality we used Eq. \ref{eq:wυ}. By the restrictions on $ε$, imposed above Eq. \ref{eq:wυ}, we can see $|z|>υ_c$. Hence, we can write
\begin{equation}
    \langle Ψ, Ψ^{\prime} \rangle = \langle U_ω^{q-p}(\boldsymbol n,  z^{-1}|\boldsymbol n|)Ψ, Ψ^{\prime} \rangle
\end{equation}
and similarly to the previous case we can repeat this process to get
\begin{equation}
\langle Ψ, Ψ^{\prime} \rangle = \langle P^{q-p}_{z,\boldsymbol n}Ψ, Ψ^{\prime} \rangle = 0. 
\end{equation}
\end{proof}
Equipped with this Lemma we can prove the Projection Theorem:
\begin{proof}[Proof of \Cref{th:rankone} ]
The quasi-local algebras of  quantum spin lattices are uniformly hyperfinite, or UHF, see \cite{glimm_certain_1960} . The GNS representation of a UHF algebra is over separable Hilbert space, since it is cyclic. The proof of this claim is exactly the same as \cite[Theorem 3.5]{glimm_certain_1960}, by using the cyclic vector $Ω_ω$. Thus every orthonormal set in $H_ω$ is countable.

Consider for  arbitrary $\boldsymbol n \in \mathbb{Z}^D$ the set
\begin{equation}
    \mathcal{K}_{\boldsymbol n} =\{ υ \in  \hat{\mathbb{R}}: {\rm rank}\,{P_{υ,\boldsymbol n}}>1. \}  
\end{equation}
For each  $υ \in \mathcal{K}_{\boldsymbol n}$ we can choose (by the axiom of choice) a non-zero $Ψ_{υ} \in P_{υ,\boldsymbol n}H_ω$ with $\norm{Ψ_υ} =1$ and (since ${\rm rank}\,{P_{υ,\boldsymbol n}}>1$) $Ψ_{υ} \bot Ω_ω$. By Lemma \ref{lem:orthogonal} we have $\langle Ψ_υ , Ψ_w \rangle=0$ for every $υ,w \in \mathcal{K}_{\boldsymbol n}$ with $υ \neq w$. Hence the cardinality of $\mathcal{K}_{\boldsymbol n}$ is the cardinality of an orthonormal set of $H_ω$, which is countable, i.e.\ of (Lebesgue) measure $0$. 

\iffalse %%%%%%%%%%%%%%%%%%%%%%%%%%%%%%%
Now consider any velocity of rational direction $υ = υ \frac{\boldsymbol n}{|\boldsymbol n|}$, $υ \in \R$, $\boldsymbol n \in \R$, and the limit as $T \to \infty$ of the operator
\begin{equation}
    \frac{1}{T} \int_0^T U_ω( \floor{υ \frac{\boldsymbol n}{|\boldsymbol n|} t}, t) \,dt  = \frac{1}{υT |\boldsymbol n|^{-1}} \int_0^{υT |\boldsymbol n|^{-1}} U_ω( \floor{x \boldsymbol n}, υ^{-1} |\boldsymbol n|x)  \,dx
\end{equation}
The limit can be taken over the integers, as the integrand is bounded and hence the integral is continuous in $T$.
\begin{equation}
   P \coloneqq \lim_{T \to \infty} \frac{1}{T} \int_0^T U_ω( \floor{υ \frac{\boldsymbol n}{|\boldsymbol n|} t}, t) \,dt = \lim_{N \to \infty} \frac{1}{N} \int_0^N U_ω( \floor{x \boldsymbol n}, υ^{-1} |\boldsymbol n|x)  \,dx
\end{equation}
We split the integral into a sum  into a sum $\int_0^1 \,dx_0 + \int_1^2 \,dx_1 + ...+ \int_{N-1}^N \,dx_{N-1}$ and in each of these integrals we change variable to $x_k=y_k+k$, with k being the lower limit of each integral:
\begin{equation}
    P= \lim_{N \to \infty} \frac{1}{N} \sum_{k=0}^{N-1} \int_0^1 U_ω( \floor{(y+k)\boldsymbol n}, (y+k)υ^{-1}|\boldsymbol n|) \,dy 
\end{equation}
now each $k n_i$ is an integer and can be moved outside the floor function. Using this fact and the group properties we get
\begin{equation}
      P= \lim_{N \to \infty} \frac{1}{N} \sum_{k=0}^{N-1} \int_0^1 U_ω( k \boldsymbol n, kυ^{-1} |\boldsymbol n| ) U_ω( \floor{y \boldsymbol n}, y υ^{-1})
\end{equation}
\fi
\end{proof}

Finally we  use von Neumann's ergodic theorem and the Projection Theorem \ref{th:rankone} to prove the Ergodicity Theorem \ref{th:maintheorem}.

\begin{proof}[Proof of Theorem \ref{th:maintheorem}]
Let $\boldsymbol q = \boldsymbol{n} / |\boldsymbol n| \in \Srat^{D-1}$ for $\boldsymbol n=(n_1,n_2,...,n_D) \in \mathbb{Z}^D$, and $A,B \in \mathfrak{U}$. For $υ>0$ (this suffices since we show the result for all $\boldsymbol n \in \Z^D$) and $\boldsymbol{υ}= υ \boldsymbol{q}=υ \frac{\boldsymbol n}{|\boldsymbol n|}$ we write:
\begin{equation} %\arraycolsep=1.4pt\def\arraystretch{2.2}
I \coloneqq  \frac{1}{T} \int_0^T ω \big( ι_{\floor{ \boldsymbol{υ} t}}τ_t (A) B \big)  \,dt 
= \frac{1}{υT| \boldsymbol n|^{-1}} \int_0^{υT| \boldsymbol n|^{-1}}  ω\big( ι_{\floor{x \boldsymbol n}} τ_{υ^{-1}|\boldsymbol n|x} (A) B\big) \,dx
\end{equation}
where we changed variable to $x=υt/|\boldsymbol n|$ and use the notation $\floor{x\boldsymbol n} = (\floor{x n_1}, ... , \floor{xn_D})$. We are interested in the limit $\lim_{T \to \infty} I $. Τhis limit can be taken over the integers since the integrand is bounded and hence the integral is continuous in $T$. We now write this expression in the GNS representation:
\begin{equation}
      I =  \frac{1}{N} \int_0^N \langle Ω_ω, U_ω\big(\floor{x \boldsymbol n},υ^{-1}|\boldsymbol n|x\big) π_ω(A)  U_ω^*\big(\floor{x \boldsymbol n},υ^{-1}|\boldsymbol n|x\big)  π_ω(B) Ω_ω \rangle \,dx
\end{equation}
\iffalse
\begin{equation}
\arraycolsep=1.4pt\def\arraystretch{2.2}
\begin{array}{*3{>{\displaystyle}lc}p{5cm}}
      I =  \lim_{ N \to \infty} \frac{1}{N} \int_0^N&& \langle Ω_ω, U_ω\big(\floor{x \boldsymbol n},υ^{-1}|\boldsymbol n|x\big) π_ω(A) \times  \\
  &&U_ω^*\big(\floor{x \boldsymbol n},υ^{-1}|\boldsymbol n|x\big)  π_ω(B) Ω_ω \rangle \,dx
    \end{array}
\end{equation}
\fi
We use invariance of $Ω_ω$ so that the integrand becomes:  $$\langle Ω_ω, π_ω(A) U_ω^*\big(\floor{x \boldsymbol n},υ^{-1}|\boldsymbol n|x\big) π_ω(B) Ω_ω \rangle.$$ We then split the integral into a sum $\int_0^1 \,dx_0 + \int_1^2 \,dx_1 + ...+ \int_{N-1}^N \,dx_{N-1}$ and in each of these integrals we change variable to $x_k=y_k+k$, with k being the lower limit of each integral:
%{\setstretch{2.4}
\begin{equation}
    I = \frac{1}{N} \sum_{k=0}^{N-1} \int_0^1 \langle Ω_ω, π_ω(A) U_ω^*\big(\floor{(y_k+k)\boldsymbol n},(y_k+k) υ^{-1}|\boldsymbol n|\big) \times \\
    π_ω(B) Ω_ω \rangle \, dy_k . \label{eq:splittingintegral}
\end{equation}%}
Now $kn_i$ is an integer and can be moved outside the floor function, and using the group properties we write:
\begin{equation}
\arraycolsep=1.4pt\def\arraystretch{2.7}
\begin{array}{*3{>{\displaystyle}lc}p{5cm}}  
    I &=&\frac{1}{N} \sum_{k=0}^{N-1} \int_0^1 \langle Ω_ω, π_ω(A) U^*_ω\big(k \boldsymbol n,kυ^{-1}|\boldsymbol n|\big) U^*_ω\big( \floor{y \boldsymbol n}, yυ^{-1}|\boldsymbol n|\big) π_ω(B) Ω_ω \rangle \,dy \\
    &=&\frac{1}{N} \sum_{k=0}^{N-1} \int_0^1 \langle Ω_ω, π_ω(A) \bigg(U^k_ω\big( \boldsymbol n,υ^{-1}|\boldsymbol n|\big) \bigg)^* U^*_ω\big( \floor{y \boldsymbol n}, yυ^{-1}|\boldsymbol n|\big) π_ω(B) Ω_ω \rangle \,dy
   \end{array}
\end{equation}
We now apply the large $N$ limit. We can calculate $\lim_{N \to \infty}I$ by using the bounded convergence theorem to exchange the limit of the sum and the integral,
 \begin{equation} 
 \begin{array}{*3{>{\displaystyle}lc}p{5cm}} 
 \lim_{N \to \infty} I= \\
     \int_0^1 \langle Ω_ω, π_ω(A) U_ω^* (\floor{y \boldsymbol n}, yυ^{-1}|\boldsymbol n|) \lim_{N \to \infty} \frac{1}{N} \sum_{n=0}^{N-1} \big(U^*_ω(\boldsymbol n,υ^{-1}|\boldsymbol n|)\big)^n  π_ω(B) Ω_ω \rangle\,dy
     \end{array}
 \end{equation}
and  using von Neumann's ergodic theorem, \cite[Theorem II.11]{simon_reed_functional_1}:
\begin{equation}
    I=\int_0^1 \langle Ω_ω, π_ω(A) U_ω^* \big(\floor{y \boldsymbol n}, yυ^{-1}|\boldsymbol n|)  P_{υ,\boldsymbol n} π_ω(B) Ω_ω \rangle\,dy.
\end{equation}
Theorem \ref{th:rankone} tells us that $P_{υ,\boldsymbol n}$ is the rank one projection on $Ω_ω$ for almost all $υ$, hence:
\begin{equation}
    I= \int_0^1 \langle Ω_ω, π_ω(A) U_ω^* \big(\floor{y \boldsymbol n}, yυ^{-1}|\boldsymbol n|\big) Ω_ω\rangle \langle Ω_ω, π_ω(B) Ω_ω  \rangle \,dy \ \ \text{ a.e.} 
\end{equation}
and finally, since $Ω_ω$ is invariant we get
\begin{equation}
    I = \langle Ω_ω, π_ω(A)  Ω_ω\rangle \langle Ω_ω, π_ω(B) Ω_ω  \rangle = ω(A) ω(B) 
\end{equation}
for almost all $υ$. 
\end{proof}
\section{Mean-square ergodicity} \label{section:meansquared}
In \cite{ampelogiannis_2023_almost} we further show that the variance, with respect to the state, of a ray averaged observables will vanish at long times.  In fact, we have that the $n-$moment will converge to the $n-$th power of the expectation value in the state. The physical  meaning of the result is that the ray-averaged observable $A$ does not fluctuate within the state $ω$ under the long time ray averaging, for almost every ray. In fact, this is solidified by our results of \cite{ampelogiannis_2024_clustering} in \Cref{chapter:higher_clustering}, where we obtain clustering for all $n-$th order connected correlations.

\begin{theorem}[Mean-square ergodicity] \label{th:meansquared}
 Consider a quantum spin dynamical system $(\mathfrak{U}, ι, τ)$  with interaction that satisfies \Cref{eq:interaction}   and a space-like clustering state $ω\in E_{\mathfrak{U}}$, \Cref{eq:space-like_clustering}. It follows that for all $A,B \in \mathfrak{U}$, any rational direction $\boldsymbol{q} \in \Srat^{D-1}$ and almost every speed $v \in \mathbb{R}:$
\begin{equation}
    \lim_{T \to \infty} \frac{1}{T^2}\int_0^{T} \int_0^{T} \ ω\big( ι_{\floor{ \boldsymbol{v} t_1}}τ_{t_1} (A)  \   ι_{\floor{ \boldsymbol{v} t_2}}τ_{t_2} (B)   \big) \,dt_1 \, dt_2= ω(A) ω(B)
\end{equation}
with $\boldsymbol{v}= v \boldsymbol{q}$. In fact, this holds as a double limit $\lim_{T,T^{\prime} \to \infty} \frac{1}{T^{\prime} T}\int_0^{T}\int_0^{T^{\prime}}  (\cdots) \, dt  \, dt^{\prime}$.
\end{theorem}

\begin{theorem} \label{th:meanN}
Under the assumptions of \Cref{th:meansquared} we have that for any $n \in \N$,
\begin{equation}
    \lim_{T \to \infty} \frac{1}{T^n} ω \bigg( \big(\int_0^T  ι_{\floor{ \boldsymbol{v} t}}τ_{t} A  \,dt  \big)^n \bigg)= \big( ω(A) \big)^n.
\end{equation}
We refer to the case $n=2$ as mean-square ergodicity.

This also holds also as a multiple limit on $T_1,T_2,\ldots, T_n$, with different observables $A_1,A_2,\ldots,A_n\in\mathfrak U$:
\begin{equation}
    \lim_{T_1,\ldots,T_n\to\infty}
    \omega\big(\prod_{j=1}^n 
    \frac1{T_j} \int_0^{T_j} dt_j \,ι_{\floor{ \boldsymbol{v} t_j}}τ_{t_j} A_j\big)= 
    \prod_{j=1}^n \omega(A_j).
\end{equation}
The integral $\int_0^T  ι_{\floor{ \boldsymbol{v} t}}τ_{t} A  \,dt  $ is well defined as a Bochner integral, see \cite[Section 3.7]{hille_functional_1996} and \Cref{appendix1}.
\end{theorem}
To be precise, the multiple limit is in the sense that $(T_1^{-1},T_2^{-1},\ldots,T_{n+1}^{-1})\to \boldsymbol 0$ with respect to the topology of $\R^{n+1}$. See \Cref{appendix:Moore-Osgood} for details on how we treat multiple limits.

\subsection{Proof} \label{section:meansquaredproof}

\begin{proof}[Proof of \Cref{th:meansquared}]
Let $\boldsymbol n \in \mathbb{Z}^D$, $A,B \in \mathfrak{U}$, $υ >0$ so that $\boldsymbol{υ}= υ \boldsymbol {n}/|\boldsymbol n|$ and consider
\begin{equation}
    I(T,T^{\prime}) = \frac{1}{T^{\prime}T} \int_0^{T^{\prime}} dt_2\int_0^{T} dt_1\, ω\big(ι_{\floor{ \boldsymbol{υ} t_1}}τ_{t_1} (A)  ι_{\floor{\boldsymbol{υ}t_2}}τ_{t_2} (B)  \big) \  , \ \  T,T^{\prime}>0.
\end{equation}

We will apply the Moore-Osgood \Cref{th:MooreOsgood} \cite[p. 140]{taylor_general_1985} to show that the double limit $\lim_{T,T^{\prime} \to \infty}I(T,T^{\prime})$ is equal to the iterated one. The Theorem states that if the iterated limit $\lim_{T^{\prime}\to\infty}(\lim_{T\to\infty}\cdot)$ exists, and the limit $\lim_{T\to\infty}\cdot$ exists uniformly in $T'$, then the double limit exists and is equal to the iterated one. We can easily calculate the iterated limit $\lim_{T^{\prime} \to \infty}\big(\lim_{T \to \infty}  I(T,T^{\prime}) \big)$ by using Theorem \ref{th:maintheorem}:
\begin{equation}
 \begin{array}{*3{>{\displaystyle}lc}p{5cm}} 
   \lim_{T^{\prime} \to \infty}\big(\lim_{T\to \infty} \frac{1}{T^{\prime}T} \int_0^{T^{\prime}} dt_2\int_0^{T} dt_1\, ω\big(ι_{\floor{ \frac{υ}{|\boldsymbol n|} t_1}\boldsymbol n}τ_{t_1} (A)  ι_{\floor{ \frac{υ}{|\boldsymbol n|} t_2}\boldsymbol n}τ_{t_2} (B)  \big) \big)\\ =
   \lim_{T^{\prime} \to \infty} \frac1{T'}
   \int_0^{T'} dt_2\,
   ω(A) ω\big(ι_{\floor{ \frac{υ}{|\boldsymbol n|} t_2}\boldsymbol n}τ_{t_2} (B)  \big)
   =ω(A)\omega(B)
   \end{array}
\end{equation}
for almost all $υ$. This shows that $\lim_{T \to \infty}$ exists pointwise in $T'$. In order to use the Moore-Osgood Theorem it remains to show that the limit on $T$ is uniform in $T'$. Consider the $T \to \infty$ limit, by following the same steps as in the proof of \Cref{th:maintheorem} in \Cref{section:ergodicityproof}. Since the integrand is bounded $I(T,T^{\prime})$ is continuous in $T$, hence the continuous limit can be changed with a discrete one. We have for all $T^{\prime}$, as in \Cref{eq:splittingintegral}
\begin{equation}
    \begin{array}{*3{>{\displaystyle}lc}p{5cm}}
    I(N,T^{\prime}) \coloneqq \\
    \frac{1}{T^{\prime}} \int_0^{T^{\prime}}dt_2 \int_0^1 dy \langle Ω_ω , π(A) \big( \frac{1}{N} \sum_{k=0}^{N-1} U^*(k \boldsymbol n, k υ^{-1}|\boldsymbol n|) \big) U^*( \floor{y\boldsymbol n}, yυ^{-1} |\boldsymbol n|) \times \\ U(\floor{υt_2},t_2) π(B) Ω_ω \rangle = \\
    \frac{1}{T^{\prime}} \int_0^{T^{\prime}}dt_2 \int_0^1 dy \langle \big( \frac{1}{N} \sum_{k=0}^{N-1} U(k \boldsymbol n, k υ^{-1}|\boldsymbol n|) \big) π(A^*) Ω_ω ,   U^*( \floor{y\boldsymbol n}, yυ^{-1} |\boldsymbol n|) \times \\ U(\floor{υt_2},t_2) π(B) Ω_ω \rangle.
    \end{array}
\end{equation}
The limit is done by using the bounded convergence theorem and von Neumann's ergodic theorem:
\begin{equation}
 \begin{array}{*3{>{\displaystyle}lc}p{5cm}}
    I(\infty,T^{\prime}) \coloneqq \lim_{N \to \infty} I(N,T') =\\
    =  \frac{1}{T^{\prime}} \int_0^{T^{\prime}}dt_2 \int_0^1 dy \langle  P_{υ,\boldsymbol n}π(A^*)Ω_ω , U^*( \floor{y\boldsymbol n}, yυ^{-1} |\boldsymbol n|) U(\floor{υt_2},t_2) π(B) Ω_ω \rangle  
\end{array}
\end{equation}
We can see
\begin{equation} \label{eq:Mean_squared_proof1}
  \begin{array}{*3{>{\displaystyle}lc}p{5cm}}
    |I(N,T^{\prime}) - I(\infty, T^{\prime})|= \\
  \bigg|   \frac{1}{T^{\prime}} \int_0^{T^{\prime}}dt_2 \int_0^1 dy \bigg\langle \big( \frac{1}{N} \sum_{k=0}^{N-1} U(k \boldsymbol n, k υ^{-1}|\boldsymbol n|) - P_{υ,\boldsymbol n}\big) π(A^*)Ω_ω ,  \\ U^*( \floor{y\boldsymbol n}, yυ^{-1} |\boldsymbol n|) U(\floor{υt_2},t_2) π(B) Ω_ω \bigg\rangle \bigg  |.
    \end{array}
\end{equation}
Using the Cauchy-Shwarz inequality and the fact that unitary operators preserve the norm of vectors,
\[
 \begin{array}{*3{>{\displaystyle}lc}p{5cm}}
    |I(N,T^{\prime}) - I(\infty, T^{\prime})| \leq  \\
   \leq \frac{1}{T^{\prime}}  \int_0^{T^{\prime}}dt_2 \int_0^1 dy \norm{\big( \frac{1}{N} \sum_{k=0}^{N-1} U(k \boldsymbol n, k υ^{-1}|\boldsymbol n|) - P_{υ,\boldsymbol n}\big) π(A^*)Ω_ω}\times\\  \norm{ U^*( \floor{y\boldsymbol n}, yυ^{-1} |\boldsymbol n|) U(\floor{υt_2},t_2) π(B) Ω_ω}\\ 
  = \norm{\big( \frac{1}{N} \sum_{k=0}^{N-1} U(k \boldsymbol n, k υ^{-1}|\boldsymbol n|) - P_{υ,\boldsymbol n}\big) π(A^*)Ω_ω}\,
   \norm{π(B) Ω_ω}.
    \end{array}
\]
Finally, for any $Ψ \in H_ω$ and for any  $ε>0$ there exists $N_0$ such that $N\geq N_0$ implies $$\norm{ \big(\frac{1}{N} \sum_{k=0}^{N-1} U^*(k \boldsymbol n, k υ^{-1}|\boldsymbol n|) - P_{υ,\boldsymbol n} \big)Ψ} <ε$$ by von Neumann's ergodic theorem. Therefore
\begin{equation}
    |I(T,T^{\prime}) - I(\infty, T^{\prime})| \leq ε \norm{π(B)Ω_ω} \ \text{ for all } T^{\prime}
\end{equation}
which shows that the limit is uniform in $T^{\prime}$.
Hence, the double limit exists and is equal to the iterated one, which completes the proof.
\end{proof}

\begin{proof}[Proof of \Cref{th:meanN}]
For simplicity and clarity of the notation, we do the case where all times are equal $T_1=T_2=\ldots=T_n$, and all operators are the same $A_1=A_2=\ldots=A_n$. The extension to the general case and the existence of the multiple limit will be discussed afterwards.

The proof is  done inductively.  Suppose  we have for $n \in \mathbb{N}$ that 
\begin{equation}
   \lim_{T \to \infty} \frac{1}{T^n} ω \bigg( \big(\int_0^T  ι_{\floor{ \boldsymbol{υ} t}}τ_{t} A  \,dt  \big)^n \bigg)= \big( ω(A) \big)^n \ , \text{ a.e. in υ}
\end{equation}
which is certainly true for $n=2$ by the Mean-Square Ergodicity Theorem. 
We will show that the same will then hold for $n+1$. In order to do so, we first note that, using the GNS representation, the Cauchy-Schwartz inequality, boundedness of the state and the $C^*$ property,
\begin{equation}\label{eq:basicineq}
    |\omega(AB)|^2 \leq \omega(AA^*)\omega(B^*B)
    \leq \norm{A}^2\omega(B^*B).
\end{equation}
We consider $  I_{n+1}(T) \coloneqq \frac{1}{T^{n+1}} ω \bigg( \big(\int_0^T  ι_{\floor{ \boldsymbol{υ} t}}τ_{t} A  \,dt  \big)^{n+1} \bigg) $ in the form
\begin{equation}
\arraycolsep=1.4pt\def\arraystretch{2.4}
  \begin{array}{*3{>{\displaystyle}lc}p{5cm}}
 I_{n+1}(T) &=&\frac{1}{T^{n+1}} ω \bigg( \prod_{j=1}^{n+1} \int_0^T ι_{\floor{\boldsymbol{υ} t_j}} τ_{t_j} A \, dt_j \bigg)\\
 &=& \frac{1}{T^{n+1}} \Big(\prod_{j=1}^{n+1}\int_0^Tdt_j\Big)\,  ω \bigg( \prod_{j=1}^{n+1} ι_{\floor{\boldsymbol{υ} t_j}} τ_{t_j} A \bigg).
 \end{array}
\end{equation}
We will show that the distance of $I_{n+1}(T)$ from $ω(A)I_n(T) $ goes to $0$ in the $T \to \infty$ limit.  We denote $\int^{(n)} = \Big(\prod_{j=1}^{n}\int_0^T dt_j\Big)$ and write:
\begin{equation} \label{eq:moment_theorem_proof_1}
     \begin{array}{*3{>{\displaystyle}l}p{5cm}}
 Δ_{n+1}(T) \coloneqq |I_{n+1(T)} - \omega(A)I_n(T)| \\
 = \big| \frac{1}{T^{n+1}} \int^{(n)} \int_0^T dt_{n+1} \,\omega\big( \prod_{j=1}^{n} ι_{\floor{ \boldsymbol{υ} t_j}}τ_{t_j} (A)\, (ι_{\floor{ \boldsymbol{υ} t_{n+1}}}τ_{t_{n+1}}(A)  - ω(A)) \big) \big |\\
 \leq 
 \frac{1}{T^{n+1}} \int^{(n)} \,\big| \omega\big( \prod_{j=1}^{n} ι_{\floor{ \boldsymbol{υ} t_j}}τ_{t_j} (A)\, \int_0^T dt_{n+1}(ι_{\floor{ \boldsymbol{υ} t_{n+1}}}τ_{t_{n+1}}(A)  - ω(A)) \big) \big |.
    \end{array}
\end{equation} 
By submultiplicativity of the operator norm and the fact that $\iota$ and $\tau$ preserve the norm,
\begin{equation}
    \norm{\prod_{j=1}^{n} ι_{\floor{ \boldsymbol{υ} t_j}}τ_{t_j} (A)}
    \leq \norm{A}^n.
\end{equation}
Therefore, from \Cref{eq:basicineq} and performing the integral $\int^{(n)}$,
\begin{equation} 
\arraycolsep=1.4pt\def\arraystretch{2.5}
     \begin{array}{*3{>{\displaystyle}l}p{5cm}}
    \Delta_{N+1}(T)^2\leq \\
    \leq\norm{A}^{2n}\frac1{T^2}
    \int_0^T dt
    \int_0^T dt'\,
    \omega\bigg(
    \big(ι_{\floor{ \boldsymbol{υ} t}}τ_{t}(A) - ω(A)\big)^*
    \big(ι_{\floor{ \boldsymbol{υ} t'}}τ_{t'}(A) - ω(A)\big)\bigg)\nonumber\\
    =
    \norm{A}^{2n}\bigg(\frac1{T^2}
    \int_0^T dt\int_0^T dt'\,
    \omega\big(
    ι_{\floor{ \boldsymbol{υ} t}}τ_{t}(A^*)ι_{\floor{ \boldsymbol{υ} t'}}τ_{t'}(A)\big) - |ω(A)|^2\bigg)
\end{array}
\end{equation}
where in the last step we used invariance of the state. We now use the  Mean-Square Ergodicity Theorem \ref{th:meansquared}, which shows that for almost every $υ$, i.e.\  $ Δ_{n+1}(T) \to 0$. This means that the limit of $I_{n+1}(T)$ coincides with the limit of
\begin{equation}
    ω(A) \frac{1}{T^N} \int_0^T \langle Ω_ω, \prod_{j=1}^{n} π_ω\big(ι_{\floor{ \boldsymbol{υ} t_j}}τ_{t_j} A \big) Ω_ω \rangle \,dt_j
\end{equation}
which is $ω(A) ω(A)^n= ω(A)^{n+1}$, for almost all $υ$, by the induction hypothesis.

In the case where the times $T_1,T_2,\ldots$ are different, the induction hypothesis states the existence of the multiple limit for $I_n(T_1,T_2,\ldots,T_n)$, which is certainly true for $n=2$ by the Mean-Square Ergodicity Theorem. Then, we note that the integral $\int^{(n)}$ was performed before taking the limit on the last time $T_{n+1}$. Therefore, we find that $\Delta_{n+1}(T_1,T_2,\ldots,T_{n+1})$ is bounded, uniformly on all $T_1,T_2,\ldots T_n$, by a value that vanishes as $T_{n+1}\to\infty$. Thus the limit $\lim_{T_{n+1}\to\infty} I_{n+1}(T_1,T_2,\ldots,T_{n+1})$ exists uniformly on $T_1,T_2,\ldots,T_n$. Since the iterated limit exists (on $T_{n+1}\to\infty$, then on $T_1,T_2,\ldots,T_n\to\infty$) by the induction hypothesis, we can apply the Moore-Osgood theorem, and the multiple limit on $I_{n+1}(T_1,T_2,\ldots,T_{n+1})$ exists and gives the stated result. Note that the Moore-Osgood \Cref{th:MooreOsgood} holds for double limits over the Cartesian product of arbitrary directed  sets $X \times Y$, see \Cref{appendix:Moore-Osgood}.

The case of different $A_1,A_2,\ldots$ is obtained from an immediate generalisation of the steps above.

\end{proof}
 \section{Almost everywhere ergodicity in KMS states} \label{section:ergodicity_kms}

The results in \cite{ampelogiannis_2023_almost} go beyond simply extending ergodicity of correlations to higher dimensions. In the case of space-like clustering KMS states, we can take advantage of the KMS property, \Cref{eq:KMS_state}, to prove much stronger results. In particular, we can show that in the GNS representation of a KMS state the ray average of an observable tends (in the strong operator topology) to the state average times the identity operator, in the long time limit, almost everywhere. The result will hold for any space-like clustering state that is also a $(τ^{\prime},β)$-KMS state for any time evolution $τ^{\prime}$ which commutes with the initial time evolution of the system $τ$.

Few remarks are in order concerning the importance of admitting $\tau'\neq \tau$. First, in the context of modern non-equilibrium physics, especially in integrable systems, KMS states with respect to different time evolutions are of fundamental importance. Indeed,  these should be identified with generalised Gibbs ensembles (GGE) \cite{rigol_2007_relaxation}, which formally have density matrices $e^{-\sum_a \beta_a Q_a}$ where $Q_a$ are some set of short-range conserved charges for the evolution Hamiltonian $H$; here $\sum_a \beta_a Q_a = H'$ would be the short-range Hamiltonian generating $\tau'$. GGEs are the states which a system is expected to reach after quantum quenches and in other non-equilibrium situations at large times. The currently most rigorous understanding of GGEs in interacting systems \cite{doyon_thermalization_2017} has not yet been connected with the KMS condition, and it would be interesting to investigate this in future works.

Second, mathematically, using the Tomita-Takesaki theory \cite{summers2005tomitatakesaki}, every faithful normal state of a von Neumann Algebra \cite[Section 2.2]{bratteli_operator_1987} is a KMS state with time evolution given by the modular group. Adapting our proofs and results to the context of von Neumann algebras, we would thus have general results for space-like clustering, faithful, normal states, with $\tau'$ the modular group. Of course, closing the $C^*$ algebra of operators on $H_\omega$ under the strong topology gives a natural von Neumann algebra to work with. This requires some additional technical work (e.g. extending our results to $W^*$ dynamical systems), and is one of the many possible avenues that this thesis sets the stage for exploring.

Finally, as an aside, making the connection with GGEs, we note that one may wish to identify the set of all space-time stationary, faithful, normal states with (at least some large subset of) GGEs, the modular group being the group generated by the particular charge $\sum_a \beta_a Q_a = H'$ for the GGE. Thus the Tomita-Takesaki theory could give a very general and mathematically useful understanding of GGEs. This is yet another open problem, the importance of which is highlighted by our work.

Our main theorem using the KMS condition is the following.
\begin{theorem} \label{th:KMS_average_strong_convergence}
Consider a quantum lattice dynamical system $(\mathfrak{U}, ι, τ)$ and a possibly different time evolution $\tau'$ such that $(\mathfrak{U}, ι, τ')$ is also a dynamical system. Suppose the interactions, both for $\tau$ that $\tau'$, satisfy \Cref{eq:interaction}, and that $\tau$ and $\tau'$ commute. Consider a   $(\tau',\beta)$-KMS state $ω_β$ that is space-like clustering for  $(\mathfrak{U}, ι, τ)$  and its respective GNS representation $(H_{ω_β},π_{ω_β},Ω_{ω_β})$. It follows that for any $A \in \mathfrak{U}$, its ray average in the GNS representation converges to the state average in the strong operator topology, for almost every ray. That is, the following limit holds in the norm (of $H_{ω_β}$)
\begin{equation}
    \lim_{T \to \infty} \frac{1}{T} \int_0^T π_{ω_β} \big( ι_{\floor{ \boldsymbol{υ} t}}τ_t A \big) \, dt \, Ψ= ω_β(A) Ψ \ , \ \ \forall Ψ \in H_{ω_β}
\end{equation}
with $\boldsymbol{υ}= υ \boldsymbol q$ of any rational direction $\boldsymbol{q} \in \Srat^{D-1}$  and almost every speed $υ\in \R$. Again, the integral is to be understrood as a Bochner integral (\Cref{appendix1}). 
\end{theorem}

\textbf{This can be understood as convergence, in the strong operator topology of the GNS Hilbert space, of the ray average to the ensemble average, along almost every space-time ray (of rational direction)}:
\begin{equation}
    \frac{1}{T} \int_0^T π_{ω_β} \big( ι_{\floor{ \boldsymbol{υ} t}}τ_t A \big) \, dt  \xrightarrow{SOT} ω_β(A) \mathds{1} \ , \text{ almost everywhere.}
\end{equation}

\begin{remark}
Note that in the case $τ^{\prime}=τ$ and of small $β$ (high temperatures) it suffices to assume that the state is only KMS, as this immediately implies space-like clustering, see the discussion above \Cref{th:maintheorem}. For higher values of $β$ where the KMS state is not unique, any KMS state that is invariant and factor will subsequently be space-like clustering.
\end{remark}

\subsection{Proof}
The proof of \Cref{th:KMS_average_strong_convergence} requires the following Lemmata.
\begin{lem} \label{lemma:6.1}
Consider the assumptions of \Cref{th:KMS_average_strong_convergence}. It follows that for any $A,B,C,D \in \mathfrak{U}$, any rational direction $\boldsymbol{q} \in \Srat^{D-1}$ of the velocity $\boldsymbol{υ}= υ \boldsymbol q$ and almost every speed  $υ \in \mathbb{R}$ 
\begin{equation} \label{eq:lemma6.11}
    \lim_{T \to \infty} \frac{1}{T} \int_0^T  ω_β( B ι_{\floor{ \boldsymbol{υ} t}}τ_t (A) D) \, dt = ω_β(A) ω_β(B \, D)
\end{equation}
and
\begin{equation} \label{eq:lemma6.12}
     \lim_{T \to \infty} \frac{1}{T^2} \int_0^T \int_0^T ω_β( B ι_{\floor{ \boldsymbol{υ} t}}τ_t (A) \, ι_{\floor{ \boldsymbol{υ} {t^{\prime}}}}τ_{t^{\prime}} (C) D) \, dt \, dt^{\prime} = ω_β(A)ω_β(C) ω_β(B \, D)
\end{equation}
\end{lem}

\begin{proof}
We first show \Cref{eq:lemma6.11} for arbitrary $A,B,D \in \mathfrak{U}$. The idea is to use the Ergodicity \Cref{th:maintheorem}, but to do this we have to change the order of elements inside $ω_β( \cdots)$. This is done by exploiting the KMS condition, which however does not necessarily hold for all elements of the algebra, but for a dense subset. Hence, we consider a sequence of analytic elements $B_n \in \mathfrak{U}_τ$ that satisfies the KMS condition, see \Cref{defn:KMS}, and approximates $B$, i.e.\ $B_n \to B$. To show \Cref{eq:lemma6.11} we have to calculate the iterated limit
\begin{equation} \label{proof:lemm611_1}
     \lim_{T \to \infty} \frac{1}{T} \int_0^T  ω_β( \lim_n B_n ι_{\floor{ \boldsymbol{υ} t}}τ_t (A) D) \, dt
\end{equation}
but to apply the KMS condition we need to do the $\lim_n$ after the $\lim_T$. For this we will use again the Moore-Osgood \Cref{th:MooreOsgood} \cite[p. 140]{taylor_general_1985} to show that the double limit exist (and is equal to the iterated ones).
Consider $I_n(T)=\frac{1}{T} \int_0^T  ω_β( B_n ι_{\floor{ \boldsymbol{υ} t}}τ_t (A) D) $. Using the KMS condition for $B_n$ the integrand becomes:
\begin{equation}
     ω_β( B_n ι_{\floor{ \boldsymbol{υ} t}}τ_t (A) D) =  ω_β(  ι_{\floor{ \boldsymbol{υ} t}}τ_t (A) D \tau'_{iβ} B_n ) .
\end{equation}
Applying the average $\frac{1}{T}\int_0^T \,dt$  and taking the limit $T \to \infty$ we get, by using the Ergodicity \Cref{th:maintheorem}:
\begin{equation}
 \lim_{T \to \infty}    \frac{1}{T}\int_0^T  ω_β( B_n ι_{\floor{ \boldsymbol{υ} t}}τ_t (A) D)  \,dt = ω_β(A) ω_β(D \tau'_{iβ}B_n)
\end{equation}
for almost every $υ$. Re-applying the KMS condition (in the opposite direction) we get  that $\lim_{T \to \infty} I_n(T)$ exists for all $n$:
\begin{equation}
    \lim_{T \to \infty} \frac{1}{T}\int_0^T  ω_β( B_n ι_{\floor{ \boldsymbol{υ} t}}τ_t (A) D)  \,dt = ω_β(A) ω_β(B_n D) \ , \ \ \forall n \in \N.
\end{equation}

It is easy to see that  $\lim_n I_n(T)$ exists and is $\frac{1}{T}\int_0^T  ω_β( B ι_{\floor{ \boldsymbol{υ} t}}τ_t (A) D)  \,dt$ \textbf{uniformly} in $T$, for instance by using the dominated convergence theorem to bring the limit inside the integral with the uniform bound $|ω_β( (B_n-B) ι_{\floor{ \boldsymbol{υ} t}}τ_t (A) D)|\leq \norm{B_n-B}\norm{A}\norm{D}$, then the Cauchy-Schwartz inequality, and finally the uniform bound $ω_β( D^*ι_{\floor{ \boldsymbol{υ} t}}τ_t (A^*A) D)\leq \norm{D}^2\norm{A}^2$.  By the Moore-Osgood theorem this means that the double limit and the two iterated limits exist and are equal. Hence the result holds for any $A,B,D \in \mathfrak{U}$ in the sense of the iterated limit:
\begin{equation}
      \lim_{T \to \infty} \frac{1}{T}\int_0^T  ω_β( \lim_nB_n ι_{\floor{ \boldsymbol{υ} t}}τ_t (A) D)  \,dt = ω_β(A) ω_β(B D) .
\end{equation}

The proof of \Cref{eq:lemma6.12} is similar. We consider again arbitrary $A,B,C,D \in \mathfrak{U}$ and sequences $A_n, B_n \in \mathfrak{U}_{\tau'}$ of analytic elements for $\tau'$ that satisfy the KMS condition and $A_n \to A$, $B_n \to B$. We first use the KMS condition to write:
\begin{equation}
     ω_β\big( B_n ι_{\floor{ \boldsymbol{υ} t}}τ_t (A_n) \, ι_{\floor{ \boldsymbol{υ} {t^{\prime}}}}τ_{t^{\prime}} (C) D\big) =  ω_β\big(  ι_{\floor{ \boldsymbol{υ} t}}τ_t (A_n) \, ι_{\floor{ \boldsymbol{υ} {t^{\prime}}}}τ_{t^{\prime}} (C) D \tau'_{iβ}(B_n)\big)
\end{equation}
apply the average $\frac{1}{T^2} \int_0^T \int_0^T \,dt \,dt^{\prime}$ and consider $\lim_{T \to \infty}$.  To calculate this limit we consider the double limit: $$\lim_{T,T^{\prime} \to \infty} \frac{1}{T^{\prime}T} \int_0^T \int_0^{T^{\prime}} ω_β(  ι_{\floor{ \boldsymbol{υ} t}}τ_t (A_n) \, ι_{\floor{ \boldsymbol{υ} {t^{\prime}}}}τ_{t^{\prime}} (C) D \tau'_{iβ}(B_n)) \, dt \, dt^{\prime}.$$ Define:
\begin{equation} \label{eq:lemma_6.1_double_limit}
    I_n(T,T^{\prime}) \coloneqq \frac{1}{T^{\prime}T} \int_0^T \int_0^{T^{\prime}} ω_β\big(  ι_{\floor{ \boldsymbol{υ} t}}τ_t (A_n) \, ι_{\floor{ \boldsymbol{υ} {t^{\prime}}}}τ_{t^{\prime}} (C) D \tau'_{iβ}(B_n)\big) \, dt \, dt^{\prime}.
\end{equation}
We intend to apply the Moore-Osgood \Cref{th:MooreOsgood}. First consider $\lim_{T}$  of $I_n(T,T^{\prime}) $. Using the dominated convergence Theorem to move the $\lim_{T}$ inside the integral, and then the Ergodicity \cref{th:maintheorem} we see that 
\begin{equation} \label{eq:lemma611_proof_limit}
    \lim_{T \to \infty} I_n(T,T^{\prime}) = \frac{1}{T^{\prime}} \int_0^{T^{\prime}} ω_β(A_n) ω_β\big( ι_{\floor{ \boldsymbol{υ} {t^{\prime}}}}τ_{t^{\prime}} (C) D \tau'_{iβ}(B_n)\big) \,dt^{\prime}.
\end{equation}
In exactly the same manner as in the proof of the Mean-Square Ergodicity theorem, in \Cref{section:meansquaredproof}, and specifically in the derivation around \Cref{eq:Mean_squared_proof1}, we can see that this limit is also uniform in $T^{\prime}$. Now, the $\lim_{T^{\prime}}$ can be done by using the KMS condition to write 
\begin{equation}
    I_n(T,T^{\prime}) =  \frac{1}{T^{\prime}T} \int_0^T \int_0^{T^{\prime}} ω_β\big( \, ι_{\floor{ \boldsymbol{υ} {t^{\prime}}}}τ_{t^{\prime}} (C) D \tau'_{iβ}(B_n)   \tau'_{iβ} ι_{\floor{ \boldsymbol{υ} t}}τ_t (A_n)\big) \, dt \, dt^{\prime}
\end{equation}
and again using dominated convergence and the ergodicity theorem we see that the limit exists and is
\begin{equation}
    \lim_{T^{\prime} \to \infty} I_n(T,T^{\prime}) = \frac{1}{T} \int_0^T ω_β(C) ω_β\big( \, ι_{\floor{ \boldsymbol{υ} {t^{\prime}}}}τ_{t^{\prime}} (C) D \tau'_{iβ}(B_n)   \tau'_{iβ} ι_{\floor{ \boldsymbol{υ} t}}τ_t (A_n) \big) \,dt.
\end{equation}

Hence, by the Moore-Osgood theorem the double limit exists and is equal to the iterated one. This implies that also the limit $\lim_{T \to \infty} I_n(T,T)$ exist and is equal to the double limit. The iterated limit is the one that we can calculate, by reapplying the ergodicity theorem in \Cref{eq:lemma611_proof_limit} 
\begin{equation}
    \lim_{T \to \infty} I_n(T,T) = ω_β(A_n) ω_β(C) ω_β\big(D \tau'_{iβ} (B_n)\big) =ω_β(A_n) ω_β(C) ω_β( B_n D ) 
\end{equation}
where in the last line we re-applied the KMS condition for $B_n$. Now, in a similar manner as before, we can see that $\lim_n I_n (T,T)$ exists uniformly in $T$. Hence, we can apply to Moore-Osgood \Cref{th:MooreOsgood} to get the result.
\end{proof}

\begin{lem} \label{lemma:6.2}
Consider the assumptions of \Cref{th:KMS_average_strong_convergence}. Then for any $A \in \mathfrak{U}$ it follows that for all $B \in \mathfrak{U}$, any rational direction $\boldsymbol{q} \in \Srat^{D-1}$ and almost every speed  $υ \in \mathbb{R}:$
\begin{equation}
\arraycolsep=1.4pt\def\arraystretch{1.8}
\begin{array}{*3{>{\displaystyle}lc}p{5cm}}
  \lim_{T \to \infty} \frac{1}{T^2}  \int_0^T \int_0^T ω_β \big( B^* \big(\,ι_{\floor{ \boldsymbol{υ} t}}τ_t (A)-ω_β(A) \mathds{1} \big)^* \big(\,ι_{\floor{ \boldsymbol{υ} {t^{\prime}}}}τ_{t^{\prime}} (A)-ω_β(A) \mathds{1} \big)  \,B \big) \,dt \,dt^{\prime} \\ = 0 
    \end{array}
\end{equation}
with $\boldsymbol{υ}= υ \boldsymbol q$.
\end{lem}

\begin{proof}
The integrand is the sum of four terms:
\begin{equation}
\arraycolsep=1.4pt\def\arraystretch{1.2}
\begin{array}{*3{>{\displaystyle}lc}p{5cm}}
    ω_β(B^* \, ι_{\floor{ \boldsymbol{υ} t}}τ_t (A^*)\, ι_{\floor{ \boldsymbol{υ} {t^{\prime}}}}τ_{t^{\prime}} (A) \, B ) - ω_β(B^*  ι_{\floor{ \boldsymbol{υ} t}}τ_t (A^*)\,  B )ω_β(A) - \\ ω_β(A) ω_β( B^* ι_{\floor{ \boldsymbol{υ} {t^{\prime}}}}τ_{t^{\prime}} (A) B) + ω_β(A^*)ω_β(A) ω_β(B^*B)
    \end{array}
\end{equation}
The result easily follows by applying \Cref{lemma:6.1} for each of these terms.
\end{proof}

Finally we can prove \Cref{th:KMS_average_strong_convergence}:
\begin{proof}[Proof of \Cref{th:KMS_average_strong_convergence}]
We have by \Cref{lemma:6.2} that  for all $A,B \in \mathfrak{U}$ and almost every speed $υ$:
\begin{equation}
\begin{array}{*3{>{\displaystyle}lc}p{5cm}}
    \lim_{T \to \infty} \frac{1}{T^2}  \int_0^T \int_0^T ω_β \big(B^*&&\big(\,ι_{\floor{ \boldsymbol{υ} t}}τ_t (A)-ω_β(A) \mathds{1} \big)^* \times \\
    &&\big(\,ι_{\floor{ \boldsymbol{υ} {t^{\prime}}}}τ_{t^{\prime}} (A)-ω_β(A) \mathds{1} \big)  \,B \big) \,dt \,dt^{\prime}  = 0 .
    \end{array}
\end{equation}
We write this in the GNS representation of $ω_β$:
\begin{equation}
\begin{array}{*3{>{\displaystyle}lc}p{5cm}}
    \lim_{T \to \infty} \frac{1}{T^2}  \int_0^T \int_0^T  \big\langle Ω_{ω_β}, && π_{ω_β} \big( B^* \big(\,ι_{\floor{ \boldsymbol{υ} t}}τ_t (A)-ω_β(A) \mathds{1} \big)^* \times \\
    &&\big(\,ι_{\floor{ \boldsymbol{υ} {t^{\prime}}}}τ_{t^{\prime}} (A)-ω_β(A) \mathds{1} \big)  \,B \big) Ω_{ω_β} \big\rangle \, dt \,dt^{\prime}=0
    \end{array}
\end{equation}
which can be written as
\begin{equation}
\arraycolsep=1.4pt\def\arraystretch{2.3}
\begin{array}{*3{>{\displaystyle}lc}p{5cm}}
     \lim_{T \to \infty}   \big\langle  &&\frac{1}{T}\int_0^T π_{ω_β}    \big(\,ι_{\floor{ \boldsymbol{υ} t}}τ_t (A)-ω_β(A) \mathds{1} \big) π_{ω_β}(B) Ω_{ω_β}  \, dt,  \\
    &&\frac{1}{T}\int_0^T π_{ω_β} \big(\,ι_{\floor{ \boldsymbol{υ} {t^{\prime}}}}τ_{t^{\prime}} (A)-ω_β(A) \mathds{1}  \big) π_{ω_β}(B) Ω_{ω_β} \, dt^{\prime} \big\rangle =0
     \end{array}
\end{equation}
where the integrals are to be understood as Bochner integrals (see \cite[Section 3.8]{hille_functional_1996}). This gives:
\begin{equation}
    \lim_{T \to \infty} \norm{ \frac{1}{T}\int_0^T π_{ω_β}  \big(\,ι_{\floor{ \boldsymbol{υ} t}}τ_t (A)-ω_β(A) \mathds{1} \big) π_{ω_β}(B)Ω_{ω_β}  \, dt} =0
\end{equation}
for any $A,B \in \mathfrak{U}$ and almost all $υ$. By cyclicity of $Ω_{ω_β}$ this statement reads
\[\arraycolsep=1.4pt\def\arraystretch{2.5}
\begin{array}{*3{>{\displaystyle}lc}p{5cm}}
      \lim_{T \to \infty} \norm{ \frac{1}{T}\int_0^T π_{ω_β} \big(  \,ι_{\floor{ \boldsymbol{υ} t}}τ_t (A)-ω_β(A) \mathds{1} \big) Ψ  \, dt} =0 \implies \\ \lim_{T \to \infty} \norm{  \frac{1}{T}\int_0^T π_{ω_β} \big(  \,ι_{\floor{ \boldsymbol{υ} t}}τ_t (A) \big) Ψ \, dt - ω_β(A) Ψ} =0 
      \end{array}
\]
for any $Ψ \in H_{ω_β}$, i.e.\ we have that $\frac{1}{T} \int_0^T π_{ω_β} \big(  \,ι_{\floor{ \boldsymbol{υ} t}}τ_t (A) \big) \, dt$ (understood as a  Bochner integral) converges in the Strong operator topology to $ω_β(A) \mathds{1}$, for almost every speed $υ$.

\end{proof}
\section{Oscillatory Ergodicity} \label{section:oscillatory_ergo}

The ray average can also be taken in such a way so as to account for oscillations, in order to show that correlations between observables cannot sustain certain types of oscillations.
The main assumption is what we call space-like non-oscillating states, which will again follow from the Lieb-Robinson bound, in factor states. The intuition is that we assume states whose correlation functions do not exhibit oscillatory behaviours inside a space-like cone:

\begin{defn}[Space-like non-oscillating state]
Consider a dynamical system $(\mathfrak{U}, ι, τ)$. A state $ω \in E_{\mathfrak{U}}$ is called space-like non-oscillating if it is space and time translation invariant and there exists a $υ_c>0$ such that for any $A,B \in \mathfrak{U}_{\rm loc}$, $\boldsymbol n \in \mathbb{Z}^D$ and $υ \in \hat{ {\mathbb{R}}}= \mathbb{R} \cup \{ -\infty, \infty \}$ with $|υ| > υ_c$ it holds that:
\begin{equation}
  \lim_N  \frac{1}{N} \sum_{m=0}^{N-1} e^{-i(\boldsymbol k \cdot \boldsymbol n - f υ^{-1} |\boldsymbol n|)m} ω \big( ι_{\boldsymbol n}^m τ_{υ^{-1}|\boldsymbol n|}^m (A) B \big) =0
\end{equation}
for all wavenumber-frequency pairs $(\boldsymbol k ,f )\in \R^D \times \R$ with $\boldsymbol k \cdot \boldsymbol n - f υ^{-1} |\boldsymbol n| \not\in 2\pi \Z$.
\label{def:phase-spacelike-ergodic}
\end{defn}

It is trivial to see space-like clustering states, \Cref{eq:space-like_clustering}, are also space-like non-oscillating. Having the assumption of a space-like non-oscillating state we can extend the Ergodicity  \Cref{th:maintheorem} to preclude oscillatory behaviours of correlations along almost every ray:
\begin{theorem}
Consider a dynamical system $(\mathfrak{U}, ι, τ)$ with interaction that satisfies \Cref{eq:interaction}, a space-like non-oscillating state  $ω\in E_{\mathfrak{U}}$ and any $A,B \in \mathfrak{U}$.  Then for every non-zero $(\boldsymbol k, f)\in \mathbb{R}^D \times \mathbb{R}$ and every rational  direction $\boldsymbol q \in \Srat^{D-1}$ of the velocity $\boldsymbol{υ}= υ \boldsymbol{q}$ it follows that  for almost all $υ\in \R$
\begin{equation}
   \lim_{T \to \infty} \frac{1}{T} \int_0^T  e^{i (\boldsymbol k\cdot \boldsymbol v   -  f) t} \bigg( ω \big( ι_{\floor{ \boldsymbol{υ}t}}τ_t (A) B \big) - ω(A) ω(B) \bigg)  \,dt = 0.
    \label{eq:frequencyav2}
\end{equation}
Note that generally this means $\lim_{T \to \infty} \frac{1}{T} \int_0^T  e^{i (\boldsymbol k\cdot \boldsymbol v   -  f) t}  ω \big( ι_{\floor{ \boldsymbol{υ}t}}τ_t (A) B \big) \,dt =0$, except for the special case $\boldsymbol k\cdot \boldsymbol v-f=0$ where the Theorem reduces to $\Cref{th:maintheorem}$.

Additionally, we have for any $n$-th moment with respect to the state:
\begin{equation} \label{eq:frequency-moments}
    \lim_{T \to \infty} \frac{1}{T^n} ω \bigg( \big(\int_0^T e^{i (\boldsymbol k\cdot \boldsymbol v   -  f) t}   ι_{\floor{ \boldsymbol{υ} t}}τ_{t} A  \,dt  \big)^n \bigg)=  
   \begin{cases}
		ω(A)^n,  & \mbox{if } \boldsymbol k \cdot \boldsymbol v - f =0 \\
		0, &  \text{otherwise}
	\end{cases}
\end{equation}
for almost all $υ \in \mathbb{R}$. \label{th:frequencyav2}. 
\end{theorem}
This  tells us that both correlations and moments cannot sustain oscillations in the long time limit, along almost every ray. 

The SOT convergence result \Cref{th:KMS_average_strong_convergence}, in the GNS representation of KMS states can also be extended to account for oscillations. The proof is very similar and is hence omitted.
\begin{theorem} \label{th:kms_strong_convergence_oscil}
    Consider a dynamical system $(\mathfrak{U}, ι, τ)$ and a possibly different time evolution $\tau'$ such that $(\mathfrak{U}, ι, τ')$ is also a dynamical system. Suppose the interactions, both for $\tau$ that $\tau'$, satisfy \Cref{eq:interaction}, and that $\tau$ and $\tau'$ commute. Consider a   $(\tau',\beta)$-KMS state $ω_β$ that is space-like non-oscillating for  $(\mathfrak{U}, ι, τ)$  and its respective GNS representation $(H_{ω_β},π_{ω_β},Ω_{ω_β})$. It follows that for any $A\in \mathfrak{U}$, any $(\boldsymbol k,f)\in \R^D \times \R$ and any rational direction $\boldsymbol q\in \Srat^{D-1}$ of the velocity $\boldsymbol{υ}=υ\boldsymbol{q}$
\begin{equation}
    \lim_{T \to \infty} \frac{1}{T} \int_0^T e^{i (\boldsymbol k\cdot \boldsymbol v   -  f) t}\bigg(π_{ω_β} \big( ι_{\floor{ \boldsymbol{υ} t}}τ_t A \big) -ω(A) \bigg) \, dt \, Ψ= 0\ , \ \ \forall Ψ \in H_{ω_β}
\end{equation}
for almost every speed $υ \in \R$.  The limit is in the norm of the GNS Hilbert space.  Note that for generic $(\boldsymbol k,f)$ this means $\lim_{T \to \infty} \frac{1}{T} \int_0^T e^{i (\boldsymbol k\cdot \boldsymbol v   -  f) t}π_{ω_β} \big( ι_{\floor{ \boldsymbol{υ} t}}τ_t A \big)  \, dt \, Ψ= 0\ $, precluding sustained oscillations of the ray averaged operator.
\end{theorem}

\chapter{Clustering of higher order connected correlations in C* dynamical systems} \label{chapter:higher_clustering}
\section*{Preface}

This chapter incorporates our paper \cite{ampelogiannis_2024_clustering}, which concerns the clustering properties of higher order  correlations. The work in \cite{ampelogiannis_2024_clustering} is largely self-contained and written in the same mathematical framework  we have used so far, but in a much more general manner. It contains general results for group actions on C$^*$ algebras, not specific to space or time translations. In \Cref{section:qsl} the main theorems are applied to clustering states of quantum lattice models, where bounds on $n-$order correlations are obtained. The main result is the decay of    correlations when one observable is moved outside the Lieb-Robinson light-cones of all others. These findings are particularly significant in light of their applications. In~\cite{ampelogiannis_2024_diffusion}, the clustering estimates developed here are used to establish rigorous lower bounds on the diffusion constant for open quantum spin chains.  This application and its implications are discussed in detail \Cref{chapter:diffusion}. 

\chapter*{Clustering of higher order connected correlations in C* dynamical systems}
\begin{center}
\vspace{1cm}

{\em Dedicated to the memory of Petros Meramveliotakis}
\vspace{1cm}

{\large Dimitrios Ampelogiannis and Benjamin Doyon}

\vspace{0.2cm}
Department of Mathematics, King's College London, Strand WC2R 2LS, UK

\end{center}

In the context of $C^*$ dynamical systems, we consider a locally compact group $G$ acting by $^*$-automorphisms on a C$^*$ algebra $\mathfrak{U}$ of observables, and assume  a state of $\mathfrak{U}$ that satisfies the clustering property with respect to a net of group elements of $G$. That is, the two-point connected correlation function vanishes in the limit on the net, when one observable is translated under the group action. Then we show that all higher order connected correlation functions (Ursell functions, or classical cumulants) and all free correlation functions (free cumulants, from free probability) vanish at the same rate in that limit. Additionally, we show that mean clustering, also called ergodicity,  extends to higher order correlations.
We then apply those results to equilibrium states of quantum spin lattice models. Under certain assumptions on the range of the interaction, high temperature Gibbs states are known to be exponentially clustering w.r.t.\ space translations. Combined with the Lieb-Robinson bound, one obtains exponential clustering for space-time translations outside the Lieb-Robinson light-cone. Therefore, by our present results, all the higher order connected and free correlation  functions will vanish exponentially under space-time translations outside the Lieb-Robinson light cone, in high temperature Gibbs states. Another consequence is that their long-time averaging over a space-time ray vanishes for almost every ray velocity.

\section{Introduction}
In this paper we are concerned with the clustering properties of multi-point connected correlations, or joint cumulants, of states of $C^*$ algebras, with respect to group actions. We show that if the 2-point connected correlation (covariance) between two observables vanishes when we move one ``infinitely far'', under the group action, then the $n$-th cumulant between $n$ observables also vanishes at the same rate. Additionally, if we average over the group action and the 2-point connected correlation vanishes in the mean, then the $n$-point connected correlation also vanishes in the mean. We show these both for classical cumulants, or Ursell functions, and free cumulants from the theory of free probability \cite{speicher_2019_notes}.

These results are particularly interesting in quantum many-body physics, for the group being that of translations along space and / or time. For instance, this includes quantum spin lattice (QSL) models with translation symmetry (such as the $\mathbb Z^d$ lattice). Indeed, in this context higher-order correlations are relevant as they characterise higher order hydrodynamics \cite{myers_2020_fluctuations,doyon_2020_fluctuations} and in particular provide bounds on diffusion coefficients \cite{doyon_2019_diffusion}.
%The latter is one application of our results which we develop in a coming work.
Higher-order correlations also serve as a measure of chaos in quantum systems via out-of-time-order correlators \cite{garcia_2022_OTOC_chaos_review,bhattacharyya_2022_quantum_chaos}, and as a measure of multi-partite entanglement \cite{zhou_2006_multiparty,pappalardi_2017_multi_entanglement}. The accuracy of the mean field approximation depends on the assumption that correlations between observables are negligible, and to improve it higher order cumulants can be used \cite{Fowler_2023_cumulant_expansion, Kramer_2015_generalisedMFT, Robicheaux_2021_beyond}, generally by a cumulant expansion method \cite{kubo_1962_cumulant, fricke_transport_1996, Kira_2008_clusterexp, Sanchez_2020_cumulant}. Our result on vanishing in the mean is especially relevant to averages along rays in space-time, as there we have recently proven the vanishing in the mean of the covariance within QSL with a large amount of generality \cite{ampelogiannis_2023_almost}. The vanishing of covariance was a crucial ingredient in the proof of the hydrodynamic projection principle for two-point functions \cite{doyon_hydrodynamic_2022,ampelogiannis_long-time_2023}, and we expect the projection principle for higher-point functions \cite{doyon_2023_mft} can be established from our present results.

The connected correlations generally considered in the above applications are the classical cumulants, which measure classical independence (as random variables) between observables. However, one can also consider different types of independence. In non-commutative probability one generally considers the notion of ``freeness'', introduced by Voiculescu who developed the theory \cite{voiculescu_1985_Symmetries,voiculescu_1986_addition,voiculescu_1987_multiplication}, and the free cumulants measure free independence, introduced by Speicher \cite{speicher_1994_multiplicative}. Recently a connection between free probability theory and the  Eigenstate Thermalization Hypothesis (ETH) was made in \cite{Pappalardi_2022_ETH_free}, where higher order correlation functions are determined by the free cumulants. Additionaly, in \cite{jindal_2024_free} quantum chaos and the decay of the out-of-time-order correlators are studied using free cumulants.  In light of this recent activity using tools from free probability in quantum statistical mechanics, we consider here both classical and free cumulants. 

The vanishing of higher-order correlation functions has been considered using cumulant expansion methods for classical and quantum  systems \cite{Ueltschi2003Cluster}. A similar result in product states of finite-volume systems extending the Lieb-Robinson bound on $n$-partite correlations was shown in \cite{Tran_2017_Lieb_Robinson_npartite}. However, we are not aware of results as general as those established here, in particular for free cumulants.

This paper is organised as follows. In \Cref{section:set-up} we discuss the precise set-up of C$^*$ dynamical systems in which we work and then in \Cref{section:cumulants} we precisely define both the classical (connected correlations) and the free  cumulants  for a state of a C$^*$algebra. In \Cref{section:results} we provide our main results, with rigorous statements. In \Cref{section:qsl}
we apply our results to quantum lattice models with either short-range or long-range interactions. Finally, the proofs of the theorems discussed in \Cref{section:results} are done in \Cref{section:proofs}.

\section{Set-up} \label{section:set-up}
We consider a unital $C^*$-algebra $\mathfrak{U}$, as the set of observables of our physical system, and a group $G$ that acts on $\mathfrak{U}$ by $*$-automorphisms $α_g$, $g \in G$. The physical system in mind can be classical ($\mathfrak{U}$ abelian) or quantum  ($\mathfrak{U}$ non-commutative). We only make the assumption that the group $G$ is locally compact; it can for example describe time evolution, $G=\R$, or space translations, $G=\Z^D$ for a spin lattice and $G=\R^D$ for continuous systems in $D$ dimensions.

\begin{defn}[$C^*$  Dynamical System] \label{defn:dynamicalsystem}
A  $C^*$ dynamical system is a triple $(\mathfrak{U},G,α)$ where $\mathfrak{U}$ is a unital $C^*$-algebra, $G$ a locally compact group, and $α$ is a representation of the group $G$ by $^*$-automorphisms of $\mathfrak U$.
\end{defn} 

We examine states $ω$ of $\mathfrak{U}$ that are clustering with respect to $G$, in the sense that the connected correlation between two observables vanishes under some limit action, e.g.\ space-translating one observable infinitely far.

\begin{defn}[Clustering] \label{defn:clustering}
A state $ω$ of a $C^*$ dynamical system  $(\mathfrak{U},G,α)$ is called $G$-clustering if there exists a net $ g_i \in G$ such that 
\begin{equation}%\label{clustering}
    \lim_i \big(ω( α_{g^i}(A) B) - ω(α_{g^i}A) ω(B) \big)= 0 , \ \forall A,B \in \mathfrak{U}.
\end{equation} 

\end{defn}

The decay of 2-point connected correlations, for space-translations, has been established rigorously in many physical systems. In classical systems see \cite{ruelle_1974_rigorous}; in particular a one-dimensional classical lattice gas satisfies the clustering property, as shown by Ruelle  \cite{Ruelle_1968_classical_gas}. For quantum spin chains (QSC), Araki \cite{Araki:1969bj} showed exponential clustering in Gibbs states, for finite-range interactions. In QSC with short-range interactions the correlations also decay exponentially in Gibbs states \cite{perez_2023_locality}. Clustering for quantum gases was established by Ginibre \cite{Gibibre_1965_quant_gases}. Cluster expansion methods can be used quite generally, to show decay of correlations for both classical and quantum systems \cite{Park_1982_cluster_expansion,Ueltschi2003Cluster}.

In the non-commutative setup, it can be the case that the algebra is asymptotically abelian under the action of $G$, i.e.\ there exists a net $g_i \in G$:
\begin{equation}
    \lim_i \norm{ [ α_{g^i} A, B] }=0 , \ \forall A,B \in \mathfrak{U}.
\end{equation}
It follows from asymptotic abelinanness that every factor state will satisfy the cluster property \Cref{clustering}, see \cite[Example 4.3.24]{bratteli_operator_1987}.  A physically relevant example of factors is that of thermal equilibrium states, described by the Kubo-Martin-Schwinger (KMS) condition, see \cite[Section 5.3]{bratteli_operator_1997}. An extremal KMS state is factor (\cite[Theorem 5.3.30]{bratteli_operator_1997}) hence it has clustering properties. 

\begin{prop}
    Consider a $C^*$ dynamical system  $(\mathfrak{U},G,α)$ that is asymptotically abelian. If $ω$ is a factor state, then ω is $G$-clustering. 
\end{prop}

\begin{corollary}
Consider a $C^*$ dynamical system  $(\mathfrak{U},G,α)$ that is asymptotically abelian and $τ_t$ is the strongly continuous group of time evolution. If $ω_β$ is an extremal $(τ,β)$-KMS state, then $ω$ is factor and hence $G$-clustering.
\end{corollary}

A concrete example is a quantum spin chain, such as the XXZ-chain: the spin chain is asymptotically abelian under space translations $ \lim_{x\to \infty} \norm{ [ \alpha_{x} A, B] }=0$ (\cite[Example 4.3.26]{bratteli_operator_1987}) and at non-zero temperature there is a unique KMS state \cite{Araki1975uniqueness} which is subsequently factor and hence clustering with respect to space translations. A continuous example is the ideal Bose gas, described by the CCR (canonical commutation relations) algebra, which is also asymptotically abelian for space translations (\cite[Example 5.2.19]{bratteli_operator_1997}). The equilibrium state of the ideal Bose gas is clustering for both space and time translations, in the single phase region (\cite[Section 5.2.5]{bratteli_operator_1997}). Of course, asymptotic abelianness is not a neccessary condition, the  ideal Fermi gas is described by the CAR (canonical anticommutation relations) algebra which is not asymptotically abelian (\cite[Section 5.2.21]{bratteli_operator_1997}), but its equilibrium states are clustering both in space and in time (\cite[Section 5.2.4]{bratteli_operator_1997}).

\section{Higher order cumulants and free cumulants} \label{section:cumulants}

\subsection{Connected correlations -- classical cumulants} \label{section:classical_cumulants}
Consider a $C^*$ algebra $\mathfrak{U}$ and a state $ω$. We start by defining the  connected correlations between $n$-observables, or joint classical cumulants, for the state $ω$. The first three cumulants are 
\begin{equation} \label{eq:example_cumulants}
 \begin{array}{*3{>{\displaystyle}lc}p{5cm}}
  c_1 ( A_1 )  &=& ω(A_1) \\
  c_2( A_1, A_2) &=& ω(A_1 A_2)-ω( A_1) ω(A_2)\\
   c_3(A_1, A_2, A_3) &=& ω(A_1 A_2 A_3) - ω(A_1)ω(A_2 A_3) - ω(A_2) ω(A_1 A_3)\\&&
   - ω(A_3) ω(A_1A_2) +2 ω(A_1)ω(A_2)ω(A_3) 
        \end{array}    
\end{equation}
for $A_1,A_2,A_3 \in \mathfrak{U}$. To give a general definition, we have to keep in mind that the observables might not commute. We define for every $n\in \N$ the functional
\begin{equation}
    ω_n( A_1, A_2, \cdots , A_n) \coloneqq ω(A_1 A_2 \cdots A_n)  \ , \ \ A_1,\cdots , A_n \in \mathfrak{U}.
\end{equation}
Then, we extend it to the multiplicative family of moments $(ω_π)_{π \in P}$, where $P = \cup_{n \in \N} P(n)$ and $P(n)$ the set of all partitions of $\{1,2,\cdots,n\}$, as follows:
\begin{equation}
    ω_π (A_1, A_2, \cdots, A_n)  \coloneqq \prod_{V \in π} ω_{|V|} (A_1, A_2, \cdots , A_n | V)
\end{equation}
where $ ω_s(A_1, A_2, \cdots , A_n | V)$ keeps the indices $i\in V$ in the correct order; for any $s\in \N$, $i_1,\cdots , i_s\in \{1,\cdots,n\}$ with $i_1 < \cdots <i_s$ and $V=\{i_1,\cdots ,i_s\}$:
\begin{equation} \label{eq:notation_ordering}
    ω_s(A_1, A_2, \cdots , A_n | V)  \coloneqq ω_s ( A_{i_1}, \cdots , A_{i_s} ).
\end{equation}
The most direct way of defining the classical cumulants of the state $ω$ is by a sum, over all partitions, of products of joint moments \cite{Speed_1983_Cumulants1}:
\begin{defn} \label{defn:classical_cumulants}
  Consider  a C$^*$ algebra  $\mathfrak{U}$ and a state $ω$. The classical cumulants of $ω$  are defined by 
  \begin{eqnarray}
%    \begin{array}{*3{>{\displaystyle}lc}p{5cm}}
    c_n (A_1, A_2,\ldots,A_n) &=& \sum_{π \in P(n)} (-1)^{|π|-1}( |π| - 1)! \prod_{V \in π} ω_{|V|} (A_1, A_2, \cdots , A_n | V)\nonumber \\
    &=&\sum_{π \in P(n)} (-1)^{|π|-1}( |π| - 1)! \, ω_π( A_1 , A_2, \ldots , A_n)
%    \end{array}
\end{eqnarray}
for $n \in \N$ and $A_1, A_2, \ldots, A_n\in\mathfrak U$.
\end{defn}

A more useful approach, is an indirect recursive definition as multilinear functionals $c_n: \mathfrak{U}^n \to \C$ that we extend to a multiplicative family $\{c_π\}_{π \in P}$:
\begin{equation}
    c_π(A_1,A_2,\cdots, A_n) = \prod_{V\in π} c_{|V|} (A_1,A_2,\cdots,A_n | V)
\end{equation}
and require that they satisfy the moments-to-cumulants formula
\begin{equation} \label{eq:moments_to_cumulants}
    ω_n(A_1,A_2, \ldots, A_n) = \sum_{π \in P(n)} c_π (A_1,\cdots,A_n).
\end{equation}

For our proofs, it will be important to use the following recursive definition of the set $P(n)$, where the recursion involves the possible sets in which the particular element $1$ lies (similarly, any other element could have been chosen):
\begin{equation}\label{eq:recursionP}
    P(n) = \{ \{ V\} \cup \pi: V\ni 1,\pi \in P(\{1,\ldots,n\}\setminus V)\}
\end{equation}
where we use the notation $P(V)$ to denote the set of all partitions of the elements of the set $V$ (in particular $P(n) = P(\{1,\ldots,n\})$).

\subsection{Free cumulants} \label{section:free_cumulants}
A double $(\mathfrak{U},ω)$ consisting of a non-commutative C$^*$ algebra $\mathfrak{U}$ and a state $ω$ forms a non-commutative probability space  \cite{speicher_2019_notes}. The free cumulants where introduced by Speicher \cite{speicher_1994_multiplicative} in the context of free probability theory. The definition is based on the non-crossing partitions, which are partitions with blocks that do not cross, in a natural diagrammatic representation where elements are ordered (say along a circle). We denote by $NC(n)$ the set of all non-crossing partitions of $\{1,2,\ldots, n\}$ and $NC= \cup_{n\in\mathbb N} NC(n)$. We give a recursive definition via the moments-to-cumulants formula, based on the discussion in \cite{speicher_2019_notes}:

\begin{defn}[Free cumulants]\label{defn:free_cumulant}
    Consider a $C^*$ algebra $\mathfrak{U}$ and a state $ω$. The free cumulants are a multiplicative family $κ_π$ over $π\in NC$, in the sense that $κ_n:\mathfrak{U}^n \to \C$ are multilinear for all $n \in \N$ and (using a notation as in Eq.~\eqref{eq:notation_ordering})
    \begin{equation} \label{eq:multiplicative_free}
    κ_π(A_1,A_2,\ldots, A_n) = \prod_{V\in π} κ_{|V|} (A_1,A_2,\ldots,A_n | V)
\end{equation}
for $n\in \N$, $π \in NC(n)$ and $A_1,A_2,\ldots,A_n \in \mathfrak{U}$, such that they satisfy
\begin{equation} \label{eq:moments_to_free_cumulants}
    ω_n(A_1,A_2, \ldots, A_n) = \sum_{π \in NC(n)} κ_π (A_1,\ldots,A_n).
\end{equation}
The $n$-th free cumulant corresponds to the maximal partition $κ_n \coloneqq κ_{\mathds{1}_n}$.
\end{defn}

See also \Cref{appendix:mobius} for a more direct (equivalent) definition. Note that the first three free cumulants are the same as the classical ones \Cref{eq:example_cumulants}, as all the partitions of $\{1,2\}$ and $\{1,2,3\}$ are non-crossing. For $n=4$ there is one crossing partition, $\{ \{1,3\}, \{2,4\} \}$, and hence for $n\geq 4$ the free cumulants are different from the classical ones.

Again, for our proofs, it will be important to use a recursive definition of the set of non-crossing partitions. Let us denote $NC(V)$ the non-crossing partitions of the set $V$, and in particular $NC(n) = NC(\{1,\ldots,n\})$. For a subset $V$ of $\{1,\ldots,n\}$, let us denote $\pi^V$ the unique partition of $V$ that is the minimal one (with the smallest number of parts) such that every part is composed of consecutive elements: every $W\in \pi^V$ is of the form $\{w_1,w_2,\ldots\}$ with $w_j = a+j\,{\rm mod}\,n$ for some $a$. Then we have
\begin{equation}\label{eq:recursionNC}
    NC(n) = \{ \{ V\} \cup \cup_j\pi_j: V\ni 1, \pi_j \in NC(W_j) \mbox{ for } \pi^{\{1,\ldots,n\}\setminus V}=\{W_1,W_2,\ldots\}\}
\end{equation}
and similarly, any other element instead of 1 could have been chosen to write the recursion. Note the difference with \eqref{eq:recursionP}: the non-crossing condition is implemented in the smaller set of partitions that combine with $V$, i.e.~those of the form $\cup_j \pi_j$: restricting to sets formed of consecutive elements of $\{1,\ldots,n\}\setminus V$ is what imposes the non-crossing condition.
\section{Main results: Clustering for $n$-th order cumulants}  \label{section:results}

Our main result is clustering for the $n$-th classical, and free, cumulants in any state that is clustering for the covariance (second cumulant), as per \Cref{defn:clustering}. In fact, all higher cumulants inherit the same rate of decay as that of the covariance.  A weaker form of clustering, ergodicity, can also be extended to higher cumulants; averaging the cumulants over the group action (e.g.\ time average) gives $0$. 

\subsection{Clustering for $n$-th order (free) cumulants} \label{section:results1}

Taking advantage of the moment-to-cumulants formula \Cref{eq:moments_to_cumulants} we can inductively show that whenever the two-point connected correlation vanishes under some limit of the group action, $ \lim_{i}  \big(ω( \alpha_{g^i} (A) B) -  ω(\alpha_{g^i} A) ω(B)\big) = 0$, then all higher cumulants will also vanish at the same limit. We show this in detail in \Cref{section:proof_1} for the free cumulants, but the proof for the classical ones is the same.
\begin{theorem} \label{th:cumulants_general}
    Let $(\mathfrak{U},G,α)$ be a C$^*$  dynamical system and $ω$ a $G$-clustering state, i.e.\  there exists a net $g^i \in G$ such that 
    \begin{equation} \label{eq:clustering_th1}
       \lim_{i}  \big(ω( α_{g^i} (A) B) -  ω(α_{g^i} A) ω(B) \big) = 0,   \ \forall A,B \in \mathfrak{U}
    \end{equation}
then for any $n\in \N$ and $A_1,A_2,\ldots , A_n \in \mathfrak{U}$ the free cumulant  vanishes 
\begin{equation} \label{eq:main}
    \lim_{i} κ_n( \alpha_{g^i}A_1 , A_2, \ldots, A_n)
    =
    \lim_{i} κ_n( A_1 , A_2, \ldots, \alpha_{g^i} A_n)= 0 .
\end{equation}
The same holds for the classical cumulants $c_n$:
\begin{equation} \label{eq:maincumu}
    \lim_{i} c_n( \alpha_{g^i}A_1 , A_2, \ldots, A_n) =
    \lim_{i} c_n( A_1 , A_2, \ldots, \alpha_{g^i}A_n)= 0 . \implies
\end{equation}
\end{theorem}

The theorem also holds for states that are clustering only for a $^*$-subalgebra of observables. Suppose   $\mathfrak{V} \subset \mathfrak{U}$ is a  $^*$-subalgebra and that 2-point clustering, \Cref{eq:clustering_th1}, holds for all $A,B \in \mathfrak{V}$. Then the $n$-th cumulants between any $A_1,A_2,\ldots,A_n \in \mathfrak{V}$ satisfy the clustering property.

Note that under the assumptions of \Cref{th:cumulants_general} it is not necessarily true that 
\begin{equation}\label{eq:cumulant_i}
 \lim_{i} κ_n( A_1 ,  \ldots, \alpha_{g^i} (A_m), \ldots, A_n) = 0 
\end{equation}
for $1<m<n$ (similarly for the classical cumulants). This is because a clustering state, as per \Cref{defn:clustering}, does not necessarily satisfy the three-element clustering property $ \lim_{i}  \big(ω(  A \alpha_{g^i}(B) C) -  ω(\alpha_{g^i} B) ω(AC)\big) = 0$. However, if we add this property as an assumption, then \Cref{th:cumulants_general} can be extended so that \Cref{eq:cumulant_i} holds for any $m=1,2,\ldots,n$. This is shown in \Cref{section:proof_1}, at the end of the proof. Note that in asymptotically abelian algebras the two element clustering property is equivalent to the three-element property (\cite[Theorems 4.3.22 \& 4.3.23]{bratteli_operator_1987}), hence all factor states satisfy the more general case.

We can also have a set of observables $\{ α_{g^i}  A_1,  α_{g^i} A_2, \ldots ,  α_{g^i} A_m\}$ translated away from another set $\{ A_{m+1}, \ldots , A_n\}$, this is shown in \Cref{section:proof_3}.

\begin{theorem} \label{th:clustering_for_groups}
Consider the assumptions of \Cref{th:cumulants_general}. For any $m < n \in N$ and $A_1, A_2 , \ldots, A_n \in \mathfrak{U}$ it follows that
\begin{equation}
  \lim_i  κ_n( α_{g^i}  A_1,  α_{g^i} A_2, \ldots ,  α_{g^i} A_m, A_{m+1}, \ldots , A_n ) =0.
\end{equation}
The same true is for classical cumulants $c_n$. %If additionally $ \lim_{i}  \big(ω(  A \alpha_{g^i}(B) C) -  ω(\alpha_{g^i} B) ω(AC)\big) = 0$, $\forall A,B,C \in \mathfrak{U}$, then the order of $α_{g^i} A_1, \ldots, A_n$  in $κ_n$ doesn't matter.
\end{theorem}
In the results that follow, we only consider the simplest case of translating one element under the group action and examining the clustering properties of joint cumulants. However, using the same techniques, these results can be generalised in the same manner as \Cref{th:clustering_for_groups}.

\subsection{Rate of decay}
Consider a clustering state $ω$ of a $C^*$ dynamical system $(\mathfrak{U},G,α)$ and a function $f:G \to \R^+$. We call $ω$ $f(g)$-clustering if the rate of decay of the covariance is $1/f(g)$ in some $^*$-subalgebra $\mathfrak{V} \subset \mathfrak{U}$. For example, KMS states in quantum lattice models are exponentially clustering in space, that is $e^{\lambda |x|}$-clustering for some $\lambda>0$ \cite{Araki:1969bj}. The proof of \Cref{th:cumulants_general} can easily be modified to show that for a $f(g)$-clustering state the $n$-th classical, and free, cumulant also decays with rate $1/f(g)$, for all $n \geq 2$. Both proofs are done in \Cref{section:proof_1}.

\begin{theorem} \label{th:decay}
    Let $(\mathfrak{U},G,α)$ be a $C^*$  dynamical system and $ω$ a f(g)-clustering state for a $*$-subalgebra $\mathfrak{V} \subset \mathfrak{U}$ , i.e.\  there exists a net $g^i \in G$ such that
    \begin{equation}
       \lim_{i}  f(g^i)\big(ω( α_{g^i}(A) B) - ω( α_{g^i}A) ω(B) \big)= 0 , \ \forall A,B \in \mathfrak{B} 
    \end{equation}
for some non-zero $f: G \to \R^+$. Then for any $n\in \N$ the free cumulants vanish at the same rate:
\begin{equation} \label{eq:maindecay}
    \lim_{i} f(g^i)κ_n( α_{g^i}A_1 , A_2, \cdots, A_n) = 0 , \ \forall A_1, A_2, \ldots, A_n \in \mathfrak{B}. 
\end{equation}
The same holds for the classical cumulants $c_n$.
\end{theorem}

Note that the theorem  includes the case $\mathfrak{V} = \mathfrak{U}$, but does not require that $\mathfrak{V} $ is norm closed. 

Additionally, bounds on the second cumulant also apply to higher order cumulants. In a similar manner one can show that $|ω( α_{g}(A) B) - ω( α_{g}A) ω(B)| \leq C_2\norm{A} \norm{B}f(g)$ implies that $|κ_n( α_{g}A_1 , A_2, \cdots, A_n)| \leq  C_n \prod_j \norm{A_j} f(g)$ where $C_n>0$. We obtain such a result for quantum lattice models in Section \cref{subsection:52}.

\subsection{Mean Clustering}
We can go further and assume a weaker clustering property called mean clustering, a form of ergodicity with respect to group actions on $C^*$-algebras (\cite[Section 4.3]{bratteli_operator_1997}). That is, instead of clustering states, with covariance that vanishes, we consider states such that the mean of the covariance over the group action goes to $0$. This is relevant with our recent results of almost everywhere ergodicity \cite{ampelogiannis_2023_almost}, that we discuss in \Cref{section:qsl}.

In general, it is not always possible to define an invariant mean over the group action.
This is possible in groups called amenable \cite{greenleaf_1969_invariant}. If $μ$ is the Haar measure of the locally compact group $G$, then one of the equivalent definitions of an amenable group is that for every compact $K\subset G$ there exists a net $U_i \subset G$, with $μ(U_i) \leq \infty$, such that
\begin{equation} \label{eq:amenable}
    μ( U_i Δ gU_i)/ μ(U_i) \to 0 , \ \forall g \in K
\end{equation}
where $gU_i = \{ h \in G : h= g u, u\in U_i \}$ is $U_i$ translated by $g$, and $A Δ B = (A \cup B) \setminus(A \cap B)$. Then, one can define an invariant mean for functions over $G$  as
\begin{equation}
    M(f) = \lim_i \frac{1}{μ(U_i)} \int_{U_i} f(g) \,dμ(g).
\end{equation}
Every locally compact abelian group is amenable and we restrict the next theorem to such groups, but locally compact amenable would still be sufficient. Time and space translations are of course described by amenable groups, such as  $G=\R^D$ which clearly satisfies \Cref{eq:amenable} for a net of balls of increasing radius.
 
\begin{theorem} \label{th:ergodicity}
    Let $(\mathfrak{U},G,α)$ be a $C^*$ dynamical system and $G$ a locally compact abelian group with Haar measure $μ$. Let $ω$ be a state such that there exists a net $U_i \subset G$ with
    \begin{equation} \label{eq:ergodicity_general}
        \lim_i \frac{1}{μ(U_i)} \int_{U_i} \big( ω( α_{g}(A) B) - ω( α_{g}A) ω(B)  \big) \,dμ(g) =0, \forall A,B \in \mathfrak{U}.
    \end{equation}
It follows that the free cumulants also have vanishing mean:
\begin{equation}
     \lim_i \frac{1}{μ(U_i)} \int_{U_i} κ_n( α_{g}A_1 , A_2, \cdots, A_n) \,dμ(g)= 0 
\end{equation}
for all $A_1,A_2,\ldots,A_n\in\mathfrak{U}$. The same holds for the classical cumulants $c_n$.
\end{theorem}

In the context of ergodicity the state $ω$ is generally considered to be $G$-invariant, that is $ω(α_gA)=ω(A)$. However, it is not necessary to assume invariance in order to prove \Cref{th:ergodicity}, so we have kept the expression as general as possible.  The proof of \Cref{th:ergodicity} is done in Section \ref{section:proof_2}.

\subsection{Banach Limit clustering}

An interesting observation is that the proofs only rely on the linearity of the limit. This allows us to consider generalised limits, such as a Banach limit, in order to extend our theorems so that whenever a Banach limit of the second cumulant vanishes, then the $n$-th cumulants also vanish for the same limit. Banach limits are used in proving a hydrodynamic projection theorem for the Euler scale correlation functions in \cite{doyon_hydrodynamic_2022,ampelogiannis_long-time_2023}, for spin chains ($D=1$) and spin lattices ($D>1$), respectively.

\begin{theorem}
    Let $(\mathfrak{U},G,α)$ be a dynamical system and $G$ a locally compactgroup . Let $ω$ be a state such that there exists a net $g_i \in G$ and a Banach Limit $\widetilde \lim$ with
    \begin{equation} \label{eq:7 }
        \widetilde{\lim_i}\,  \big( ω( α_{g^i}(A) B) - ω( α_{g}A) ω(B)  \big)  =0, \ \forall A,B \in \mathfrak{U}.
    \end{equation}
It follows that the free cumulants also vanish at the same limit
\begin{equation}
    \widetilde {\lim_i}\, κ_n( α_{g^i}A_1 , A_2, \cdots, A_n)  = 0, \ \forall A_1,A_2,\ldots,A_n \in \mathfrak{U}.
\end{equation}
The same holds for the classical cumulants.
\end{theorem}

\section{Application to QSL: space-like clustering from Lieb-Robinson bound} \label{section:qsl}
We consider a quantum spin lattice (QSL)  with either short-range interactions (exponentially decaying or finite range) or long-range (power-law decaying). The $C^*$-algebra of observables is a quasi-local algebra $\mathfrak{U}= \overline{\cup_{Λ\subset \Z^D} \mathfrak{U}_{Λ}}$ over $\Z^D$ and we have the groups of space translations $ι_x$ and time evolution $τ_t$, acting as $^*$-automorphisms on $\mathfrak{U}$.  We  assume that $τ$ is such that $τ_t ι_{\boldsymbol x} = ι_{\boldsymbol x} τ_t  $, $\forall t \in \R$, $\boldsymbol x \in \Z^D$, i.e.\ time evolution is homogeneous.  We call the triple $(\mathfrak{U},ι,τ)$ a quantum lattice $C^*$ dynamical system. See \cite[Section 6.2]{bratteli_operator_1997} for the detailed construction. Note that existence of the infinite volume dynamics for long-range interactions has been established, see \cite{nachtergaele_quasi-locality_2019}.

A central result in the context of QSL is the Lieb-Robinson bound, which  exponentially bounds the effect of any time-evolved observable outside a light-cone. The Lieb-Robinson bound was initially shown for finite range interactions \cite{Lieb:1972wy} and later extended to short range  interactions, see for example \cite{nachtergaele_2006_LR}. In the case of long-range interactions, the bound was extended by Hastings and Koma \cite{Hastings_2006_spectralgap}, but with a light-cone velocity that diverges with distance. However, in \cite{Chen_2019_finite_scrambling,Kuwahara_2020_linear_LR}, under the assumption of a power-law decaying interaction $\sim 1/r^a$ with $a>2D+1$, where $D$ is the lattice dimension, it is shown that one obtains a linear light-cone. Here, we are not interested in the specifics of the Lieb-Robinson bounds, but in the existance of the linear light cone and the resulting asymptotic abelianness for space-time translations outside of it. To summarise, for short-range interactions or power-law decaying with exponent  $a>2D+1$, there exists a $υ_{LR}>0$, called the Lieb-Robinson velocity, and a $λ>0$ such that for any local observables $A,B$ 
\begin{equation}  
      \norm{[ τ_t (A) , B]} \leq \begin{cases}
 L_{A,B} \exp\{-λ( \dist ( A,B) - υ_{LR}|t|)\},  &\text{for short-range}\\
 \\[1pt]
\displaystyle L^{\prime}_{A,B} \frac{ |t|^{D+1} }{(\dist (A,B) -υ_{LR}|t|)^{a-D} }, &\text{for long-range}
\end{cases}\label{eq:liebrobinsonbound}
\end{equation}
where $\dist(A,B)$ is the distance between their supports and $L_{A,B}$, $L^{\prime}_{A,B}$ depend on the norms and support sizes of $A,B$. 

This section is divided into three subsections. In \ref{susbsection:51} we apply results from the previous section to obtain clustering and exponential clustering, for higher order cumulants, for space-time translations outside the Lieb-Robinson light cone, in physically relevant states. We also discuss ergodicity results for higher order connected correlations, applying results from \cite{ampelogiannis_2023_almost}.  In \ref{subsection:52} we consider an inverse power-law bound on second order connected correlations and obtain a bound for $n$-th order connected correlations $c_n(A_1(x_1,t_1),\ldots, A_n(x_n,t_n))$ at different (not necessarily ordered) times. This bound requires a three element clustering property, for $ω \big( A B(x,t) C) - ω(AC) ω(B(x,t))$, which comes as a consequence of the Lieb-Robinson bound in states that satisfy the usual two element clustering for $ω(A(x,t) B) -ω(A(x,t))ω(B)$. This is discussed in \ref{section:three-element-clustering}. These results are complemented by Appendices \ref{appendix:lemma} and \ref{appendix:space-like}.

\subsection{Clustering for space-like translations and almost everywhere ergodicity} \label{susbsection:51}

The Lieb-Robinson bound  implies that $\mathfrak{U}$ is asymptotically abelian for space-time translations along any space-time ray that is outside the Lieb-Robinson light cone (\cite[Theorem 4.2]{ampelogiannis_2023_almost}):
\begin{equation} \label{space_like_abelianness}
    \lim_{\boldsymbol{x} \to \infty} \norm{ [ι_{\boldsymbol x} τ_{|\boldsymbol{x}|υ^{-1}}A,B]} \to 0     , \ υ> υ_{LR}, \ A,B \in \mathfrak{U}.
\end{equation}
More specifically, this is asymptotic abelianness of the algebra $\mathfrak{U}$ with respect to the group of space-time translations along a ray $G_{υ,\boldsymbol x} \coloneqq  \{ (r\boldsymbol x, rυ^{-1}|\boldsymbol x|) : r \in \Z \} \subset \Z^D \times \R$, for any velocity with $|\boldsymbol{v}|>v_{LR}$ and $\boldsymbol{x} \in \Z^D$.

Therefore one concludes that factor states are clustering with respect to space-time translations outside the Lieb-Robinson light-cone:
\begin{equation}
    \lim_{\boldsymbol{x} \to \infty} \big( ω( ι_{\boldsymbol{x}} τ_{|\boldsymbol{x}|υ^{-1}}A B) - ω(ι_x τ_{|\boldsymbol{x}|υ^{-1}}A) ω(B) \big) =0 , \ υ> υ_{LR}, \ A,B \in \mathfrak{U}.
\end{equation} 
This is discussed in detail in \cite{ampelogiannis_2023_almost} for factor invariant states and short-range interactions, but generalising the proof to factor states and long-range is immediate. This is because long-range of the interaction does not change \Cref{space_like_abelianness}, and the proofs of clustering of factor states from asymptotic abelianness, on which  \cite{ampelogiannis_2023_almost} relies, are done in  (\cite[Theorem 8]{kastler_invariant_1966}), (\cite[Example 4.3.24]{bratteli_operator_1987}) without requiring  invariance of the state.  Therefore, by \Cref{th:cumulants_general} the higher order classical and free cumulants also satisfy the clustering property in factor states:
\begin{corollary} \label{eq:qsl_clustering}
Consider a quantum lattice $C^*$ dynamical system $(\mathfrak{U},ι,τ)$ with either short-range (finite range or exponentially decaying) or power-law decaying ($a>2D+1$)  interaction and a factor state $ω$ of $\mathfrak{U}$. It follows,  by  \cite[Theorem 4.2]{ampelogiannis_2023_almost}, that $ω$ is space-like clustering 
and consequently the $n$-th free and classical cumulants are also space-like clustering:  
    \begin{equation}
    \lim_{t \to \infty} c_n ( ι_{\floor{ \boldsymbol{v}t}}τ_t A_1, A_2, \ldots, A_n)  =  0 , \ \forall \boldsymbol{v}\in \R^D: |\boldsymbol{v}| > υ_{LR}  , \forall A_1,A_2,\ldots,A_n \in \mathfrak{U}
\end{equation}
\end{corollary}

In the above corollary, and below, we use the notation $\floor{\boldsymbol{a}} \coloneqq ( \floor{a_1},...,\floor{a_D})$ for any vector $\boldsymbol{a}=(a_1,a_2,...,a_D) \in \R^D$.

Additionally, in the case of short-range interactions, if we consider exponentially space-clustering states, such as high temperature KMS states, we can combine the exponential bound on the commutator with  space-clustering to obtain clustering for space-like translations, as shown in \cite[Theorem C.1]{ampelogiannis_long-time_2023}. Thus, by \Cref{th:decay}, we ultimately get exponential space-like clustering for all cumulants in the short-range case.

\begin{corollary} \label{th:space-like-clustering}
Consider a quantum lattice $C^*$ dynamical system $(\mathfrak{U},ι,τ)$ with short-range (finite range or exponentially decaying) interaction and an exponentially space clustering  state $ω$ of $\mathfrak{U}$. It follows, by \cite[Theorem C.1]{ampelogiannis_long-time_2023}, that $ω$ is exponentially space-like clustering for some $λ>0$:
\begin{equation}
    \lim_{t \to \infty} e^{λt} \big( ω\big( ι_{\floor{ \boldsymbol{v}t}}τ_t(A) B \big) - ω(ι_{\floor{ \boldsymbol{v}t}}τ_tA) ω(B) \big) = 0 , \forall \boldsymbol{v}\in \R^D: |\boldsymbol{v}| > υ_{LR} ,\ A,B \in \mathfrak{U}_{\rm loc}
\end{equation}
Consequently, all $n$-th cumulants are clustering w.r.t space-like translations:
\begin{equation}
    \lim_{t \to \infty}e^{λt} c_n (  ι_{\floor{ \boldsymbol{v}t}}τ_tA_1, A_2, \ldots, A_n)  = 0  ,\forall \boldsymbol{v}\in \R^D: |\boldsymbol{v}| > υ_{LR}
\end{equation}
for all $A_1, A_2,\ldots , A_n \in \mathfrak{U}_{\rm loc}$, and the same is true for the free cumulants.
\end{corollary}

A weaker form of clustering, ergodicity, can also be useful in the study of quantum lattice models. A recent result in (short-range) quantum lattice models is almost-everywhere ergodicity \cite{ampelogiannis_2023_almost}, which shows that for almost every $υ \in \R$ the long-time averaging of connected correlations over a space-time ray vanishes,
\begin{equation}
    \lim_{T \to \infty} \frac{1}{T} \int_0^T  ω \bigg( ι_{\floor{ \boldsymbol{v}t}}τ_t (A) B \bigg)  \,dt = ω(A) ω(B). %\label{eq:maintheorem}
\end{equation}
Combining this result with \Cref{th:ergodicity} we show:

\begin{corollary} \label{th:almost-everywhere-clustering}
Consider a quantum lattice $C^*$ dynamical system $(\mathfrak{U},ι,τ)$ with short-range (finite range or exponentially decaying) translation invariant interactions  and a factor $ι,τ$-invariant state $ω$ of $\mathfrak{U}$. For almost every $υ\in \R$, and every lattice direction $\boldsymbol{q}= \frac{\boldsymbol{x}}{|\boldsymbol{x}|}$, $\boldsymbol{x}\in \Z^D$, the long-time averaging of any $n$-th cumulant (or any $n$-th free cumulant), over a space-time ray with velocity $\boldsymbol{v}=υ \boldsymbol{q}$, will vanish:
\begin{equation}
    \lim_{T \to \infty} \frac{1}{T} \int_0^T  c_n \bigg( ι_{\floor{ \boldsymbol{v}t}}τ_t A_1, A_2,\ldots,  A_n \bigg)  \,dt = 0.
\end{equation}
for all $A_1, A_2, \ldots, A_n \in \mathfrak{U}$. 
\end{corollary} 

A few remarks are in order as to which physically relevant states satisfy the above  corollaries \ref{eq:qsl_clustering}, \ref{th:space-like-clustering}. In short-range interaction quantum spin chains, $D=1$, we have two cases. For finite range interactions any KMS state at non-zero temperature satisfies exponential clustering, by the result of Araki \cite{Araki:1969bj}. For exponentially decaying interactions Araki's result has been extended, but only for high enough temperature \cite{perez_2023_locality}. At higher dimensions, it is shown that high temperature KMS states  are exponentially clustering \cite{frohlich_properties_2015}. These cases fall into the single phase regimes of QSL and will satisfy Corollary \ref{th:space-like-clustering}. For lower temperatures, where multiple phases exist\footnote{The KMS state is not unique, but the KMS states of fixed temperature form a convex set.}, clustering, not necessarily exponential, still holds in space-invariant extremal KMS states, as these are factor (\cite[Theorem 5.3.30]{bratteli_operator_1997}), hence these will satisfy \Cref{eq:qsl_clustering}. Note that Corollaries \ref{eq:qsl_clustering}, \ref{th:space-like-clustering} hold for interactions that are not necessarily translation invariant, while  Corollary \ref{th:almost-everywhere-clustering} requires translation invariance and that the state is space and time invariant, which is automatically true for KMS states  in the single phase regime, otherwise it has to be added as an assumption at lower temperatures.

\subsection{Bound on $n-$th order connected correlations} \label{subsection:52}

In applications, it might be useful to have a more explicit expression of clustering, in terms of a bound. For example, see \cite{doyon_2019_diffusion}, where a Lieb-Robinson type bound on the $n$-th order cumulants is used to obtain a lower bound on diffusion. Our goal here is to obtain such a bound. In the previous results in QSL we only had one element separated from all others, but what is the relevant space-time distance that bounds correlations when we want to talk about the connected correlation $   c_n\big( A_1(x_1,t_1) , \ldots , A_n(x_n,t_n) \big)$?  We will see that $ c_n\big( A_1(x_1,t_1) , \ldots , A_n(x_n,t_n) \big)$ is controlled by the max-min of the  distances $z_{ij} \coloneqq \dist(A_i(x_i),A_j(x_j))$, whenever $t_k < υ^{-1} \max_i \min_j \{ z_{ij} \}$  with $υ>υ_{LR}$, for all $k$, so that whenever at least one observable is outside the light-cones of all others, the cumulant goes to zero.

For clarity we use the notation $A(x,t) \coloneqq ι_x τ_t A$ for $A\in \mathfrak{U}$. Assuming that $ω \big( A B(x,t) C) \\- ω(AC) ω(B(x,t))$ is bounded by $1/(\dist(B(x),AC))^p$, we will show that higher order connected correlations (and free cumulants) $ c_n\big( A_1(x_1,t_1) , \ldots , A_n(x_n,t_n) \big)$ are bounded by $1/(max_i \{ min_j \{ z_{ij} \} \})^{(p-rD)}$, where $D$ is the lattice dimension and $r$ the polynomial degree of the dependence of clustering on the support sizes of local observables, see \Cref{defn:sizeable}.  The proof is done in \Cref{section:proof_4}. Note that the assumed three-element cluster property will be true in any state that is clustering in space with rate $1/(1+x)^{p+rD}$, $p>1$, which will be space-like clustering (see \Cref{appendix:space-like}) with rate $1/(1+x)^p$ as a consequence of the Lieb-Robinson bound. This is discussed in \Cref{section:three-element-clustering}.

\begin{theorem} \label{th:n-order-lieb-robinson}
    Consider a short-range quantum lattice $C^*$ dynamical system $(\mathfrak{U},ι,τ)$ and a state  $ω$ for which there exist  $p,r>1$ with $p-rD>1$, such that for every $A,B,C\in \mathfrak{U }_{\rm loc}$,$υ>υ_{LR}$, $x\in \Z^D , t \in -υ^{-1}[ -\dist(B(x),AC), \dist(B(x),AC) ]$:
\begin{equation} \label{eq:three-element-LR-clustering}
   |ω \big( A B(x,t) C) - ω(AC) ω(B(x,t))| \leq  \frac{C_2(B,AC)}{\big(1+ \dist(B(x), AC)\big)^p}   
%  \leq C_2(A_1,\ldots,A_n) \frac{1}{ (1+ \min_i \{z_{mi} \} )^p} \text{, for any } m\leq n
\end{equation}
where $C_2(A,B)=u\norm{A} \norm{B} P_r( |Λ_A|, |Λ_B|)$ with $u>0$ constant that depends on $υ$, $P_r$ a polynomial of degree $r$ and $|Λ_A |$ the size of the support of $A$.

It follows that for any $n \in \N$ the $n$-th joint cumulant of $A_1,\ldots,A_n \in \mathfrak{U}_{loc}$ satisfies the following clustering property for every $υ > υ_{LR}$:
\begin{equation}  \label{eq:th:n-order-lieb-robinson}
    c_n\big( A_1(x_1,t_1) , \ldots , A_n(x_n,t_n) \big) \leq  \frac{C_n(A_1,\ldots, A_n)}{\big (1+ \max_i \min_j \{\dist(A_i(x_i), A_j(x_j) \} \big)^{p-rD}}
\end{equation}
for $x_1,\ldots,x_n \in \Z^D$, $\displaystyle  t_1,\ldots,t_n \in [-υ^{-1}z+1,υ^{-1}z-1]$, where $ z \coloneqq \max_i \min_j \{\dist(A_i(x_i), A_j(x_j) \}$ and $ C_n(A_1,\ldots, A_n)$ is a function of the $C_2$ and $υ$. This is also true for the free cumulants.
\end{theorem} 

To show \Cref{eq:th:n-order-lieb-robinson} from \Cref{eq:three-element-LR-clustering} we need the following technical lemma. Its proof is discussed in \Cref{appendix:lemma}.
\begin{lem} \label{lemma}
    Consider the assumptions of \Cref{th:n-order-lieb-robinson}. For any $n,m\in \N$ with $m \leq n$, $A_1,\ldots,A_n \in \mathfrak{U}_{\rm loc}$, $x_1,\ldots, x_n \in \Z^D$ and $υ> υ_{LR}$, it follows that 
    \begin{eqnarray}
      \big | ω \big( A_1(x_1,t_1) \cdots  A_n(x_n,t_n) \big) - ω(A_m(x_m,t_m) ) ω\big( \prod_{j\neq m} A_j(x_j,t_j) \big) \big|\\  
        \leq C^{\prime}_2(A_1,\ldots,A_n) \frac{1}{ (1+ \min_i \{z_{mi} \} )^{p-rD}} 
    \end{eqnarray}
for $\displaystyle  t_1,\ldots,t_n \in υ^{-1}[-\min_{i \neq m} \{ z_{mi} \}, \min_{i \neq m} \{ z_{mi} \}] $, where  $  z_{mi} \coloneqq \dist( A_m(x_m), A_i(x_i) )$.
\end{lem}

If the state $ω$ is exponentially clustering, then it satisfies the Theorem for every $p>1$, thus its higher-order correlations will also be exponentially clustering. The case of long-range interactions will yield a similar result, that will be affected by the degree $a$ of power-law decay.

\subsection{Three element clustering property} \label{section:three-element-clustering}
 
Consider the general set-up introduced in \Cref{section:set-up}. A state $ω$ satisfies the three-element clustering property when  
\begin{equation}
  \lim_i \big(  ω( A α_{g^i}(B) C) - ω(AC) ω(a_{g^i}B \big) =0  ,\, \forall A,B,C \in \mathfrak{U}
\end{equation}
This property is important for proving more general clustering properties of higher order connected correlations, as was the case in \Cref{section:results1}. It is established that in an asymptotically abelian algebra the two-element and three-element clustering properties are equivalent, see  \cite[Section 4.3.2]{bratteli_operator_1987}. This is shown by considering the limit
\begin{equation}
    \lim_i ω( A [α_{g^i}B,C] ) = 0 
\end{equation}
which is $0$ by asymptotic abelianness. 

Consider now the set-up of QSL with a short-range interaction. We will show that a state satisfying space-like clustering (between two observables)  will satisfy the three-element clustering property with the same rate. Suppose $ω$ is a state that is space-like $q$-clustering for some $q>1$; for $υ>υ_{LR}$, $A,B\in \mathfrak{U}_{\rm loc}$ and any $x \in \Z^D$, $t \in υ^{-1} [-\dist(A(x),B), \dist(A(x),B)]$  we have the bound:
\begin{equation} \label{eq:qsl_spacelike_clustering}
        |ω( A(x,t) B) - ω(A) ω(B) | \leq C_2(A,B) \frac{1}{(1+ \dist(A(x),B))^{q}} 
\end{equation}
See Appendix \ref{appendix:space-like} for more details on how one obtains a space-like clustering bound. By the Lieb-Robinson bound we also get
\begin{equation}
    |ω(A [B(x,t),C)])| \leq \norm{A} \norm{[B(x,t),C)]} \leq L_{B,C} \norm{A}e^{-λ( \dist(B(x),C)- υ_{LR}|t|)} 
\end{equation}
and we also have that
\begin{eqnarray}
    |ω(A [B(x,t),C)])|  &\geq& |ω(A B(x,t) C) - ω(AC)ω(B) |  
    - |ω(AC B(x,t) ) -ω(AC)ω(B) |
\end{eqnarray}
Combining the above two inequalities:
\begin{eqnarray}
     |ω(A B(x,t) C) - ω(AC)ω(B) | &\leq&  |ω(AC B(x,t) ) -ω(AC)ω(B) | 
     + L_{B,C} \norm{A}e^{-λ( \dist(B(x),C)- υ_{LR}|t|)} 
\end{eqnarray}
and the first term in the right-hand side is bounded by \Cref{eq:qsl_spacelike_clustering}, while the exponential decay is dominated by the polynomial. Therefore we obtain a three-point $q$-clustering property, with the same rate, of the form
\begin{eqnarray}
     |ω(A B(x,t) C) - ω(AC)ω(B) | \leq \frac{ C^{\prime}_2(B,AC) } {(1+\dist(B(x),AC))^{q}}
\end{eqnarray}
for $υ>υ_{LR}$, $A,B,C\in \mathfrak{U}_{\rm loc}$ and any $x \in \Z^D$, $t \in υ^{-1} [-\dist(B(x),AC), \dist(B(x),AC)]$.
\section{Proofs} \label{section:proofs}

\subsection{Proof of  \Cref{th:cumulants_general} and \Cref{th:decay}} \label{section:proof_1}
\begin{proof}
    We prove  \Cref{th:decay}, as  \Cref{th:cumulants_general} follows by choosing $f(g) = 1$.
    We will prove the claim by induction. For $n=2$ the (scaled) second cumulant $f(g^i)κ_2( α_{g^i} A, B) = f(g^i)\,(ω(α_{g^i}( A )B) - ω(α_{g^i}A) ω(B))$ does indeed vanish by the assumption. Suppose \Cref{eq:maindecay} is true for every $m\leq n$, for some $n\in \N$. We want to show that this implies \Cref{eq:maindecay} for $n+1$.
 
    Consider the moments-to-cumulants formula \Cref{eq:moments_to_cumulants}:
    \begin{equation} \label{eq:proof0}
        ω_{n+1}(α_{g^i}A_1,A_2, \cdots, A_{n+1}) = \sum_{π \in NC(n+1)} κ_π (α_{g^i}A_1,\cdots,A_{n+1}).
    \end{equation}
On the r.h.s. the sum $\sum_{π \in NC(n+1)}$ has only one maximal partition of size $1$, the partition $\{1,2,\cdots ,n+1\}$. The maximal partition corresponds to $κ_{n+1}$, which we want to show vanishes. We write
\begin{equation} \label{eq:rhs}
 \begin{array}{*3{>{\displaystyle}lc}p{5cm}}
    r.h.s. \coloneqq \sum_{π \in NC(n+1)} κ_π (α_{g^i}A_1,\cdots,A_{n+1}) &=&κ_{n+1} (α_{g^i}A_1,\cdots,A_{n+1})
    + \sum_{π \in NC(n+1), |π|\geq 2} κ_π (α_{g^i}A_1,\cdots,A_{n+1})
    \end{array}
\end{equation}
and then rearrange the terms in \Cref{eq:proof0}:
\begin{equation} \label{eq:proof0.1}
 \begin{array}{*3{>{\displaystyle}lc}p{5cm}}
 κ_{n+1}(α_{g^i}A_1,\cdots,A_{n+1}) = ω_{n+1}(α_{g^i}A_1,A_2, \cdots, A_{n+1}) 
    - \sum_{π \in NC(n+1), |π|\geq 2} κ_π (α_{g^i}A_1,\cdots,A_{n+1}).
    \end{array}
\end{equation}          
The set $K=\{π \in NC(n+1), |π|\geq 2\}$ is a union of two disjoint sets: those partitions where the element $1$ is a singleton $K_1=\{π \in NC(n+1), |π|\geq 2 \wedge \{1\}\in π\}$ and those were it is in a block of size at least $2$, $K_2=\{π \in NC(n+1), |π|\geq 2 \wedge (\exists V \in π : 1\in V, |V|\geq 2)\}$.

Hence:
\begin{equation}\label{eq:proof1}
 \begin{array}{*3{>{\displaystyle}lc}p{5cm}}
    \sum_{π \in NC(n+1), |π|\geq 2} κ_π (α_{g^i}A_1,\cdots,A_{n+1}) =\ \sum_{π\in K_1} κ_π (α_{g^i}A_1,\cdots,A_{n+1}) 
    +\ \sum_{π\in K_2} κ_π (α_{g^i}A_1,\cdots,A_{n+1}).
       \end{array} 
\end{equation}
We note that, as a consequence of the general recursion relation \eqref{eq:recursionNC} on non-crossing partitions, we have the recursion relation $K_1= \{ \{ \{1\}\}\cup \pi: \pi\in NC(2,\cdots,n+1)\}$, where $NC(2,\cdots,n+1)$ is the set of non-crossing paritions of $\{2, 3,\cdots, n+1\}$. Hence, as $κ_π$ is completely determined by the $κ_n$, \Cref{eq:multiplicative_free}, we obtain 
\begin{equation} \label{eq:proof2}
 \begin{array}{*3{>{\displaystyle}lc}p{5cm}}
    \sum_{π\in K_1} κ_π (α_{g^i}A_1,\cdots,A_{n+1}) &=& κ_1(α_{g^i}A_1) \sum_{π\in NC(2,\cdots,n+1)}   κ_π (A_2,\cdots,A_{n+1})\\
    &=&ω(α_{g^i}A_1)  \sum_{π\in NC(2,\cdots,n+1)}   κ_π (A_2,\cdots,A_{n+1}).
 \end{array}
\end{equation}
The second term in \Cref{eq:proof1} is:
\begin{equation}
 \begin{array}{*3{>{\displaystyle}lc}p{5cm}} \label{eq:proof1_sum2}
    \sum_{π\in K_2} κ_π (α_{g^i}A_1,\cdots,A_{n+1}) &=& \sum_{π\in K_2} \prod_{V \in π} κ_{|V|} (α_{g^i}A_1,\cdots,A_{n+1}|V). \\
 %   &=& \sum_{π\in K_2} \prod_{V \in π} κ_{|V|} (ι_xA_1,\cdots,A_{n+1}|V)
     \end{array}
\end{equation}
Since in every parition  $π \in K_2$ we have $1$ in a block of size at least $2$, every product in the above sum will contain a $V$ with $1\in V$ and $|V|\geq 2$ but $|V|\leq n$, since $|π|\geq 2$. That is, the product will contain a term $κ_m(α_{g^i}A_1, A_{i_{s_1}}, \cdots A_{i_{s_m}})$, $i_{s_1}< \cdots <i_{s_m}$, with $2\leq m+1\leq n$. By the induction hypothesis 
\begin{equation}
    \lim_{i} f(g^i) κ_m(α_{g^i}A_1, A_{i_{s_1}}, \cdots A_{i_{s_m}}) =0 , \forall 2\leq m+1\leq n \text{ and } i_{s_1}< \cdots <i_{s_m}.
\end{equation}
Therefore,
\begin{equation}
    \lim_{i}  f(g^i)\sum_{π\in K_2} κ_π (α_{g^i}A_1,\cdots,A_{n+1}) = 0.
\end{equation}
Thus, taking the limit of \Cref{eq:proof0.1} and using the moments-to-cumulants formula 
\begin{equation} \label{eq:lim_rhs}
 \begin{array}{*3{>{\displaystyle}lc}p{5cm}}
   && \lim_{i}f(g^i) κ_{n+1}(α_{g^i}A_1,\cdots,A_{n+1}) = \lim_i f(g^i)\bigg( ω_{n+1}( α_{g^i} A_1, A_2,\ldots,A_{n+1}) 
   \\ &-& ω( α_{g^i} A_1)  \sum_{π\in NC(2,\cdots,n+1)}   κ_π (A_2,\cdots,A_{n+1}) \bigg) \\
   &=&  \lim_i f(g^i)\bigg( ω_{n+1}( α_{g^i} A_1, A_2,\ldots,A_{n+1}) - ω( α_{g^i} A_1) ω_{n}(A_2,\ldots,A_{n+1} )\bigg) \\
   &=&\lim_i f(g^i) \bigg( ω\big( α_{g^i} (A_1) \prod_{j=2}^{n+1} A_j\big) - ω( α_{g^i} A_1) ω(\prod_{j=2}^{n+1} A_j)\bigg) \\
   &=& 0
\end{array}
\end{equation}
where the limit is zero by $f(g)$-clustering of the second cumulant between the observables $α_{g^i}A_1$ and $\prod_{i=2}^{n+1} A_i$. This concludes the proof. The proof is identical for $ \lim_{i} κ_n( A_1 ,  \ldots, α_{g^i}A_n) = 0$.

If we additionally assume that  $\lim_{i}  \big(ω(  A \alpha_{g^i}(B) C) -  ω(α_{g^i} B) ω(AC)\big) = 0$, then we can show that $ \lim_{i} κ_n( A_1 ,  \ldots, α_{g^i} (A_m), \ldots, A_n) \\= 0 $, for any $1<m<n$. Follow the same steps as above, the sum over $K_1$ will again cancel out in the limit by the additional clustering property. Each partition in the sum over $K_2$, \Cref{eq:proof1_sum2}, will contain a block $V$ with $m\in V$ and at least one other element
\begin{equation}
        \sum_{π\in K_2} κ_π (A_1 ,  \ldots, \alpha_{g^i} (A_m), \ldots, A_n)= \sum_{π\in K_2} \prod_{V \in π} κ_{|V|} (A_1 ,  \ldots, \alpha_{g^i} (A_m), \ldots, A_n|V).
\end{equation}
And since $m$ can be in any of the first $i\leq m$ positions in $κ_{|V|}$, we have to modify the induction hypothesis so that \\$ \lim_{i} κ_l( A_1 ,  \ldots, \alpha_{g^i} (A_m), \ldots, A_l) = 0 $ for every $l\leq n$ and all $m<l$, and then show that it's also true for $n+1$. This induction hypothesis implies that the sum over $K_2$ vanishes and concludes the proof. 
\end{proof} 

\subsection{Proof of \Cref{th:clustering_for_groups} } \label{section:proof_3}
The proof is done inductively in $m,n$. We define the proposition $P(m,n)$ as follows, $P(n,m):$
\begin{equation}
    (\forall A_1,  \ldots, A_n \in \mathfrak{U}) [\lim_i κ_n(  α_{g^i} (A_1),  α_{g^i} (A_2),\ldots,  α_{g^i} (A_m), A_{m+1}, \ldots, A_n) =0]
\end{equation}

By \Cref{th:cumulants_general}, $P(1,n)$ is true for every $n\geq 2$. We will prove inductively that $P(n-1,n) \implies P(n,n+1)$, hence $P(n,n+1)$ will be true for all $n$. Afterwards, we will prove that $(\forall m<n)P(m,n) \implies (\forall m<n)P(m,n+1)$ \footnote{This means that if $P(m,n)$ is true for all $m<n$, then $P(m,n+1)$ holds for $m<n$}.  These two implications show that $P(m,n)$ is true for any $n,m$, with $m<n$, since a base case $P(1,2)$ is true, $P(1,n)$ for all $n$ is true and $P(m,m+1)$  for every $m$. Then all intermediate cases $P(m,n)$ with $1<m <n-1$ follow from $(\forall m<n)P(m,n) \implies (\forall m<n)P(m,n+1)$. The figure bellow illustrates how we navigate the double induction for the proposition $P(m,n)$.
%%%%%%%%%%%%%%%%%%%%%%%%%%%%%%%%%%%%%%%%%%%%%%%%%%%%%%%%%%%%%%%%%%%%%%%%%%%%%%%%%%%%%%%%%%%%%%%%%%%%%%
{
\setlength{\arraycolsep}{10pt} % Default value: 6pt
\renewcommand{\arraystretch}{2} % Default value: 1
\begin{minipage}{0.5\textwidth}
\[
\begin{NiceMatrix}
    (1,2)     &  (1,3)    & (1,4)  & (1,5)  & \ldots   \\
      \cross  & (2,3 )    & (2,4)  & (2,5)   &           \\
      \cross  &   \cross  & (3,4)  & (3,5)   &           \\
      \cross  &   \cross  & \cross &   (4,5)  &      \ldots   
\CodeAfter
\begin{tikzpicture}
\begin{scope} [->, thick]
\draw (1-1) -- (1-2);
\draw (1-2) -- (1-3);
\draw (1-3) -- (1-4);
%\draw (1-4) -- (1-5);
{\color{red}\draw (1-1) -- (2-2) ; 
\draw (2-2) -- (3-3) ;
\draw (3-3) -- (4-4) ;
%\draw (4-4) -- (5-5) ;
}
{\color{blue}
\draw (2-2.east) to (2-3.west) ;
\draw (1-2.east) to [bend right=45] (2-3.west) ;
\draw (1-1.east) to (2-3.west) ;
%\draw (3-4.east) to (3-5.west);
}
\end{scope}
\end{tikzpicture}
\end{NiceMatrix}
\]
\end{minipage}
\begin{minipage}{0.5\textwidth}
$\rightarrow$ \text{by } $P(1,n) \implies P(1,n+1) $  \\
\\
 {\color{blue}$\rightarrow$  \text{by } $(\forall m<n)P(m,n) \implies \\ (\forall m<n)P(m,n+1)$} \\
 \\
{\color{red}$\rightarrow$ \text{by } $P(n-1,n) \implies P(n,n+1) $}
\end{minipage}
}

\begin{center}
Figure 1: Implications used in double induction for $P(m,n)$
\end{center}

%%%%%%%%%%%%%%%%%%%%%%%%%%%%%%%%%%%%%%%%%%%%%%%%%%%%%%%%%%%%%%%%%%%%%%%%%%%%%%%%%%%%%%%%%%%%%%%%%%%%%

In the spirit of the proof of \Cref{th:cumulants_general},  we write
\begin{equation}  \label{eq:proof2_1}
 \begin{array}{*3{>{\displaystyle}lc}p{5cm}}
  κ_{n}^{(m)} \coloneqq κ_{n}(α_{g^i}A_1, α_{g^i}A_2,\ldots, α_{g^i}A_m, A_{m+1},\ldots,A_{n}) =ω_{n}(α_{g^i}A_1, \ldots, A_{n}) 
    - \sum_{π \in NC(n), |π|\geq 2} κ_π (α_{g^i}A_1,\cdots,A_{n})
    \end{array}
\end{equation}      
and we separate $\{π\in NC(n), |π|\geq 2\}$ into two disjoint subsets. The first subset contains all the partitions $π\in NC(n)$ with all $i \in \{1,\ldots,m\}$ belonging to different blocks from every $j\in \{m+1,\ldots,n\}$. The second subset is the complement of the first, where the $i$'s and the $j$'s mix.
\begin{equation}
    K_1 = \{π\in NC(n): |π| \geq 2, \forall V \in π : V \cap M = \emptyset \vee \ V\cap N = \emptyset \}
\end{equation}
where $M= \{1,2,\ldots,m\}$ and $N=\{m+1, \ldots,n\}$, and
\begin{equation}
  K_2= \{π\in NC(n): |π| \geq 2, \exists V \in π : V \cap M \neq \emptyset \wedge \ V\cap N \neq \emptyset \}
\end{equation}
Then \Cref{eq:proof2_1} becomes
\begin{equation}
 \begin{array}{*3{>{\displaystyle}lc}p{5cm}}
    κ_{n}^{(m)} =  ω_{n}(α_{g^i}A_1, \ldots, A_{n})   - \sum_{π \in K_1} κ_π (α_{g^i}A_1,\cdots,A_{n}) 
    - \sum_{π \in K_2} κ_π (α_{g^i}A_1,\cdots,A_{n})
    \end{array}
\end{equation}
First, note that $κ_π = \prod_{V \in π} κ_{|V|}$, so the sum over $K_1$ factorises into sums over partitions of $N$ and $M$ (note in particular that because $M$ and $N$ are each composed of contiguous sites, the non-crossing condition indeed factorises):
\begin{equation}
     \sum_{π \in K_1} κ_π = \sum_{π \in NC(1,\ldots,m)} κ_π (α_{g^i}A_1,\ldots ,α_{g^i}A_m) \sum_{π \in NC(m+1,\ldots,n)} κ_π (A_{m+1},\ldots ,A_{n}) 
\end{equation}
and by the moments to cumulants formula this is equal to a product of the respective joint moments:
\begin{equation}
     \sum_{π \in K_1} κ_π = ω(α_{g^i}A_1,\ldots ,α_{g^i}A_m) ω(A_{m+1},\ldots ,A_{n})
\end{equation}
which combined with the term $ ω_{n}(α_{g^i}A_1, \ldots, A_{n})$ will vanish in the limit by two point clustering. 

Using this, let us first prove $P(n,n+1)$,$\forall n \in \N$ by induction; suppose that $P(l-1,l)$ is true for every $l\leq n$. Then, take the limit of $κ^{(n)}_{n+1}$ in \Cref{eq:proof2_1} where the limit of the sum over $K_1$ vanishes, and we are left with $K_2$:
\begin{equation}
    \lim_i κ^{(n)}_{n+1}  =\lim_i\sum_{π \in K_2} \prod_{V \in π} κ_{|V|} (α_{g^i}A_1,\cdots, α_{g^i}A_n, A_{n+1}|V)
\end{equation}
In any partition $π\in K_2$, by definition of $K_2$, the block $V_1$ of $π$ that contains $n+1$ must also contain at least one element of $\{1,\ldots,n\}$, but not all because $π \in K_2$ has at least two blocks. Therefore the product will have a cumulant $κ_l^{(l-1)}$ of some order $l<n+1$ and $l-1$ translated observables. By the induction hypothesis the limit of every term in the sum above must vanish. Therefore $P(n,n+1)$ is true.

Finally, it remains to show $P(m,n) \implies P(m,n+1)$. Consider again by \Cref{eq:proof2_1} the limit of  $κ_{n+1}^{(m)}$ and assume the induction hypothesis that $P(m,n)$ is true for any $l \leq n$ and any $m< l$. The limit for the terms in $K_1$ vanishes, while the limit of the terms in $K_2$ vanishes by the induction hypothesis: the cumulants in the sum are of order less than $n+1$ and at most $m$ translated observables pair with non-translated ones. Therefore $(\forall m <n)P(m,n) \implies (\forall m <n)P(m,n+1)$, and this concludes the proof.

\subsection{Proof of higher order mean clutering} \label{section:proof_2}

The proof closely follows that of \Cref{th:cumulants_general}. By assumption, for $n=2$ we have:

\begin{equation}
    \lim_i \frac{1}{μ(U_i)} \int_{U_i} κ_2( α_{g}A , B) \,dμ(g) = 0  , \ \forall A,B \in \mathfrak{U}.
\end{equation}
We assume this holds for every $m \leq n$, for some $n >2$, and proceed to prove the theorem by induction. Consider \Cref{eq:proof0.1} and instead of taking its limit we take the limit of the avrage $\lim_i \frac{1}{μ(U_i)} \int_{U_i}$ and follow the same steps as before. We split the sum over $π\in NC(N+1)$, $|π|\geq 2$ into $K_1$ and $K_2$ and use \Cref{eq:proof2}

\begin{equation} 
\renewcommand{\arraystretch}{2.5}
 \begin{array}{*3{>{\displaystyle}l}p{5cm}}
   & \lim_{i}  \frac{1}{μ(U_i)} \int_{U_i} κ_{n+1}(α_{g}A_1,\cdots,A_{n+1}) dμ(g) =
    \lim_i  \frac{1}{μ(U_i)} \int_{U_i} \bigg( ω_{n+1}( α_{g} A_1, A_2,\ldots,A_{n+1})  \\
   &- ω( α_{g} A_1)  \sum_{π\in NC(2,\cdots,n+1)}   κ_π (A_2,\cdots,A_{n+1}) \bigg) \,dμ(g) 
    -  \lim_i  \frac{1}{μ(U_i)} \int_{U_i}\sum_{π\in K_2} κ_π (α_{g}A_1,\cdots,A_{n+1}) \,dμ(g) \end{array}
\end{equation}

By the induction hypothesis and the same reasoning as before, the sum over $K_2$ is zero, while the first term is also zero by the assumption of mean clustering for $n=2$ between $A_1$ and $B= A_2 A_3 \ldots A_{n+1}$.

\subsection{Proof of n-th order Lieb-Robinson bound} \label{section:proof_4}
Consider $c_n(A_1(x_1,t_1),\ldots,A_n(x_n,t_n))$, write $z_{ij}= |x_i-x_j| $ and  $z \coloneqq \max_i \min_j \{ z_{ij} \}$. Following the proof of \Cref{th:cumulants_general}, we write the $n-$th cumulant as in \Cref{eq:proof0.1}. We single out the observable that is furthest from the rest. Let $1\leq m,l \leq n$ be such that  the values $i=m,\,j=l$ achieve the max-min of $z_{ij}$ (in particular, $z_{ml} = z$), and suppose $m$ is the one such that $\min_j \{ z_{mj} \} = z$\footnote{This has to be true for at least one of $m$ or $l$, since they achieve the max-min, then for at least one of the two their minimum distance from all others has to be the maximum of all other minimum distances.}. We then proceed to split the partitions with $|π| \geq 2$ into $K_1$ that contains the partitions in which $m$ is a singleton, and $K_2$, as in the original proof:
\renewcommand{\arraystretch}{1.5}   
\begin{equation} \label{eq:proof4_1}
 \begin{array}{*3{>{\displaystyle}lc}p{5cm}}
 && c_{n}(A_1(x_1,t_1),\ldots,A_n(x_n,t_n)) = ω_{n+1}(A_1(x_1,t_1),\ldots,A_n(x_n,t_n)) \\
    &-& ω(A_m(x_m,t_m))  \sum_{π\in P( \{1,\ldots,n\} \setminus \{m\})}   c_π (A_1(x_1,t_1),\ldots, \hat{A_m}, \ldots,A_n(x_n,t_n)) 
    - \sum_{π\in K_2}   c_π (A_1(x_1,t_1),\ldots, A_n(x_n,t_n))
    \end{array}
\end{equation}
where $K_2=\{π \in P(n), |π|\geq 2 \wedge (\exists V \in π : m\in V, |V|\geq 2)\}$.
The first two terms in the right-hand side will be bounded by the clustering assumption for  $\displaystyle  t_1,\ldots,t_n \in υ^{-1}[-\min_{i \neq m} \{ z_{mi} \}, \min_{i \neq m} \{ z_{mi} \}] $, using \Cref{lemma}, by
\begin{equation}
     \frac{C_2(A_1,\ldots,A_n) }{(1+min_i \{z_{mi}\})^{p-rD}} =\frac{C_2(A_1,\ldots,A_n) }{(1+z)^{p-rD}} 
\end{equation}
and by definition of $m$ we have $min_i \{z_{mi}\}= z$.
The terms in the sum over $K_2$ are products of cumulants of order at most $n-1$. Consider an arbitrary term $π\in K_2$, with $π = \{ V_1, \ldots , V_{|π|} \}$
\begin{equation} \label{eq:proof4_2}
  \prod_{V_i \in π}  κ_{|V_i|} (A_1(x_1,t_1),\ldots,A_n(x_n,t_n) |V_i ) 
\end{equation}
By definition of $K_2$, for every $π\in K_2$ there is a $V^{\prime}  \in π$ with $|V^{\prime} | \geq 2$ and $m \in V^{\prime} $; $V^{\prime}  = \{i_1 \ldots , i_{|V^{\prime} |} \} \ni m$. Thus, in \Cref{eq:proof4_2} we have a term  
\begin{equation} \label{eq:proof4_3}
   I_{V^{\prime}}\coloneqq κ_{|V^{\prime}|} ( A_{i_1} (x_{i_1},t_{i_1}), \ldots , A_m( x_m,t_m) , \ldots,  A_{i_{|V^{\prime} |}}(x_{i_{|V^{\prime} |}},t_{i_{|V^{\prime} |}}) )
\end{equation}
We assume the theorem is true for $1,2,\ldots, n-1$ and show by induction that it's true for $n$. In \Cref{eq:proof4_3} the cumulant is of order at most $n-1$, hence 
\begin{equation}
\begin{array}{*3{>{\displaystyle}lc}p{5cm}}
     I_{V^{\prime}}&\leq& C_{|V^{\prime}|}(A_{i_1}, \ldots, A_{i{|V^{\prime}|}} )  \frac{1}{ (1+ \max_{i \in V^{\prime}} \min_{j  \in V^{\prime}} \{ z_{ij} \} )^{p-rD}} \\
    &\leq& C_{|V^{\prime}|}(A_{i_1}, \ldots, A_{i{|V^{\prime}|}} )  \frac{1}{ (1+ z)^{p-rD}} .
        \end{array}
\end{equation}
Here we used $\max_{i \in V^{\prime}} \min_{j  \in V^{\prime}} \{ z_{ij} \} \geq \min_{j  \in V^{\prime}} \{ z_{mj} \}\geq \min_{j  \in \{1,\ldots,n\}}\{ z_{mj} \} = z_{ml} = z$.
The rest of the  terms in the product \ref{eq:proof4_2} will give a similar bound:
\begin{equation}
    I_{V_i} \leq C_{|V_i|}(A_j: j \in V_i) \frac{1}{ (1+ \max_{i \in V_i} \min_{j  \in V_i} \{ z_{ij} \} )^{p-rD} } \ \leq C_{|V_i|}(A_i : i \in V_i)  
\end{equation}
where $ C_{|V_i|}(A_j: j \in V_i)= C_{|V_i|}(A_{i_1},\ldots, A_{i_{|V_i|}})$ for the indices in $V_i$. Therefore the term in \Cref{eq:proof4_2} will be bounded by a product of the constants
\begin{equation}
    C_n({A_1,\ldots, A_n} )\coloneqq \prod_{V_i \in π} C_{|V_i|}(A_j : j \in V_i)
\end{equation}
times $\frac{1}{(1+z)^{p-rD}}$, which concludes the proof.

\section{Conclusion}
In this paper we discussed the clustering properties of higher order connected correlations, in the general setting of group actions on C$^*$ dynamical systems. We showed that the $n$-th order joint classical cumulants, and the free cumulants, follow the same clustering properties as those of the second order connected correlations. The proofs were based on induction and utilisation of the moments-to-cumulants formula. We then applied these results to quantum spin lattice models, where we also obtained a bound on the $n$-th order connected correlation $c_n(A_1(x_1,t_1),\ldots,A_n(x_n,t_n))$ outside the Lieb-Robinson light-cone, with respect to the intermediate distance $\max_i \min_j \{\dist(A_i(x_i), A_j(x_j)) \}$. An example of physically relevant states that will satisfy our results are high temperature Gibbs states of short-range QSL; these are well known to be exponentially clustering in space, and also exponentially clustering for space-time translations outside the Lieb-Robinson light cone \cite{ampelogiannis_long-time_2023}.

We expect that our results can be applied to obtain rigorous lower bounds on diffusion coefficients (in QSL), by using the results in \cite{doyon_2019_diffusion}. Another possible application can be found in establishing a hydrodynamic projection principle for higher order correlations \cite{doyon_2023_mft}. Finally, the clustering properties for free cumulants may be relevant to the recent connection between Free probability theory and the Eigenstate Thermalisation Hypothesis \cite{Pappalardi_2022_ETH_free,jindal_2024_free}. 

\section*{Acknowledgments}
This work has benefited from discussions with Theodoros Tsironis. BD is supported by EPSRC under the grant ``Emergence of hydrodynamics in many-body systems: new rigorous avenues from functional analysis", ref.~EP/W000458/1.  DA is supported by a studentship from EPSRC. 

%\section*{Conflict of Interest}
%The authors have no conflicts to disclose.

\begin{appendices}

\chapter{Multi-point clustering property} \label{appendix:lemma}
Here we give a proof of \Cref{lemma}, which follows the same ideas as the proofs of space-like clustering from space clustering and the Lieb-Robinson bound in \cite[Theorem 8.5]{doyon_hydrodynamic_2022} and \cite[Appendix C]{ampelogiannis_long-time_2023}. Consider local $A_1,\ldots,A_n \in \mathfrak{U}_{\rm loc}$ and $x_1,\ldots,x_n \in \Z^D$, $t_1,\ldots,t_n \in \R$ and the quantity
\begin{equation} \label{eq:appendixA_S}
     S \coloneqq ω \big( A_1(x_1,t_1)  \ldots  A_n(x_n,t_n) \big) - ω(A_m(x_m,t_m) ) ω\big( \prod_{j\neq m} A_j(x_j,t_j) \big)
\end{equation}
where $j\neq m$ indicates $j=1,\ldots,n$ except $m$. The proof relies on approximating the time evolved operators by local ones, by using the Lieb-Robinson bound, and then applying the assumption \Cref{eq:three-element-LR-clustering}. We use the expression \cite[Corollary 3.1]{sims_2011_LR}   for the Lieb-Robnson bound between local $A,B$, there exist $L,υ_{LR}>0$\footnote{L depends on the lattice structure and $υ_{LR}$ on the interaction} independent of $A,B$:
\begin{equation}\label{eq:liebrobinsonbound_detailed}
    \norm{[ τ_t (A) , B]} \leq L \norm{A}\norm{B} \min \{|Λ_A|,|Λ_B|\} \exp{-λ( \dist ( A,B) - υ_{LR}|t|)} 
\end{equation}
We then use \cite[Corollary 4.4]{nachtergaele_quasi-locality_2019} and the Lieb-Robinson bound to obtain  for any local $A\in \mathfrak{U}_{Λ_A}$, supported on $Λ_A \subset Z^D$, an approximation of its time evolution $A(t)$ by a  sequence of local observables $A^\nu(t) \in \mathfrak{U}_{Λ_\nu}$ supported on:
\begin{equation} \label{eq:approximation_support}
    Λ_\nu \coloneqq \cup_{x \in Λ_A}B_{x}(\nu), \, \nu=1,2,3, \dots
\end{equation}
where $B_x(\nu)$ is the $D-$ball of radius $\nu$ around $x$:
\begin{equation} \label{eq:localapproximation}
    \norm{A^\nu(t) - A(t)} \leq  L \norm{A}  |Λ_A|  \exp\{-λ(\nu- υ_{LR}|t|)\} 
\end{equation}
with $\norm{A^{\nu}(t)}=\norm{A(t)}=\norm{A}$.

We approximate all observables $A_i(x_i,t_i)$, except $A_m$, by their local approximations, denoted by $A_{i}^{\nu_i}$ for simplicity:
\begin{equation} 
    \norm{A_{i}^{\nu_i} - A_i(x_i,t_i)} \leq  L \norm{A}  |Λ_A|  \exp\{-λ(\nu- υ_{LR}|t_i|)\} ,\, \nu_i=1,2,\ldots
\end{equation}
for $i \in \{1,2,\ldots, n \} \setminus\{m\}$.

Consider now $μ\in \N$ and the quantity of interest approximated by local observables:
\begin{equation}
  S_μ \coloneqq  ω( \prod_{i=1}^{m-1} A_{i}^{μ} \, A_m(x_m,t_m) \, \prod_{i=m+1}^{n} A_{i}^{μ}) - ω(A_m(x_m,t_m)) ω(\prod_{i\neq m} A_{i}^{μ})
\end{equation}
where $\lim_{μ \to \infty} S_μ=S$ is what we have to bound, in order to prove the Lemma. Now let $\nu\in \N$, that we will specify later, with $\nu < μ$ and write:
\begin{equation} \label{eq:appendixA_1}
\begin{array}{*3{>{\displaystyle}lc}p{5cm}}
  S_μ &=& ω\bigg( \prod_{i=1}^{m-1} ( A_{i}^{μ} + A_{i}^{\nu}- A_{i}^{\nu})   \, A_m(x_m,t_m) \, \prod_{i=m+1}^{n} (A_{i}^{μ}+A_{i}^{\nu}-A_{i}^{\nu} ) \bigg) 
  - ω(A_m(x_m,t_m)) ω\big(\prod_{i\neq m} (A_{i}^{μ}+A_{i}^{\nu} - A_{i}^{\nu}) \big) \\
  &=& ω( \prod_{i=1}^{m-1} A_{i}^{\nu} \, A_m(x_m,t_m) \, \prod_{i=m+1}^{n} A_{i}^{\nu}) - ω(A_m(x_m,t_m)) ω(\prod_{i\neq m} A_{i}^{\nu}) \\
  &+& ω\big(\sum_{\substack{a_1,\ldots,a_n\in \{0,1\}^n \\ \neq (1,\ldots,1)}}\prod_{i=1}^n ζ(i,a_i)\big)  -  ω(A_m(x_m,t_m)) ω\big(\prod_{i\neq m} (A_{i}^{μ} - A_{i}^{\nu}) \big)  
\end{array}
\end{equation}
where $ζ(m,0)= ζ(m,1)= A_m(x_m,t_m)$, while $ζ(i,0)= A_i^{μ}- A_{i}^{\nu}$ and $ζ(i,1)=A_{i}^{\nu}$ for all $i \neq m$. The sum is over all $n-$tuples $a_1,\ldots,a_n$ with $a_i\in \{0,1\}$, except $a_i=1$ for all $i$, indicating whether the distributive property of the product picks $A_i^μ - A_{i}^{\nu}$ or  $A_i^{\nu}$.  The last two terms in \Cref{eq:appendixA_1} are bounded as:
\begin{equation}
  |  ω(A_m(x_m,t_m)) ω\big(\prod_{i\neq m} (A_{i}^{μ} - A_{i}^{\nu}) \big)  | \leq \norm{A_m} \norm{\prod_{i\neq m} (A_{i}^{μ} - A_{i}^{\nu})}
\end{equation}
and
\begin{equation} \label{eq:appendixA_2}
\begin{array}{*3{>{\displaystyle}lc}p{5cm}}
   \big| \sum_{\substack{a_1,\ldots,a_n\in \{0,1\}^n \\ \neq (1,\ldots,1)}} ω\big(\prod_{i=1}^n ζ(i,c_i)\big) \big| \leq \sum_{\substack{a_1,\ldots,a_n\in \{0,1\}^n \\ \neq (1,\ldots,1)}} \norm{\prod_{i=1}^n ζ(i,c_i)}
\end{array}
\end{equation}
for all $ν,μ$.
These are both  exponentially small by \Cref{eq:localapproximation} whenever $ν> υ_{LR}|t_i|$. In particular \Cref{eq:appendixA_2}, we have  $\norm{ζ(i,1)}=\norm{A_{i}^{\nu}} \leq \norm{A_i}$ and (using $\nu < μ$)
\begin{equation} \label{eq:appendixA_3}
    \norm{ζ(i,0)} = \norm{ A_i^{μ} - A_{i}^{\nu}}   \leq  L \norm{A_i}  |Λ_{A_i}|  \exp\{-λ( \nu - υ_{LR}|t_i|)\} 
\end{equation}
and there is always a term $ζ(i,0)$ in each product since $(a_1,\ldots, a_n) \neq (1,\ldots,1)$. 

The two terms in the second line of \Cref{eq:appendixA_1} are of the form $ω(A B(x,t) C)- ω(B(x,t))ω(C)$, with $A,B,C$ local, hence we can use our main clustering assumption:

\begin{equation}\label{eq:appendixA_T1}
\begin{array}{*3{>{\displaystyle}lc}p{5cm}}
    S_{\nu} &\coloneqq& |ω\big( \prod_{i=1}^{m-1} A_{i}^{\nu} \,  A_m(x_m,t_m) \prod_{j=m+1}^n  A_{j}^{\nu} \big)- ω(A_m(x_m,t_m) ) ω\big( \prod_{j\neq m} A_{j}^{\nu}\big) | \\ &\leq& C_2(A_m, \prod_{j\neq m} A_j^{\nu})\frac{1}{\big(1+ \dist(A_m(x_m), \prod_{j\neq m} A_{j}^{\nu})\big)^p} 
     \end{array}
\end{equation}
The constant $C_2(A,B)$ is a polynomial, of degree $r$, of the sizes of the supports of $A,B$. For simplicity we will assume that $C_2(A,B)= u \norm{A} \norm{B} |Λ_A|^r |Λ_B|^r$. The support of $\prod_{j\neq m} A_j^{\nu}$ is the union of the supports of each observable in the product, hence:
\begin{equation} \label{eq:appendixA_estimate1}
     C_2(A_m, \prod_{j\neq m} A_j^{\nu}) \leq u \prod_{j=1}^n \norm{A_j} |Λ_{A_m}|^r \sum_{i\neq m} | Λ_{A_i^{\nu}}|^r
\end{equation}
And by definition of $A_i^{\nu}$, its support given by \Cref{eq:approximation_support}  is
\begin{equation} \label{eq:appendixA_estimate2}
    |Λ_{A_i^{\nu}}| \leq |Λ_{A_i} | \, |B_0(\nu)| \leq B_D |Λ_{A_i}| \nu^D
\end{equation}
Where $|B_0(\nu)|$ the size of the $D-$ball of radius $\nu$, which is a polynomial of degree $D$, hence bounded by $B_D \nu^D$, with $B_D>0$ a constant that depends on the lattice dimension.

The last thing to estimate is $\dist(A_m(x_m), \prod_{j\neq m} A_{j}^{\nu})$. The support of $A_{j}^{\nu}$ is the set that contains all points of the support of $A_j(x_j)$ and $D-$balls of radius $\nu$ around all those points.  Since the support of the product of observables is the union of their supports,  the support of $\prod_{j\neq m} A_{j}^{\nu}$ will be equal to the support of $(\prod_{j\neq m} A_{j}(x_j))^{\nu}$:
\begin{equation}
    \rm{supp}(\prod_{j\neq m} A_{j}^{\nu}) = \bigcup_{j\neq m}  \bigcup_{x \in \supp(A_j(x_j))}B_{x}(\nu) = \bigcup_{x \in \supp(\prod_{j\neq m}A_j(x_j))}B_{x}(\nu) \coloneqq A^{\nu}
\end{equation}
Therefore, as $A^{\nu}$ extends a radius $\nu$ around $\prod_{j\neq m}A_j(x_j)$, a simple geometric argument gives
\begin{equation}
    \dist(A_m(x_m),  \prod_{j \neq m} A_j^{\nu}) \geq \dist(A_m(x_m), \prod_{j \neq m} A_j(x_j)) - \nu 
\end{equation}
Again, since the support of the product is the union of the supports, we get
\begin{equation}\label{eq:appendixA_dist}
    \dist(A_m(x_m),  \prod_{j\neq m} A_j^{\nu}) \geq \min_{j\neq m}\{\dist(A_m(x_m),  A_j(x_j)) \} - \nu 
\end{equation}
With this in mind, we now specify  $\nu$ to be:
\begin{equation}
    \nu = \lfloor ε \min_{i\neq m} \{ \dist (A_m(x_m), A_i(x_i) )\} \rfloor
\end{equation}
for some $0<ε<1$ constant. Note that the estimate \Cref{eq:appendixA_3} now requires
\begin{equation}
    \lfloor ε \min_{i\neq m} \{ \dist (A_m(x_m), A_i(x_i) )\} \rfloor > υ_{LR} |t_j| , \, j=1,\ldots,n
\end{equation}
hence our final bound will be valid for times in the compact set:
\begin{equation}
    t_j \in  [ -  υ^{-1}\min_{i\neq m} \{ \dist (A_m(x_m), A_i(x_i) ) \}-1,  υ^{-1} \min_{i\neq m} \{ \dist (A_m(x_m), A_i (x_i) )\}-1] 
\end{equation}
where $υ= υ_{LR}/ε > υ_{LR}$. 
\begin{remark}
    Note that we have the freedom to choose any $0<ε<1$, and this affects two things. If we choose $ε$ near $1$ we will have a loose clustering bound, but for a large compact set of times, while $ε$ near $0$ yields a tighter bound, but only for short times.
\end{remark}

With this choice of $\nu$, and  using $\lfloor x \rfloor \leq x$,  \Cref{eq:appendixA_dist} gives
\begin{equation} \label{eq:appendixA_estimate3}
\begin{array}{*3{>{\displaystyle}lc}p{5cm}}
   \dist(A_m(x_m),  \prod_{j\neq m} A_j^{\nu}) \geq  (1-ε) \min_{j\neq m} \{ \dist (A_m(x_m), A_j(x_j) )\} 
\end{array}
\end{equation}
We denote $z \coloneqq \min_{j \neq m}\{ \dist(A_m(x_m), A_j(x_j) ) \}$. With this, we now return to \Cref{eq:appendixA_T1} and use \Cref{eq:appendixA_estimate1}, \Cref{eq:appendixA_estimate2} and  \Cref{eq:appendixA_estimate3}:
\begin{equation} \label{eq:appendixA_6}
    S_{\nu} \leq  u \prod_{j=1}^n \norm{A_j} |Λ_{A_m}|^r \sum_{i\neq m} |B_D Λ_{A_i}|^r \frac{ (εz)^{rD} }{(1+(1-ε)z)^p}
\end{equation}
for all $μ \in \N$ with $μ$ large enough.  
The rest of the proof is trivial. To summarize, in \Cref{eq:appendixA_1} the last two terms are bounded exponentially for times $t_i$ in the compact intervals specified above; the exponential bound is dominated by the power-law decay $\frac{1}{(1+ z)^q}$ for any $q>1$. The first two terms in \Cref{eq:appendixA_1} are bounded by $\frac{1}{(1+z)^{p-rD}}$, by \Cref{eq:appendixA_6}. These bounds are uniform in $μ$, hence we can take $μ \to \infty$ in  \Cref{eq:appendixA_1} and conclude the proof.

\chapter{Space-like clustering bound} \label{appendix:space-like}
Consider  the set-up of QSL with a short-range interaction and a state that is $p$-clustering in space
\begin{equation} \label{eq:qsl_p-clustering}
    |ω( A(x) B) - ω(A) ω(B) | \leq k_2(A,B) \frac{1}{\big(1+\dist(A(x),B)) \big)^p} 
\end{equation}
To obtain clustering for space-time translations, from clustering in space, we need to control the growth of $k_2(A,B)$ with respect to the support sizes of the observables. We call such states sizably clustering:

\begin{defn}[Sizable clustering] \label{defn:sizeable}
    A state $ω$ of a QSL dynamical system $(\mathfrak{U},ι,τ)$ is called $r$-sizably $p$-clustering if it satisfies \Cref{eq:qsl_p-clustering} with
    \begin{equation}
    k_2(A,B) = k \norm{A} \norm{B} |Λ_A|^r |Λ_B|^r, \, k>0, \, r \geq 1
\end{equation}
where $Λ_A$ denotes the support of $A\in \mathfrak{U}_{\rm loc}$, i.e.\ the smallest subset of $\Z^D$ such that $A\in \mathfrak{U}_{Λ_A}$.  
\end{defn}
Using the Lieb-Robinson bound, we can approximate time-evolved observables by local ones \cite{nachtergaele_quasi-locality_2019} and  obtain space-like $q$ clustering for $q=p-rD$. See for example \cite[Theorem 8.5]{doyon_hydrodynamic_2022}. We can obtain for $υ>υ_{LR}$, $A,B\in \mathfrak{U}_{\rm loc}$ and any $x \in \Z^D$, $t \in υ^{-1} [-\dist(A(x),B)+1, \dist(A(x),B)+1]$ we have the bound:
\begin{equation} 
        |ω( A(x,t) B) - ω(A) ω(B) | \leq C_2(A,B) \frac{1}{(1+ \dist(A(x),B))^{p-rD}} 
\end{equation}
Similar proofs are done in \cite[Appendix C]{ampelogiannis_2023_almost} and \cite[Appendix C]{ampelogiannis_long-time_2023} for exponential clustering.

 \chapter{Defining cumulants by Mobius functions} \label{appendix:mobius}
Consider \Cref{defn:classical_cumulants} for the classical cumulants. The set of partitions $P=\cup P(n)$ is partially ordered by inclusion and one can define a Mobius function over any  locally finite partially ordered set \cite{Speed_1983_Cumulants1}.  It is then noted that the coefficients in \Cref{defn:classical_cumulants} correspond to the value of Mobius function of $P$: 
\begin{equation}
    μ_P(π, \mathds{1}_n) = (-1)^{|π|-1}( |π| - 1)!
\end{equation}
where  $\mathds{1}_n$ is the maximal partition of $\{ 1 ,2 , \ldots, n \}$ \cite{Speed_1983_Cumulants1}. The cumulants of \Cref{defn:classical_cumulants} can then be equivalently defined by the Mobius function of the partition lattice:
\begin{equation}
    c_π (A_1, A_2, \ldots, A_n) \coloneqq \sum_{σ \in P(n), σ \leq π} μ_P(σ,π) ω_σ (A_1, A_2, \ldots, A_n) 
\end{equation}

The free cumulants are then defined similarly, by the Mobius function of the partially ordered set of non-crossing partitions $NC$ (ordered by inclusion):
\begin{equation} \label{eq:classical_cumulants_mobius}
    κ_π (A_1, A_2, \ldots, A_n) \coloneqq \sum_{σ \in NC(n), σ \leq π} μ_{NC}(σ,π) ω_σ (A_1, A_2, \ldots, A_n) 
\end{equation}
It follows that $κ_π$, $π\in NC$, and $c_π$, $π\in P$, form  multiplicative families, determined by the n-th cumulants, which are the ones that correspond to the maximal partition $\mathds{1}_n = \{ \{1,2,\ldots,n\} \}$ of $\{ 1, 2, \ldots,n\}$:
\begin{equation} \label{eq:free_cumulants_mobius}
    κ_n \coloneqq κ_{\mathds{1}_n} , \  c_n \coloneqq c_{\mathds{1}_n}
\end{equation}

In both \Cref{eq:classical_cumulants_mobius} and \Cref{eq:free_cumulants_mobius}, we can apply the respective Mobius inversion \cite{rota_1964_mobius} to obtain the moments-to-cumulants formulae.
\end{appendices}

%\bibliographystyle{ieeetr}
%\bibliography{references}

%

\chapter{Hydrodynamic projections} \label{chapter:hydroproj}
In hydrodynamics it is posited that the dynamics of the system, microscopically described by myriads of degrees of freedom, are effectively reduced to  a handful of emergent variables that evolve slowly over time. A striking example of this idea is the Boltzmann-Gibbs principle \cite{spohn_large_1991,demasi_mathematical_2006,kipnis_1988_scaling}, in the context of hydrodynamic linear response theory: correlations due to local perturbations are carried by long-lived modes that slowly propagate at hydrodynamic velocities. Indeed, the dominant correlations between a person's vocal chords and another's eardrum are carried by sound waves. The principle holds in extended Hamiltonian quantum models, including integrable ones, see for examples \cite{doyon_drude_2017,doyoncorrelations,denardis_2022_correlation_ghd}. Intuitively, this should require the loss of information at large scales; a property of mixing, or ergodicity of the system. It is then natural to ask, what are the strict mathematical conditions for the emergence of the Boltzmann-Gibbs principle, and how is this related to ergodicity? 

In this chapter, we discuss a hydrodynamic projection Theorem, based on our work in \cite{ampelogiannis_long-time_2023}, to which a significant part of this Ph.D. project was devoted, by utilising almost-everywhere ergodicity. This is the main and most interesting application of the relaxation properties of the previous chapter, and constitutes a universal expression of the Boltzmann-Gibbs principle. Hydrodynamic projections were initially considered in the renowned works of Mori \cite{mori_transport_1965} and Zwanzig \cite{zwanzig_statistical_1961}, where integration over ``fast modes" gave rise to a projection mechanism over ``slow modes", see also \cite{spohn_large_1991,demasi_mathematical_2006,kipnis_1988_scaling}. Conserved quantities play a very important role in this problem, as noted early on \cite{th_brox_equilibrium_1984}; in generic (chaotic) systems they  are assumed to be the energy, momentum and number of particles, while in integrable systems there are extensively many \cite{doyon_drude_2017,doyon_lecture_2020}, see the review \cite{denardis_2022_correlation_ghd}. Our main theorem shows that (a) there is a certain fluid-cell average under which a projection indeed occurs (see \cite{doyoncorrelations,denardis_2022_correlation_ghd,delvecchio2021hydro} for discussions of fluid-cell means); (b) Given a quantum spin lattice $C^*$ dynamical system, we can always define  the space of extensive conserved quantities $\mathcal Q$, which exists universally; (c) $\mathcal Q$ may in general be state-dependent. It remains an important problem to show that $\mathcal Q$ is spanned by the conventional extensive conserved quantities in chaotic systems. In fact, proving chaoticity is a non-trivial task that is currently out of reach, as it requires one to establish the absence of non-trivial {\em extensive} charges \cite{Prosen_1999_ergodic,prosen_1998_quantum_invariants,doyon_thermalization_2017,ampelogiannis_long-time_2023, doyon_hydrodynamic_2022}. There are, nevertheless, exciting partial results, showing the absence of charges with local densities (a subset of extensive charges) in certain models \cite{shiraishi_2019_absence_local,chiba_2023_ising_nonintegrability,shiraishi2024absence,park_2024_nonintegrability}. These results are still not sufficient, as they do not prove the absence of quasi-local conserved charges, which is an important unsolved problem in mathematical physics.

Our main Theorem \ref{hydroprojection} and its proof in \cite{ampelogiannis_long-time_2023} are new results,  extending some of the results of \cite{doyon_hydrodynamic_2022} to higher dimensions and to short-range interaction. Going further than that, we importantly generalise the theory to also describe oscillatory behavrious. The proofs presented here, for $D>1$, are not simple generalisations of the respective ones in $D=1$ \cite{doyon_hydrodynamic_2022}, and require the construction of appropriate Hilbert spaces, by quotient spaces of $\mathfrak{U}_{\rm loc}$, in order to recast the $D$-dimensional problem into a $1-$dimensional problem.  

What is a ``hydrodynamic projection"? See for instance the short non-rigorous discussion in \cite[Sec 2.6]{doyon_lecture_2020} as well as the aforementioned review \cite{denardis_2022_correlation_ghd}  for a modern introduction to this idea. The main point of the hydrodynamic projection theorem, that we will prove in generality in short-range quantum spin lattices, is that the Euler-scale correlation functions can be fully evaluated only by knowing the full set of conserved extensive charges of the system, and the ``projection" onto these charges of the local operators involved in the correlation function. The physical idea is that the initial, dynamically complicated disturbance quickly relaxes and projects, at the Euler scale of long time and large distances, onto the conserved extensive quantities, that then carry correlations.

But there are a number of concepts in this statement that need clarifying and appropriate mathematical definitions. What is an Euler-scale correlation function? It heuristically involves the ballistic scaling of space and time, $\bm x, t$ large with $\bm x/t$ finite, but it requires a precise definition. What are the conserved extensive charges, and how do we define the full set of them? One needs a precise definition of an extensive quantity, and then of a conserved extensive quantity (or charge). We will define these as certain Hilbert spaces, in a way that is similar to, but different from, the GNS construction. Then, what does it mean to project a local operator onto such a conserved extensive quantity? This will be from the Hilbert space structure. Our main hydrodynamic projection theorem will make the connection between all these notions, the result being in agreement with the physical intuition.

For the Hilbert spaces $\mathcal{H}_0$ of extensive quantities, to every local observable $A\in \mathfrak{U}_{\rm loc}$ we will associate, in a manner similar to the GNS construction, a respective ``extensive quantity", which we will denote as $ΣA$, defined as an appropriate equivalence class of the observable $A$. As the notation suggests, $ΣA$ will have the interpretation as the sum of all translates, ``$ΣA= \sum_{\bm x \in \Z^D}  A(\bm x)$", of the observable $A$, and $A$ as a density of $ΣA$. The density of an extensive observable is not unique -- one can add ``total derivatives" for instance -- and $ΣA$ can be seen as the equivalence class of all densities of the extensive quantity of which $A$ is a density. $ΣA$ will indeed contain all spatial translates of $A$. Naturally, the infinite sum itself, $\sum_{\bm x \in \Z^D}  A(\bm x)$, does not converge in $\mathfrak U$; but we will have convergence in a different topology. Note that, unlike the GNS construction where the observables of the C$^*$ algebra are represented by operators acting on a Hilbert space, here instead we will form a Hilbert space of extensive observables -\textit{ not a representation as operators}. On this Hilbert space, the $C^*$-algebra time evolution $\tau_t$ induces a time evolution $\tau^{0}_t$, which is unitary. The conserved extensive charges $\mathcal{Q}_0$ are just the elements of $\mathcal{H}_0$ that are invariant under $\tau^0_t$. This is a closed subspace of $\mathcal{H}_0$, thus allowing us to define an orthogonal projection $\mathbb{P} : \mathcal{H}_0 \to \mathcal{Q}_0$ of extensive quantities to the conserved ones.
 
The objects of interest are the time-averaged, long-wavelength Fourier transform of two-point connected correlations:
\begin{equation} \label{eq:spacetime}
    S_{ΣA,ΣB} (\bm{\kappa}) \coloneqq \widetilde{\lim_{T \to \infty}} \frac{1}{T} \int_0^T \dd t \, \sum_{\bm x \in \Z^D}e^{\ri \boldsymbol\kappa\cdot \boldsymbol x/t} (A (\bm x, t),B)
\end{equation}
where $(A,B) = \omega(A^\dag B) - \omega(A^\dag)\omega(B)$ is the sesquilinear connected correlation. We do not know how to show the existence of the limit $\lim_{T \to \infty}$, but we can show that $\sum_{\boldsymbol x\in\Z^D}\,e^{\ri \boldsymbol\kappa\cdot \boldsymbol x/t} (A(\boldsymbol x,t), B)$ is uniformly bounded. This allows us to use the notion of a Banach limit $\widetilde{\lim_{T \to \infty}}$, see \cite[Chapter III.7]{conwayFunctionalAnalysis2007} and \cite[Appendix A]{doyon_hydrodynamic_2022}. We can also show that the result is indeed a function not merely of $A,\,B$, but in fact of the equivalence classes $\Sigma A,\,\Sigma B$, as the notation suggests; and that it is continuous, in both variables, with respect to the norm $||\cdot||_0$ on $\mathcal H_0$. The essence of the hydrodynamic projection is that inside correlation functions every extensive observable will project to a conserved charge, irrespectively of the choice of a Banach limit:
\begin{equation}\label{hydroprojstandard}
    S_{\Sigma A,\Sigma B}(\boldsymbol\kappa) = S_{\mathbb{P}(\Sigma A),\mathbb{P}(\Sigma B)}(\boldsymbol\kappa).
\end{equation}
This is an expression of the aforementioned Boltzmann-Gibbs principle: the reduction of the number of degrees of freedom at large space-time separations by projections over hydrodynamic modes.  This occurs thanks almost-everywhere ergodicity, as will be made apparent in the proof of Theorem \ref{hydroprojection} in \Cref{section:hydroproj_proof} (see \cite{doyon_hydrodynamic_2022} for proof in D=1).

This is the first general, rigorous result concerning the Boltzmann-Gibbs principle in deterministic interacting systems of arbitrary dimensions. The principle is applicable to the large class of quantum lattices, going beyond interacting particle systems conventionally studied in statistical physics and hydrodynamics. It provides further support to the idea that the basic principles of hydrodynamics hold independently from the details of the microscopic dynamics.

Surprisingly, the hydrodynamic projection can be generalised to describe oscillatory behaviours: to any frequency $f \in \R$, and wavenumber $\bm k \in \R^D$.  We show that the  projection still takes place for $\bm k$-extensive quantities, formally $\Sigma^{\boldsymbol k} A = \sum e^{-\ri {\boldsymbol k}\cdot {\boldsymbol x}} A( x)$, which will project to $(f,\bm k)$-conserved charges, defined as those $\Sigma^{\boldsymbol k} A$ such that their time evolution take the form $ \Sigma^{\boldsymbol k} A (t) =e^{\ri ft} \Sigma^{\boldsymbol k} A$.

Oscillatory hydrodynamic projection describes oscillatory behaviours that emerge at large space and time separations in correlation functions. One in general expects that $(f,\bm k)$-conserved charges only exist for certain frequency-wavenumber pairs $(f,\bm k)$, depending on the specific model.  A simple example is the free fermionic lattice, with some dispersion relation $E(\boldsymbol k)$. In this system, creation and annihilation operators at momentum $\boldsymbol k$ are $(E(\boldsymbol k),\boldsymbol k)$-extensive conserved quantities, see the discussion in \cite[Section 4]{ampelogiannis_long-time_2023}.

This chapter largely incorporates Section 3 of our work \cite{ampelogiannis_long-time_2023} and is organised as follows. \Cref{section:hydroproj_assumptions} establishes the framework and our main assumptions. In \Cref{section:extensive_charge} we rigorously construct Hilbert-spaces of extensive quantities by appropriate quotient spaces of the algebra of local observables $\mathfrak{U}_{\rm loc}$. Subsequently, we define time evolution to act on these spaces, leading to an unambiguous definition of the extensive conserved charges. The same procedure is followed to define oscillatory charges. In \Cref{section:drude} we state a hydrodynamic projection for the familiar Drude weight, and then in \Cref{section:hydroproj} we present a fully general hydrodynamic projection Theorem, for the Euler-scale correlations, which also takes account of oscillatory behaviours. Finally, \Cref{section:hydroproj_proof} contains the proof of our main theorem. 

\section{Assumptions} \label{section:hydroproj_assumptions}
Throughout this chapter we consider a short-range quantum spin lattice dynamical system $(\mathfrak{U},ι,τ)$, \Cref{def:short_range_qsl}. For clarity, we will use the notation $A(\bm x,t)$ for space-time translations of observables $ι_{\bm x}τ_t A$. We also denote $A(\bm x)=A(\bm x,0)$, while we keep the notation $τ_tA$ for pure time translations. 
The system will be in a space-time invariant state $ω$, which will be assumed to have clustering properties with respect to space translations. The almost-everywhere ergodicity theorems shown in \cite{ampelogiannis_2023_almost}, and discussed in \Cref{chapter:ergo}, require the state $ω$ to be clustering in space, in particular $\lim_{\boldsymbol x\to\infty}\big( ω(  A(\bm x) B)  - ω( A)ω( B) \big)=0$ for any $A,B\in\mathfrak U$. Almost-everywhere ergodicity holds no matter how fast the connected correlations of observables decay at large spatial separations.  For the hydrodynamic projection to occur stronger clustering assumptions need to be imposed. In particular, clustering of correlations at large space separations will need to happen faster than $|\bm x|^p$, for $p$ large enough.

In order for the time evolution to be well defined as a unitary group action on the Hilbert spaces of extensive quantities, we need to
assume that the time translations of local observables  cluster in a uniform enough manner. First, we require that any element $A$ in ${\rm span} \{ τ_t A : A \in  \mathfrak{U}_{\rm loc}, t \in \mathbb{R} \}$ is approximated by a sequence $σ_nA$ of local elements. We will show that this follows from the Lieb-Robinson bound. Additionally, we require that for any local $A,B$ the set of pairs $\{ (\sigma_n A, \sigma_n B ) \}$ is uniformly clustering in space, as per \Cref{eq:uniformclustering}.

Combining spatial clustering  of the state $ω$ with the Lieb-Robinson bound, one can obtain space-like $p_c$ clustering, defined as follows:
\begin{defn} \label{defn:spacelikeclustering}
The dynamical system $(\mathfrak{U},ι,τ)$ in the state $ω$ is called space-like $p_c$-clustering  with velocity $υ_c$, if
\begin{enumerate}
    \item $\forall A \in \hat{ \mathfrak{U}}_{\rm loc} = {\rm span} \{ τ_t A : A \in  \mathfrak{U}_{\rm loc}, t \in \mathbb{R} \}$ there exists a sequence $ \sigma_n A \in \mathfrak{U}_{\rm loc}$, $n \in \mathbb{N}$, such that 
$\lim σ_n A = A$. For any $A \in  \mathfrak{U}_{\rm loc}$ we define $\sigma_n A = A$, $\forall n$. \label{spacelike1}
    \item  $\forall A,B \in \hat{ \mathfrak{U}}_{\rm loc}$ the set of pairs $\{ (\sigma_n A, \sigma_n B ) \}$ is uniformly $p$-clustering for some $p>p_c$. \label{spacelike2}
    \item $\forall A,B \in \mathfrak{U}_{\rm loc}$ there exist $p> p_c$, $0<V<υ_c$ and $C_{A,B}>0$ such that
\begin{equation}
\big| ω( A(\bm x,t)  B ) - ω(A)ω(B)  \big|\leq  \frac{C_{A,B}}{ ( |\boldsymbol x| +1)^p} \label{eq:spacelikeclustering}
\end{equation}
for all $\boldsymbol x\in\Z^D$, $|\boldsymbol x|\geq V|t|$. \label{spacelike3}
\end{enumerate}
Similarly, we define exponential space-like clustering by the same conditions, with Condition 2 replaced by uniform exponential clustering and \Cref{eq:spacelikeclustering} replaced by  $\big| ω( A(\bm x,t) , B) - ω(A)ω(B)  \big|\leq C_{A,B} {\rm e}^{-λ |\bm x|}$, for some $λ>0$. 
\end{defn}

\section{Spaces of extensive quanties} \label{section:extensive_charge}
We start by constructing Hilbert spaces of extensive quantities from the dynamical system $(\mathfrak{U},ι,τ)$. For each wavenumber $\bm k\in \mathbb{R}^D$ we define the positive-semidefinite sesquilinear form 
\begin{equation}\label{inner}
    \langle A,B\rangle_{\boldsymbol k} = \sum_{\boldsymbol x\in \Z^D}\,
    \re^{\ri \boldsymbol k \cdot \boldsymbol x} ( A(\boldsymbol x), B)
        \quad A,B\in\mathfrak U_{\rm loc},
\end{equation}
where we recall that $(A,B) = \omega(A^\dag B) - \omega(A^\dag)\omega(B)$ is the sesquilinear connected correlation. We define the equivalence relation $A \sim^{\bm k} A^{\prime}$ on $\mathfrak{U}_{\rm loc}$  by $\langle A'-A,A'-A\rangle_{\boldsymbol k}=0$. The Hilbert spaces of extensive quantities are the norm-completion of the quotient spaces  formed of the set of equivalence classes of $\sim^{\bm k}$. That is, the equivalence class of $A \in \mathfrak{U_{\rm loc}}$ is $Σ^{\bm k} A \coloneqq \{ A^{\prime}\in \mathfrak{U_{\rm loc}} : A^{\prime} \sim^{\bm k}A \}$. This is to be understood as the $\bm k$-extensive observable associated to the density $A$, and it contains all other densities $A'$ that lead to the same $\bm k$-extensive observable. For instance, it is clear that $A' = e^{\ri \bm k\cdot\bm x'} A(\bm x')\in Σ^{\bm k} A$ for any fixed $\bm x'\in\Z^D$: the $\bm k$-translate of a density $A$ still is a density for the same extensive observable. Formally, we can associate $Σ^{\bm k} A$ with the infinite series ``$Σ^{\bm k} A = \sum_{\bm x \in \Z^D} e^{-i \bm k \cdot \bm x} A(\bm x)$", the ``total'' $A$ on the full quantum lattice, as is intuitively suggested by the \eqref{inner}. Of course this series does not converge within $\mathfrak{U}$; the quantity $Σ^{\bm k} A$ is defined as an equivalence class. The set of equivalence classes is $\mathcal{V}_{\bm k} \coloneqq \mathfrak{U_{\rm loc}} ~/ \sim^{\bm k} = \{ Σ^{\bm k}A : A \in \mathfrak{U_{\rm loc}}\}$ and its Cauchy completion gives the Hilbert space of $\bm k$-extensive observables  $\mathcal{H}_{\boldsymbol k} = \overline{\mathcal{V}}_{\bm k}$.
\begin{remark}
The quantity $Σ^{\bm k} A$ is not to be confused with the spatially-averaged quantity \\ $lim_{Λ \to \Z^D}\frac{1}{|Λ|}\sum_{\bm x \in Λ} A(\bm x)$. Here, we are studying the total, extensive observable itself\\ $lim_{Λ \to \Z^D}\sum_{\bm x \in Λ} A(\bm x)$, as an element of some Hilbert space. The information that is kept of the extensive observable, as encoded within the Hilbert space inner product, are its convergent connected correlation functions with local observables; while of course the spatially-averaged quantities have zero connected correlation functions with local observables. Note that connected correlation functions of conserved extensive observables with local observables have the physical interpretation as susceptibilities. For example, consider a chain with spin $1/2$ at each site. Suppose that the Hamiltonian conserved the total spin  $\sum_{x \in \Z} σ_3 (x)$. Then, in a thermal state $\omega$ with inverse temperature $\beta$ and magnetic field $h$, we have $\partial\omega(A)/\partial h = \beta \langle \sigma_3,A\rangle_{0}$. Similarly, a staggered magnetic field would correspond to the case $k = \pi$.
\end{remark}
 
 \begin{remark}
 As noted in the introduction of this Section, the construction of the Hilbert spaces $\mathcal{H}_{\bm k}$ should not be confused with the GNS representation. The GNS representation is a representation of the elements of a C$^*$ algebras by bounded operators $B(H_ω)$ acting on a Hilbert space $H_ω$. Instead, we directly form Hilbert spaces by Cauchy completion of the quotient spaces $\mathfrak{U}_{loc}/ \sim^{\bm k}$; these are not representations of the C$^*$ algebra.
 \end{remark}
 %remove weakly conv comment, give example eg. \sum σ_3(x) , stress that it is different than average, susceptibilities
 
 In order to examine the long time dynamics of these extensive observables we have to extend the action of time evolution $τ_t,\,t\in\R$ to the Hilbert spaces $\mathcal H_{\boldsymbol k}$. This is done rigorously in \cite[Section 5.3]{doyon_hydrodynamic_2022} for $D=1$ (spin chains) and can be immediately extended to arbitrary dimension $D$. The key in order to be able to do this is the uniform clustering condition 2 in \Cref{defn:spacelikeclustering}. Thus, we can  show that time evolution acts as a unitary operator $τ_t^{\bm k}$ on $\mathcal H_{\boldsymbol k}$. For lightness of notation, we omit the superscript and simply write $τ_t$ for the unitary action of time evolution on these spaces.
 
 With this construction, we can define the subspace of conserved extensive charges as those elements of $\mathcal H_{\boldsymbol k}$
 that are invariant under $τ_t$:
 \begin{equation}
     \mathcal{Q}_{\bm k} = \{ q \in \mathcal{H}_{\bm k} : τ_t q = q, \forall t \in \R \}.
 \end{equation}
 In fact, we can go one step further and define the subspace of $(f,\boldsymbol k)$-extensive conserved quantities (or $f$-oscillatory $\bs k$-extensive charges) as:
\begin{equation}\label{Qfk}
\mathcal Q_{(f,\boldsymbol k)} = \{ a\in \mathcal H_{\boldsymbol k}: τ_t  a = e^{-\ri f t}  a ,\forall t\in\R\}.
\end{equation}

It is immediate that $\mathcal Q_{(f,\boldsymbol k)}$ is a closed subspace of $\mathcal{H}_{\bm k}$, hence there is an orthogonal projection
\begin{equation} \label{eq:conserved_projection}
    \mathbb P_{(f,\boldsymbol k)}: \mathcal{H}_{\boldsymbol k} \to \mathcal Q_{(f,\boldsymbol k)}.
\end{equation}

 \section{Hydrodynamic projection: the Drude Weight} \label{section:drude}
We believe it is instructive to begin with a projection formula for the well-known Drude weight, before giving the more general theorem. The Drude weights are transport coefficients, often used in condensed matter physics to classify materials as conductors or insulators, as they  measure the effective electron density contributing to DC conductivity. In the context of quantum many-body systems, the Drude weights quantify ballistic transport.

Consider local observables $A,B$ and define the Drude weight $D_{A,B}$ by the Kubo formula
\begin{equation} \label{eq:drude1}
    D_{A,B} =  \lim_{T \to \infty} \frac{1}{T}\int_0^T \sum_{x \in \Z^D} (A(x,t),B)
\end{equation}
As a consequence of \Cref{hydroprojection} that the Drude weight satisfies the projection formula:
\begin{equation}
    D_{A,B} = D_{\mathbb{P}A, \mathbb{P}B}
\end{equation}
 where $\mathbb{P} = \mathbb{P}_{(0,0)}$ is the projection \Cref{eq:conserved_projection} onto the space of extensive conserved charges.  

 Going further, in \cite{ampelogiannis_long-time_2023} we define the oscillatory Drude weight at frequency $f$ and wavenumber $k$ as
 \begin{equation}
    \mathsf D_{\Sigma^{\boldsymbol k} A,\Sigma^{\boldsymbol k} B}^{(f,\boldsymbol k)} := \lim_{T\to\infty} \frac1T \int_0^T \dd t\,\sum_{\boldsymbol x\in\Z^D} \re^{\ri \boldsymbol k \cdot \boldsymbol x - \ri f t} ( A(\boldsymbol x,t), B)
\end{equation}
As a consequence of \Cref{hydroprojection}, we obtain a projection onto the space of oscillatory charges:
\begin{equation}
    \mathsf D_{A ,B}^{(f,\boldsymbol k)} = \mathsf D_{\mathbb P_{(f,\boldsymbol k)}A,\mathbb P_{(f,\boldsymbol k)}B}^{(f,\boldsymbol k)} 
\end{equation}

Obtaining these projection formulas for the Drude weight is in fact much easier than the general theorem we give below, by a  simple application of von Neumann's mean ergodic theorem and application of the dominated convergence theorem. This is done in \cite[Section 6]{doyon_hydrodynamic_2022}.

\section{Hydrodynamic projection theorem} \label{section:hydroproj}
The correlation functions with $(f,\boldsymbol k)$-fluid-cell averaging are given by
\begin{equation} \label{spacetime}
    S_{\Sigma^{\boldsymbol k} A,\Sigma^{\boldsymbol k} B}^{(f,\boldsymbol k)}(\boldsymbol\kappa) := \widetilde{\lim_{T\to\infty}} \frac1T \int_0^T \dd t\,
    \sum_{\boldsymbol x\in\Z^D}\,\re^{\ri \boldsymbol k \cdot \boldsymbol x - \ri f t}\re^{\ri \boldsymbol\kappa\cdot \boldsymbol x/t} ( A(\boldsymbol x,t), B).
\end{equation}
Note how one extracts, thanks to the time integral and factor $e^{-ift}$, the time-oscillatory behaviour of the correlation function with frequency $f$. Note also how the full wavenumber is $\bs k + \bs\kappa/t$, representing, in the large-time limit, a long-wavelength modulation of a $\bs k$-oscillatory factor; this extracts the space-oscillatory behaviour with wavenumber $\bs k$. 

The limit on $T$ is in general a Banach limit \cite[Chapter III.7]{conwayFunctionalAnalysis2007}, \cite[Appendix A]{doyon_hydrodynamic_2022}. The result will hold irrespectively of the choice of Banach limit, and all the aspects of the proof regarding the Banach limit are the same as in $D=1$, as shown in \cite[Section 6]{doyon_hydrodynamic_2022}. As the notation implies, the result only depends on the equivalence classes $\Sigma^{\boldsymbol k} A,\Sigma^{\boldsymbol k} B$, which we'll denote with the respective lowercase letters $a, b$. In fact this defines a continuous sesquilinear form on $\mathcal H_{\bm k}$. The hydrodynamic projection theorem is rigorously stated as follows:  

\begin{theorem}[Hydrodynamic Projection]\label{hydroprojection}
Consider a dynamical system $(\mathfrak{U},ι,τ)$ with interactions satisfying \Cref{eq:interaction}, in an $r$-sizeably $D(r+1)$-clustering  state $ω$, as per \Cref{def:power_clustering_space}.  For every frequency-wavenumber pair $(f,\bm k) \in \R \times \R^D$, rational vector $\boldsymbol\kappa\in \R^D : \exists r\in\R \,|\,r\boldsymbol \kappa \in\Z^D$, and any $\bm k$-extensive elements $a,b \in \mathcal H_{\boldsymbol k}$,
\begin{equation}\label{proj}
    S_{a,b}^{(f,\boldsymbol k)}(\boldsymbol\kappa) =
    S_{\mathbb P_{(f,\boldsymbol k)}a,\mathbb P_{(f,\boldsymbol k)}b}^{(f,\boldsymbol k)}(\boldsymbol\kappa).
\end{equation}
Specifically for the simple case $(f,\bm k)=(0, \bm 0)$ we have
\begin{equation}
     S_{a,b} (\bm \kappa) =
    S_{\mathbb Pa,\mathbb P   b}(\bm \kappa).
\end{equation}
\end{theorem}

\section{Proof} \label{section:hydroproj_proof}

This section is contains the main part of the proof  in \cite{ampelogiannis_long-time_2023}. The main body of the proof is presented here for completeness, while minor details are omitted.

The key idea for the proof of the hydrodynamic projection formula in quantum spin lattices is this: By using an appropriate geometric construction, we recast the $D$-dimensional problem of hydrodynamic projection into the a 1-dimensional problem. This is done by identifying the summation
over space coordinates in a plane perpendicular to the wavenumber direction, with a sesquilinear form and its new associated Hilbert space, which is to play the role of the sesquilinear correlation
    in an effective one-dimensional problem. Once this is done, the proof follows that of \cite{doyon_hydrodynamic_2022} done for $D=1$. The schematics of our proof is as follows: We assume a dynamical system in a $p$-clustering state $ω$, \Cref{def:power_clustering_space} and with sufficiently fast decaying interactions, \Cref{eq:interaction}, so that the Lieb-Robinson bounds, \Cref{eq:liebrobinsonbound}, holds. We combine $p$-clustering with the Lieb-Robinson bound (proof in Appendix \ref{appC}) in order to obtain space-like $p_c$-clustering, \Cref{defn:spacelikeclustering}. This leads to the special Property \ref{PROPERTY}, below, for the dynamical system. Using this property we can prove Lemma \ref{thelemma}, which forms the basis for the hydrodynamic projection formula, Theorem \ref{hydroprojection}.

The most important difference from the $D=1$ case is the new geometric constructon, which we now explain. We define the rational unit sphere in $D$-dimensional space as $\mathbb S_{\mathbb Q}^{D-1} = \{\boldsymbol \kappa/|\boldsymbol \kappa|:\boldsymbol \kappa\in\Z^D\}$. We also denote the set of vectors in rational directions in $\R^D\setminus \{\boldsymbol 0\}$ by $\R^D_{\mathbb Q} = \{\boldsymbol \kappa \in \R^D\setminus\{\bs 0\} : \exists r\in\R \,|\,r\boldsymbol \kappa \in\Z^D\}$. Clearly, by definition, $\mathbb S_{\mathbb Q}^{D-1}\subset \R^D_{\mathbb Q}$. For any such vector $\boldsymbol \kappa\in \R^D_{\mathbb Q}$, there is a unique unit vector $\widehat{\boldsymbol \kappa}' = \pm\boldsymbol \kappa/|\boldsymbol \kappa|\in \mathbb S_{\mathbb Q}^{D-1}$ that lies on the half-unit sphere: such that $\widehat \kappa_1'> 0$, or $\widehat \kappa_1'= 0,\,\widehat \kappa_2'>0$, or $\cdots$, or $\widehat \kappa_1'=0,\ldots,\widehat \kappa_{D-1}'=0,\, \widehat \kappa_D'>0$. Then, given such a unit vector $\widehat{\boldsymbol \kappa}'$, we denote by $\widehat{\boldsymbol \kappa} = r \widehat{\boldsymbol \kappa}'$ the ``smallest" integer vector associated to it, where $r>0$ is the smallest positive number such that $\widehat{\boldsymbol \kappa} \in \mathbb Z^D$ ($\widehat{\boldsymbol \kappa}$ exists because $\widehat{\boldsymbol \kappa}'\in \R^D_{\mathbb Q}$).  Clearly $\widehat{r\boldsymbol \kappa} = \widehat{\boldsymbol \kappa}$ for every $r\in\R$, and in fact, $\widehat{\boldsymbol \kappa}$ is an element of $\Z^D$ that identifies the ``rational ray" on which $\boldsymbol \kappa$ lies: there is a bijection between the set of $\widehat{\boldsymbol \kappa}$'s and the set of $\{r\boldsymbol \kappa:r\in\R\}$'s for $\boldsymbol \kappa\in\R^D_{\mathbb Q}$. Below we will refer to $\widehat{\boldsymbol \kappa}$ as a rational ray, and it will be important that it lies in $\Z^D$.
For any rational ray $\widehat{\boldsymbol \kappa}$, we may build its perpendicular plane in $\Z^D$: a sublattice of $\Z^D$ that is isomorphic to $\Z^{D-1}$, and whose vectors have vanishing vector dot product with $\widehat{\boldsymbol \kappa}$. Given $\widehat{\boldsymbol \kappa}$, there is a set (possibly empty) of zero components $\zeta = \{i: \widehat \kappa_i = 0\}$ and its complement (always non-empty) $\bar\zeta = \{1,\ldots,D\}\setminus \zeta = \{j_1,\ldots,j_d\}$ for unique $1\leq d = |\b\zeta|\leq D$ and $j_1<\ldots<j_d$. We construct the vectors
\begin{equation}  \begin{array}{*3{>{\displaystyle}lc}p{5cm}}
    \boldsymbol h^{(i)}\in\Z^D&:&
    h^{(i)}_m = \delta_{i,m} \quad (m\in\{1,\ldots,D\},\,i\in\zeta)\n
    && h^{(j_l)}_{j_l} = \widehat \kappa_{j_{l+1}},\ 
    h^{(j_l)}_{j_{l+1}} = -\widehat \kappa_{j_l},\ 
    h^{(j_l)}_{m} = 0 \quad\\ && (l\in \{1,\ldots,d-1\}, m\in \{1,\ldots,D\}\setminus\{j_l,j_{l+1}\}).\n
\end{array}\end{equation}  
This defines $D-1$ vectors: all $\boldsymbol h^{(i)}, i=1,\ldots,D$, except $\boldsymbol h^{(j_d)}$ which has not been defined. If $j_d \neq D$, then $D\in\zeta$,     and for convenience, in this case, we simply define $\boldsymbol h^{(j_d)} = \boldsymbol h^{(D)}$ as well as $\t\zeta = \zeta \setminus \{D\}\cup \{j_d\}$ (otherwise $\t\zeta = \zeta$). Thus we may always concentrate on $\boldsymbol h^{(i)}:i=1,\ldots,D-1$.

By construction, the set $\{\boldsymbol h^{(i)}:i=1,\ldots,D-1\}$ is linearly independent, and satisfy $\boldsymbol h^{(i)} \cdot \widehat{\boldsymbol \kappa} = \boldsymbol h^{(i)}  \cdot \boldsymbol \kappa = 0$. The perpendicular plane is $\Z^D\supset \mathbb H_{\widehat{\boldsymbol \kappa}}^{D-1} \coloneqq {\rm span}_{\Z}(\boldsymbol h^{(1)},\ldots,\boldsymbol h^{(D-1)})\simeq \Z^{D-1}$, i.e.\ $\mathbb H_{\widehat{\boldsymbol \kappa}}^{D-1}$ is the $\mathbb{Z}$-module freely generated by the $\bm h^{(i)}$. The isomorphism between $\mathbb{Z}^{D-1}$ and $\mathbb H_{\widehat{\boldsymbol \kappa}}^{D-1} $, as $\Z$-modules, is clear by  construction, in particular
\begin{equation}
     \mathbb H_{\widehat{\boldsymbol \kappa}}^{D-1} = \{ x_i\boldsymbol h^{(i)}:{\boldsymbol x}\in\Z^{D-1}\}
\end{equation}
(with implied summation over repeated indices). Of course, we could have taken away the greatest common divisor of the components of $\boldsymbol h^{(i)}$ in order to have ``denser" planes, but this is not necessary in the following construction.

By stacking the perpendicular planes parallely, we obtain a sublattice of $\Z^D$ which is itself isomorphic to $\Z^D$, defined as the $\Z$-module freely generated by the set $\{ \bm h^{(i)},\widehat{\bm \kappa} \}$:
\begin{equation}
    \Z_{\widehat{\boldsymbol \kappa}}^D \coloneqq \{x_i\boldsymbol h^{(i)}+z\widehat{\boldsymbol \kappa}:(\boldsymbol x,z)\in\Z^D\}.
\end{equation}
%is isomorphic to $\Z^D$, because the set  $\{ \boldsymbol h^{(i)},\widehat{\boldsymbol \kappa}\}$ is linearly independent.
In-between the points of this lattice lie ``fundamental cells". That is, consider the region $R_{\widehat{\boldsymbol \kappa}} =\{x_i\boldsymbol h^{(i)}+z\widehat{\boldsymbol \kappa} : x_i\in[0,1)\forall i,\,z\in[0,1)\}\subset\R^D$ and the cell in $\Z^D$ given by
\begin{equation}
    \Lambda_{\widehat{\boldsymbol \kappa}} = \Z^D\cap R_{\widehat{\boldsymbol \kappa}} = 
    \Z^D \cap \{x_i\boldsymbol h^{(i)}+z\widehat{\boldsymbol \kappa}
    : x_i\in[0,1)\forall i,\,z\in[0,1)\}.
\end{equation}
Then the union of all unit cells shifted by the sublattice gives back $\Z^D$:
\begin{equation}\label{ZDdecomp}
    \Z^D = \bigcup_{\boldsymbol r\in \Z_{\widehat{\boldsymbol \kappa}}^D} \Big(\Lambda_{\widehat{\boldsymbol \kappa}} + \boldsymbol r\Big).
\end{equation}
Indeed, by linear independence and the fact that the Jacobian $J$ is finite and nonzero, every $\boldsymbol u\in\R^D$ may be written in a unique way as $\boldsymbol u = x_i\boldsymbol h^{(i)}+z\widehat{\boldsymbol \kappa}$ for some $(\boldsymbol x,z)\in\R^D$, and thus $\boldsymbol u = \floor{x_i}\boldsymbol h^{(i)}+\floor{z}\widehat{\boldsymbol \kappa} + \boldsymbol\lambda$ where $\boldsymbol\lambda \in R_{\widehat{\boldsymbol \kappa}}$. This holds in particular for $\bm u\in\Z^D$, and since $\floor{x_i}\boldsymbol h^{(i)}+\floor{z}\widehat{\boldsymbol \kappa}\in\Z^D_{\widehat{\boldsymbol \kappa}}$ it gives \eqref{ZDdecomp}.

We define for any wavenumber $\boldsymbol \kappa\in\R^D_{\mathbb Q}$ and $\boldsymbol \kappa'\in\R^D$, some positive-semidefinite sesquiliniear forms, and associated Hilbert spaces, in a way similar to that done above: first by summation over the perpendicular planes $\mathbb H^{D-1}_{\widehat{\boldsymbol \kappa}}$, and second by summation over the full sublattices $\Z^D_{\widehat{\boldsymbol \kappa}}:= \{x_i\bs h^{(i)} + z\widehat{\bs \kappa}:(\bs x,z)\in\Z^D\}$. Let
\begin{equation} \label{lemma_sesquiliniar}
( A , B )^{\widehat {\boldsymbol \kappa},\perp} \coloneqq \sum_{\boldsymbol x \in \mathbb H_{\widehat{\boldsymbol \kappa}}^{D-1}} ( A(\bm x) , B),\quad
( A , B )^{\widehat {\boldsymbol \kappa}}_{\boldsymbol \kappa'} \coloneqq \sum_{\boldsymbol x \in \Z_{\widehat{\boldsymbol \kappa}}^{D}} 
e^{\ri \boldsymbol \kappa'\boldsymbol x}
(  A( \bm x) , B), \quad A,B \in \mathfrak U_{\rm loc}.
\end{equation}
Note that these are well defined by clustering of the connected correlator $(A,B)=ω(A^\dagger B)-ω(A^\dagger )ω(B)$. Similarly to the construction of the spaces $\mathcal{H}_{\boldsymbol \kappa}$, we define the Hilbert spaces $\mathcal{H}^{\widehat {\boldsymbol \kappa},\perp}$ and $\mathcal H^{\widehat {\boldsymbol \kappa}}_{\boldsymbol \kappa'}$ as the norm completions of $\mathcal{V}^{\widehat {\boldsymbol \kappa},\perp}\coloneqq \mathfrak U_{\rm loc}/ \sim^{\widehat {\boldsymbol \kappa},\perp}$ and $\mathcal V^{\widehat {\boldsymbol \kappa}}_{\boldsymbol \kappa'}\coloneqq \mathfrak U_{\rm loc}/ \sim^{\widehat {\boldsymbol \kappa}}_{\boldsymbol \kappa'}$, respectively, where $A \sim^{\widehat {\boldsymbol \kappa},\perp} A^{\prime} \Leftrightarrow ( A - A^{\prime}, A-A^{\prime} )^{\widehat {\boldsymbol \kappa},\perp}=0$ and $A\sim^{\widehat {\boldsymbol \kappa}}_{\boldsymbol \kappa'} A' \Leftrightarrow ( A - A^{\prime}, A-A^{\prime} )^{\widehat {\boldsymbol \kappa}}_{\boldsymbol \kappa'} = 0$. The equivalence class of $A \in \mathfrak{U}_{\rm loc}$ is denoted by the respective lowercase $a$ and it is immediate that $( \cdot , \cdot )^{\widehat {\boldsymbol \kappa},\perp}$ and $(\cdot,\cdot)^{\widehat {\boldsymbol \kappa}}_{\boldsymbol \kappa'}$ are inner products on their respective Hilbert spaces, and thus satisfies the Cauchy-Schwarz inequality. See \cite[Appendix D]{ampelogiannis_long-time_2023} for proofs of basic properties. It can be established that $ι_x \coloneqq \iota_{x \widehat{\boldsymbol \kappa}}$ ($x\in\Z$) and $τ_t$ ($t\in\R$) act unitarily on $\mathcal{H}^{\widehat {\boldsymbol \kappa},\perp}$ as representations of the groups $\Z$ and $\R$ respectively, and that $τ_t$ ($t\in\R$) act unitarily on $\mathcal H^{\widehat {\boldsymbol \kappa}}_{\boldsymbol \kappa'}$ as a representation of the group $\R$. See also \cite[Appendix E]{ampelogiannis_long-time_2023}  We denote space-time translations of elements $a \in \mathcal{H}^{\hat{\kappa},\perp}$ by $ι_z τ_t a$, time translations of $a \in \mathcal{H}^{\bm \hat \kappa}_{\bm \kappa^{\prime}}$ by $τ_t a$, while keeping the notation $A(\bm x,t)$ for $A \in \mathfrak{U}$.
\begin{property}\label{PROPERTY}
Consider a dynamical system $(\mathfrak{U},ι,τ)$ (\Cref{defn:dynamicalsystem}) that is space-like $D+1$-clustering (\Cref{defn:spacelikeclustering}). For every $\boldsymbol \kappa \in \R^{D}_{\mathbb Q}$, it follows that:
\begin{enumerate}
\item $\forall\,a,b \in \mathcal{H}^{\widehat {\boldsymbol \kappa},\perp}$, for almost all $v \in \mathbb{R}$,
\begin{equation}
\lim_{T \to \infty} \frac{1}{T} \int_0^T  ( ι_{\lfloor vt \rfloor}τ_t a, b )^{\widehat {\boldsymbol \kappa},\perp} \dd t=0.
\end{equation} \label{property1}
\item $\forall\, a,b \in  \mathcal{V}^{\widehat {\boldsymbol \kappa},\perp}$, there exist $T>0$ (independent of $\widehat{\boldsymbol \kappa}$) and a Lebesgue measurable function $f : \mathbb{R} \mapsto \mathbb{R}_+$ such that:
\begin{equation}
  | ( ι_{  \floor{vt}} τ_{t} a,b )^{\widehat {\boldsymbol \kappa},\perp}| \leq f(v) \ \forall \  v \in \mathbb{R}, \ t\geq T \text{ satisfying } \int_{-\infty}^{\infty} \dd v \,( |v| +1) f(v) < \infty.
\end{equation}\label{property2}
\end{enumerate} \label{property}
\end{property} 
\begin{proof}
Property \ref{PROPERTY}.\ref{property1} is a consequence of the almost-everywhere ergodicity theorem, \cite[Theorem 3.2]{ampelogiannis_2023_almost}.  One can see this by proving that $( ι_{\lfloor vt \rfloor}τ_t a, b )^{\widehat {\boldsymbol \kappa},\perp}$ is space-like ergodic, \cite[Definition 3.1]{ampelogiannis_2023_almost}, see  \cite[Appendix B]{ampelogiannis_long-time_2023} for the general idea. That is, first, note that our assumptions immediately imply that the state $ω$ is space-like ergodic:
\begin{equation}
  \lim_N  \frac{1}{N} \sum_{m=0}^{N-1} \bigg( A\big( m \bm n , m υ^{-1} |\bm n|\big) , B\bigg) = 0  \ , υ>υ_{LR} \ , A,B \in \mathfrak{U}_{\rm loc}.
\end{equation}
Consider $a,b \in \mathcal{V}^{\bm{\hat{\kappa}},\perp}$ and class representatives $A,B \in \mathfrak{U}_{\rm loc}$ respectively. Then,
\begin{equation}
\label{eq63}
  \lim_N  \frac{1}{N} \sum_{m=0}^{N-1} \big( ι_{\floor{m z}} τ_{mυ^{-1}z}a,b \big)^{\hat{\bm \kappa}, \perp} =
  \lim_N  \frac{1}{N} \sum_{m=0}^{N-1}  \sum_{\bm x \in \mathbb{H}_{\widehat{\boldsymbol \kappa}}^{D-1}} \bigg(A \big( \floor{m z}\bm{\hat{\kappa}} + \bm x, mυ^{-1}z\big),B \bigg).
\end{equation}
Using space-like clustering we can see for any $z\in \R$, $υ>υ_c$ and  $A,B \in \mathfrak{U}_{\rm loc}$
\begin{equation}
    \Big| \frac{1}{N} \sum_{m=0}^{N}
    \bigg( A \big(\floor{m z}\bm{\hat{\kappa}} + \bm x, mυ^{-1}z\big)
    ,B \bigg) \Big| \leq \frac{1}{N} \sum_{m=0}^{N} \frac{c}{ (\sqrt{\floor{mz}^2|\bm{\hat{\kappa}}|^2 + |\bm x|^2 } +1)^p}
\end{equation}
where the summand is $N$-independent, $\bs x$-summable, and uniformly bounded by a $m$-summable function, allowing us to apply the dominated convergence theorem in order to move $\lim_N$ inside the sum $\sum_{\bm x \in \mathbb{H}_{\widehat{\boldsymbol \kappa}}^{D-1}}$ in Eq.~\eqref{eq63}. Thus, 
\begin{equation}
     \lim_N  \frac{1}{N} \sum_{m=0}^{N-1} \big( ι_{\floor{m z}} τ_{mυ^{-1}z}a,b
     \big)^{\hat{\bm \kappa}, \perp} =0 , \text{ for $υ>υ_{LR}$ }.
\end{equation}
Applying the almost-everywhere ergodicity theorem for the system $(\mathcal{H}^{\widehat {\boldsymbol \kappa},\perp},ι,τ)$ (see Appendix \cite[Appendix B]{ampelogiannis_long-time_2023}) we get Property \ref{property1}.

For Property \ref{property}.\ref{property2}: Choose a $V$ as per the definition of space-like clustering, Def.~\ref{defn:spacelikeclustering}, choose $T>0$, and consider all $t\geq T$. Note that the choice of $T$ and $V$ does not depend on $\widehat{\boldsymbol \kappa}$.

First consider the case $|v|\geq V/|\widehat{\boldsymbol \kappa}|+T^{-1}$. Note that $(|v|t-1)|\widehat{\boldsymbol \kappa}| \geq Vt$ and $(|v|t-1)|\widehat{\boldsymbol \kappa}|\geq (|v|T-1)|\widehat{\boldsymbol \kappa}|\geq VT$.  We have, for some $p>D+1$ (each step is explained below):
\begin{equation}  \begin{array}{*3{>{\displaystyle}lc}p{5cm}}
    | ( ι_{  \floor{vt}} τ_{t} a,b )^{\widehat {\boldsymbol \kappa},\perp} |
    &\leq &
    \sum_{{\boldsymbol x}\in\Z^{D-1}}
    \Big|
    \big(  A(\floor{vt}{\widehat{\boldsymbol \kappa}}
    + x_i\boldsymbol h^{(i)}, t), B \big)
    \Big|
    \n
    &\leq &
    \sum_{{\boldsymbol x}\in\Z^{D-1}}
    \frc c{\lt(\sqrt{(|v|t-1)^2|\widehat{\boldsymbol \kappa}|^2
    +| x_i\boldsymbol h^{(i)}|^2}+1\rt)^p}
    \n
    &\leq& \sum_{{\boldsymbol x}\in\Z^{D-1}}
    \frc c{\lt((|v|t-1)^2|\widehat{\boldsymbol \kappa}|^2
    +| x_i\boldsymbol h^{(i)}|^2\rt)^{p/2}}.
\end{array}\end{equation}  
 Note that the right-hand side never gets infinite, since $\hat{ \bm \kappa}$ is fixed and $(|v|t-1)|\widehat{\boldsymbol \kappa}|\geq VT>0$. In the second line we used the fact that
\begin{equation}  \begin{array}{*3{>{\displaystyle}lc}p{5cm}}
|\floor{vt}\widehat{\boldsymbol \kappa}+ x_i\boldsymbol h^{(i)}|^2 
    &=&
    |\floor{vt}|^2\,|\widehat{\boldsymbol \kappa}|^2  +|x_i\boldsymbol h^{(i)}|^2 \quad \mbox{(by orthogonality and the $L^2$ norm)}\n
    &\geq& (|v|t-1)^2|\widehat{\boldsymbol \kappa}|^2 +| x_i\boldsymbol h^{(i)}|^2
    \quad \mbox{(as $\floor{vt}> vt-1$ and $|vt-1|\geq ||v|t-1|$)}\n
    &\geq & (Vt)^2 \quad \mbox{(as $(|v|t-1)|\widehat{\boldsymbol \kappa}| \geq Vt$)}
\end{array}\end{equation}  
which allowed us to use space-like $p$-clustering. We now use the fact that $\sum_{\boldsymbol x\in\Z^{D-1}} F(\boldsymbol x) = \int \dd^{D-1}x\,F(\floor{\boldsymbol x})$, hence we'll have to deal with $\big|\,\floor{ x_i} \bm h^{(i)} \, \big|$. The idea is to do the following transformation:
\begin{equation}\label{eq:chgy}  \begin{array}{*3{>{\displaystyle}lc}p{5cm}} 
    |x_i\boldsymbol h^{(i)}|^2 &=& \sum_{m\in\t\zeta}
    |x_m|^2 + |x_{j_1} \widehat \kappa_{j_2}|^2
    + |x_{j_2}\widehat \kappa_{j_3}- x_{j_1}\widehat \kappa_{j_1}|^2
    + \ldots
    + |x_{j_{d-1}}\widehat k_{j_d}- x_{j_{d-2}}\widehat \kappa_{j_{d-2}}|^2\n
    &=& \sum_{m=1}^{D-1} |y_m|^2 = |\boldsymbol y|^2,\quad
    y_m = \lt\{\ba{ll}
    x_m & (m\in\t\zeta)\\
    x_{j_1} \widehat \kappa_{j_2} & (m = j_1)\\
    x_{j_l}\widehat \kappa_{j_{l+1}} - x_{j_{l-1}}\widehat \kappa_{j_{l-1}} & (m = j_l \in\{j_2,\ldots,j_{d-1}\}).
    \ea\rt.
\end{array}\end{equation}  

The  finite Jacobian of the change of variable above is (nonzero)
\begin{equation}
    J = \Big|\frc{\p \boldsymbol y}{\p\boldsymbol x}\Big|
    = \prod_{l=1}^{d-1} |\widehat \kappa_{j_{l+1}}|,\quad 0<J<\infty.
\end{equation}

%First, recall that for any two vectors $\boldsymbol a,\,\boldsymbol b$, we have $|\boldsymbol a + \boldsymbol b| \geq |\boldsymbol a| - |\boldsymbol b|$ and $|\boldsymbol a + \boldsymbol b| \leq |\boldsymbol a| + |\boldsymbol b|$.
We have by the triangle inequality:
\begin{equation}  \begin{array}{*3{>{\displaystyle}lc}p{5cm}}
    |\floor{x_i}\boldsymbol h^{(i)}| = |(x_i-\varep_i)\boldsymbol h^{(i)}|
    \geq |x_i\boldsymbol h^{(i)}| - |\varep_i\boldsymbol h^{(i)}|
    \geq |x_i\boldsymbol h^{(i)}| - h
\end{array}\end{equation}  
where we used $\floor{x_i}=x_i -ε_i$ for some $ε_i\in[0,1)$, and where $h = \sum_i|\boldsymbol h^{(i)}|$. Hence, doing the transformation $\bm x \mapsto \bm y$, defined in Eq. \eqref{eq:chgy}:
\begin{equation}  \begin{array}{*3{>{\displaystyle}lc}p{5cm}}
     | ( ι_{  \floor{vt}} τ_{t} a,b )^{\widehat {\boldsymbol \kappa},\perp} | &\leq& \int_{\R^{D-1}} \dd^{D-1} x\,
    \frc c{\big((|v|T-1)^2|\widehat{\boldsymbol \kappa}|^2+(  | x_i\boldsymbol h^{(i)}| - h)^2\big)^{p/2}}
    \n
    &= &
   J^{-1} \int_{\R^{D-1}} \dd^{D-1} y\,
    \frc c{((|v|T-1)^2|\widehat{\boldsymbol \kappa}|^2+(|\boldsymbol y|-h)^2)^{p/2}}
    \n
    &= &
    J^{-1}((|v|T-1)|\widehat{\boldsymbol \kappa}|)^{D-1-p} \int_{\R^{D-1}} \dd^{D-1} y\,
    \frc c{\lt(1+\lt(|\boldsymbol y|-\frc h{(|v|T-1)|\widehat{\boldsymbol \kappa}|}\rt)^2\rt)^{p/2}}
    \n
    &\leq &
   cIJ^{-1}((|v|T-1)|\widehat{\boldsymbol \kappa}|)^{D-1-p}\quad
    (\mbox{$I$ is defined in \eqref{defI}}).
\end{array}\end{equation}  
In the third line we used the fact that $(|v|T-1)|\widehat{\boldsymbol \kappa}|>0$. In the final line, we used $(|v|T-1)|\widehat{\boldsymbol \kappa}|\geq VT$ and, for every $\ell\in(0, h/VT]$,
\begin{equation}  \begin{array}{*3{>{\displaystyle}lc}p{5cm}}
    \lefteqn{\int_{\R^{D-1}} \dd^{D-1} y\,
    \frc 1{\lt(1+\lt(|\boldsymbol y|-\ell\rt)^2\rt)^{p/2}}} \n
    &\leq&
    \int_{\R^{D-1},|\boldsymbol y|\geq\frc{h}{VT}} \dd^{D-1} y\,
    \frc 1{\lt(1+\lt(|\boldsymbol y|-\frc{h}{VT}\rt)^2\rt)^{p/2}}
    +
    \int_{\R^{D-1},|\boldsymbol y|<\frc{h}{VT}} \dd^{D-1} y\,1
   \  =:\  I\label{defI}
\end{array}\end{equation}  
where $I$, as defined by the right-hand side of the inequality, is finite, as $p>D+1>D-1$, and only depends on $h/(VT)$ and $D$.

For $|v|< V/|\widehat{\boldsymbol \kappa}|+T^{-1}$, we instead use the fact that 
\begin{equation}
    |( ι_{  \floor{vt}} τ_{t} a,b )^{\widehat {\boldsymbol \kappa},\perp} |
    \leq ||a||^{\widehat {\boldsymbol \kappa},\perp}\,
    ||b||^{\widehat {\boldsymbol \kappa},\perp}
\end{equation}
by the Cauchy-Schwartz inequality and space-time translation invariance.

Thus we set
\begin{equation}
    f(v) = \lt\{\ba{ll}
    c IJ^{-1}
    ((|v|T-1)|\widehat{\boldsymbol \kappa}|)^{D-1-p}
    & (|v|\geq V/|\widehat{\boldsymbol \kappa}|+T^{-1}) \\
    ||a||^{\widehat {\boldsymbol \kappa},\perp}\,
    ||b||^{\widehat {\boldsymbol \kappa},\perp}
    & (|v|< V/|\widehat{\boldsymbol \kappa}|+T^{-1}).
    \ea\rt.
\end{equation}
We see that, for $p>D+1$, this indeed satisfies the right properties, and in particular this lower bound on $p$ is necessary for the integral of $(|v|+1)f(v)$ to exist on $\R$.
\end{proof}

From this, we obtain the following crucial lemma:
\begin{lem}\label{thelemma}
If the dynamical system satisfies Property \ref{property}, then $\forall A,B \in \mathfrak{U}_{\rm loc}$, $\forall s \in \mathbb{R}$, there exists $T_0>0$ such that for every ${\boldsymbol \kappa}\in \mathbb{R}^D_{\mathbb Q}$, the following holds:
\begin{equation}\label{basiclem}
\lim_{T \to \infty} \frac{1}{T} \int_{T_0}^T \dd t \,g(t) = 0 \ , \ \ g(t)= \sum_{{\boldsymbol x} \in \mathbb{Z}^D} \big( e^{\ri{\boldsymbol \kappa}{\boldsymbol x}/t} -e^{\ri{\boldsymbol \kappa}{\boldsymbol x}/(t+s)} \big) \big(A(\bm x,t), B \big).
\end{equation} 
\end{lem}

\begin{proof}
First we write
\begin{equation}
    g(t) = G_0(t) - G_s(t),\quad
    G_s(t) = \sum_{{\boldsymbol x} \in \mathbb{Z}^D} e^{\ri{\boldsymbol \kappa}{\boldsymbol x}/(t+s)} \big(A(\bm x,t), B \big).
\end{equation}
Let us first simplify $G_s(t)$ for arbitrary $s$. Writing (in a unique fashion) $\boldsymbol x = \boldsymbol r + \boldsymbol \lambda$ where $\boldsymbol r \in \Z^D_{\widehat{\boldsymbol \kappa}}$ and $\bm \lambda\in\Lambda_{\widehat{\boldsymbol \kappa}}$, we get
\begin{equation}
    G_s(t) = \sum_{\boldsymbol r\in \Z^{D}_{\widehat{\boldsymbol \kappa}}}
    e^{\ri {\boldsymbol \kappa}\boldsymbol r/(t+s) } \big( \t A_{t+s}(\bm r, t), B \big)
    = (\tau_t \t a_{t+s},b)^{\widehat{\boldsymbol \kappa}}_{\boldsymbol \kappa/(t+s)}
\end{equation}
where (recall that $\Lambda_{\widehat{\boldsymbol \kappa}}$ is finite)
\begin{equation}
    \t A_{u} = \sum_{\boldsymbol\lambda\in\Lambda_{\widehat{\boldsymbol \kappa}}}
    e^{\ri  \boldsymbol \kappa
    \boldsymbol\lambda /u}\iota_{\boldsymbol \lambda}A
\end{equation}
and the lowercase $\t a_u,b\in \mathcal V^{\widehat {\boldsymbol k}}_{\boldsymbol \kappa/(t+s)}$ are the respective equivalence classes of $\t A_u,B$ (see \Cref{lemma_sesquiliniar} and discussion below). We note that
\begin{equation}  \begin{array}{*3{>{\displaystyle}lc}p{5cm}}
    \lim_{t\to\infty} |G_s(t) - (τ_t \t a_{\infty},b)^{\widehat{\boldsymbol \kappa}}_{\boldsymbol \kappa/(t+s)}|
    &\leq& \lim_{t\to\infty}
    ||τ_t(\t a_{t+s} - \t a_{\infty})||^{\widehat{\boldsymbol \kappa}}_{\boldsymbol \kappa/(t+s)}
    ||b||^{\widehat{\boldsymbol \kappa}}_{\boldsymbol \kappa/(t+s)}\n
    &\leq& \lim_{t\to\infty}
    {\rm max}\{
    |\sin(\boldsymbol \kappa
    \boldsymbol\lambda /(2(t+s)))|
    :{\boldsymbol\lambda \in \Lambda_{\widehat{\boldsymbol \kappa}}}\}
    \, ||\t a_{\infty}||^{\widehat{\boldsymbol \kappa}}_{\boldsymbol \kappa/(t+s)}
    ||b||^{\widehat{\boldsymbol \kappa}}_{\boldsymbol \kappa/(t+s)}\n
    &=& 0
\end{array}\end{equation}  
where we used the fact that $\lim_{\boldsymbol \kappa'\to\boldsymbol 0} ||c||_{\boldsymbol \kappa'}^{\widehat{\boldsymbol \kappa}} = ||c||_{\boldsymbol 0}^{\widehat{\boldsymbol \kappa}}$ exists for any $c$. Therefore, using the form $\boldsymbol r = x_i\boldsymbol h^{(i)}+z\widehat{\boldsymbol \kappa} \in \Z^D_{\widehat{\boldsymbol \kappa}}$,
\begin{equation}  \begin{array}{*3{>{\displaystyle}lc}p{5cm}}
    \lim_{T \to \infty} \frac{1}{T} \int_{T_0}^T \dd t \,g(t) &=&
    \lim_{T \to \infty} \frac{1}{T} \int_{T_0}^T \dd t \,\Big(
    (τ_t \t a_{\infty},b)^{\widehat{\boldsymbol \kappa}}_{\boldsymbol \kappa/t}
    -
    (τ_t \t a_{\infty},b)^{\widehat{\boldsymbol \kappa}}_{\boldsymbol \kappa/(t+s)}\Big)\n
    &=&
    \lim_{T \to \infty} \frac{1}{T} \int_{T_0}^T \dd t \,
    \sum_{z\in\Z}
    \Big(
    e^{\ri |{\boldsymbol \kappa}| |\widehat{\boldsymbol \kappa}|z/t }- 
    e^{\ri |{\boldsymbol \kappa}| |\widehat{\boldsymbol \kappa}|z/(t+s) }\Big) ( ι_{z} τ_t  \t a_{\infty}, b )^{\widehat {\boldsymbol \kappa},\perp}.
\end{array}\end{equation}  
At this point, we have managed to recast the problem into an effectively one-dimensional problem. The proof now broadly follows that in $D=1$, \cite{doyon_hydrodynamic_2022}. In order to show the Lemma, i.e\ show $\lim_{T \to \infty} \frac{1}{T}\int_{T_0}^{T} \dd t g(t)=0$, we want to commute the limit and the time integral past the summation $\sum_{z \in \Z}$ in $g(t)$. The result would then follow by applying Property \ref{PROPERTY}.\ref{property1}. We proceed to uniformly (in $t$) bound the summand in $g(t)$. As in the proof of Lemma 6.6 in \cite{doyon_hydrodynamic_2022}, we write (denoting $κ=|\bm \kappa| | \bm{ \hat \kappa|}$):
\begin{equation}
 \begin{array}{*3{>{\displaystyle}lc}p{5cm}}
    g(t) &=& \sum_{z \in \Z} 2i \exp{\frac{iκz}{2} \big( \frac{1}{t} +\frac{1}{t+s}\big)} \operatorname{sin}\big[ \frac{κz}{2} \big( \frac{1}{t} - \frac{1}{t+s} \big) \big] ( ι_z τ_t \t a_{\infty},b)^{\bm{\hat \kappa},\perp} \\
    &=& \sum_{υ \in t^{-1} \Z} \frac{i κυ}{2} \exp{\frac{iκυ}{2} \big( 1+\frac{t}{t+s}\big)} \frac{2t}{κυ}\operatorname{sin}\big[ \frac{κυ}{2t} \big( t - \frac{t^2}{t+s} \big) \big] ( ι_{υt} τ_t \t a_{\infty},b)^{\bm{\hat \kappa},\perp}\\
    &= &\int_{\R} i κ υ_t \exp{\frac{ i κ υ_t}{2}\big(1 + \frac{t}{t+s})} 
     \frac{2t}{ κ υ_t}  \operatorname{sin}\big(\frac{κ υ_t}{2t}  \big(t - \frac{t^2}{t+s}\big)\big) ( ι_{\floor{υt}} τ_t \t a_{\infty} ,b)^{\bm{\hat \kappa}, \perp}\, \dd υ
\end{array}
\end{equation}
where $υ_t= t^{-1} \floor{ υt}$. Having established Property \ref{PROPERTY}, the rest of the proof follows exactly as in \cite[Lemma 6.6]{doyon_hydrodynamic_2022}.
\iffalse
\begin{equation}
\sum_{\boldsymbol x\in \Z^{D-1}}\sum_{z\in\Z}
    e^{\ri |{\boldsymbol \kappa}| |\widehat{\boldsymbol \kappa}|z/(t+s) } \langle ι_{x_i\boldsymbol h^{(i)}+z\widehat{\boldsymbol \kappa}} τ_t  \t a_{t+s}, b \rangle
\end{equation}
\fi

\end{proof}

Finally, having shown Lemma \ref{basiclem}, we have for all $A,B \in \mathfrak{U}_{\rm loc}$, any rational vector ${\boldsymbol \kappa} \in \mathbb{R}^D$, $s \in \mathbb{R}$ and any choice of Banach Limit $\widetilde{\lim_{t \to \infty}}$:

\begin{equation}
\begin{array}{*3{>{\displaystyle}lc}p{5cm}}
0&=& \widetilde{\lim_{t \to \infty}}\frac{1}{T}\int_0^T \bigg( \langle τ_t a,b \rangle_{{\boldsymbol \kappa}/t} - \langle τ_t a,b \rangle_{{\boldsymbol \kappa}/(t+s)} \bigg) \,\dd t \\
&=& \widetilde{\lim_{t \to \infty}}\frac{1}{T}\int_0^T \langle τ_t a,b \rangle_{{\boldsymbol \kappa}/t} \, \dd t- \widetilde{\lim_{t \to \infty}}\frac{1}{T}\int_0^T \langle τ_t a,b \rangle_{{\boldsymbol \kappa}/(t+s)}  \,\dd t
\end{array}
\end{equation}
where $a,b \in \mathcal{V}_{\bm \kappa/t}$ are the respective equivalence classes (see \Cref{inner} and below).
We then proceed exactly as in the proof  \cite[Theorem 6.7]{doyon_hydrodynamic_2022}, which completes the proof of the Hydrodynamic projection theorem for any dimension $D$ and for zero wavenumber and frequency.

The proofs can easily be generalised to all frequencies and wavelengths, as all the bounds remain essentially unchanged.  We consider different representations of the groups of space and time translations on the C$^*$-algebra $\mathfrak{U}$, for each frequency-wavenumber pair, defined as
\begin{equation}
    \t {ι}_{\bm x} \coloneqq e^{\ri \bm k \bm x} ι_{\bm x} , \ \t {τ_t} \coloneqq e^{\ri f t} τ_t
    \quad \mbox{for}\  \bs k\in\R^D, f\in\R.
\end{equation}
These form representations of the groups $\Z^D$, $\R$ respectively, by bijective linear maps on $\mathfrak{U}$. They are not $^*$-automorphisms like $ι_{x}$ and $τ_{t}$. However, linearity and unitarity on the Hilbert spaces defined from the state are all that is used in our proofs (and of course the group representation properties). Hence we can relax the Definition of a dynamical system, \Cref{defn:dynamicalsystem}, to a triplet $(\mathfrak{U},\t{ι},\t{τ})$ where $\t{τ}$ is a strongly continuous representation of the group $\R$ by bijective linear maps $\{ \t{τ}_t : \mathfrak{U} \xrightarrow{\sim} \mathfrak{U} \}_{t \in \R}$, and $\t{ι}$ is a representation of the translation group $\Z^D$ by bijective linear maps $\{ \t{ι}_{\boldsymbol x}: \mathfrak{U} \xrightarrow{\sim} \mathfrak{U} \}_{\boldsymbol x \in \Z^D}$, and the state is required to have the invariance property $(\t\iota_{\bs x}\t\tau_t(A),\t\iota_{\bs x}\t\tau_t(B)) = (A,B)$.
It is clear that if  uniform $p$-clustering holds for a  subset of elements in $\mathfrak{U}_{\rm loc}$ under
$\iota_{\boldsymbol x}$, then it also holds under $\t\iota_{\boldsymbol x}$. Under this new represenations of space translations we can construct the Hilbert space of extensive quantities as
\begin{equation}
    \t{\mathcal H}_{\bs 0} = \mathcal H_{\bs k}.
\end{equation}
Further, it is immediate that the Lieb-Robinson bound also holds for $\t\tau_t = e^{\ri ft}\tau_t$.  Thus, by the Lieb-Robinson bound, in this case $\t\tau_t$ is also a one-parameter unitary group on $\t{\mathcal H}_{\bs 0}$. 
We then can define the space of $\t\tau_t$-invariants $\t{\mathcal Q}$,
\begin{equation}
    \t{\mathcal Q} = \{a\in\mathcal H_{\bs k}:
    \tau_t a = e^{-\ri f t}a\;\forall\;
    t\in\R\},
\end{equation}
and the orthogonal projection
\begin{equation}
    \t{\mathbb P}:\t{\mathcal H}_{\bs 0} \to \t{\mathcal Q}.
\end{equation}

Finally, it is easily seen (by $p$-clustering and the Lieb-Robison bound) that if the dynamical system $(\mathfrak U,\iota,\tau)$ and state $ω$ is space-like $p_c$ clustering according to \Cref{defn:spacelikeclustering}, then so is $(\mathfrak U,\t\iota,\t\tau)$, $ω$.  From there on, the results we proved above also hold for these new quantities:
\begin{equation}
\t S_{a,b} ({\boldsymbol \kappa}) = \lim_{t\to \infty} \frac1T \int_0^Te^{-\ri f t}\langle  τ_t a,b \rangle_{\boldsymbol k + {\boldsymbol \kappa}/t}	 \, \dd t.
\end{equation}

\chapter{Diffusion bound in open chaotic nearest neighbor spin chains} \label{chapter:diffusion}

The results of the previous chapter concerned the Euler scale of hydrodynamics and the emergence of the ballistic modes, characterised by the extensive conserved quantities, which carry correlations. In quantum spin chains $D=1$, our results can be used to obtain the Euler equations, as shown in \cite{doyon_hydrodynamic_2022}. This corresponds to the first order of the hydrodynamic equation we discussed in \Cref{section:intro_hydro}.  What happens if we go to finer scales? Is there a way of rigorously characterising diffusion?  As argued heuristically in \Cref{section:intro_hydro}, diffusion occurs within a derivative expansion of the current, accounting for the principle that the
system tends towards entropy maximization. In a long-wavelength state $\langle \ldots \rangle$ we argued: 
    \begin{equation} \label{eq:intro_current_expansion2}
    \overline j_i(x,t) =  j_i^{\rm av}(   \overline q(x,t)) - \frac12 \sum_k \mathcal{D}_i^k( \overline q(x,t)) \partial_x \overline q_k(x,t) + \mathcal{O} ( \partial^2_x)
\end{equation}
which in combination with the conservation laws, yields the hydrodynamic equation \Cref{eq:hydrodynamic_eq}. The question we ask is how to gain access to the diffusion matrix $\mathcal{D}_i^k$, what are the microscopic quantities necessary to describe it, and can we show it to be non-zero, whereby diffusive transport occurs?

 Besides the diffusion integral kernel in hard-rod models \cite{boldrighini_1997_hard_rod}, this remains largely unsolved, with very few rigorous results that establish lower bounds on the strength of diffusion. A crucial first step is to determine rigorous lower bounds on the strength of the diffusive spread of response functions, via the celebrated Green-Kubo formula for the Onsager matrix and Einstein's relation connecting it to the diffusion parameters (see e.g.~\cite{spohn_large_1991,DeNardis_Doyon_2019_diffusion_GHD}). The Onsager matrix is manifestly non-negative in parity-time-symmetric dynamics \cite{DeNardis_Doyon_2019_diffusion_GHD}, leading to non-negative entropy production; however showing that it is strictly positive is a difficult problem.  One such bound on spin diffusion at infinite temperature, obtained by using special quadratically extensive charges, was obtained in \cite{Prosen_2014_diffusion_bound}, and a second bound by the curvature of the Drude weight in \cite{Medenjak_2017_diffusion_bound}. Both rely on the assumed absence of ballistic transport of spin and focus on integrable models, where the charge is explicitly constructed.

In general, there is a mechanism for diffusive effects, coming from large-scale fluctuations of the system's emergent hydrodynamic modes. This happens if hydrodynamic modes exhibit {\em interacting ballistic transport} -- more precisely, the eigenvalues of the flux Jacobian \Cref{eq:intro_flux}, which are the hydrodynamic velocities generalising the sound velocity in gases, are non-constant functions of the state. Indeed, in such cases, it is expected that fluctuations of the ballistic trajectories induced by initial-state macroscopic, thermodynamic fluctuations \cite{doyon_2023_hydro_longrange,doyon_2023_ballistic_fluctuations}, will lead to spreading that may contribute to diffusion \cite{Medenjak_Yoshimura_2020_Diffusion_bound,hubner_2024_diffusive} and even anomalous effects \cite{gopalakrishnan_2024_universality,krajnik_2022_anomalous_current_fl,yoshimura_2024_anomalous_current,gopalakrishnan_2024_nongaussian_dif}. A proof of this principle, which generalizes the ideas of \cite{Prosen_2014_diffusion_bound}, is found in \cite{doyon_2019_diffusion} where the Onsager matrix is bounded rigorously by a projection formula onto quadratically extensive charges constructed from ballistic modes and representing initial-state fluctuations. This works in one-dimensional systems under the assumption of strong-enough clustering properties of correlations. This principle is important: although chaotic Hamiltonian chains do not admit ballistic transport of spin or energy, as follows from a general result of Kobayashi and Watanabe \cite{kobayashi_2022_vanishing_currents}, as soon as the Hamiltonian structure is broken -- such as in open chains (i.e.~with incoherent processes) or unitary quantum circuits (i.e.~with discrete time) -- interacting ballistic transport may arise. However, the bound in \cite{doyon_2019_diffusion} is not explicitly non-zero, the presence of symmetries can in fact force it to be zero, and it relies on clustering assumptions that were until recently unproven.

In this chapter, based on our work in \cite{ampelogiannis_2024_diffusion}, we construct a family of interacting, translation-invariant, nearest-neighbor spin-$1/2$ open (Lindbladian) infinite-length quantum chains, which conserve total magnetization, and for which we show an explicitly positive bound on the strength of linear-response spin diffusion. The models accounts for spin hopping, both coherently and incoherently, on nearest neighbors, along with phase decoherence effects. For the notion of diffusion to be meaningful in a Lindbladian system, it is natural to impose two properties which suggest that spin transport be of ``hydrodynamic type". First, we require magnetization $M = \sum_{x\in\Z} \sigma_x^3$ to be {\em strongly conserved} (generating a strong symmetry \cite{buca_2012_lindblad_transport}). This implies that we have a local spin conservation law. Second, we require the two-site local incoherent processes to satisfy a {\em local detailed-balance condition} (related to quantum detailed balance \cite{Alicki_1976_detailed_balance,fangola_2007_generator_quantum_markov,carlen_2017_gradient}). We show that this makes Gibbs states with density matrix $e^{\mu M}$ stationary. Thus one may expect emergent hydrodynamics for the local conserved spin thanks to local thermalization. Transport is generically not reversible in Lindbladian systems, but we show that local detailed balance also implies the existence of a (different) time-reverse dynamics that is well defined on local observables in the infinite chain. The family of models we consider is the most general family of translation-invariant, nearest-neighbor spin-1/2 open chains solving these two requirements.

For this family, we show that the spin-spin Onsager coefficient, which we define below, is bounded by projection onto quadratic charges, and that the bound is strictly positive if and only if  incoherent transport processes generate a non-zero current. When this happens, spin is an interacting ballistic mode as, we show, there is a spin current in Gibbs states that depends quadratically on the average local magnetization $\mathsf s$, leading to a hydrodynamic velocity $v(\mathsf s)$ that depends on the $\mathsf s$; the bound is proportional to $(v'(\mathsf s))^2$ (Eq.~\Cref{boundjprime} below). Physically, then, the bound accounts for the spreading effects of initial-state macroscopic fluctuations, the mechanism described above. We establish the bound by applying the result of \cite{doyon_2019_diffusion}, which we make fully rigorous by adapting the proof in order to use recent results on clustering of $n$-th order correlations \cite{ampelogiannis_2024_clustering} presented in \Cref{chapter:higher_clustering}. 
%Finally, we show that for certain choices of parameters, the Lindbladian evolution is in fact {\em reversible}, and the irreversibility diffusion strength vanishes. Thus, for this sub-family of models, we have established a strictly positive normal spin diffusion strength.
The techniques will apply to any Lindbladian quantum system with strong spin conservation and local detailed balance, and we believe the results will generically hold.

\section{Overview}\label{section:diffusnion_overview}

Because of the lack of time reversal (more precisely, parity-time-reversal) symmetry, constructing a manifestly non-negative quantity that measures diffusion requires special care. We construct a {\em normal diffusion strength} $\mathfrak L_{\rm norm}(t)$ measuring the linear-response diffusive spread of forward spin propagation at time $t$, and an {\em irreversibility diffusion strength} $\mathfrak L_{\rm irr}(t)$ measuring the diffusive spread of the mismatch, due to irreversibility, obtained by performing forward then backward evolution by time $t$:
\begin{align}
    \mathfrak L_{\rm norm}(t) &=
    \frc1{t} \sum_{x\in\Z} (x^2-(v(\mathsf s)t)^2) \langle
    \tau_t^*(s(x)),s(0)\rangle^c\\ 
    \mathfrak L_{\rm irr}(t)&=
    \frc1{2t} \sum_{x\in\Z} x^2\langle \tau_{t}\tau^*_t( s(x)),s(0)\rangle^c.
\end{align}
\nomenclature[cor]{$\langle A,B\rangle^c$}{ Connceted correlation between $A$ and $B$ in the state $ω$: $ω(ΑB)-ω(A)ω(B)$}
Here $\langle\cdot,\cdot\rangle^c$ is the connected correlation function in the Gibbs state $e^{\mu M}$ with average magnetization density $\mathsf s = \Tr e^{\mu M}\sigma^3_0/\Tr e^{\mu M}$; $A(x,t) = \tau_t^*(A(x))$ is the Heisenberg-picture forward Lindbladian time evolution of the local operator $A(x)$; $\tau_t$ is the backward time evolution; $v(\mathsf s)$ is the hydrodynamic spin velocity in the Gibbs state; and $s(x)=\sigma_x^3$ is the spin operator at $x$. Note that in the state with zero chemical potential ($\mathsf s=0$), the hydrodynamic velocity vanishes $v(0)=0$. Intuitively, the irreversibility diffusion strength encodes the entropy increase that may occur in the ``forgotten" environment.
Then we show, with full mathematical rigor and without extra assumption, in the general, explicit family of translation-invariant, nearest-neighbor spin-1/2 open chains with strong magnetization conservation and detailed balance, Eqs.~\eqref{time_evolution}, \eqref{hamiltonian_density}, \eqref{jump_operators}, that:
\begin{equation}\label{boundintro}
    \mathfrak L = \liminf_{t\to\infty} \Big(\mathfrak L_{\rm norm}(t) - \mathfrak L_{\rm irr}(t)\Big) \geq
    \frc{(\chi v'(\mathsf s))^2}{8v_{\rm LR}}
\end{equation}
where $\chi$ is the magnetic susceptibility and $v_{\rm LR}$ is the Lieb-Robinson velocity. In the family of models considered, expressions for $\chi$, $v'(\mathsf s)$ and $v_{\rm LR}$ are given in \eqref{chiexplicit}, \eqref{fluxcurvature}, and \eqref{eq:LR_velocity} with \eqref{Vtilde}, respectively. We expect the bound in the form \eqref{boundintro} to be universal for chaotic models with strong spin conservation and local detailed balance. Note that the limits $\lim_{t\to\infty} \mathfrak L_{\rm norm}(t)$ and $\lim_{t\to\infty}\mathfrak L_{\rm irr}(t)$ may in fact exist, but we have not proven this; in this case, the result is for the difference of these limits. The proof is obtained by naturally adapting the standard Green-Kubo formula, for spin diffusion, to Lindbladian evolution:
\begin{align}
   & \mathfrak L = \liminf_{T\to\infty} \frc1{T}\int_0^T\dd t \int_0^T\dd t'\,\sum_{x\in\Z}
    \langle j^-(x,t),j^-(0,t')\rangle^c, \label{Onsager} \\
    &\hspace{6cm}j^-(x,t) = j(x,t) - vs(x,t) .\label{eq:proejcted_spin_current}
\end{align}
We use quadratic charge projections in order to lower-bound this quantity, and we show that it indeed decomposes into the normal and irreversible contributions as above.

The usefulness of the above results relies on the idea that there are no other extensive conserved quantities coupling to the spin current -- as otherwise, additional ballistic spreading is expected to occur and the quantity $\mathfrak L$ is infinite. In its most general form, the Onsager matrix is
\begin{align}
   & \mathfrak L_{mn} = \liminf_{T\to\infty} \frc1{T}\int_0^T\dd t \int_0^T\dd t'\,\sum_{x\in\Z}
    \langle j_m^-(x,t),j_n^-(0,t')\rangle^c, \label{Onsagerintro} \\
    &\hspace{6cm}A^-(x,t) = (1 - \mathbb P) A(x,t)\label{eq:projected_operator}
\end{align}
where $j_m$ are the currents, and $\mathbb P$ is the projector onto the Hilbert space $\mathcal H_1$ of extensive conserved quantities \cite{doyon_hydrodynamic_2022} (we show that this Hilbert space exists for Lindbladian evolution with local detailed balance). As usual, this definition requires the knowledge of all extensive conserved quantities, which is difficult to establish in general, and \eqref{Onsager} with \eqref{eq:proejcted_spin_current} is obtained by assuming that the spin chain is chaotic, in the sense that $\mathcal H_1$ be one-dimensional, spanned by $M$ only (there is only the spin current $j=j_1$). More precisely, we assume that {\em spin transport be chaotic}: the spin current projects only onto $M$. Proving chaoticity is a non-trivial task that is currently out of reach, as it requires one to establish the absence of other extensive charges \cite{Prosen_1999_ergodic,prosen_1998_quantum_invariants,doyon_thermalization_2017,ampelogiannis_long-time_2023, doyon_hydrodynamic_2022}. In Hamiltonian systems there are exciting partial results, showing the absence of charges with local densities (a subset of extensive charges) in certain models \cite{shiraishi_2019_absence_local,chiba_2023_ising_nonintegrability,shiraishi2024absence,park_2024_nonintegrability}. In our understanding it is widely expected that generic interacting spin chains be chaotic.

\section{Open quantum spin chain}
We consider a quantum spin-$1/2$ chain ($D=1$) described by a quasi-local C$^*$-algebra $\mathfrak{U}$, as described in \Cref{section:quantum_spin_lattice}, with local spaces $\C^2$.
 We consider a homogeneous Lindbladian evolution where the generator of the dynamics is defined for each finite $Λ\subset \Z$ and contains both Hamiltonian nearest-neighbor interactions and dissipative two-site terms. The local Hamiltonian, for finite $Λ\subset \Z$, is
\begin{equation}
   H_Λ = \sum_{x\in Λ} h(x), \quad h(x) \in \mathfrak{U}_{\rm loc} .
\end{equation} 
The dissipative part is described by local quantum jump (or Lindblad) operators $L_i(x)\in \mathfrak{U}_{\rm loc}$, $x\in \Z$ for $i$ in some finite index set. On local observables $A\in \mathfrak{U}_{Λ}$ the Lindbladian acts as\footnote{This is the standard form of the Lindbladian in the Heisenberg representation, but written in terms of commutators -- this makes clear that its action is well defined on local operators.}
\begin{equation} \label{time_evolution}
    \mathcal L_{Λ}^*(A) = \partial_t A(t) =  i[H_Λ,A] 
     + \frc12\sum_{x\in Λ,\, i}\big(L_i(x)^\dag [A,L_i(x)] + [L_i(x)^\dag,A]L_i(x)\big). \no
\end{equation}
The thermodynamic limit on the generator $\lim_{Λ \to \Z}\mathcal L_\Lambda^* = \mathcal L^*$ exists on $\mathfrak U_{\rm loc}$; and that of the dynamics
\begin{equation}
    \tau^*_t = \lim_{Λ\to\Z} e^{t\mathcal L^* _\Lambda}
\end{equation}
exists on $\mathfrak U$ and is a strongly continuous completely positive dynamical semigroup  \cite{nachtergaele_2011_LR_irreversible}. 

We require strong spin conservation: the Hamiltonian density, and each quantum jump, are all required to preserve the total spin of the quantum chain,
\begin{equation}\label{strongconserved}
    [M,h(x)]= [M,L_{i}(x)] = 0,\quad M= \sum_{x\in\Z} \sigma^3_x.
\end{equation}
Here and below, for any $A\in \mathfrak{U}_{\rm loc}$ we write $[M,A(x)] = \lim_{\Lambda\to \Z} [M_\Lambda,A(x)]$ with $M_\Lambda= \sum_{x\in\Lambda} \sigma^3_x$. By  spin conservation, the Hamiltonian density and quantum jump operators take the general forms
\begin{equation} \label{hamiltonian_density}
    h(x) = \alpha\sigma^+_x\sigma^-_{x+1}
    + \bar\alpha \sigma^-_x\sigma^+_{x+1}
    + \beta\sigma^3_{x}\mathds{1}_{x+1}
    + \gamma\sigma^3_x\sigma^3_{x+1},
\end{equation}
and
\begin{equation} \label{jump_operators}
    L_i(x) = a_i\sigma^+_x\sigma^-_{x+1}
    + b_i \sigma^-_x\sigma^+_{x+1}
    + c_i\sigma^3_{x}\mathds{1}_{x+1}
    + d_i \mathds{1}_{x}\sigma^3_{x+1}
    + e_i\sigma^3_x\sigma^3_{x+1}
\end{equation}
where $α\in\C$, $\beta,γ\in \R$, $a_i,b_i,c_i,d_i,e_i \in \C$ are constants, which we will refer to as interaction parameters. Here, $σ_x^1$,$σ_x^2$,$σ_x^3 \in \mathfrak{U}_{ \{x \}}$ are the x, y and z Pauli matrices respectively acting on $\mathfrak{U} _{\{x \} }$ and $σ^{\pm}= \frac12(σ^1 \pm i σ^2)$.  Note that for the Hamiltonian density it is not necessary to have the term $\sigma^3_{x+1}$ as under $\sum_{x\in\Z}$ in $H$ this is redundant; while for the Lindblad operator $L_i$ there is no redundancy. Observe that for $|a_i|>|b_i|$ the $i$th process leads to incoherent spin transport towards the left, while for $|a_i|<|b_i|$ it is towards the right; and the sum of terms for $c_i,d_i,e_i$ represent various types of incoherent local dephasing.

Time evolution for Hamiltonian systems satisfies the Lieb-Robinson bound \cite{Lieb:1972wy}, which can be extended to Lindbladian dynamics \cite{poulin_2010_LR_markovian,nachtergaele_2011_LR_irreversible}:
\begin{equation}
    \norm{[A(x,t),B] }\leq C_{A,B} e^{-λ(x-υ_{\rm LR}|t|)}.
\end{equation}
The Lieb-Robinson velocity $υ_{\rm LR}$ for \eqref{time_evolution} can be estimated by the results \cite{nachtergaele_2011_LR_irreversible,sims_2011_LR}:
\begin{equation} \label{eq:LR_velocity}
    υ_{\rm LR} = 4eζ(2) V , \quad V=2 \norm{h(x)} + 2 \sum_i \norm{L_i(x)}
\end{equation}
where $ζ$ is the Riemann function, and we can estimate for \eqref{hamiltonian_density}, \eqref{jump_operators}:
\begin{equation}\label{Vtilde}
    V \leq \t V = 2( 2|α|+|β|+|γ|) + 2 \sum_i( |a_i|+|b_i|+|c_i|+|d_i|+|e_i|).
\end{equation}

We assume the system to be in a Gibbs state for the total spin, at chemical potential $μ$:

\begin{equation} \label{gibbs_state}
    \omega_\mu(A) = \lim_{Λ \to \Z} \frc{\Tr_{\prod_{i\in\Lambda} \C^2}( e^{\mu M_Λ}A)}{\Tr_{\prod_{i\in\Lambda} \C^2}(e^{\mu M_Λ})}, \quad A \in \mathfrak{U}
\end{equation}

Clearly, this state satisfies a strong clustering property, for every $A\in\mathfrak U_\Lambda$, $B\in\mathfrak U_{\Lambda'}$ with $\Lambda\cap \Lambda'=\emptyset$:
\begin{equation} \label{eq:strong_custering_mu}
    \omega_\mu(AB) = \omega_\mu(A)\omega_\mu(B)
\end{equation}

\subsection{Local detailed-balance assumption}
We assume the quantum jump operators to satisfy the local detailed-balance condition:
\begin{equation} \label{detailed_balance}
    \sum_i [L_{i}(x),L_{i}(x)^\dagger] = o(x+1)-o(x)
\end{equation}
for some ``detailed-balance current" operator $o(x)\in \mathfrak U_{\rm loc}$. Note that by strong spin conservation and locality of $o(x)$, we have $[M,o(x)]=0$ for all $x$. The local detailed-balance condition, along with strong spin conservation, guarantees that the Gibbs state $\omega_\mu$ is stationary, see \cite[Section 2]{ampelogiannis_2024_diffusion}. By a straightforward calculation we find that the local detailed-balance condition is equivalent to
\begin{align}\label{condition}
    \sum_i a_i(\b d_i - \b c_i) = \sum_i \b b_i(d_i-c_i),
\end{align}
Quantum detailed balance conditions have been discussed in \cite{Alicki_1976_detailed_balance,fangola_2007_generator_quantum_markov,carlen_2017_gradient}.

Another important consequence of the detailed-balance condition, is that is allows us to define a backward time-evolution by the adjoint of the Lindbladian. Consider the local Lindbladian defined by exchanging $L_i(x)\leftrightarrow L_i(x)^\dagger$ and changing the sign of the Hamiltonian,
\begin{equation}
    \label{time_evolution_conjugate}
    \mathcal L(A) = -i[H,A] 
     + \frc12\sum_{x\in \Z,\, i}\big(L_i(x) [A,L_i^\dag(x)] + [L_i(x),A]L_i^\dag(x)\big)
\end{equation}
Then, this gives rise to a strongly continuous completely positive dynamical semigroup 
\begin{equation}
    \tau_t=\lim_{\Lambda\to\Z} e^{t\mathcal L_\Lambda}
\end{equation} 
on $\mathfrak U$ with a Lieb-Robinson bound with \eqref{eq:LR_velocity}. Explicitly, $\mathcal L$ is obtained from $\mathcal L^*$ by exchanging $b_i\leftrightarrow \b a_i$ and $c_i,d_i,e_i\leftrightarrow \b c_i, \b d_i,\b e_i$ in \eqref{jump_operators}.  We interpret $\tau_t$ as the {\em backward time evolution} of the system; there is also a notion of extensive conserved quantities for backward time evolution. Note that $\mathcal L$ {\em does not} take the form the Lindbladian usually takes on density matrices; by contrast, the form above indeed makes sense on local operators in the infinite-volume setting. We show in \cite[Appendix B]{ampelogiannis_2024_diffusion} with certain choices of parameters, $\mathcal L = -\mathcal L^*$, in which case $\tau_t = \tau_{-t}^*$ and time evolution is unitary on $\mathcal H_k$: the dynamics is reversible.

\subsection{Chaoticity Assumption}
In general, the relevant degrees of freedom for the hydrodynamic scale are the extensive conserved quantities, formally $Q_n = \sum_{x \in \Z}q_n(x)$, afforded by the microscopic dynamics, $\mathcal{L}^*(Q_n)=0$. With $q_n(x,t)$, $j_n(x,t)$ conserved densities and their currents, we have conservation laws $\partial_t q_n(x,t) + j_n(x,t)- j_n(x-1,t)=0$; it is those that control hydrodynamics.

Systems with Lindbladian \eqref{time_evolution} may admit an infinity of conserved quantities (e.g.~if it is integrable, which may happen also in the presence of incoherent processes \cite{ziolkowska_2020_yang_baxter,essler_2020_integrablility_lindbladian}), or a finite number (e.g.~if it is chaotic). In many integrable models, one knows how to write down the conserved charges iteratively, but proving that this set is complete, and no other charges exist, remains a conjecture.
Likewise, proving that a system is chaotic remains an open problem. In our understanding, it is however expected that non-chaotic models are of measure $0$ in the space of interactions. As such, we make the assumption that the magnetisation $M$ is the only conserved charge of our system.

%%%We therefore assume that the only conserved extensive quantity of \eqref{time_evolution} is the magnetization $Q_1 = M= \sum_{x \in \Z} σ^3_x$; that is $\mathcal Q = {\rm span}(M)$. More precisely, in fact, we only need to assume that the spin current $j$ (as an element of $\mathcal H_1$) projects onto $M$; physically, we assume that spin transport be chaotic. We emphasize that this assumption is only used in addressing the projection operator in the Green-Kubo formula \eqref{Onsagerintro}. Once the formula is written for the projection onto magnetization, our lower bound result is rigorous. Note that $M$ is also conserved by the backward time evolution $\tau_t$.

\section{Onsager coefficient}
The spin current $j(x)$ for spin density $s(x) = \sigma_x^3$ is defined by the conservation law:
\begin{equation} \label{eq:spin_conservation}
\mathcal L^*(s(x)) + j(x)-j(x-1) = 0
\end{equation}
and decomposes into a Hamiltonian (coherent) and Lindbladian (incoherent) parts, $j(x) = j^{\rm H}(x) + j^{\rm L}(x)$. Both can be straightforwardly evaluated, but we will only need the incoherent part. A straightforward calculation gives:
\begin{align}
    j^{\rm L}(x) &= \sum_i\Big(
    2|b_i|^2 P_{x}^+P_{x+1}^- - 2|a_i|^2 P_x^-P_{x+1}^+ \n & \quad + \big(2a_i(\b d_i-\b c_i) + \b e_i a_i - e_i \b b_i\big)\sigma_x^+\sigma_{x+1}^-
    \n & \quad + \big(2\b a_i(d_i-c_i) + e_i \b a_i - \b e_i b_i\big)\sigma_x^-\sigma_{x+1}^+
    \Big) \label{Lcurrent}
\end{align}
where $P^\pm_x = (1\pm\sigma_x^3)/2$ are local projections onto positive (up) and negative (down) spins. Our goal is to estimate the diffusive transport coefficient \eqref{Onsager} for $j(x)$.

With the chaoticity assumption, formula \eqref{Onsagerintro} simplifies to ($\mathfrak L = \mathfrak L_{11}$):
\begin{align}
   & \mathfrak L = \liminf_{T\to\infty} \frc1{T}\int_0^T\dd t \int_0^T\dd t'\,\sum_{x\in\Z}
    \langle j^-(x,t),j^-(0,t')\rangle^c, \label{Onsager1} \\
    &\hspace{5cm} j^-(x,t) = j(x,t) - s(x,t) \chi^{-1} \langle M,j(0,0)\rangle^c \label{eq:proejcted_spin_current_chi}
\end{align}
where
\begin{equation}
    \chi = \langle M,s(0,0)\rangle^{\rm c}
\end{equation}
is the magnetic susceptibility.

What is the interpretation of formula \eqref{Onsager1}?

Technically diffusion occurs within a derivative expansion of the current, accounting for the principle that the system tends towards entropy maximization, as discussed in \Cref{section:intro_hydro}. In a long-wavelength state $\langle\cdots\rangle$, one writes phenomenologically $\langle j(x,t)\rangle = \mathsf j(\langle s(x,t)\rangle) -\frac12 \mathfrak D(\langle s(x,t)\rangle)\partial_x \langle s(x,t)\rangle + \ldots$ where $\mathsf j(\mathsf s) = \omega_\mu(j)$ for $
\mu$ such that $\mathsf s = \omega_\mu(s)$. It is a simple matter to evaluate the current and velocity by an explicit calculation of the trace,  as it factorises into the individual sites, by \Cref{eq:strong_custering_mu}, and we obtain
\begin{equation}
    \mathsf j = \omega_\mu(j^{\rm L}(x)) = \sum_i \frc{|b_i|^2-|a_i|^2}{2\cosh^2\mu}.
\end{equation}
As $\mathsf s = \omega_\mu(\sigma_x^3) = \tanh\mu$, we can express the current as a function of the spin density, obtaining
\begin{equation}\label{js}
    \mathsf j(\mathsf s) = 
    \sum_i(|b_i|^2-|a_i|^2)
    \frc{1-\mathsf s^2}2.
\end{equation}
Linear response arguments for the (phenomenologically justified) diffusive-scale hydrodynamic equation $\p_t \mathsf s(x,t) + v(\mathsf s(x,t)) \p_x \mathsf s(x,t)  = \frc12\p_x (\mathfrak D(\mathsf s(x,t)) \p_x \mathsf s(x,t))$ then give an equation for the correlation $\langle s(x,t),s(0,0)\rangle^c$, solved as a Gaussian centered around the velocity $v(\mathsf s)= d \mathsf j(\mathsf s)/d \mathsf s$:
\begin{equation}\label{diffusiveform}
    \langle s(x,t),s(0,0)\rangle^c \sim \frc{χ}{\sqrt{2\pi \mathfrak D_{\rm norm} t}}\, e^{-(x-vt)^2/(2\mathfrak D_{\rm norm}t)}\qquad \mbox{(hydrodynamic linear response)}
\end{equation}
where we recall that $A(x,t) = \tau^*_t(A(x))$ is the Heisenberg-picture time evolution of the local observable $A(x)\in \mathfrak{U}_{\rm loc}$. Here we have renamed $\mathfrak D\to\mathfrak D_{\rm norm}$ in order to distinguish it from the ``irreversibility diffusion" discussed below; this is {\em normal} diffusion.

Recall that we have different forward and backward time evolutions. According to hydrodynamic intuition, both should lead to diffusion, thus there should be diffusion associated to irreversibility as well. Time evolution $\tau_t$ leads to the opposite average current $-\mathsf j(\mathsf s)$, Eq.~\eqref{js}, thus the opposite hydrodynamic velocity. Evolving forward, then backward in time, and phenomenologically assuming diffusive extension around the end-point of the resulting ballistic trajectory, we thus expect, in analogy to hydrodynamic linear response,
\begin{equation}\label{diffusiveirr}
    \langle \tau_t \tau_t^*s(x),s(0,0)\rangle^c \sim \frc{χ}{\sqrt{4\pi \mathfrak D_{\rm irr} t}}\, e^{-x^2/(4\mathfrak D_{\rm irr}t)}\qquad \mbox{(hydrodynamic linear response).}
\end{equation}
This is {\em irreversibility} diffusion.

Using the relation
\begin{equation}
    v(\mathsf s) = \frc{d \mathsf j}{d\mu} \Big/ \frc{d\mathsf s}{d\mu} = \chi^{-1}\langle M,j(0,0)\rangle^c
\end{equation}
we write from \eqref{eq:proejcted_spin_current_chi}
\begin{equation}
    j^-(x,t) = j(x,t) - v s(x,t)
\end{equation}
and \eqref{Onsager} (or \eqref{Onsager1}) with \eqref{eq:proejcted_spin_current} follows. Using conservation laws, we obtain the physical meaning of \eqref{Onsager}. Following the steps of \cite[App A]{DeNardis_Doyon_2019_diffusion_GHD} (the steps can be made fully rigorous, using discrete-space integration by parts $\sum_x x \p_x j(x) = -\sum_x j(x)$, $\sum_x x^2 \p_x j(x) = -\sum_x (1+2x)j(x)$, with $\p_x j(x) = j(x)-j(x-1)$, and the Lieb-Robinson bound and clustering of correlation functions for boundary terms to vanish), we find that it decomposes into (the infimum limit of) the difference
\begin{equation}
    \mathfrak L := \liminf_{t\to\infty} (\mathfrak L_{\rm norm}(t)-\mathfrak L_{\rm irr}(t))
\end{equation}
of a normal linear-response diffusion strength
\begin{align}
    \mathfrak L_{\rm norm}(t) &=
    \frc1{t} \sum_{x\in\Z} (x^2-(vt)^2) \langle
    s(x,t),s(0,0)\rangle^c\\ &= 
    \frc1t \sum_{x\in\Z} (x^2 - (vt)^2) \frc{d}{d\mu}\frc{\Tr \Big(e^{M + \mu s(0)} s(x,t)\Big)}{\Tr e^{M + \mu s(0)} }\Bigg|_{\mu=0},
\end{align}
and an irreversibility diffusion strength
\begin{align}
    \mathfrak L_{\rm irr}(t)&=
    \frc1{2t} \sum_{x\in\Z} x^2\langle \tau_{t}\tau^*_t( s(x)),s(0,0)\rangle^c
    \\ &= 
    \frc1{2t} \sum_{x\in\Z} x^2\frc{d}{d\mu}\frc{\Tr \Big(e^{M + \mu s(0)} \tau_{t}\tau^*_t(s(x))\Big)}{\Tr e^{M + \mu s(0)} }\Bigg|_{\mu=0}
\end{align}
the linear response of the spin evolved forward, and then backward, for the same time. If the correlation functions satisfy hydrodynamic diffusive equations, then \eqref{diffusiveform} and \eqref{diffusiveirr} hold and we have Einstein relations for the limits $t\to\infty$:
\begin{equation}
    \mathfrak L_{\rm norm}(\infty) = \mathfrak D_{\rm norm}\chi,\quad \mathfrak L_{\rm irr}(\infty) = \mathfrak D_{\rm irr}\chi\quad
    \mbox{(hydrodynamic linear response)},
\end{equation}
which provides the physical meaning of $\mathfrak L$. With the choices of parameters making the dynamics reversible, see \cite[Appendix B]{ampelogiannis_2024_diffusion}, we have $\tau_t = \tau_{-t}^*$ and therefore $\mathfrak L_{\rm irr}=0$.

\section{Lower bound on the Onsager coefficient}
To prove that the system exhibits at least diffusive transport, it suffices to find a lower, strictly positive, bound for $\mathfrak{L}$. In \cite{doyon_2019_diffusion}  a general bound is obtained in one-dimensional systems, under the assumption of clustering of $n$-th order connected correlations under space-time translations, which until recently remained without a proof. In finite range quantum spin chains the Gibbs states for nonzero temperatures are exponentially clustering \cite{Araki:1969bj}, and the state $\omega_\mu$ is manifestly clustering because of ultra-locality of $M$. Combined with the Lieb-Robinson bound this implies space-like clustering for two-point connected correlations, \cite[Appendix C]{ampelogiannis_long-time_2023}. Our results in \Cref{chapter:higher_clustering} establish  clustering of $n-$th order cumulants from two-point clustering, which is slightly different from the assumed clustering in \cite{doyon_2019_diffusion}. In particular, applying \Cref{th:n-order-lieb-robinson} in the case of exponential clustering, we find that the $n$-th order connected correlations of $ω_μ$ satisfy:
\begin{equation} \label{eq:clustering}
\arraycolsep=1.4pt\def\arraystretch{1.5}
\begin{array}{*3{>{\displaystyle}l}p{5cm}}
    \langle  A_1(x_1,t_1), A_2(x_2,t_2), \ldots ,A_n(x_n,t_n) \rangle^c \leq C_{A_1,\ldots,A_n}  
    e^{ -\nu z}, \\
     \text{where } z=\max_i \min_j \{ \dist(A_i(x_i),A_j(x_j))\}
    \end{array}
\end{equation}
for any $n\in \N$, $A_1,\ldots,A_n \in \mathfrak{U_{\rm loc}}$, $x_1,\ldots,x_n\in \Z$ and $t_1,\ldots ,t_n \in [-υ^{-1}z+1, υ^{-1}z-1]$, where $C_{A_1,\ldots,A_n},\nu>0 $ constants \footnote{The constant $C_{A_1,\ldots,A_n}$  depends linearly on the norms of the observables and polynomially on the sizes of their supports}. Having established \Cref{eq:clustering}, we can slightly modify the proof of \cite[Theorem 5.1]{doyon_2019_diffusion}, see \cite[Appendix A]{ampelogiannis_2024_diffusion}, to rigorously obtain the bound\footnote{Note that there is a difference of a factor of $2$ with \cite[Theorem 5.1]{doyon_2019_diffusion}, this is because in the present set-up the Onsager coefficient \eqref{Onsager1} is defined as an integral over $t\geq 0$ instead of  $\R$}:
\begin{equation} \label{bound}
    \mathfrak{L} \geq \mathfrak L_{\rm lower} := \frac{|\langle M ,M,j^-\rangle^c|^2}{8υ_{\rm LR} (\langle M,s \rangle^c)^2}.
\end{equation}
Here $M=\sum_x{σ_x^3}$ is the magnetisation, $s=σ^3_0$ the spin density\footnote{The result is irrespective of the site $x\in \Z$ of the density}, $υ_{LR}$ the Lieb-Robinson velocity \eqref{eq:LR_velocity} and $j^-$ is the projection of the local spin currect \eqref{eq:proejcted_spin_current}. If the three point connected correlation $\langle M,M,j^-\rangle^c$ is non-zero, then the Onsager coefficient is strictly positive. We show the following:

\begin{theorem} \label{theorem:1}
    Consider a quantum spin chain with Lindbladian evolution \eqref{time_evolution} with nearest neighbor Hamiltonian interaction \eqref{hamiltonian_density} and Lindblad operators \eqref{jump_operators}, the spin current \eqref{eq:spin_conservation}, and the associated Onsager coefficient \eqref{Onsager}. It follows that  for all interaction parameters $α\in \C$, $β,γ \in \R$, $a_i,b_i,c_i,d_i,e_i \in \C$ satisfying \eqref{condition} and $\sum_i(|a_i|^2 - |b_i|^2)\neq 0$, the Onsager coefficient is strictly positive $\mathfrak{L}\geq\mathfrak{L}_{\rm lower} > 0$. 
\end{theorem}
\begin{proof}
We proceed to calculate the bound \eqref{bound} for \eqref{hamiltonian_density}, \eqref{jump_operators}. Consider the connected correlation:
\begin{equation} \label{eq:numerator}
    \langle M, M,j^-\rangle^c = \langle M,M, j \rangle^c - χ^{-1}\langle M, M, s\rangle^c \langle M, j \rangle^c.
\end{equation}
As $\langle M,a\rangle^c = d\mathsf a/dμ$ and $\langle M,M,a\rangle^c = d^2\mathsf a/dμ^2$ for any observable $a$ with average $\mathsf a=\omega_μ(a)$, changing variable $μ\to \mathsf s=\omega_μ(s)$, the bound \eqref{bound} can then be written  in terms of the flux curvature $\mathsf j''(\mathsf s) = v'(\mathsf s)$ as
\begin{equation}\label{boundjprime}
    \mathfrak L_{\rm lower} = \frc{(\chi \mathsf j'')^2}{8v_{\rm LR}},\quad
    ' = \frac{d}{d\mathsf s}
\end{equation}
(this is valid in general in chaotic systems). Therefore, as
\begin{equation}\label{chiexplicit}
    χ = \frc1{(\cosh\mu)^{2}}>0,
\end{equation}
diffusion is strictly positive if and only if the flux curvature is non-zero (see also \cite[Eq 1]{doyon_2019_diffusion}). This above formula shows that if the hydrodynamic mode has state-dependent velocity $v(\mathsf s)$, then the above bound is strictly positive. With a state-dependent velocity, the hydrodynamic trajectory of the mode is affected by macroscopic fluctuations of the initial state, leading to diffusive effects. We interpret the bound above as arising from this diffusion mechanism.

The bound \eqref{bound} can be calculated by elementary means. First, note that by the general result of \cite{kobayashi_2022_vanishing_currents} the Hamiltonian part $j^H$ of the spin current \eqref{eq:spin_conservation}, vanishes in the Gibbs state $\omega_\mu(j^{\rm H}(x)) = 0$. Thus our bound vanishes in generic Hamiltonian spin chain, both for energy and spin: the lack of ballistic transport does not allow the above diffusion mechanism to apply. However the incoherent processes lead to non-vanishing current and non-zero diffusion strength.

Indeed, we calculate the flux curvature 
\begin{equation}\label{fluxcurvature}
    v'(\mathsf s) = \mathsf j''(\mathsf s) = \sum_i(|a_i|^2-|b_i|^2)
\end{equation}
resulting in a lower bound for the spin-spin Onsager coefficient:
\begin{equation}\label{lowerexplicit}
   \mathfrak L_{\rm lower} = \frac{  \big(  \sum_i(|a_i|^2-|b_i|^2)  \big)^2 }{ 32eζ(2) \tilde{V} \cosh^4(\mu)}    
\end{equation}
where $\t V$ is defined in \eqref{Vtilde}. We note that as soon as there is incoherent spin transport, $\sum_i|a_i|^2\neq \sum_i|b_i|^2$, spin is an interacting ballistic mode as according to \eqref{js} the hydrodynamic velocity $v = \mathsf j'$ depends on the state, and we have a strictly positive diffusion bound. This shows \Cref{theorem:1}.
\end{proof}

\chapter{Conclusion and way forward}

In this thesis, we addressed the problem of rigorously establishing the emergence of hydrodynamics from the microscopic dynamics of quantum many-body systems, specifically focusing on quantum spin lattice models. This forms an important part of Hilbert's sixth problem of axiomatising the laws of physics. 

We emphasized the significance of the algebraic approach to quantum statistical mechanics, which provides the appropriate mathematical framework for addressing this problem. This approach enables the rigorous treatment of systems directly in the thermodynamic limit and equips us with a range of valuable results, such as Lieb-Robinson bounds and clustering of correlations in thermal states. 

Using these properties, we obtained an ergodic property which applies to high temperature thermal states, called almost-everywhere ergodicity;  the long-time averaging of
any observable A over a space-time ray converges, in the strong
operator topology of the  (GNS) representation, to its expectation in the state. This holds for almost every speed with respect to the Lebesgue measure, and
all rational directions on the lattice. This property indicates that a lot of information, initially encoded in the microscopic dynamics, is lost at larger scales. Since the results are general, for any $D$-dimensional quantum lattice model, they show that almost-everywhere ergodicity does not require particular properties of the dynamics, such as ``chaos". 
The loss of information at large scales of space-time can be formalised by the phenomenon of hydrodynamic projection: intuitively, the projection of observables onto conserved quantities at large scales. By applying almost-everywhere ergodicity, we have shown that the principle of hydrodynamic projection -- by which correlation functions, at large wavelengths and long times, project onto the extensive conserved quantities admitted by the model -- holds in every short-range quantum spin lattice, by unambiguously constructing Hilbert-spaces of conserved extensive quantities, in a manner similar to the GNS construction. Importantly, our ergodicity and hydrodynamic projection results can be formulated to take oscillatory behaviours into account.
Going to finer scales of hydrodynamics, we  established   a lower bound on spin diffusion in chaotic, translation-invariant, nearest-neighbor open quantum spin-$1/2$ chain. For this purpose, we developed new results on clustering of $n-$th order connected correlations.

These results underline the universality of hydrodynamic principles and act as a proof-of-principle that large-scale behavirous can indeed be seen as emerging from microscopic dynamics, with mathematical proof. At the same time they open many doors for further studies. Firstly, almost-everywhere ergodicity and the hydrodynamic projection theorem are agnostic to the specific details of the interaction, and we believe they can be extended to classical and quantum gases, and systems with disorder. Additionally, given that Lieb-Robinson bounds apply to systems with long-range (invserse power-law decay) interactions, we also expect our results to hold in such cases. Further, it would be of great interest to have an explicit construction of the space of extensive conserved quantities, defined in \cref{section:extensive_charge}, in specific models, such as a free fermion chain where one knows how to write down the full set of conserved charges. This raises another problem; the importance of proving that the set of conserved quantities is complete. Can we, for example, show that there are no conserved charges in a chaotic system that preserves only the Hamiltonian? Physically it is expected that there exist no further charges, but can this be established rigorously, especially in the case of quasi-local ones? This thesis sets the stage for all these questions, and highlights their significance.

In conclusion, this thesis underscores the broad universality of hydrodynamic principles,
provides a framework for establishing them rigorously and sets the stage for tackling  many open
problems leading to the ultimate goal of proving the hydrodynamic equations.
An important takeaway  is that it is extremely fruitful to extract the relevant many-body physics out of equilibrium by analysing the operator algebra of local observables already in the thermodynamic limit. This approach contrasts with methods reliant on how the spectrum of large Hamiltonians behaves during the thermodynamic limit in quantum systems, or on kinetic equations in classical models. The algebraic method allows one to directly concentrate on the small part of the Hilbert space  that is significant in the thermodynamic limit. We hope that our work lays the foundation for further studies on the problem of emergence, guided by mathematical rigor.

% ********************************** Back Matter *******************************
% Backmatter should be commented out, if you are using appendices after References
%\backmatter

% ********************************** Bibliography ******************************
\begin{spacing}{0.9}

% If you would like to use BibLaTeX for your references, pass `custombib' as
% an option in the document class. The location of 'reference.bib' should be
% specified in the preamble.tex file in the c   ustombib section.
% Comment out the lines related to natbib above and uncomment the following line.

%\printbibliography[heading=bibintoc, title={References}]

\end{spacing}

% ********************************** Appendices ********************************

\begin{appendices} % Using appendices environment for more functunality

\chapter{Remark on integration} \label{appendix1}
Throughout this work we deal with ray averaged observables $\frac{1}{T} \int_0^T τ_t ι_{\floor{\boldsymbol{υ}t}} A \, \dd t$, where $A \in \mathfrak{U}$ and $(\mathfrak{U},ι,τ)$ is a dynamical system. This integral is to be understood in the Bochner sense \cite{hille_functional_1996}. It is easy to see that the function $τ_t ι_{\floor{\boldsymbol{υ}t}} A$ is Bochner integrable \cite[Definition 3.7.3]{hille_functional_1996} in any compact subset of $\R$,  as it can be approximated by the countable valued functions (simple functions):
\begin{equation}
    A_n(t) =τ_{\frac{\lfloor nt \rfloor }{n}}ι_{\lfloor \boldsymbol{υ} t \rfloor}  A
\end{equation}
 Indeed, $τ_t$ is strongly continuous and $\lim_n \frac{\lfloor nt \rfloor }{n}=t$. Hence, $$\lim_n \norm{A_n(t)-τ_t ι_{\floor{\boldsymbol{υ}t}} A}=0 $$ for all $t$, for any $A \in \mathfrak{U}$. The integral is defined in any interval $[0,T]$ of $\R$, uniquely and independently of the choice of $A_n(t)$ as
\begin{equation}
    \int_0^T τ_t ι_{\floor{\boldsymbol{υ}t}} A \, \dd t \coloneqq \lim_{n} \int_0^T    A_n(t) \, \dd t
\end{equation}
where the integral of $A_n(t)$ is defined as a sum over its (countable) values, in the usual way. The Bochner integral calculated inside a linear functional can be pulled outside, by \cite[Theorem 3.7.12]{hille_functional_1996}, in particular for any state $ω$ we have $ω\big(  \int_0^T τ_t ι_{\floor{\boldsymbol{υ}t}} A \, \dd t \big)= \int_0^T ω\big( τ_t ι_{\floor{\boldsymbol{υ}t}} A \big) \, \dd t$.

Likewise, in the GNS representation  $\int_0^T π\big(τ_t ι_{\floor{\boldsymbol{υ}t}} A  \big)\, \dd t$ is defined, and the two definitions are compatible in the sense that the integral commutes with $π$, i.e.\ $π(\int_0^T τ_t ι_{\floor{\boldsymbol{υ}t}} A  \, \dd t) = \int_0^T π\big(τ_t ι_{\floor{\boldsymbol{υ}t}} A \big)\, \dd t$,  by \cite[Theorem 3.7.12]{hille_functional_1996}. Finally, the integral $\int_0^T π\big(τ_t ι_{\floor{\boldsymbol{υ}t}} A  \big) Ψ \, \dd t$ is well defined and equal to  $\int_0^T π\big(τ_t ι_{\floor{\boldsymbol{υ}t}} A  \big) \, \dd t Ψ$, again by \cite[Theorem 3.7.12]{hille_functional_1996} and considering $ π\big(τ_t ι_{\floor{\boldsymbol{υ}t}} A  \big) Ψ$ as the evaluation map from $π(\mathfrak{U})$ to $H_ω$.

\chapter{Multiple Limits and the Moore-Osgood Theorem} \label{appendix:Moore-Osgood}

The Moore-Osgood Theorem is a general tool that lets us deal with double limits over arbitrary sets, given an appropriate limiting notion on these sets (called direction). We give an overview of  \cite[p. 139]{taylor_general_1985}. We assume a non-empty set $X$ and a family of subsets $\mathcal{N}$ of $X$ called a \textbf{direction on $X$ }such that $N_1, N_2 \in \mathcal{N}$ implies that there exists a $N_3 \in \mathcal{N}$ such that $N_3 \subset N_1 \cap N_2$. For example, in our Theorems we deal with  $\lim_{T \to \infty}$ which corresponds to the direction $\mathcal{T} = \big \{ \{ T \in \R :  T\geq m \}$, $m \in \N \big\}$.

A pair $(g,\mathcal{N})$ of a  function $g:X \to Z$, where $Z$ a metric space, and a direction $\mathcal{N}$ is called a directed function. We say that\textbf{ $(g,\mathcal{N})$  converges to $x_0 \in Z$, and write $(g,\mathcal{N}) \to x_0$}, if for each neighborhood $U$ of $x_0$ there exists a $N$ in $\mathcal{N}$ such that $g(N)\subset U$.  If $X,Y$ are sets with directions $\mathcal{M}, \mathcal{N}$, respectively, then the family of sets $\big\{M \times N$, $M \in \mathcal{M}$, $N \in \mathcal{N} \big\}$ is a direction in $X \times Y$, denoted $\mathcal{M \times N}$. The double limit of a function $f: X \times Y \to Z$, where $(Z,d)$ is a metric space, is given by the convergence of $(f , \mathcal{M} \times \mathcal{N})$. It is easy to see that $( f(x, \cdot) , \mathcal{N})$
and $(f(\cdot, y) , \mathcal{M})$ are directed functions. We say that $( f(x, \cdot) , \mathcal{N})$ converges to $g(x)$ \textbf{uniformly}, if for every $ε>0$ there exists a $N \in \mathcal{N}$ s.t. $d\big( f(x,y) , g(x) \big) < ε$ for $x\in X$ and $y \in N$.

The notion of the multiple limits $\lim_{T_1,\ldots,T_{n} \to \infty} I(T_1, T_2\ldots,T_n)$ in the main text is to be understood as convergence of the directed function $(I,\mathcal{T}^n)$. It is clear that the multiple limit $\lim_{T_1,\ldots,T_{n} \to \infty}$ coincides with the double limit $\lim_{(T_1,\ldots,T_{n-1}) , T_n \to \infty}$ over $\R^{n-1} \times \R$. One can then take advantage of the Moore-Osgood Theorem:
\begin{theorem}[Moore-Osgood] \label{th:MooreOsgood}
 Consider a complete metric space $Z$, the sets $X,Y$ with directions $\mathcal{M}, \mathcal{N}$ respectively  and a function $f: X \times Y \to Z$. Denote by $\mathcal{P}$  the direction $\mathcal{M} \times \mathcal{N}$. If there exist a function $g(x)$ s.t. $( f(x, \cdot), \mathcal{N}) \to g(x)$ uniformly on $x \in X$, and a function $h(y)$ s.t. $ ( f(\cdot,y) , \mathcal{M} ) \to h(y)$ for any $y \in Y$, then $(f,\mathcal{P})$ converges to a $z \in Z$ and $(g,\mathcal{M}) \to z$, $(h,\mathcal{N})\to z$.
\end{theorem}

\chapter{Space-like clustering proof}\label{appC}
Throughout this appendix we use the notation $ι_{\bm x}τ_tA$ for space-time translations of observables $A\in \mathfrak{U}$, in order to maintain visual clarity of expressions. We use a slightly different version of the Lieb-Robinson bound than that of the main text, which is of little consequence to the final result. We prove the following proposition:
\begin{prop}
    Consider a dynamical system $(\mathfrak{U},ι,τ)$ with interaction satisfying \Cref{eq:interaction}, in an exponentially clustering state (\Cref{defn:exp_clustering_space}). It follows that the system is space-like exponentially clustering with respect to $ω$, \Cref{defn:spacelikeclustering}, for the Lieb-Robison velocity $υ_{LR}$.
\end{prop}

 First, we show the first condition of \Cref{defn:spacelikeclustering}. 
The Lieb-Robinson bound allows us to approximate the time evolved observables by local ones, by projecting the time evoluted $τ_t(A)$, $A \in \mathfrak{U}_{loc}$ onto local ones $σ_{Λ}(τ_t (A) )$ supported on finite $Λ \subset \Z^D$. This is done by using the result \cite[Corollary 4.4]{nachtergaele_quasi-locality_2019} and satisfies the first condition of \Cref{defn:spacelikeclustering}:
\begin{lem} \label{lem:localapprox}
Let $A\in \mathfrak{U}$ and consider a finite $Λ \subset \mathbb{Z}^D$. If there is an $ε>0$ such that 
\begin{equation}
\norm{ [ A, B] }\leq ε \norm{A}  \norm{B}  \ , \ \ \forall B \in \mathfrak{U}_{ \mathbb{Z}^D \setminus Λ}
\end{equation}
then we can approximate $A$ by a strictly local $σ_Λ (A) \in \mathfrak{U}_Λ$:
\begin{equation}
\norm{σ_Λ (A) - A } \leq 2ε \norm{A}.
\end{equation} 
\end{lem}

This Lemma states that if an observable $A \in \mathfrak{U}$ almost commutes with every $B\in \mathfrak{U}_{\Z^D \setminus Λ}$ supported outside a finite $Λ$, then it can be well approximated by a local $σ_Λ(A) \in \mathfrak{U}_Λ$. Combined with the Lieb-Robinson bound  we can prove (this is also described in \cite[Chapter 4.3]{naaijkens_quantum_2017}):
\begin{prop} \label{propo:LRapproximation}
    Consider a dynamical system with exponentially decaying interactions, the time evolution $τ_t A$ of a local $A \in \mathfrak{U}_Λ$ and the finite sets $Λ_r = \cup_{\boldsymbol x \in Λ}B_{\boldsymbol x}(r)$, $r=1,2,3, \dots$, where $B_{\boldsymbol x}(r)$ is the ball of radius $r>0$ around $\boldsymbol x$, i.e.\ $Λ_r$ is $Λ$ extended by a distance $r$ around all of its points. Then, we can approximate $τ_t A$ by the local $σ_{Λ_r}(τ_t A) \in \mathfrak{U}_{Λ_r}$ :
    \begin{equation}
        \norm{ σ_{Λ_r} (τ_t(A)) - τ_t(A)} \leq 2ε_r \norm{A}
    \end{equation}
    with $ε_r = C |Λ|N^{2|Λ|} \exp{-λ(r- υ_{LR}|t|)}$, $r=1,2,3 \dots$.
\end{prop}
Using this result we can show the first condition of space-like clustering:
\begin{proof}[Proof of \ref{defn:spacelikeclustering}.\ref{spacelike1}]
The projection $σ$ allows us to define the sequences $σ_r(A) \coloneqq σ_{Λ_r}A$ required by condition 1 of  \Cref{defn:spacelikeclustering}. 
%\begin{proof}[Proof of space-like clustering - \Cref{propo:space-like}]
 Consider $\boldsymbol n \in \Z^D$ and $A,B  \in \mathfrak{U}_{loc}$ with supports
\begin{equation}
    \supp(A) \coloneqq Λ_A \ , \ \ \supp(B) \coloneqq Λ \text{ and } \supp(ι_{\boldsymbol n} A) \coloneqq Λ_A+\boldsymbol n
\end{equation}
%We denote by $a,b \in \mathcal{V}$ the respective class representatives of $A,B$.
Let $r \in \N$ and by \Cref{propo:LRapproximation} consider the local approximation of $τ_t B$ by $σ_{Λ_r}(τ_t B)$, supported in $Λ_r = \cup_{\boldsymbol x \in Λ} B_{\boldsymbol x}(r)$:
\begin{equation}
\norm{ σ_{Λ_r} τ_t B - τ_t B} \leq 2C |Λ| N^{2|Λ|} \norm{B} \exp\big(  -λ(r-υ_{LR} |t|) \big) \label{ineq:localapproximationB}
\end{equation}
Using this we can approximate the time evolution of observables by a local sequence $σ_{r} (τ_t A) \coloneqq σ_{Λ_r} (τ_t A)$ so that $\lim_{r} σ_{r} (τ_t A) = τ_t A$. 
\end{proof}

We show the uniformity condition 2 of the elements $σ_r (τ_t A)$ in the end of the appendix, we first proceed to show the third condition of \Cref{defn:spacelikeclustering}.

\begin{proof}[Proof of \ref{defn:spacelikeclustering}.\ref{spacelike2}]
Consider the quantity $I \coloneqq | ω( ι_{\boldsymbol n}(A) σ_{Λ_r}(τ_t B)) - ω( A) ω( σ_{Λ_r}(τ_t B)) |$, $r \in \N$ where we are interested in the large $r$ limit.  By linearity and the triangle inequality, it holds for any $r,l \in \N$ :
\begin{equation} \label{inequality1}
\begin{array}{*3{>{\displaystyle}lc}p{5cm}}
I  &=& \big| ω\bigg( ι_{\boldsymbol n}(A) \big(σ_{Λ_r}(τ_t B) + σ_{Λ_l}(τ_t B)  - σ_{Λ_l}(τ_t B) \big)\bigg)  - ω( A) ω\big ( σ_{Λ_r}(τ_t B)+σ_{Λ_l}(τ_t B)  - σ_{Λ_l}(τ_t B)\big ) \big| \\
&\leq& \big| ω\big( ι_{\boldsymbol n}(A) σ_{Λ_l}(τ_t B) \big) -ω( A)  ω( σ_{Λ_l}(τ_t B))\big|+  \\
&+&\big| ω\big( ι_{\boldsymbol n}(A) \big(σ_{Λ_r}(τ_t B)  - σ_{Λ_l}(τ_t B)\big) \big) -ω( A)  ω\big( σ_{Λ_r}(τ_t B)  - σ_{Λ_l}(τ_t B)\big) \big|
\end{array} 
\end{equation}

The idea is to control the first part using exponential clustering and the second one using the approximation of time evolution. We now estimate the quantities
\begin{equation}
     T_1=\big| ω\big(ι_{\boldsymbol n}(A) σ_{Λ_l}(τ_t B) \big) - ω( A ) ω(σ_{Λ_l}(τ_t B))\big|
\end{equation} and 
\begin{equation}
     T_2=\big| ω\big( ι_{\boldsymbol n}(A) \big(σ_{Λ_r}(τ_t B)  - σ_{Λ_l}(τ_t B)\big) \big ) - ω(  A ) ω\big( σ_{Λ_r}(τ_t B)  - σ_{Λ_l}(τ_t B)\big ) \big|
\end{equation} 
Consider $r$ large enough and 
\begin{equation}
    l= \floor{ε \dist (Λ_A+\boldsymbol n, Λ)} + \floor{ε \diam (Λ_A \cup Λ)} + 2 \ \text{ for some } \frac12<ε<1 \label{eq:l}
\end{equation}
We first estimate $T_1$, using exponential clustering 
\begin{equation}
    T_1 \leq k_{Λ_A+\boldsymbol n, Λ_l}e^{-λ \dist(Λ_A+\boldsymbol n, Λ_l)} \label{ineq:T_1}
\end{equation}
Where we have the following bound bound for $k(Λ_A+\boldsymbol n, Λ_l)$:
\begin{equation}
   k_{Λ_A+\boldsymbol n, Λ_l} \leq u \norm{A} \norm{ σ_{Λ_l}( τ_t B)}  |Λ_A|^r |Λ_l|^r
\end{equation}
Obviously $|Λ_A+\boldsymbol n|=|Λ_A|$ and $|Λ_l| \leq |B_{\boldsymbol 0}(l)||Λ|$. ,where  $|B_{\boldsymbol 0}(l)|$ is the number of lattice points in the D-ball of radius $l$, which is a polynomial in $l$ of degree $D$, hence it can be bounded by $K l^D$ for some constant $K>0$, $\forall l \geq 1$. Using this, the chosen value for $l$ (\Cref{eq:l}) and the triangle inequality for $\dist(Λ_A +\boldsymbol n, Λ_l)$:
\begin{equation}
    \begin{array}{*3{>{\displaystyle}lc}p{5cm}}
 
   |Λ_l| &\leq& |Λ|K( \floor{ε \dist ( Λ_A +\boldsymbol n , Λ)}+ \floor{ε \diam(Λ_A \cup Λ)}+2)^D \\
   &\leq& |Λ|K ε^D \big( \dist(Λ_A + \boldsymbol n, Λ) + \diam( Λ_A \cup Λ) +4\big)^D \\
    &\leq& |Λ|K ε^D \big( |\boldsymbol n| + \dist(Λ_A, Λ) + \diam( Λ_A \cup Λ) +4\big)^D
    \end{array}
\end{equation}
Since $ (|\boldsymbol n| + \dist(Λ_A, Λ) + \diam( Λ_A \cup Λ) +2)^D$ is a polynomial in $|\boldsymbol n|$ of degree $D$ it can be bounded by $L|\boldsymbol n|^D$ for some constant $L>0$ (depending on $\dist(Λ_A,Λ)$, $\diam(Λ_A \cup Λ)$), $\forall |\boldsymbol n| \geq 1$.  Hence:
\begin{equation}
    |Λ_l| \leq |Λ|KL ε^{D} |\boldsymbol n|^D
\end{equation}
By the explicit construction of the projection $σ$ in \cite{nachtergaele_quasi-locality_2019} it holds that $\norm{σ_{Λ_l}C} \leq \norm{C}$, $\forall C\in \mathfrak{U}$. Additionally, since $τ_t$ is a $^*$-automorphism for all $t \in \R$, it is norm preserving, hence
\begin{equation}
    \norm{σ_{Λ_l}(τ_tB)} \leq \norm{B}
\end{equation}
Next, we can estimate $\dist(Λ_A +\boldsymbol n, Λ_l)$ with respect to $|\boldsymbol n|$ by simple geometric arguements:
\begin{equation}
        \dist(Λ_A+\boldsymbol n, Λ_l) \geq \dist(Λ_A +\boldsymbol n, Λ)-l \geq (1-ε) \dist(Λ_A +\boldsymbol n, Λ) - ε \diam(Λ_A \cup Λ)-2
\end{equation}
and
\begin{equation}
 \begin{array}{*3{>{\displaystyle}lc}p{5cm}}
    \dist(Λ_A+\boldsymbol n, Λ) &=& \min \{ |\boldsymbol y-\boldsymbol z| : \boldsymbol y \in Λ_A+\boldsymbol n, \boldsymbol z \in Λ \} \\
        &=& \min \{ |\boldsymbol y+\boldsymbol n-\boldsymbol z| : \boldsymbol y \in Λ_A, \boldsymbol z \in Λ \} \\
        &\geq& \min \{ |\boldsymbol n| - |\boldsymbol y-\boldsymbol z| :\boldsymbol y \in Λ_A, \boldsymbol z \in Λ \} \\
        &=& |\boldsymbol n| - \max\{ |\boldsymbol y-\boldsymbol z| : \boldsymbol y \in Λ_A ,\boldsymbol z \in Λ \} \\
        &\geq& |\boldsymbol n| - \diam(Λ_A \cup Λ)
    \end{array} \label{eq:distΛvecn}
\end{equation}
Putting these together we get
\begin{equation}
    e^{-λ \dist(Λ_A +\boldsymbol n,Λ)} \leq e^{2\lambda}e^{λ \diam(Λ_A \cup Λ)} e^{-λ(1-ε)|\boldsymbol n|}
\end{equation}
and combining all the estimates for the terms in $T_1$, inequality \eqref{ineq:T_1} becomes
\begin{equation}
    T_1 \leq u e^{2\lambda}K^rL^r |Λ_A|^r|Λ|^r ε^{rD} \norm{A}\norm{B} e^{λ \diam(Λ_A \cup Λ)} |\boldsymbol n|^{rD} e^{-λ(1-ε)|\boldsymbol n|} \label{ineq:T1}
\end{equation}
To estimate $T_2$, we use the triangle inequality and note that $ω$ has norm $1$:
\begin{equation}
    T_2 \leq  \norm{ι_{\boldsymbol n}A} \norm{σ_{Λ_r}(τ_t B)  - σ_{Λ_l}(τ_t B)}  +  \norm{A} \norm{σ_{Λ_r}(τ_t B)  - σ_{Λ_l}(τ_t B)}  \label{ineq:T2}
\end{equation}
where $ \norm{ι_{\boldsymbol n}A}=\norm{A}$. Using the approximation of $τ_tB$, inequality \eqref{ineq:localapproximationB}, we get
\begin{equation}
    T_2 \leq 2\norm{A}  2C |Λ| N^{2|Λ|} \norm{B} \big( \exp \{ -λ(r-υ_{LR} |t|)\} +  \exp\{-λ(l-υ_{LR} |t|)\}\big)
\end{equation}

We consider a compact subset $T\coloneqq ε^{\prime} υ^{-1}_{LR} [-|\boldsymbol n|,|\boldsymbol n|]$ for some $0<ε^{\prime}<ε$, and let $t \in T$. Then, for the value \eqref{eq:l} of $l$, since $\floor{y} \geq y-1, \forall y>0$, we have:
\begin{equation}
\begin{array}{*3{>{\displaystyle}lc}p{5cm}}
 l-υ_{LR} |t|  &\geq& ε \dist(Λ_A +\boldsymbol n,Λ)-1+ε\diam(Λ_A \cup Λ)-1+2 - ε^{\prime}|\boldsymbol n| \\
 &\geq& ε |\boldsymbol n| -ε \diam(Λ_A \cup Λ) +ε\diam(Λ_A \cup Λ)- ε^{\prime}|\boldsymbol n|  \\
 &=& (ε-ε^{\prime})|\boldsymbol n| >0 \label{ineq:lvtestimate}
 \end{array}
\end{equation}
where in the second line we used relation \eqref{eq:distΛvecn}.  Hence:
\begin{equation}\label{ineq:lvtestimateprime}
    \exp\{-λ(l-υ_{LR}|t|)\}  \leq \exp\{-λ(ε-ε^{\prime})|\boldsymbol n|\} \ , \ \ \forall t \in T, 0<ε^{\prime}<ε<1
\end{equation}
and  also $\lim_r \exp\{-λ(r-υ_{LR}|t|)\} =0$, $\forall t \in T$. We can now take $\lim_r$ of inequality \ref{inequality1}, where the left-hand side becomes $\lim_r I = |ω(  ι_{\boldsymbol n}A τ_tB ) - ω(  A ) ω(  B )|$ by continuity. The right-hand side is bounded by the estimates \eqref{ineq:T1}, \eqref{ineq:T2}, \eqref{ineq:lvtestimateprime}, to get:
{
\begin{equation}
 \begin{array}{*3{>{\displaystyle}lc}p{5cm}}
  |ω(  ι_{\boldsymbol n}A τ_tB ) - ω( A ) ω(  B ) | \leq \\
  ue^{2\lambda}K^rL^r|Λ_A|^r|Λ|^r ε^{rD} \norm{A} \norm{B} e^{λ\diam(Λ_A \cup Λ)}|\boldsymbol n|^{rD} e^{-λ(1-ε)|\boldsymbol n|} \\
  + 4 C N^{2|Λ|} |Λ| \norm{A} \norm{B} e^{-λ(ε-ε^{\prime}) |\boldsymbol n|}
    \end{array} 
\end{equation}
}
Concluding the proof, note that for any $0<\tilde{λ}<λ(1-ε)$ there exists a large enough $G>0$ s.t. $|\boldsymbol n|^{rD} e^{-\lambda (1-ε)|\boldsymbol n|} \leq G e^{-\tilde{λ}|\boldsymbol n|}$. Hence we arrive at the result by defining $μ \coloneqq \min(\tilde{λ}, λ(ε-ε^{\prime})) >0$ and
\begin{equation}
    c= \norm{A} \norm{B} |Λ| (u e^{-2}K^rL^r |Λ_A|^r |Λ|^{r-1}ε^{rD}e^{λ\diam(Λ_A \cup Λ)}G + 4 CN^{2|Λ|})
\end{equation}
to get that there exist $c>0$ and a $μ>0$ s.t.
\begin{equation}
    | ω( ι_{\boldsymbol n}A τ_t B ) - ω(  A ) ω(  B ) | \leq c e^{-μ|\boldsymbol n|}
\end{equation}
for all $\boldsymbol n \in \Z^D$ and for any $t \in T = ε^{\prime} υ_{LR} [ - |\boldsymbol n|, |\boldsymbol n| ]$, with any choice $0<ε^{\prime}<ε$ and $1/2<ε<1$. Hence, for any $υ=υ_{LR}/ε^{\prime}>υ_{LR}$ and choosing $t= - υ^{-1}|\boldsymbol n|$, we have
\begin{equation}
    |ω(  ι_{\boldsymbol n}A τ_{-υ^{-1}|\boldsymbol n|} B )- ω(  A ) ω(  B) | \leq c e^{-μ|\boldsymbol n|} \ , \forall \boldsymbol n \in \Z^D \label{eq:spacelikeergodicproof}
\end{equation}
and taking advantage of the time invariance of $ω$, we have established exponential space-like clustering, and as a consequence space-like $p$-clustering for all $p$.
\end{proof}

It remains to prove the uniformity condition, i.e.\ that the set $\{ (σ_n τ_t a , σ_m τ_s b) : n,m \in \N \}$ is uniformly clustering, \Cref{eq:uniformclustering}.

\begin{proof}[Proof of \ref{defn:spacelikeclustering}.\ref{spacelike3}]
We start with $I \coloneqq | ω( ι_{\boldsymbol n}(A) σ_{Λ_r}(τ_t B)) - ω( A) ω( σ_{Λ_r}(τ_t B)) |$ and choose $l= \dist(Λ_A+\boldsymbol n, Λ)$. If $r< l$ then we bound $I$ by the space clustering property, as in \Cref{ineq:T_1}:
\begin{equation}
     I_{r<l} \leq u |Λ_A|^r |Λ_l|^r \norm{A} \norm{B}  e^{-λ \dist(Λ_A+\boldsymbol n, Λ_r)} 
\end{equation}
and $e^{-λ \dist(Λ_A+\boldsymbol n, Λ_r)} \leq e^{-λ|\boldsymbol n|} e^{λ \diam (Λ_A \cup Λ) - λ l}= e^{-2λ|\boldsymbol n|} e^{2λ \diam (Λ_A \cup Λ) } $. Thus, we get
\begin{equation}
    I_{r < l} \leq u |Λ_A|^r |Λ|^r \norm{A} \norm{B} e^{2λ \diam (Λ_A \cup Λ) }  e^{-2λ|\boldsymbol n|} 
\end{equation}

Now, if $r>l$ we proceed as in \Cref{inequality1} to get
\begin{equation}
 \begin{array}{*3{>{\displaystyle}lc}p{5cm}}
    I &\leq& I_{r<l} + 2\norm{A}  2C |Λ| N^{2|Λ|} \norm{B} \big( \exp\{ -λ(r-υ_{LR} |t|)\} +  \exp\{-λ(l-υ_{LR} |t|)\}\big) \\
    &\leq& I_{r<l} +4\norm{A}  2C |Λ| N^{2|Λ|} \norm{B} \exp\{-λ(l- υ_{LR}t)\} \\
    &\leq& I_{r<l} +4\norm{A}  2C |Λ| N^{2|Λ|} \norm{B}  e^{υ_{LR}t} e^{-λ \dist(Λ_A+ \boldsymbol n,Λ) } \\
    &\leq& I_{r<l} +4\norm{A}  2C |Λ| N^{2|Λ|} \norm{B}  e^{υ_{LR}t} 
    e^{λ \diam( Λ_A \cup Λ)}  e^{-λ |\boldsymbol n| }
    \end{array}
\end{equation}

Hence, we can overall obtain a uniform in $r$ exponential clustering. Similarly we can repeat the process for the replacement $ι_{\boldsymbol n} A \rightarrow σ_{r^{\prime}}  τ_{s} ι_{\boldsymbol n} A$, which will finally yield uniform exponential clustering for the set $\{( σ_{r^{\prime}}  τ_{s}  a , σ_{r}  τ_{t}  b)  : r,r^{\prime} \in \N \}$ for any $t,s \in \R$. This obviously implies uniform $p$-clustering for all $p$.
\end{proof}

\chapter{Von Neumann's mean ergodic theorem}

\begin{theorem} \label{th:neumann}
   Let $U$ be a unitary operator on a Hilbert sapce $H$ and $P$ the orthogonal projection onto $ker(U-\mathds{1})$: the space invariant under $U$. Then for any $Ψ\in H$ von Neumann showed that 
    \begin{equation}
        \lim_{N} \frac{1}{N} \sum_{n=0}^N U^nΨ = PΨ
    \end{equation}
\end{theorem}
\end{appendices}

% To use the conventional natbib style referencing  
% Bibliography style previews: http://nodonn.tipido.net/bibstyle.php
% Reference styles: http://sites.stat.psu.edu/~surajit/present/bib.htm

\bibliographystyle{unsrt}
\cleardoublepage
\bibliography{references} % Path to your References.bib file

@article{buca_2012_lindblad_transport,
	title = {A note on symmetry reductions of the {Lindblad} equation: transport in constrained open spin chains},
	volume = {14},
	issn = {1367-2630},
	url = {https://dx.doi.org/10.1088/1367-2630/14/7/073007},
	doi = {10.1088/1367-2630/14/7/073007},
	number = {7},
	journal = {New Journal of Physics},
	author = {Buca, Berislav and Prosen, Tomaz},
	month = jul,
	year = {2012},
	note = {Publisher: IOP Publishing},
	pages = {073007},
}

@book{landau_1987_fluid,
	title = {Fluid mechanics: {Volume} 6},
	volume = {6},
	publisher = {Elsevier},
	author = {Landau, Lev Davidovich and Lifshitz, Evgenii Mikhailovich},
	year = {1987},
}

@misc{hannani_2021_hydro_harmonic,
	title = {Hydrodynamic limit for a disordered quantum harmonic chain},
	url = {https://arxiv.org/abs/2011.07552},
	author = {Hannani, Amirali},
	year = {2021},
	note = {arXiv: 2011.07552 [math-ph]},
}

@article{bernardin_2019_hydro_chain,
	title = {Hydrodynamic {Limit} for a {Disordered} {Harmonic} {Chain}},
	volume = {365},
	issn = {1432-0916},
	url = {https://doi.org/10.1007/s00220-018-3251-4},
	doi = {10.1007/s00220-018-3251-4},
	number = {1},
	journal = {Communications in Mathematical Physics},
	author = {Bernardin, Cédric and Huveneers, François and Olla, Stefano},
	month = jan,
	year = {2019},
	pages = {215--237},
}

@article{Klobas:2018ldp,
	title = {Time-{Dependent} {Matrix} {Product} {Ansatz} for {Interacting} {Reversible} {Dynamics}},
	volume = {371},
	issn = {1432-0916},
	url = {https://doi.org/10.1007/s00220-019-03494-5},
	doi = {10.1007/s00220-019-03494-5},
	number = {2},
	journal = {Communications in Mathematical Physics},
	author = {Klobas, Katja and Medenjak, Marko and Prosen, Tomaz and Vanicat, Matthieu},
	month = oct,
	year = {2019},
	pages = {651--688},
}

@article{even_2014_hydrolimit,
	title = {Hydrodynamic {Limit} for a {Hamiltonian} {System} with {Boundary} {Conditions} and {Conservative} {Noise}},
	volume = {213},
	issn = {1432-0673},
	url = {https://doi.org/10.1007/s00205-014-0741-1},
	doi = {10.1007/s00205-014-0741-1},
	number = {2},
	journal = {Archive for Rational Mechanics and Analysis},
	author = {Braxmeier-Even, Nadine and Olla, Stefano},
	month = aug,
	year = {2014},
	pages = {561--585},
}

@article{pavel_2003_holo_hydro,
	title = {Holography and hydrodynamics: diffusion on stretched horizons},
	volume = {2003},
	issn = {1126-6708},
	url = {https://dx.doi.org/10.1088/1126-6708/2003/10/064},
	doi = {10.1088/1126-6708/2003/10/064},
	number = {10},
	journal = {Journal of High Energy Physics},
	author = {{Pavel Kovtun} and {Dam T. Son} and {Andrei O. Starinets}},
	month = nov,
	year = {2003},
	pages = {064},
}

@article{jeon_1996_qft_hydro,
	title = {From quantum field theory to hydrodynamics: {Transport} coefficients and effective kinetic theory},
	volume = {53},
	url = {https://link.aps.org/doi/10.1103/PhysRevD.53.5799},
	doi = {10.1103/PhysRevD.53.5799},
	number = {10},
	journal = {Physical Review D: Particles and Fields},
	author = {Jeon, Sangyong and Yaffe, Laurence G.},
	month = may,
	year = {1996},
	note = {Number of pages: 0
Publisher: American Physical Society},
	pages = {5799--5809},
}

@article{lucas_2018_electrons_graphene,
	title = {Hydrodynamics of electrons in graphene},
	volume = {30},
	issn = {0953-8984},
	url = {https://dx.doi.org/10.1088/1361-648X/aaa274},
	doi = {10.1088/1361-648X/aaa274},
	number = {5},
	journal = {Journal of Physics: Condensed Matter},
	author = {Lucas, Andrew and Fong, Kin Chung},
	month = jan,
	year = {2018},
	note = {Publisher: IOP Publishing},
	pages = {053001},
}

@article{torre_2015_viscosity_graphene,
	title = {Nonlocal transport and the hydrodynamic shear viscosity in graphene},
	volume = {92},
	url = {https://link.aps.org/doi/10.1103/PhysRevB.92.165433},
	doi = {10.1103/PhysRevB.92.165433},
	number = {16},
	journal = {Physical Review B},
	author = {Torre, Iacopo and Tomadin, Andrea and Geim, Andre K. and Polini, Marco},
	month = oct,
	year = {2015},
	note = {Number of pages: 11
Publisher: American Physical Society},
	pages = {165433},
}

@article{narozhny_2015_hydro_graphene,
	title = {Hydrodynamics in graphene: {Linear}-response transport},
	volume = {91},
	url = {https://link.aps.org/doi/10.1103/PhysRevB.91.035414},
	doi = {10.1103/PhysRevB.91.035414},
	number = {3},
	journal = {Physical Review B},
	author = {Narozhny, B. N. and Gornyi, I. V. and Titov, M. and Schütt, M. and Mirlin, A. D.},
	month = jan,
	year = {2015},
	note = {Number of pages: 20
Publisher: American Physical Society},
	pages = {035414},
}

@article{muller_2009_graphene,
	title = {Graphene: a nearly perfect fluid},
	volume = {103},
	url = {https://link.aps.org/doi/10.1103/PhysRevLett.103.025301},
	doi = {10.1103/PhysRevLett.103.025301},
	number = {2},
	journal = {Physical Review Letters},
	author = {Müller, Markus and Schmalian, Jörg and Fritz, Lars},
	month = jul,
	year = {2009},
	note = {Number of pages: 4
Publisher: American Physical Society},
	pages = {025301},
}

@article{moll_2016_graphene,
	title = {Evidence for hydrodynamic electron flow in {PdCoO2}},
	volume = {351},
	url = {https://doi.org/10.1126/science.aac8385},
	doi = {10.1126/science.aac8385},
	number = {6277},
	urldate = {2025-04-28},
	journal = {Science},
	author = {Moll, Philip J. W. and Kushwaha, Pallavi and Nandi, Nabhanila and Schmidt, Burkhard and Mackenzie, Andrew P.},
	month = mar,
	year = {2016},
	note = {Publisher: American Association for the Advancement of Science},
	pages = {1061--1064},
}

@misc{pines_1967_theory,
	title = {The theory of quantum liquids, vol. 1: {Normal} fermi liquids},
	publisher = {American Institute of Physics},
	author = {Pines, David and Nozières, Phillippe and Chang, Howard HC},
	year = {1967},
}

@article{kadanoff_1963_hydrodynamics,
	title = {Hydrodynamic equations and correlation functions},
	volume = {24},
	issn = {0003-4916},
	url = {https://www.sciencedirect.com/science/article/pii/0003491663900782},
	doi = {10.1016/0003-4916(63)90078-2},
	journal = {Annals of Physics},
	author = {Kadanoff, Leo P and Martin, Paul C},
	month = oct,
	year = {1963},
	pages = {419--469},
}

@article{corry_1997_hilbert_axiomatisation,
	title = {David {Hilbert} and the axiomatization of physics (1894–1905)},
	volume = {51},
	issn = {1432-0657},
	url = {https://doi.org/10.1007/BF00375141},
	doi = {10.1007/BF00375141},
	number = {2},
	journal = {Archive for History of Exact Sciences},
	author = {Corry, Leo},
	month = jun,
	year = {1997},
	pages = {83--198},
}

@book{kolmogorov_foundations,
	title = {Foundations of the theory of probability: {Second} english edition},
	author = {{Andrey Kolmogorov}},
}

@article{hugenholtz_1972_why_algebras,
	title = {The how, why and wherefore of {C}*-algebras in statistical mechanics},
	journal = {Festkörperprobleme 12: Plenary Lectures of the Professional Groups “Semiconductor Physics”,“Magnetism”,“Low Temperature Physics”“Thermodynamics and Statistics”, of the German Physical Society Freudenstadt, April 10–14, 1972},
	author = {Hugenholtz, Nicolaas Marinus},
	year = {1972},
	note = {Publisher: Springer},
	pages = {641--647},
}

@article{haag_1967_equilibrium,
	title = {On the equilibrium states in quantum statistical mechanics},
	volume = {5},
	number = {3},
	journal = {Communications in Mathematical Physics},
	author = {Haag, Rudolf and Hugenholtz, Nicolaas Marinus and Winnink, Marinus},
	year = {1967},
	note = {Publisher: Springer},
	pages = {215--236},
}

@article{martin_1959_theory,
	title = {Theory of many-particle systems. {I}},
	volume = {115},
	number = {6},
	journal = {Physical Review},
	author = {Martin, Paul C and Schwinger, Julian},
	year = {1959},
	note = {Publisher: APS},
	pages = {1342},
}

@article{kubo_1957_statistical,
	title = {Statistical-mechanical theory of irreversible processes. {I}. {General} theory and simple applications to magnetic and conduction problems},
	volume = {12},
	number = {6},
	journal = {Journal of the physical society of Japan},
	author = {Kubo, Ryogo},
	year = {1957},
	note = {Publisher: The Physical Society of Japan},
	pages = {570--586},
}

@book{greiner2012thermodynamics,
	title = {Thermodynamics and statistical mechanics},
	publisher = {Springer Science \& Business Media},
	author = {Greiner, Walter and Neise, Ludwig and Stöcker, Horst},
	year = {2012},
}

@inproceedings{robinson_1967_extremal,
	title = {Extremal invariant states},
	volume = {6},
	booktitle = {Annales de l'institut henri poincaré. {Section} a, physique théorique},
	author = {Robinson, Derek W and Ruelle, David},
	year = {1967},
	note = {Number: 4},
	pages = {299--310},
}

@book{ruelle_1974_rigorous,
	title = {Statistical mechanics: {Rigorous} results},
	publisher = {World Scientific},
	author = {Ruelle, David},
	year = {1974},
}

@book{kipnis_1988_scaling,
	title = {Scaling limits of interacting particle systems},
	volume = {320},
	isbn = {3-540-64913-1},
	publisher = {Springer Science \& Business Media},
	author = {Kipnis, Claude and Landim, Claudio},
	year = {1998},
}

@article{rigol_2007_relaxation,
	title = {Relaxation in a {Completely} {Integrable} {Many}-{Body} {Quantum} {System}: {An} {Ab} {Initio} {Study} of the {Dynamics} of the {Highly} {Excited} {States} of {1D} {Lattice} {Hard}-{Core} {Bosons}},
	volume = {98},
	url = {https://link.aps.org/doi/10.1103/PhysRevLett.98.050405},
	doi = {10.1103/PhysRevLett.98.050405},
	number = {5},
	journal = {Physical Review Letters},
	author = {Rigol, Marcos and Dunjko, Vanja and Yurovsky, Vladimir and Olshanii, Maxim},
	month = feb,
	year = {2007},
	pages = {050405},
}

@book{simon_reed_functional_1,
	series = {Methods of {Modern} {Mathematical} {Physics}},
	title = {I: {Functional} {Analysis}},
	isbn = {978-0-08-057048-8},
	url = {https://books.google.gr/books?id=rpFTTjxOYpsC},
	publisher = {Elsevier Science},
	author = {Reed, M. and Simon, B.},
	year = {1981},
}

@article{pancotti_2020_quantum,
	title = {Quantum east model: {Localization}, nonthermal eigenstates, and slow dynamics},
	volume = {10},
	url = {https://link.aps.org/doi/10.1103/PhysRevX.10.021051},
	doi = {10.1103/PhysRevX.10.021051},
	number = {2},
	journal = {Physical Review X},
	author = {Pancotti, Nicola and Giudice, Giacomo and Cirac, J. Ignacio and Garrahan, Juan P. and Bañuls, Mari Carmen},
	month = jun,
	year = {2020},
	note = {Number of pages: 21
Publisher: American Physical Society},
	pages = {021051},
}

@article{keyserlingk_2018_operator,
	title = {Operator hydrodynamics, otocs, and entanglement growth in systems without conservation laws},
	volume = {8},
	url = {https://link.aps.org/doi/10.1103/PhysRevX.8.021013},
	doi = {10.1103/PhysRevX.8.021013},
	number = {2},
	journal = {Physical Review X},
	author = {von Keyserlingk, C. W. and Rakovszky, Tibor and Pollmann, Frank and Sondhi, S. L.},
	month = apr,
	year = {2018},
	note = {Number of pages: 19
Publisher: American Physical Society},
	pages = {021013},
}

@article{khemani_2018_operator,
	title = {Operator spreading and the emergence of dissipative hydrodynamics under unitary evolution with conservation laws},
	volume = {8},
	url = {https://link.aps.org/doi/10.1103/PhysRevX.8.031057},
	doi = {10.1103/PhysRevX.8.031057},
	number = {3},
	journal = {Physical Review X},
	author = {Khemani, Vedika and Vishwanath, Ashvin and Huse, David A.},
	month = sep,
	year = {2018},
	note = {Number of pages: 25
Publisher: American Physical Society},
	pages = {031057},
}

@article{rakovszky_2018_diffusive,
	title = {Diffusive hydrodynamics of out-of-time-ordered correlators with charge conservation},
	volume = {8},
	url = {https://link.aps.org/doi/10.1103/PhysRevX.8.031058},
	doi = {10.1103/PhysRevX.8.031058},
	number = {3},
	journal = {Physical Review X},
	author = {Rakovszky, Tibor and Pollmann, Frank and von Keyserlingk, C. W.},
	month = sep,
	year = {2018},
	note = {Number of pages: 28
Publisher: American Physical Society},
	pages = {031058},
}

@article{nahum_2018_operator,
	title = {Operator spreading in random unitary circuits},
	volume = {8},
	url = {https://link.aps.org/doi/10.1103/PhysRevX.8.021014},
	doi = {10.1103/PhysRevX.8.021014},
	number = {2},
	journal = {Physical Review X},
	author = {Nahum, Adam and Vijay, Sagar and Haah, Jeongwan},
	month = apr,
	year = {2018},
	note = {Number of pages: 30
Publisher: American Physical Society},
	pages = {021014},
}

@article{neumann_1929_beweis,
	title = {Beweis des {Ergodensatzes} und {desH}-{Theorems} in der neuen {Mechanik}},
	volume = {57},
	issn = {0044-3328},
	url = {https://doi.org/10.1007/BF01339852},
	doi = {10.1007/BF01339852},
	number = {1},
	journal = {Zeitschrift für Physik},
	author = {Neumann, J. v.},
	month = jan,
	year = {1929},
	pages = {30--70},
}

@article{nachtergaele_2011_local_approx,
	title = {Local {Approximation} of {Observables} and {Commutator} {Bounds}},
	issn = {978-3-0348-0530-8},
	doi = {10.1007/978-3-0348-0531-5_8},
	author = {Nachtergaele, Bruno and Scholz, Volkher and Werner, Reinhard},
	month = mar,
	year = {2011},
}

@article{foss-feig_2015_nearly_linear,
	title = {Nearly linear light cones in long-range interacting quantum systems},
	volume = {114},
	url = {https://link.aps.org/doi/10.1103/PhysRevLett.114.157201},
	doi = {10.1103/PhysRevLett.114.157201},
	number = {15},
	journal = {Physical Review Letters},
	author = {Foss-Feig, Michael and Gong, Zhe-Xuan and Clark, Charles W. and Gorshkov, Alexey V.},
	month = apr,
	year = {2015},
	note = {Number of pages: 5
Publisher: American Physical Society},
	pages = {157201},
}

@article{Roberts_2016_butterfly,
	title = {Lieb-robinson bound and the butterfly effect in quantum field theories},
	volume = {117},
	url = {https://link.aps.org/doi/10.1103/PhysRevLett.117.091602},
	doi = {10.1103/PhysRevLett.117.091602},
	number = {9},
	journal = {Physical Review Letters},
	author = {Roberts, Daniel A. and Swingle, Brian},
	month = aug,
	year = {2016},
	note = {Number of pages: 6
Publisher: American Physical Society},
	pages = {091602},
}

@article{Haah_2021_qalgorithm,
	title = {Quantum algorithm for simulating real time evolution of lattice hamiltonians},
	volume = {52},
	issn = {1095-7111},
	url = {http://dx.doi.org/10.1137/18M1231511},
	doi = {10.1137/18m1231511},
	number = {6},
	journal = {SIAM Journal on Computing},
	author = {Haah, Jeongwan and Hastings, Matthew B. and Kothari, Robin and Low, Guang Hao},
	month = jan,
	year = {2021},
	note = {Publisher: Society for Industrial \& Applied Mathematics (SIAM)},
	pages = {FOCS18--250--FOCS18--284},
}

@article{Tran_2019_qsimulation,
	title = {Locality and digital quantum simulation of power-law interactions},
	volume = {9},
	url = {https://link.aps.org/doi/10.1103/PhysRevX.9.031006},
	doi = {10.1103/PhysRevX.9.031006},
	number = {3},
	journal = {Physical Review X},
	author = {Tran, Minh C. and Guo, Andrew Y. and Su, Yuan and Garrison, James R. and Eldredge, Zachary and Foss-Feig, Michael and Childs, Andrew M. and Gorshkov, Alexey V.},
	month = jul,
	year = {2019},
	note = {Number of pages: 22
Publisher: American Physical Society},
	pages = {031006},
}

@article{polkovnikov_2011_nonequilibrium,
	title = {Colloquium: {Nonequilibrium} dynamics of closed interacting quantum systems},
	volume = {83},
	url = {https://link.aps.org/doi/10.1103/RevModPhys.83.863},
	doi = {10.1103/RevModPhys.83.863},
	number = {3},
	journal = {Reviews of Modern Physics},
	author = {Polkovnikov, Anatoli and Sengupta, Krishnendu and Silva, Alessandro and Vengalattore, Mukund},
	month = aug,
	year = {2011},
	note = {Number of pages: 0
Publisher: American Physical Society},
	pages = {863--883},
}

@article{Bose_2007_qtransfer,
	title = {Quantum communication through spin chain dynamics: an introductory overview},
	volume = {48},
	issn = {0010-7514},
	url = {https://doi.org/10.1080/00107510701342313},
	doi = {10.1080/00107510701342313},
	number = {1},
	journal = {Contemporary Physics},
	author = {Bose, Sougato},
	month = jan,
	year = {2007},
	note = {Publisher: Taylor \& Francis},
	pages = {13--30},
}

@article{Burrell_2007_LR_disorder,
	title = {Bounds on the speed of information propagation in disordered quantum spin chains},
	volume = {99},
	url = {https://link.aps.org/doi/10.1103/PhysRevLett.99.167201},
	doi = {10.1103/PhysRevLett.99.167201},
	number = {16},
	journal = {Physical Review Letters},
	author = {Burrell, Christian K. and Osborne, Tobias J.},
	month = oct,
	year = {2007},
	note = {Number of pages: 4
Publisher: American Physical Society},
	pages = {167201},
}

@article{nachtergaele_2006_propagation,
	title = {Propagation of {Correlations} in {Quantum} {Lattice} {Systems}},
	volume = {124},
	issn = {1572-9613},
	url = {https://doi.org/10.1007/s10955-006-9143-6},
	doi = {10.1007/s10955-006-9143-6},
	number = {1},
	journal = {Journal of Statistical Physics},
	author = {Nachtergaele, Bruno and Ogata, Yoshiko and Sims, Robert},
	month = jul,
	year = {2006},
	pages = {1--13},
}

@article{Lieb_1961_Lieb-Schultz-Mattis,
	title = {Two soluble models of an antiferromagnetic chain},
	volume = {16},
	number = {3},
	journal = {Annals of Physics},
	author = {Lieb, Elliott and Schultz, Theodore and Mattis, Daniel},
	year = {1961},
	note = {Publisher: Elsevier},
	pages = {407--466},
}

@article{Hastings_2004_lieb-schultz-mattis,
	title = {Lieb-{Schultz}-{Mattis} in higher dimensions},
	volume = {69},
	url = {https://link.aps.org/doi/10.1103/PhysRevB.69.104431},
	doi = {10.1103/PhysRevB.69.104431},
	number = {10},
	journal = {Physical Review B},
	author = {Hastings, M. B.},
	month = mar,
	year = {2004},
	note = {Number of pages: 13
Publisher: American Physical Society},
	pages = {104431},
}

@article{gogolin_equilibration_2016,
	title = {Equilibration, thermalisation, and the emergence of statistical mechanics in closed quantum systems},
	volume = {79},
	issn = {0034-4885},
	url = {http://dx.doi.org/10.1088/0034-4885/79/5/056001},
	doi = {10.1088/0034-4885/79/5/056001},
	number = {5},
	journal = {Reports on Progress in Physics},
	author = {Gogolin, Christian and Eisert, Jens},
	month = apr,
	year = {2016},
	pages = {056001},
}

@article{perez_2023_locality,
	title = {Locality {Estimates} for {Complex} {Time} {Evolution} in {1D}},
	volume = {399},
	issn = {1432-0916},
	url = {https://doi.org/10.1007/s00220-022-04573-w},
	doi = {10.1007/s00220-022-04573-w},
	number = {2},
	journal = {Communications in Mathematical Physics},
	author = {Pérez-García, David and Pérez-Hernández, Antonio},
	month = apr,
	year = {2023},
	pages = {929--970},
}

@article{chen2023speed,
	title = {Speed limits and locality in many-body quantum dynamics},
	volume = {86},
	number = {11},
	journal = {Reports on Progress in Physics},
	author = {Chen, Chi-Fang Anthony and Lucas, Andrew and Yin, Chao},
	year = {2023},
	note = {Publisher: IOP Publishing},
	pages = {116001},
}

@misc{yoshimura_2024_anomalous_current,
	title = {Anomalous current fluctuations from {Euler} hydrodynamics},
	url = {https://arxiv.org/abs/2406.20091},
	author = {Yoshimura, Takato and Krajnik, Ziga},
	year = {2024},
	note = {arXiv: 2406.20091 [cond-mat.stat-mech]},
}

@article{krajnik_2022_anomalous_current_fl,
	title = {Exact anomalous current fluctuations in a deterministic interacting model},
	volume = {128},
	url = {https://link.aps.org/doi/10.1103/PhysRevLett.128.160601},
	doi = {10.1103/PhysRevLett.128.160601},
	number = {16},
	journal = {Physical Review Letters},
	author = {Krajnik, Ziga and Schmidt, Johannes and Pasquier, Vincent and Ilievski, Enej and Prosen, Tomaz},
	month = apr,
	year = {2022},
	note = {Number of pages: 6
Publisher: American Physical Society},
	pages = {160601},
}

@article{carlen_2017_gradient,
	title = {Gradient flow and entropy inequalities for quantum {Markov} semigroups with detailed balance},
	volume = {273},
	issn = {0022-1236},
	url = {https://www.sciencedirect.com/science/article/pii/S0022123617301878},
	doi = {10.1016/j.jfa.2017.05.003},
	number = {5},
	journal = {Journal of Functional Analysis},
	author = {Carlen, Eric A. and Maas, Jan},
	month = sep,
	year = {2017},
	pages = {1810--1869},
}

@article{fangola_2007_generator_quantum_markov,
	title = {{GENERATORS} {OF} {DETAILED} {BALANCE} {QUANTUM} {MARKOV} {SEMIGROUPS}},
	volume = {10},
	issn = {0219-0257},
	url = {https://doi.org/10.1142/S0219025707002762},
	doi = {10.1142/S0219025707002762},
	number = {03},
	urldate = {2025-01-28},
	journal = {Infinite Dimensional Analysis, Quantum Probability and Related Topics},
	author = {FAGNOLA, FRANCO and UMANITÀ, VERONICA},
	month = sep,
	year = {2007},
	note = {Publisher: World Scientific Publishing Co.},
	pages = {335--363},
}

@article{Alicki_1976_detailed_balance,
	title = {On the detailed balance condition for non-hamiltonian systems},
	volume = {10},
	issn = {0034-4877},
	url = {https://www.sciencedirect.com/science/article/pii/003448777690046X},
	doi = {10.1016/0034-4877(76)90046-X},
	number = {2},
	journal = {Reports on Mathematical Physics},
	author = {Alicki, Robert},
	month = oct,
	year = {1976},
	pages = {249--258},
}

@article{gopalakrishnan_2024_nongaussian_dif,
	title = {Non-{Gaussian} diffusive fluctuations in {Dirac} fluids},
	volume = {121},
	url = {https://doi.org/10.1073/pnas.2403327121},
	doi = {10.1073/pnas.2403327121},
	number = {50},
	urldate = {2025-01-28},
	journal = {Proceedings of the National Academy of Sciences},
	author = {Gopalakrishnan, Sarang and McCulloch, Ewan and Vasseur, Romain},
	month = dec,
	year = {2024},
	note = {Publisher: Proceedings of the National Academy of Sciences},
	pages = {e2403327121},
}

@article{gopalakrishnan_2024_universality,
	title = {Distinct universality classes of diffusive transport from full counting statistics},
	volume = {109},
	url = {https://link.aps.org/doi/10.1103/PhysRevB.109.024417},
	doi = {10.1103/PhysRevB.109.024417},
	number = {2},
	journal = {Physical Review B},
	author = {Gopalakrishnan, Sarang and Morningstar, Alan and Vasseur, Romain and Khemani, Vedika},
	month = jan,
	year = {2024},
	note = {Number of pages: 10
Publisher: American Physical Society},
	pages = {024417},
}

@misc{sims_2011_LR,
	title = {Lieb-{Robinson} {Bounds} and {Quasi}-locality for the {Dynamics} of {Many}-{Body} {Quantum} {Systems}},
	url = {https://doi.org/10.1142/9789814350365_0007},
	urldate = {2024-07-18},
	publisher = {WORLD SCIENTIFIC},
	author = {Sims, Robert},
	month = may,
	year = {2011},
	doi = {10.1142/9789814350365_0007},
}

@article{nardis_2019_diffusion_ghd,
	title = {Diffusion in generalized hydrodynamics and quasiparticle scattering},
	volume = {6},
	url = {https://scipost.org/10.21468/SciPostPhys.6.4.049},
	doi = {10.21468/SciPostPhys.6.4.049},
	journal = {SciPost Physics},
	author = {Nardis, Jacopo De and Bernard, Denis and Doyon, Benjamin},
	year = {2019},
	note = {Publisher: SciPost},
	pages = {049},
}

@article{doyon_2023_ballistic_fluctuations,
	title = {Ballistic macroscopic fluctuation theory},
	volume = {15},
	url = {https://scipost.org/10.21468/SciPostPhys.15.4.136},
	doi = {10.21468/SciPostPhys.15.4.136},
	journal = {SciPost Physics},
	author = {Doyon, Benjamin and Perfetto, Gabriele and Sasamoto, Tomohiro and Yoshimura, Takato},
	year = {2023},
	note = {Publisher: SciPost},
	pages = {136},
}

@article{doyon_2023_hydro_longrange,
	title = {Emergence of hydrodynamic spatial long-range correlations in nonequilibrium many-body systems},
	volume = {131},
	url = {https://link.aps.org/doi/10.1103/PhysRevLett.131.027101},
	doi = {10.1103/PhysRevLett.131.027101},
	number = {2},
	journal = {Physical Review Letters},
	author = {Doyon, Benjamin and Perfetto, Gabriele and Sasamoto, Tomohiro and Yoshimura, Takato},
	month = jul,
	year = {2023},
	note = {Number of pages: 7
Publisher: American Physical Society},
	pages = {027101},
}

@article{essler_2020_integrablility_lindbladian,
	title = {Integrability of one-dimensional {Lindbladians} from operator-space fragmentation},
	volume = {102},
	url = {https://link.aps.org/doi/10.1103/PhysRevE.102.062210},
	doi = {10.1103/PhysRevE.102.062210},
	number = {6},
	journal = {Physical Review E: Statistical Physics, Plasmas, Fluids, and Related Interdisciplinary Topics},
	author = {Essler, Fabian H. L. and Piroli, Lorenzo},
	month = dec,
	year = {2020},
	note = {Number of pages: 7
Publisher: American Physical Society},
	pages = {062210},
}

@article{ziolkowska_2020_yang_baxter,
	title = {Yang-{Baxter} integrable {Lindblad} equations},
	volume = {8},
	url = {https://scipost.org/10.21468/SciPostPhys.8.3.044},
	doi = {10.21468/SciPostPhys.8.3.044},
	journal = {SciPost Physics},
	author = {Ziolkowska, Aleksandra A. and Essler, Fabian H.L.},
	year = {2020},
	note = {Publisher: SciPost},
	pages = {044},
}

@article{kobayashi_2022_vanishing_currents,
	title = {Vanishing and nonvanishing persistent currents of various conserved quantities},
	volume = {129},
	url = {https://link.aps.org/doi/10.1103/PhysRevLett.129.176601},
	doi = {10.1103/PhysRevLett.129.176601},
	number = {17},
	journal = {Physical Review Letters},
	author = {Kobayashi, Hirokazu and Watanabe, Haruki},
	month = oct,
	year = {2022},
	note = {Number of pages: 6
Publisher: American Physical Society},
	pages = {176601},
}

@article{poulin_2010_LR_markovian,
	title = {Lieb-robinson bound and locality for general markovian quantum dynamics},
	volume = {104},
	url = {https://link.aps.org/doi/10.1103/PhysRevLett.104.190401},
	doi = {10.1103/PhysRevLett.104.190401},
	number = {19},
	journal = {Physical Review Letters},
	author = {Poulin, David},
	month = may,
	year = {2010},
	note = {Number of pages: 4
Publisher: American Physical Society},
	pages = {190401},
}

@misc{nachtergaele_2011_LR_irreversible,
	title = {Lieb-robinson bounds and existence of the thermodynamic limit for a class of irreversible quantum dynamics},
	url = {https://arxiv.org/abs/1103.1122},
	author = {Nachtergaele, Bruno and Vershynina, Anna and Zagrebnov, Valentin A.},
	year = {2011},
	note = {arXiv: 1103.1122 [math-ph]},
}

@article{boldrighini_1997_hard_rod,
	title = {One-{Dimensional} {Hard}-{Rod} {Caricature} of {Hydrodynamics}: “{Navier}–{Stokes} {Correction}” for {Local} {Equilibrium} {Initial} {States}},
	volume = {189},
	issn = {1432-0916},
	url = {https://doi.org/10.1007/s002200050218},
	doi = {10.1007/s002200050218},
	number = {2},
	journal = {Communications in Mathematical Physics},
	author = {Boldrighini, C. and Suhov, Y.M.},
	month = nov,
	year = {1997},
	pages = {577--590},
}

@article{park_1982_clusterexpansion,
	title = {The cluster expansion for classical and quantum lattice systems},
	volume = {27},
	issn = {1572-9613},
	url = {https://doi.org/10.1007/BF01011092},
	doi = {10.1007/BF01011092},
	number = {3},
	journal = {Journal of Statistical Physics},
	author = {Park, Yong Moon},
	month = mar,
	year = {1982},
	pages = {553--576},
}

@article{spohn_2014_nonlinear,
	title = {Nonlinear {Fluctuating} {Hydrodynamics} for {Anharmonic} {Chains}},
	volume = {154},
	issn = {1572-9613},
	url = {https://doi.org/10.1007/s10955-014-0933-y},
	doi = {10.1007/s10955-014-0933-y},
	number = {5},
	journal = {Journal of Statistical Physics},
	author = {Spohn, Herbert},
	month = mar,
	year = {2014},
	pages = {1191--1227},
}

@article{Prosen_1999_ergodic,
	title = {Ergodic properties of a generic nonintegrable quantum many-body system in the thermodynamic limit},
	volume = {60},
	url = {https://link.aps.org/doi/10.1103/PhysRevE.60.3949},
	doi = {10.1103/PhysRevE.60.3949},
	number = {4},
	journal = {Physical Review E: Statistical Physics, Plasmas, Fluids, and Related Interdisciplinary Topics},
	author = {Prosen, Tomaz},
	month = oct,
	year = {1999},
	note = {Number of pages: 0
Publisher: American Physical Society},
	pages = {3949--3968},
}

@article{prosen_1998_quantum_invariants,
	title = {Quantum invariants of motion in a generic many-body system},
	volume = {31},
	issn = {0305-4470},
	url = {https://dx.doi.org/10.1088/0305-4470/31/37/004},
	doi = {10.1088/0305-4470/31/37/004},
	number = {37},
	journal = {Journal of Physics A: Mathematical and General},
	author = {{Tomaz Prosen}},
	month = sep,
	year = {1998},
	pages = {L645},
}

@misc{ampelogiannis_2024_diffusion,
	title = {Rigorous bound on diffusion for chaotic spin chains},
	author = {Ampelogiannis, Dimitrios and Doyon, Benjamin},
	note = {In review at Physical Review Letters},
}

@article{denardis_2022_correlation_ghd,
	title = {Correlation functions and transport coefficients in generalised hydrodynamics},
	volume = {2022},
	issn = {1742-5468},
	url = {https://dx.doi.org/10.1088/1742-5468/ac3658},
	doi = {10.1088/1742-5468/ac3658},
	number = {1},
	journal = {Journal of Statistical Mechanics: Theory and Experiment},
	author = {De Nardis, Jacopo and Doyon, Benjamin and Medenjak, Marko and Panfil, Miłosz},
	month = jan,
	year = {2022},
	note = {Publisher: IOP Publishing and SISSA},
	pages = {014002},
}

@misc{hubner_2024_diffusive,
	title = {Diffusive hydrodynamics from long-range correlations},
	url = {https://arxiv.org/abs/2408.04502},
	author = {Hübner, Friedrich and Biagetti, Leonardo and Nardis, Jacopo De and Doyon, Benjamin},
	year = {2024},
	note = {arXiv: 2408.04502 [cond-mat.stat-mech]},
}

@article{Wang_2020_tightening_LR,
	title = {Tightening the lieb-robinson bound in locally interacting systems},
	volume = {1},
	url = {https://link.aps.org/doi/10.1103/PRXQuantum.1.010303},
	doi = {10.1103/PRXQuantum.1.010303},
	number = {1},
	journal = {PRX Quantum},
	author = {Wang, Zhiyuan and Hazzard, Kaden R.A.},
	month = sep,
	year = {2020},
	note = {Number of pages: 24
Publisher: American Physical Society},
	pages = {010303},
}

@article{chiba_2023_ising_nonintegrability,
	title = {Proof of absence of local conserved quantities in the mixed-field {Ising} chain},
	volume = {109},
	url = {https://link.aps.org/doi/10.1103/PhysRevB.109.035123},
	doi = {10.1103/PhysRevB.109.035123},
	number = {3},
	journal = {Physical Review B},
	author = {Chiba, Yuuya},
	month = jan,
	year = {2024},
	note = {Number of pages: 15
Publisher: American Physical Society},
	pages = {035123},
}

@article{shiraishi_2019_absence_local,
	title = {Proof of the absence of local conserved quantities in the {XYZ} chain with a magnetic field},
	volume = {128},
	issn = {0295-5075},
	url = {https://dx.doi.org/10.1209/0295-5075/128/17002},
	doi = {10.1209/0295-5075/128/17002},
	number = {1},
	journal = {Europhysics Letters},
	author = {Shiraishi, Naoto},
	month = nov,
	year = {2019},
	note = {Publisher: EDP Sciences, IOP Publishing and Società Italiana di Fisica},
	pages = {17002},
}

@article{shiraishi2024absence,
	title = {Absence of {Local} {Conserved} {Quantity} in the {Heisenberg} {Model} with {Next}-{Nearest}-{Neighbor} {Interaction}},
	volume = {191},
	issn = {1572-9613},
	url = {https://doi.org/10.1007/s10955-024-03326-4},
	doi = {10.1007/s10955-024-03326-4},
	number = {9},
	journal = {Journal of Statistical Physics},
	author = {Shiraishi, Naoto},
	month = sep,
	year = {2024},
	pages = {114},
}

@misc{park_2024_nonintegrability,
	title = {Graph theoretical proof of nonintegrability in quantum many-body systems : {Application} to the {PXP} model},
	url = {https://arxiv.org/abs/2403.02335},
	author = {Park, HaRu K. and Lee, SungBin},
	year = {2024},
	note = {arXiv: 2403.02335 [cond-mat.stat-mech]},
}

@article{Medenjak_2017_diffusion_bound,
	title = {Lower bounding diffusion constant by the curvature of drude weight},
	volume = {119},
	url = {https://link.aps.org/doi/10.1103/PhysRevLett.119.080602},
	doi = {10.1103/PhysRevLett.119.080602},
	number = {8},
	journal = {Physical Review Letters},
	author = {Medenjak, Marko and Karrasch, Christoph and Prosen, Tomaz},
	month = aug,
	year = {2017},
	note = {Number of pages: 5
Publisher: American Physical Society},
	pages = {080602},
}

@article{Prosen_2014_diffusion_bound,
	title = {Lower bounds on high-temperature diffusion constants from quadratically extensive almost-conserved operators},
	volume = {89},
	url = {https://link.aps.org/doi/10.1103/PhysRevE.89.012142},
	doi = {10.1103/PhysRevE.89.012142},
	number = {1},
	journal = {Physical Review E: Statistical Physics, Plasmas, Fluids, and Related Interdisciplinary Topics},
	author = {Prosen, Tomaz},
	month = jan,
	year = {2014},
	note = {Number of pages: 5
Publisher: American Physical Society},
	pages = {012142},
}

@misc{ampelogiannis_2024_clustering,
	title = {Clustering of higher order connected correlations in {C}$^{\textrm{*}}$ dynamical systems},
	url = {https://arxiv.org/abs/2405.09388},
	author = {Ampelogiannis, Dimitrios and Doyon, Benjamin},
	year = {2024},
	note = {arXiv: 2405.09388 [math-ph]},
}

@article{Medenjak_Yoshimura_2020_Diffusion_bound,
	title = {Diffusion from convection},
	volume = {9},
	url = {https://scipost.org/10.21468/SciPostPhys.9.5.075},
	doi = {10.21468/SciPostPhys.9.5.075},
	journal = {SciPost Physics},
	author = {Medenjak, Marko and Nardis, Jacopo De and Yoshimura, Takato},
	year = {2020},
	note = {Publisher: SciPost},
	pages = {075},
}

@article{DeNardis_Doyon_2019_diffusion_GHD,
	title = {Diffusion in generalized hydrodynamics and quasiparticle scattering},
	volume = {6},
	url = {https://scipost.org/10.21468/SciPostPhys.6.4.049},
	doi = {10.21468/SciPostPhys.6.4.049},
	journal = {SciPost Physics},
	author = {Nardis, Jacopo De and Bernard, Denis and Doyon, Benjamin},
	year = {2019},
	note = {Publisher: SciPost},
	pages = {049},
}

@article{doyon_2019_diffusion,
	title = {Diffusion and {Superdiffusion} from {Hydrodynamic} {Projections}},
	volume = {186},
	issn = {1572-9613},
	url = {https://doi.org/10.1007/s10955-021-02863-6},
	doi = {10.1007/s10955-021-02863-6},
	number = {2},
	journal = {Journal of Statistical Physics},
	author = {Doyon, Benjamin},
	month = jan,
	year = {2022},
	pages = {25},
}

@article{doyon_2023_mft,
	title = {Ballistic macroscopic fluctuation theory},
	volume = {15},
	url = {https://scipost.org/10.21468/SciPostPhys.15.4.136},
	doi = {10.21468/SciPostPhys.15.4.136},
	journal = {SciPost Physics},
	author = {Doyon, Benjamin and Perfetto, Gabriele and Sasamoto, Tomohiro and Yoshimura, Takato},
	year = {2023},
	note = {Publisher: SciPost},
	pages = {136},
}

@article{Hastings_2006_spectralgap,
	title = {Spectral {Gap} and {Exponential} {Decay} of {Correlations}},
	volume = {265},
	issn = {1432-0916},
	url = {https://doi.org/10.1007/s00220-006-0030-4},
	doi = {10.1007/s00220-006-0030-4},
	number = {3},
	journal = {Communications in Mathematical Physics},
	author = {Hastings, Matthew B. and Koma, Tohru},
	month = aug,
	year = {2006},
	pages = {781--804},
}

@article{Chen_2019_finite_scrambling,
	title = {Finite speed of quantum scrambling with long range interactions},
	volume = {123},
	url = {https://link.aps.org/doi/10.1103/PhysRevLett.123.250605},
	doi = {10.1103/PhysRevLett.123.250605},
	number = {25},
	journal = {Physical Review Letters},
	author = {Chen, Chi-Fang and Lucas, Andrew},
	month = dec,
	year = {2019},
	note = {Number of pages: 5
Publisher: American Physical Society},
	pages = {250605},
}

@article{Kuwahara_2020_linear_LR,
	title = {Strictly linear light cones in long-range interacting systems of arbitrary dimensions},
	volume = {10},
	url = {https://link.aps.org/doi/10.1103/PhysRevX.10.031010},
	doi = {10.1103/PhysRevX.10.031010},
	number = {3},
	journal = {Physical Review X},
	author = {Kuwahara, Tomotaka and Saito, Keiji},
	month = jul,
	year = {2020},
	note = {Number of pages: 12
Publisher: American Physical Society},
	pages = {031010},
}

@article{park_1995_clustering_gibbs,
	title = {Uniqueness and clustering properties of {Gibbs} states for classical and quantum unbounded spin systems},
	volume = {80},
	issn = {1572-9613},
	url = {https://doi.org/10.1007/BF02178359},
	doi = {10.1007/BF02178359},
	number = {1},
	journal = {Journal of Statistical Physics},
	author = {Park, Yong Moon and Yoo, Hyun Jae},
	month = jul,
	year = {1995},
	pages = {223--271},
}

@article{nachtergaele_2006_LR,
	title = {Lieb-{Robinson} {Bounds} and the {Exponential} {Clustering} {Theorem}},
	volume = {265},
	issn = {1432-0916},
	url = {https://doi.org/10.1007/s00220-006-1556-1},
	doi = {10.1007/s00220-006-1556-1},
	number = {1},
	journal = {Communications in Mathematical Physics},
	author = {Nachtergaele, Bruno and Sims, Robert},
	month = jul,
	year = {2006},
	pages = {119--130},
}

@article{Fowler_2023_cumulant_expansion,
	title = {Determining the validity of cumulant expansions for central spin models},
	volume = {5},
	url = {https://link.aps.org/doi/10.1103/PhysRevResearch.5.033148},
	doi = {10.1103/PhysRevResearch.5.033148},
	number = {3},
	journal = {Physical Review Research},
	author = {Fowler-Wright, Piper and Arnardóttir, Kristín B. and Kirton, Peter and Lovett, Brendon W. and Keeling, Jonathan},
	month = sep,
	year = {2023},
	note = {Number of pages: 9
Publisher: American Physical Society},
	pages = {033148},
}

@article{rota_1964_mobius,
	title = {On the foundations of combinatorial theory {I}. {Theory} of {Möbius} {Functions}},
	volume = {2},
	issn = {1432-2064},
	url = {https://doi.org/10.1007/BF00531932},
	doi = {10.1007/BF00531932},
	number = {4},
	journal = {Zeitschrift für Wahrscheinlichkeitstheorie und Verwandte Gebiete},
	author = {Rota, Gian -Carlo},
	month = jan,
	year = {1964},
	pages = {340--368},
}

@incollection{greenleaf_1969_invariant,
	title = {Invariant means on topological groups and their applications},
	booktitle = {Van nostrand mathematical studies series, no. 16},
	publisher = {Van Nostrand Reinhold Company},
	author = {Greenleaf, Frederick},
	year = {1969},
}

@misc{speicher_2019_notes,
	title = {Lecture notes on "free probability theory"},
	author = {Speicher, Roland},
	year = {2019},
	note = {arXiv: 1908.08125 [math.OA]},
}

@article{Tran_2017_Lieb_Robinson_npartite,
	title = {Lieb-{Robinson} bounds on {\textless}span class="nocase"{\textgreater}n{\textless}/span{\textgreater}-partite connected correlation functions},
	volume = {96},
	url = {https://link.aps.org/doi/10.1103/PhysRevA.96.052334},
	doi = {10.1103/PhysRevA.96.052334},
	number = {5},
	journal = {Physical Review A: Atomic, Molecular, and Optical Physics},
	author = {Tran, Minh Cong and Garrison, James R. and Gong, Zhe-Xuan and Gorshkov, Alexey V.},
	month = nov,
	year = {2017},
	note = {Number of pages: 8
Publisher: American Physical Society},
	pages = {052334},
}

@inproceedings{voiculescu_1985_Symmetries,
	address = {Berlin, Heidelberg},
	title = {Symmetries of some reduced free product {C}*-algebras},
	isbn = {978-3-540-39514-0},
	booktitle = {Operator {Algebras} and their {Connections} with {Topology} and {Ergodic} {Theory}},
	publisher = {Springer Berlin Heidelberg},
	author = {Voiculescu, Dan},
	editor = {Araki, Huzihiro and Moore, Calvin C. and Stratila, Serban-Valentin and Voiculescu, Dan-Virgil},
	year = {1985},
	pages = {556--588},
}

@article{voiculescu_1986_addition,
	title = {Addition of certain non-commuting random variables},
	volume = {66},
	issn = {0022-1236},
	url = {https://www.sciencedirect.com/science/article/pii/0022123686900625},
	doi = {10.1016/0022-1236(86)90062-5},
	number = {3},
	journal = {Journal of Functional Analysis},
	author = {Voiculescu, Dan},
	month = may,
	year = {1986},
	pages = {323--346},
}

@article{voiculescu_1987_multiplication,
	title = {{MULTIPLICATION} {OF} {CERTAIN} {NON}-{COMMUTING} {RANDOM} {VARIABLES}},
	volume = {18},
	issn = {03794024, 18417744},
	url = {http://www.jstor.org/stable/24714784},
	number = {2},
	urldate = {2024-04-30},
	journal = {Journal of Operator Theory},
	author = {VOICULESCU, DAN},
	year = {1987},
	note = {Publisher: Theta Foundation},
	pages = {223--235},
}

@article{pappalardi_2017_multi_entanglement,
	title = {Multipartite entanglement after a quantum quench},
	volume = {2017},
	issn = {1742-5468},
	url = {https://dx.doi.org/10.1088/1742-5468/aa6809},
	doi = {10.1088/1742-5468/aa6809},
	number = {5},
	journal = {Journal of Statistical Mechanics: Theory and Experiment},
	author = {Pappalardi, Silvia and Russomanno, Angelo and Silva, Alessandro and Fazio, Rosario},
	month = may,
	year = {2017},
	note = {Publisher: IOP Publishing and SISSA},
	pages = {053104},
}

@article{zhou_2006_multiparty,
	title = {Multiparty correlation measure based on the cumulant},
	volume = {74},
	url = {https://link.aps.org/doi/10.1103/PhysRevA.74.052110},
	doi = {10.1103/PhysRevA.74.052110},
	number = {5},
	journal = {Physical Review A: Atomic, Molecular, and Optical Physics},
	author = {Zhou, D. L. and Zeng, B. and Xu, Z. and You, L.},
	month = nov,
	year = {2006},
	note = {Number of pages: 8
Publisher: American Physical Society},
	pages = {052110},
}

@article{garcia_2022_OTOC_chaos_review,
	title = {Out-of-time-order correlators and quantum chaos},
	journal = {arXiv preprint arXiv:2209.07965},
	author = {García-Mata, Ignacio and Jalabert, Rodolfo A and Wisniacki, Diego A},
	year = {2022},
}

@article{bhattacharyya_2022_quantum_chaos,
	title = {Towards the web of quantum chaos diagnostics},
	volume = {82},
	issn = {1434-6052},
	url = {https://doi.org/10.1140/epjc/s10052-022-10035-3},
	doi = {10.1140/epjc/s10052-022-10035-3},
	number = {1},
	journal = {The European Physical Journal C},
	author = {Bhattacharyya, Arpan and Chemissany, Wissam and Haque, S. Shajidul and Yan, Bin},
	month = jan,
	year = {2022},
	pages = {87},
}

@misc{jindal_2024_free,
	title = {Generalized free cumulants for quantum chaotic systems},
	author = {Jindal, Siddharth and Hosur, Pavan},
	year = {2024},
	note = {arXiv: 2401.13829 [cond-mat.stat-mech]},
}

@article{Pappalardi_2022_ETH_free,
	title = {Eigenstate thermalization hypothesis and free probability},
	volume = {129},
	url = {https://link.aps.org/doi/10.1103/PhysRevLett.129.170603},
	doi = {10.1103/PhysRevLett.129.170603},
	number = {17},
	journal = {Physical Review Letters},
	author = {Pappalardi, Silvia and Foini, Laura and Kurchan, Jorge},
	month = oct,
	year = {2022},
	note = {Number of pages: 6
Publisher: American Physical Society},
	pages = {170603},
}

@article{Kira_2008_clusterexp,
	title = {Cluster-expansion representation in quantum optics},
	volume = {78},
	url = {https://link.aps.org/doi/10.1103/PhysRevA.78.022102},
	doi = {10.1103/PhysRevA.78.022102},
	number = {2},
	journal = {Physical Review A: Atomic, Molecular, and Optical Physics},
	author = {Kira, M. and Koch, S. W.},
	month = aug,
	year = {2008},
	note = {Number of pages: 26
Publisher: American Physical Society},
	pages = {022102},
}

@article{fricke_transport_1996,
	title = {Transport {Equations} {Including} {Many}-{Particle} {Correlations} for an {Arbitrary} {Quantum} {System}: {A} {General} {Formalism}},
	volume = {252},
	issn = {0003-4916},
	url = {https://www.sciencedirect.com/science/article/pii/S0003491696901426},
	doi = {10.1006/aphy.1996.0142},
	number = {2},
	journal = {Annals of Physics},
	author = {Fricke, Jens},
	month = dec,
	year = {1996},
	pages = {479--498},
}

@article{Sanchez_2020_cumulant,
	title = {Cumulant expansion for the treatment of light–matter interactions in arbitrary material structures},
	volume = {152},
	issn = {0021-9606},
	url = {https://doi.org/10.1063/1.5138937},
	doi = {10.1063/1.5138937},
	number = {3},
	urldate = {2024-04-24},
	journal = {The Journal of Chemical Physics},
	author = {Sánchez-Barquilla, M. and Silva, R. E. F. and Feist, J.},
	month = jan,
	year = {2020},
	pages = {034108},
}

@article{Robicheaux_2021_beyond,
	title = {Beyond lowest order mean-field theory for light interacting with atom arrays},
	volume = {104},
	url = {https://link.aps.org/doi/10.1103/PhysRevA.104.023702},
	doi = {10.1103/PhysRevA.104.023702},
	number = {2},
	journal = {Physical Review A: Atomic, Molecular, and Optical Physics},
	author = {Robicheaux, F. and Suresh, Deepak A.},
	month = aug,
	year = {2021},
	note = {Number of pages: 12
Publisher: American Physical Society},
	pages = {023702},
}

@article{kubo_1962_cumulant,
	title = {Generalized {Cumulant} {Expansion} {Method}},
	volume = {17},
	issn = {0031-9015},
	url = {https://doi.org/10.1143/JPSJ.17.1100},
	doi = {10.1143/JPSJ.17.1100},
	number = {7},
	urldate = {2024-04-24},
	journal = {Journal of the Physical Society of Japan},
	author = {Kubo, Ryogo},
	month = jul,
	year = {1962},
	note = {Publisher: The Physical Society of Japan},
	pages = {1100--1120},
}

@article{Kramer_2015_generalisedMFT,
	title = {Generalized mean-field approach to simulate the dynamics of large open spin ensembles with long range interactions},
	volume = {69},
	issn = {1434-6079},
	url = {https://doi.org/10.1140/epjd/e2015-60266-5},
	doi = {10.1140/epjd/e2015-60266-5},
	number = {12},
	journal = {The European Physical Journal D},
	author = {Krämer, Sebastian and Ritsch, Helmut},
	month = dec,
	year = {2015},
	pages = {282},
}

@article{myers_2020_fluctuations,
	title = {Transport fluctuations in integrable models out of equilibrium},
	volume = {8},
	url = {https://scipost.org/10.21468/SciPostPhys.8.1.007},
	doi = {10.21468/SciPostPhys.8.1.007},
	journal = {SciPost Physics},
	author = {Myers, Jason and Bhaseen, M. J. and Harris, Rosemary J. and Doyon, Benjamin},
	year = {2020},
	note = {Publisher: SciPost},
	pages = {007},
}

@article{doyon_2020_fluctuations,
	title = {Fluctuations in {Ballistic} {Transport} from {Euler} {Hydrodynamics}},
	volume = {21},
	issn = {1424-0661},
	url = {https://doi.org/10.1007/s00023-019-00860-w},
	doi = {10.1007/s00023-019-00860-w},
	number = {1},
	journal = {Annales Henri Poincaré},
	author = {Doyon, Benjamin and Myers, Jason},
	month = jan,
	year = {2020},
	pages = {255--302},
}

@article{ampelogiannis_2023_almost,
	title = {Almost {Everywhere} {Ergodicity} in {Quantum} {Lattice} {Models}},
	volume = {404},
	copyright = {All rights reserved},
	issn = {1432-0916},
	url = {https://doi.org/10.1007/s00220-023-04849-9},
	doi = {10.1007/s00220-023-04849-9},
	number = {2},
	journal = {Communications in Mathematical Physics},
	author = {Ampelogiannis, Dimitrios and Doyon, Benjamin},
	month = dec,
	year = {2023},
	pages = {735--768},
}

@article{speicher_1994_multiplicative,
	title = {Multiplicative functions on the lattice of non-crossing partitions and free convolution},
	volume = {298},
	issn = {1432-1807},
	url = {https://doi.org/10.1007/BF01459754},
	doi = {10.1007/BF01459754},
	number = {1},
	journal = {Mathematische Annalen},
	author = {Speicher, Roland},
	month = jan,
	year = {1994},
	pages = {611--628},
}

@article{Speed_1983_Cumulants1,
	title = {{CUMULANTS} {AND} {PARTITION} {LATTICES} 1},
	volume = {25},
	issn = {0004-9581},
	url = {https://doi.org/10.1111/j.1467-842X.1983.tb00391.x},
	doi = {10.1111/j.1467-842X.1983.tb00391.x},
	number = {2},
	urldate = {2024-04-02},
	journal = {Australian Journal of Statistics},
	author = {Speed, T. P.},
	month = feb,
	year = {1983},
	note = {Publisher: John Wiley \& Sons, Ltd},
	pages = {378--388},
}

@article{Park_1982_cluster_expansion,
	title = {The cluster expansion for classical and quantum lattice systems},
	volume = {27},
	issn = {1572-9613},
	url = {https://doi.org/10.1007/BF01011092},
	doi = {10.1007/BF01011092},
	number = {3},
	journal = {Journal of Statistical Physics},
	author = {Park, Yong Moon},
	month = mar,
	year = {1982},
	pages = {553--576},
}

@article{Gibibre_1965_quant_gases,
	title = {Reduced {Density} {Matrices} of {Quantum} {Gases}. {II}. {Cluster} {Property}},
	volume = {6},
	issn = {0022-2488},
	url = {https://doi.org/10.1063/1.1704276},
	doi = {10.1063/1.1704276},
	number = {2},
	urldate = {2024-03-27},
	journal = {Journal of Mathematical Physics},
	author = {Ginibre, Jean},
	month = feb,
	year = {1965},
	pages = {252--262},
}

@article{Ruelle_1968_classical_gas,
	title = {Statistical mechanics of a one-dimensional lattice gas},
	volume = {9},
	issn = {1432-0916},
	url = {https://doi.org/10.1007/BF01654281},
	doi = {10.1007/BF01654281},
	number = {4},
	journal = {Communications in Mathematical Physics},
	author = {Ruelle, D.},
	month = dec,
	year = {1968},
	pages = {267--278},
}

@misc{summers2005tomitatakesaki,
	title = {Tomita-takesaki modular theory},
	author = {Summers, Stephen J.},
	year = {2005},
	note = {arXiv: math-ph/0511034 [math-ph]},
}

@article{ampelogiannis_long-time_2023,
	title = {Long-{Time} {Dynamics} in {Quantum} {Spin} {Lattices}: {Ergodicity} and {Hydrodynamic} {Projections} at {All} {Frequencies} and {Wavelengths}},
	issn = {1424-0661},
	url = {https://doi.org/10.1007/s00023-023-01304-2},
	doi = {10.1007/s00023-023-01304-2},
	journal = {Annales Henri Poincaré},
	author = {Ampelogiannis, Dimitrios and Doyon, Benjamin},
	month = may,
	year = {2023},
}

@book{bratteli_operator_1987,
	series = {Operator {Algebras} and {Quantum} {Statistical} {Mechanics}},
	title = {Operator {Algebras} and {Quantum} {Statistical} {Mechanics} 1: {C}*- and {W}*-{Algebras}. {Symmetry} {Groups}. {Decomposition} of {States}},
	isbn = {978-3-540-17093-8},
	publisher = {Springer Science \& Business Media},
	author = {Bratteli, O. and Robinson, D.W.},
	year = {1987},
	lccn = {86027877},
}

@book{bratteli_operator_1997,
	series = {Operator {Algebras} and {Quantum} {Statistical} {Mechanics}},
	title = {Operator {Algebras} and {Quantum} {Statistical} {Mechanics} 2: {Equilibrium} {States} {Models} in {Quantum} {Statistical} {Mechanics}},
	isbn = {978-3-540-17093-6},
	publisher = {Springer Science \& Business Media},
	author = {Bratteli, O. and Robinson, D.W.},
	year = {1997},
}

@book{Khinchin,
	address = {New York},
	title = {Mathematical {Foundations} of {Statistical} {Mechanics}},
	publisher = {Dover Publications},
	author = {Khinchin, A. I.},
	year = {1949},
}

@article{fidaleo2014Nonconventional,
	title = {Nonconventional ergodic theorems for quantum dynamical systems},
	volume = {17},
	issn = {0219-0257},
	url = {https://doi.org/10.1142/S021902571450009X},
	doi = {10.1142/S021902571450009X},
	number = {02},
	urldate = {2023-01-10},
	journal = {Infinite Dimensional Analysis, Quantum Probability and Related Topics},
	author = {Fidaleo, Francesco},
	month = jun,
	year = {2014},
	note = {Publisher: World Scientific Publishing Co.},
	pages = {1450009},
}

@article{ollahydro,
	title = {Hydrodynamical limit for a {Hamiltonian} system with weak noise},
	volume = {155},
	issn = {1432-0916},
	url = {https://doi.org/10.1007/BF02096727},
	doi = {10.1007/BF02096727},
	number = {3},
	journal = {Communications in Mathematical Physics},
	author = {Olla, S. and Varadhan, S. R. S. and Yau, H. T.},
	month = aug,
	year = {1993},
	pages = {523--560},
}

@article{Araki1975uniqueness,
	title = {On uniqueness of {KMS} states of one-dimensional quantum lattice systems},
	volume = {44},
	journal = {Communications in Mathematical Physics},
	author = {Araki, Huzihiro},
	year = {1975},
	pages = {1--7},
}

@unpublished{delvecchio2021hydro,
	title = {The hydrodynamic theory of dynamical correlation functions in the {XX} chain},
	author = {Del Vecchio Del Vecchio, Giuseppe and Doyon, Benjamin},
	year = {2021},
	note = {arXiv: 2111.08420 [math-ph]},
}

@article{nachtergaele_quasi-locality_2019,
	title = {Quasi-locality bounds for quantum lattice systems. {I}. {Lieb}-{Robinson} bounds, quasi-local maps, and spectral flow automorphisms},
	volume = {60},
	issn = {0022-2488},
	url = {https://doi.org/10.1063/1.5095769},
	doi = {10.1063/1.5095769},
	number = {6},
	urldate = {2021-04-15},
	journal = {Journal of Mathematical Physics},
	author = {Nachtergaele, Bruno and Sims, Robert and Young, Amanda},
	month = jun,
	year = {2019},
	note = {Publisher: American Institute of Physics},
	pages = {061101},
}

@book{naaijkens_quantum_2017,
	address = {Cham},
	series = {Lecture {Notes} in {Physics}},
	title = {Quantum {Spin} {Systems} on {Infinite} {Lattices}: {A} {Concise} {Introduction}},
	volume = {933},
	isbn = {978-3-319-51456-7 978-3-319-51458-1},
	shorttitle = {Quantum {Spin} {Systems} on {Infinite} {Lattices}},
	url = {http://link.springer.com/10.1007/978-3-319-51458-1},
	urldate = {2021-07-20},
	publisher = {Springer International Publishing},
	author = {Naaijkens, Pieter},
	year = {2017},
	doi = {10.1007/978-3-319-51458-1},
}

@article{doyon_thermalization_2017,
	title = {Thermalization and {Pseudolocality} in {Extended} {Quantum} {Systems}},
	volume = {351},
	issn = {1432-0916},
	url = {https://doi.org/10.1007/s00220-017-2836-7},
	doi = {10.1007/s00220-017-2836-7},
	number = {1},
	journal = {Communications in Mathematical Physics},
	author = {Doyon, Benjamin},
	month = apr,
	year = {2017},
	pages = {155--200},
}

@misc{buca2021local,
	title = {Local hilbert space fragmentation and out-of-time-ordered crystals},
	author = {Buca, Berislav},
	year = {2021},
	note = {arXiv: 2108.13411 [cond-mat.stat-mech]},
}

@misc{gunawardana2021dynamical,
	title = {Dynamical l-bits in {Stark} many-body localization},
	author = {Gunawardana, Thivan and Buca, Berislav},
	year = {2021},
	note = {arXiv: 2110.13135 [cond-mat.dis-nn]},
}

@article{doyon_hydrodynamic_2022,
	title = {Hydrodynamic {Projections} and the {Emergence} of {Linearised} {Euler} {Equations} in {One}-{Dimensional} {Isolated} {Systems}},
	volume = {391},
	issn = {1432-0916},
	url = {https://doi.org/10.1007/s00220-022-04310-3},
	doi = {10.1007/s00220-022-04310-3},
	number = {1},
	journal = {Communications in Mathematical Physics},
	author = {Doyon, Benjamin},
	month = apr,
	year = {2022},
	pages = {293--356},
}

@article{doyon_lecture_2020,
	title = {Lecture notes on generalised hydrodynamics},
	doi = {10.21468/SciPostPhysLectNotes.18},
	journal = {SciPost Phys. Lect. Notes},
	author = {Doyon, Benjamin},
	year = {2020},
}

@article{doyoncorrelations,
	title = {Exact large-scale correlations in integrable systems out of equilibrium},
	volume = {5},
	url = {https://scipost.org/10.21468/SciPostPhys.5.5.054},
	doi = {10.21468/SciPostPhys.5.5.054},
	number = {5},
	journal = {SciPost Phys.},
	author = {Doyon, Benjamin},
	year = {2018},
	pages = {54},
}

@book{spohn_large_1991,
	title = {Large scale dynamics of interacting particles},
	isbn = {3-642-84371-9},
	publisher = {Springer-Verlag},
	author = {Spohn, Herbert},
	year = {1991},
}

@book{demasi_mathematical_2006,
	title = {Mathematical methods for hydrodynamic limits},
	publisher = {Springer},
	author = {DeMasi, Anna and Presutti, Errico},
	year = {2006},
}

@article{hashimoto_out--time-order_2017,
	title = {Out-of-time-order correlators in quantum mechanics},
	volume = {2017},
	issn = {1029-8479},
	url = {https://doi.org/10.1007/JHEP10(2017)138},
	doi = {10.1007/JHEP10(2017)138},
	number = {10},
	journal = {Journal of High Energy Physics},
	author = {Hashimoto, Koji and Murata, Keiju and Yoshii, Ryosuke},
	month = oct,
	year = {2017},
	pages = {138},
}

@article{maldacena_bound_2016,
	title = {A bound on chaos},
	volume = {2016},
	issn = {1029-8479},
	url = {https://doi.org/10.1007/JHEP08(2016)106},
	doi = {10.1007/JHEP08(2016)106},
	number = {8},
	journal = {Journal of High Energy Physics},
	author = {Maldacena, Juan and Shenker, Stephen H. and Stanford, Douglas},
	month = aug,
	year = {2016},
	pages = {106},
}

@article{hosur_chaos_2016,
	title = {Chaos in quantum channels},
	volume = {2016},
	issn = {1029-8479},
	url = {https://doi.org/10.1007/JHEP02(2016)004},
	doi = {10.1007/JHEP02(2016)004},
	number = {2},
	journal = {Journal of High Energy Physics},
	author = {Hosur, Pavan and Qi, Xiao-Liang and Roberts, Daniel A. and Yoshida, Beni},
	month = feb,
	year = {2016},
	pages = {4},
}

@article{shenker_black_2014,
	title = {Black holes and the butterfly effect},
	volume = {2014},
	issn = {1029-8479},
	url = {https://doi.org/10.1007/JHEP03(2014)067},
	doi = {10.1007/JHEP03(2014)067},
	number = {3},
	journal = {Journal of High Energy Physics},
	author = {Shenker, Stephen H. and Stanford, Douglas},
	month = mar,
	year = {2014},
	pages = {67},
}

@article{buca2019nonstationary,
	title = {Non-stationary coherent quantum many-body dynamics through dissipation},
	volume = {10},
	issn = {2041-1723},
	url = {https://doi.org/10.1038/s41467-019-09757-y},
	doi = {10.1038/s41467-019-09757-y},
	number = {1},
	journal = {Nature Communications},
	author = {Buca, Berislav and Tindall, Joseph and Jaksch, Dieter},
	month = apr,
	year = {2019},
	pages = {1730},
}

@article{doyon_drude_2017,
	title = {Drude {Weight} for the {Lieb}-{Liniger} {Bose} {Gas}},
	volume = {3},
	url = {https://ui.adsabs.harvard.edu/abs/2017ScPP....3...39D},
	doi = {10.21468/SciPostPhys.3.6.039},
	journal = {SciPost Physics},
	author = {Doyon, Benjamin and Spohn, Herbert},
	year = {2017},
	pages = {039},
}

@article{mori_transport_1965,
	title = {Transport, {Collective} {Motion}, and {Brownian} {Motion}*)},
	volume = {33},
	issn = {0033-068X},
	url = {https://doi.org/10.1143/PTP.33.423},
	doi = {10.1143/PTP.33.423},
	number = {3},
	urldate = {2021-12-07},
	journal = {Progress of Theoretical Physics},
	author = {Mori, Hazime},
	month = mar,
	year = {1965},
	pages = {423--455},
}

@article{Lieb:1972wy,
	title = {The finite group velocity of quantum spin systems},
	volume = {28},
	issn = {1432-0916},
	url = {https://doi.org/10.1007/BF01645779},
	doi = {10.1007/BF01645779},
	number = {3},
	journal = {Communications in Mathematical Physics},
	author = {Lieb, Elliott H. and Robinson, Derek W.},
	month = sep,
	year = {1972},
	pages = {251--257},
}

@article{Boldrighini_HardRodGas,
	title = {One-dimensional hard rod caricature of hydrodynamics},
	volume = {31},
	issn = {1572-9613},
	url = {https://doi.org/10.1007/BF01019499},
	doi = {10.1007/BF01019499},
	number = {3},
	journal = {Journal of Statistical Physics},
	author = {Boldrighini, C. and Dobrushin, R. L. and Sukhov, Yu. M.},
	month = jun,
	year = {1983},
	pages = {577--616},
}

@book{conwayFunctionalAnalysis2007,
	address = {New York, NY},
	series = {Graduate {Texts} in {Mathematics}},
	title = {A {Course} in {Functional} {Analysis}},
	volume = {96},
	isbn = {978-1-4419-3092-7 978-1-4757-4383-8},
	url = {http://link.springer.com/10.1007/978-1-4757-4383-8},
	urldate = {2021-07-25},
	publisher = {Springer New York},
	author = {Conway, John B.},
	year = {2007},
	doi = {10.1007/978-1-4757-4383-8},
}

@article{Araki:1969bj,
	title = {Gibbs states of a one dimensional quantum lattice},
	volume = {14},
	issn = {1432-0916},
	url = {https://doi.org/10.1007/BF01645134},
	doi = {10.1007/BF01645134},
	number = {2},
	journal = {Communications in Mathematical Physics},
	author = {Araki, Huzihiro},
	month = jun,
	year = {1969},
	pages = {120--157},
}

@article{Ueltschi2003Cluster,
	title = {Cluster expansions and correlation functions},
	journal = {arXiv: Mathematical Physics},
	author = {Ueltschi, Daniel},
	year = {2003},
}

@article{medenjak2020rigo,
	title = {Rigorous bounds on dynamical response functions and time-translation symmetry breaking},
	volume = {9},
	url = {https://scipost.org/10.21468/SciPostPhys.9.1.003},
	doi = {10.21468/SciPostPhys.9.1.003},
	number = {1},
	journal = {SciPost Phys.},
	author = {Medenjak, Marko and Prosen, Tomaz and Zadnik, Lenart},
	year = {2020},
	note = {Publisher: SciPost},
	pages = {3},
}

@misc{buca2020quantum,
	title = {Quantum many-body attractors},
	author = {Buca, Berislav and Purkayastha, Archak and Guarnieri, Giacomo and Mitchison, Mark T. and Jaksch, Dieter and Goold, John},
	year = {2020},
	note = {arXiv: 2008.11166 [quant-ph]},
}

@article{zwanzig_statistical_1961,
	title = {Statistical mechanics of irreversibility},
	volume = {3},
	journal = {Lectures in theoretical physics},
	author = {Zwanzig, Robert},
	year = {1961},
	pages = {106--141},
}

@article{th_brox_equilibrium_1984,
	title = {Equilibrium {Fluctuations} of {Stochastic} {Particle} {Systems}: {The} {Role} of {Conserved} {Quantities}},
	volume = {12},
	url = {https://doi.org/10.1214/aop/1176993225},
	doi = {10.1214/aop/1176993225},
	number = {3},
	journal = {The Annals of Probability},
	author = {{Th. Brox} and {H. Rost}},
	month = aug,
	year = {1984},
	pages = {742--759},
}

@article{goldstein_long-time_2010,
	title = {Long-time behavior of macroscopic quantum systems},
	volume = {35},
	issn = {2102-6467},
	url = {https://doi.org/10.1140/epjh/e2010-00007-7},
	doi = {10.1140/epjh/e2010-00007-7},
	number = {2},
	journal = {The European Physical Journal H},
	author = {Goldstein, S. and Lebowitz, J. L. and Tumulka, R. and Zanghì, N.},
	month = nov,
	year = {2010},
	pages = {173--200},
}

@article{goldstein_normal_2009,
	title = {Normal {Typicality} and von {Neumann}'s {Quantum} {Ergodic} {Theorem}},
	volume = {466},
	doi = {10.1098/rspa.2009.0635},
	journal = {Proceedings of The Royal Society A Mathematical Physical and Engineering Sciences},
	author = {Goldstein, Sheldon and Lebowitz, Joel and Mastrodonato, Christian and Tumulka, Roderich and Zanghi, Nino},
	month = jul,
	year = {2009},
}

@inproceedings{hilbert_mathematical_1902,
	title = {Mathematical {Problems}},
	url = {https://doi.org/10.1090/S0002-9904-1902-00923-3},
	doi = {10.1090/S0002-9904-1902-00923-3},
	publisher = {Bull. Amer. Math. Soc. 8 (1902), 437-479},
	author = {Hilbert, David R.},
	year = {1902},
}

@article{kastler_invariant_1966,
	title = {Invariant states in statistical mechanics},
	volume = {3},
	url = {https://doi.org/},
	number = {3},
	journal = {Communications in Mathematical Physics},
	author = {Kastler, Daniel and Robinson, Derek W.},
	month = jan,
	year = {1966},
	pages = {151--180},
}

@book{hille_functional_1996,
	series = {American {Mathematical} {Society}: {Colloquium} publications},
	title = {Functional {Analysis} and {Semi}-groups},
	isbn = {978-0-8218-1031-6},
	url = {https://books.google.gr/books?id=xn-EQyIpYegC},
	publisher = {American Mathematical Society},
	author = {Hille, E. and Phillips, R.S.},
	year = {1996},
	note = {Issue: v. 31, pt. 1},
}

@book{simon_statistical_2014,
	series = {Princeton {Legacy} {Library}},
	title = {The {Statistical} {Mechanics} of {Lattice} {Gases}, {Volume} {I}},
	isbn = {978-1-4008-6343-3},
	url = {https://books.google.gr/books?id=BUAABAAAQBAJ},
	publisher = {Princeton University Press},
	author = {Simon, B.},
	year = {2014},
}

@article{glimm_certain_1960,
	title = {On a {Certain} {Class} of {Operator} {Algebras}},
	volume = {95},
	issn = {00029947},
	url = {http://www.jstor.org/stable/1993294},
	doi = {10.2307/1993294},
	number = {2},
	urldate = {2021-07-26},
	journal = {Transactions of the American Mathematical Society},
	author = {Glimm, James G.},
	year = {1960},
	note = {Publisher: American Mathematical Society},
	pages = {318--340},
}

@book{taylor_general_1985,
	title = {General theory of functions and integration},
	publisher = {Courier Corporation},
	author = {Taylor, Angus Ellis},
	year = {1985},
}

@article{frohlich_properties_2015,
	title = {Some properties of correlations of quantum lattice systems in thermal equilibrium},
	volume = {56},
	issn = {0022-2488},
	url = {https://doi.org/10.1063/1.4921305},
	doi = {10.1063/1.4921305},
	number = {5},
	urldate = {2021-04-17},
	journal = {Journal of Mathematical Physics},
	author = {Fröhlich, Jürg and Ueltschi, Daniel},
	month = may,
	year = {2015},
	note = {Publisher: American Institute of Physics},
	pages = {053302},
}
% *************************************** Index ********************************
\printthesisindex % If index is present

\end{document}